\pdfoutput = 1
\documentclass[12pt]{scrartcl}
\usepackage[utf8]{inputenc}

\areaset{158mm}{238mm}
\usepackage{setspace}
\usepackage[all]{xypic}
\usepackage{scrlayer-scrpage}

\usepackage{pdflscape}

\usepackage{multirow}
\usepackage{amsmath}
\usepackage{mathtools}
\usepackage{mathrsfs}
\usepackage{tikz}
\usepackage{amssymb}
\usepackage[unicode=true,linktoc=all]{hyperref}
\usepackage{cleveref}
\usepackage{cite}
\usepackage{bm}
\usepackage{slashed}

\usepackage{comment}

\usepackage{tikz,pgf}
\usetikzlibrary{shapes}
\usetikzlibrary{calc}
\usetikzlibrary{decorations.pathmorphing}
\usetikzlibrary{decorations.pathreplacing,shapes.misc}
\usetikzlibrary{positioning}
\usetikzlibrary{decorations.markings}
\tikzset{->-/.style={decoration={
  markings,
  mark=at position .5 with {\arrow{>}}},postaction={decorate}}}
\tikzset{-<-/.style={decoration={
  markings,
  mark=at position .5 with {\arrow{<}}},postaction={decorate}}}

\newtheorem{defi}{Definition}[section]

\usepackage{xcolor}
  \definecolor{rblue}{RGB}{81, 49, 193}
  \definecolor{rorange}{RGB}{255, 147, 40}
  \definecolor{rgreen}{RGB}{176, 233, 0}

\renewcommand{\tilde}{\widetilde}

\newcommand{\fsu}{\mathfrak{su}}
\newcommand{\fso}{\mathfrak{so}}
\newcommand{\fsp}{\mathfrak{sp}}
\newcommand{\fg}{\mathfrak{g}}
\newcommand{\ff}{\mathfrak{f}}
\newcommand{\fe}{\mathfrak{e}}

\newcommand\be{\begin{equation}}
\newcommand\ee{\end{equation}}
\renewcommand{\hat}{\widehat}

\numberwithin{equation}{section}

\title{Back to Heterotic Strings \\ on ALE Spaces }
\subtitle{Part II -- Geometry of T-dual Little Strings}

\author{Michele Del Zotto$^{\dagger\sharp}$, Muyang Liu $^\sharp$, \\ and Paul-Konstantin Oehlmann$^{\dagger\lozenge}$
\\
	\footnotesize\slshape$^\sharp$ Department of Mathematics, Uppsala University, Uppsala, Sweden\\
	\footnotesize\slshape$^\dagger$ Department of Physics and Astronomy, Uppsala University, Uppsala, Sweden\\
\footnotesize\slshape$^{\lozenge}$ 	Department of Physics, Northeastern University, Boston, USA \\
}
\date{}

\begin{document}
\maketitle

\vspace{-15.5cm}
\vspace{14.5cm}
\paragraph{\hspace{.9cm}\large{Abstract}}
\vspace{-1cm}
\begin{abstract}
\noindent This work is the second of a series of papers devoted to revisiting the properties of Heterotic string compactifications on ALE spaces. In this project we study the geometric counterpart in F-theory of the T-dualities between Heterotic ALE instantonic Little String Theories (LSTs) extending and generalising previous results on the subject by Aspinwall and Morrison. Since the T-dualities arise from a circle reduction one can exploit the duality between F-theory and M-theory to explore a larger moduli space, where T-dualities are realised as inequivalent elliptic fibrations of the same geometry. As expected from the Heterotic/F-theory duality the elliptic F-theory Calabi-Yau we consider admit a nested elliptic K3 fibration structure. This is central for our construction: the K3 fibrations determine the flavor groups and their global forms, and are the key to identify various T-dualities. We remark that this method works also more generally for LSTs arising from non-geometric Heterotic backgrounds. We study a first example in detail: a particularly exotic class of LSTs which are built from extremal K3 surfaces that admit flavor groups with maximal rank 18. We find all models are related by a so-called T-hexality (i.e. a 6-fold family of T-dualities) which we predict from the inequivalent elliptic fibrations of the extremal K3.

\end{abstract}

\vfill{}
--------------------------

November 2022

\thispagestyle{empty}

\newpage

\tableofcontents 
\addtocontents{toc}{\protect\setcounter{tocdepth}{2}}

\section{Introduction}
This work is the second of a series of papers revisiting the properties of Heterotic string compactifications on ALE spaces. The main motivation for these studies is that by combining recent progress about six-dimensional theories \cite{Heckman:2013pva,DelZotto:2014hpa,Ohmori:2014kda,
Heckman:2015bfa,
Bhardwaj:2015oru,
Mekareeya:2017jgc,
Dierigl:2020myk,
Bhardwaj:2019hhd,Fazzi:2022hal}
with our improved understanding of their continuous 2-group symmetries \cite{Cordova:2018cvg,Cordova:2020tij,DelZotto:2020sop},\footnote{\ We stress that there have been lots of recent developments about the geometric origin of generalised symmetries \cite{Gaiotto:2014kfa} for 5d theories with a geometric engineering in M-theory (see e.g. \cite{Morrison:2020ool,Albertini:2020mdx,Bhardwaj:2020phs,BenettiGenolini:2020doj,Cvetic:2021sxm,Apruzzi:2021nmk,Damia:2022bcd}), and in particular their 2-group structures mixing a continuous 0-form symmetry with a finite 1-form symmetry, found in \cite{Apruzzi:2021vcu} and then further explored by \cite{Apruzzi:2021mlh,Genolini:2022mpi,Bhardwaj:2022scy,DelZotto:2022fnw,DelZotto:2022joo,Cvetic:2022imb}. We expect all these results have interesting applications to further constrain 6d LST T-dualities via the M-theory moduli space -- in combination with the dimensional reduction of the corresponding 6d defect groups \cite{DelZotto:2015isa}. We plan to explore these aspects in future work \cite{delZottoAppear2}.} we find new insights on several open questions on the subject.

\medskip

In our previous work \cite{DelZotto:2022ohj} we have revisited the Heterotic ALE instanton little strings and the corresponding T-dualities \cite{Aspinwall:1996vc,Aspinwall:1997ye,Blum:1997mm,Intriligator:1997dh,Hanany:1997gh,Brunner:1997gf}. While the little string theories (LSTs) of heterotic $Spin(32)/\mathbb Z_2$ ALE instantons are well-known\footnote{\ Those models have been obtained from orbifolds of the theory of N $Spin(32)/\mathbb Z_2$ NS5 branes \cite{Sagnotti:1987tw,Bianchi:1990yu,Blum:1997mm,Intriligator:1997dh}.}, the LSTs governing Heterotic $E_8 \times E_8$ ALE instantons were still relatively mysterious. We have completely determined the latter thanks to 6d conformal matter methods, extending and generalising previous beautiful results in the literature about them \cite{Aspinwall:1997ye,Font:2017cya,Font:2017cmx} -- the gap which we have filled in \cite{DelZotto:2022ohj} is the dependence of the 6d LST instantons on choices of monodromies at infinity for the Heterotic $E_8\times E_8$ gauge group, parameterised by pairs of group morphisms $(\boldsymbol{\mu}_1, \boldsymbol{\mu}_2)$, $\boldsymbol{\mu}_{a} \in \text{Hom}(\Gamma_{\mathfrak g},E_8)$. The various possible $(\boldsymbol{\mu}_1, \boldsymbol{\mu}_2)$ are part of the data defining an instanton on the ALE space of type $\mathbb C^2/\Gamma_{\mathfrak g}$, where $\Gamma_\mathfrak{g}$ is a finite subgroup of $SU(2)$. Hence one expects to obtain different 6d LSTs corresponding to each pair, which we have determined thanks to recent progress in understanding 6d orbi-instanton theories \cite{Heckman:2015bfa,Frey:2018vpw,Mekareeya:2017jgc,Fazzi:2022hal} (see also \cite{Giacomelli:2022drw}). As a consistency check of our results, in \cite{DelZotto:2022ohj} we have exploited T-duality with the known $Spin(32)/\mathbb Z_2$ ALE instanton LSTs: since T-dual pairs of theories must have matching flavor symmetry ranks, 5d Coulomb branch dimensions, and continuum 2-group structure constants \cite{Cordova:2020tij,DelZotto:2020sop}, we confirmed our prediction on the structure of the $E_8\times E_8$ ALE instanton LSTs by finding a matching T-dual $Spin(32)/\mathbb Z_2$ for each example. Moreover, the above data seemed sufficiently powerful to predict T-duality between pairs of LSTs and, as a consequence, we ended up with several new families of conjectural equivalences among these models \cite{DelZotto:2022ohj}. 

\medskip

For ALE spaces of type $A_k$ and $D_k$ these conjectures can be tested by exploiting dual brane realisations in Type I$^\prime$ (which we did in \cite{DelZotto:2022ohj}, building on \cite{Blum:1997mm,Intriligator:1997dh,Hanany:1997gh,Brunner:1997gf}). However, for ALE spaces of types $E_6$, $E_7$ and $E_8$, a perturbative superstring description is lacking (which goes hand in hand with the lack of an ADHM construction for instanton moduli spaces of exceptional gauge groups \cite{Atiyah:1978ri}). Therefore, in order to check our conjectures for the ALE spaces of exceptional type, we must resort to a different method which was pioneered by Aspinwall and Morrison \cite{Aspinwall:1996vc,Aspinwall:1997ye}, as well as developed in later works \cite{Bhardwaj:2015oru,Font:2017cya,Font:2017cmx}. Namely, one realises the corresponding 6d (1,0) LSTs via geometric engineering them in F-theory and then exploits the M/F-theory duality to probe the 5d moduli spaces of the relevant 6d LSTs reduced on a circle. It is thanks to this geometrization of T-duality that we can probe some of the conjectures of \cite{DelZotto:2022ohj} and explicitly demonstrate their validity, thus confirming our predictions. This is the main aim of this paper.

\medskip

Here we recap some features of the geometric origin of the T-dualities from F-theory, which are relevant for our paper. Recall that the data of an elliptic fibration\footnote{\ More generally, a genus one fibration is sufficient, but in this paper we will not study these more general cases that should correspond to twisted T-dualities.} is needed to define a 6d (1,0) F-theory background:
\begin{equation}
    \begin{array}{rl} T^2 \rightarrow & X \\ & \downarrow \pi \\ & B_2 \end{array}\, 
\end{equation}
where the complex structure parameter of the torus fiber $T^2$ is interpreted as an axio-dilaton coupling in IIB \cite{Vafa:1996xn, Morrison:1996pp,Morrison:1996xf}, $X$ is a Calabi-Yau (CY) threefold and $B_2$ is a two-dimensional K\"ahler complex surface. The resulting 6d theory strictly depends on the property of $\pi$, hence we denote it $\mathcal T_{F/(X,\pi)}$ in this introduction. By the M/F-theory duality, one has that the KK theory corresponding to the circle reduction of $\mathcal T_{F/(X,\pi)}$ is given by  M-theory on $X$, namely
\begin{equation}
   D_{S^1} \mathcal T_{F/(X,\pi)} = \mathcal T_{M/X}\,.
\end{equation}
In particular, on the RHS one can probe the whole M-theory moduli space of $X$, which is much larger than the F-theory moduli space, as it does not depend on $\pi$. In particular, if $X$ admits an inequivalent elliptic fibration
\begin{equation}
    \begin{array}{rl} T^2 \rightarrow & X \\ & \downarrow \widetilde{\pi} \\ & \widetilde{B_2} \end{array}
\end{equation}
then the resulting 5d KK theory will have an inequivalent 6d uplift to the theory $\mathcal T_{F/(X,\widetilde{\pi})}$, which is often very different from $\mathcal T_{F/(X,\pi)}$. This is a geometrically realised T-duality between these two 6d theories.\footnote{\ We stress that the constructions of geometrically realised T-dualities in F-theory presented in \cite{Bhardwaj:2015oru} and in \cite{DelZotto:2020sop,Bhardwaj:2022ekc} are slightly more general than the one we review here: in the former, flops of $X$ are included in the M-theory moduli (indeed each $X$ corresponds just to a chamber of the extended Kähler cone which is the actual 5d extended Coulomb branch for the KK theory, hence flop transitions are indeed allowed physically), while in the latter pair of references both flops and genus one fibrations as opposed to elliptic fibrations are included (which lead to the so-called twisted T-dualities). These refinements of the geometrization of T-duality are not needed to probe our conjectures in paper \cite{DelZotto:2022ohj}, hence we refrain from mentioning them in this work.} In order to probe our conjectures, we need to find explicit CY threefolds corresponding to the F-theory construction of the Heterotic ALE $E_8 \times E_8$ instantons with nontrivial choices of monodromies at infinity that explicitly break $E_8$ to $F(\boldsymbol{\mu}_a)$, the commutant subgroup of $\boldsymbol{\mu}_a(\Gamma_\fg)$ in $E_8$. For trivial choices of monodromies preserving the whole $E_8$, this was done by Aspinwall and Morrison \cite{Aspinwall:1997ye}, and some examples of threefolds corresponding to non-trivial monodromies triggering small breakings were also studied in \cite{Font:2017cya,Font:2017cmx}. Here, we formulate a general picture based on the Heterotic/F-theory duality. On top of being elliptically fibered, the Calabi-Yau in our constrcution are also K3 fibered, which is the hallmark of the fact that these are F-theory duals of Heterotic compactifications. Thanks to this other fibration of the form
\begin{equation}
    \begin{array}{rl} K3 \rightarrow & X \\ & \downarrow \xi \\ & C \end{array}
\end{equation}
where $C$ is a non-compact complex curve, we can explicitly realise the global symmetry algebra of the 6d LST, corresponding to the Lie algebra of $F(\boldsymbol{\mu}_1) \times F(\boldsymbol{\mu_2})$, as the Picard lattice of the K3 fiber of $\xi$. In the T-dual, we see how the same Picard lattice is compatible with an inequivalent decomposition, which is induced by the other elliptic structure of $X$ and realises the flavor symmetry of the expected T-dual model. This description gives a finer more physically sound generalisation of the Kulikov description of K3 degenerations. A perspective is already implicit in the works of \cite{Lee:2021qkx}.  Viewing the curve $C$ as the family parameter, we see that the LST is localised at a point over $C$, where all the 6d degrees of freedom are coalesced in a generalised singularity. Resolving these geometries gives access to the generalised 6d quivers for the resulting 6d LSTs, which we use to confirm our analysis and probe, with F-theory techniques, our conjectures in \cite{DelZotto:2022ohj}. Interestingly, the F-theory geometry and the corresponding Mordell-Weil groups are powerful enough to constrain the global form of the resulting flavor symmetry too. We stress here that each LST class corresponds to a different geometry and in this paper we are constructing these geometries as explicitly as possible exploiting techniques of toric geometry, we are not going to demonstrate all the field theoretical results discussed in \cite{DelZotto:2022ohj}. The purpose of this paper is to test the predictions of \cite{DelZotto:2022ohj} with F-theory geometry with a special emphasis on those which are dual to Heterotic instantons along exceptional ALE spaces and exhibit exceptional flavor symmetries.

\medskip

The approach based on K3 fibrations has applications beyond proving our conjectures: we show explicitly that it can give predictions on T-dualities for non-geometric phases of the Heterotic string, which is a further result that we include in this work. In particular we find a close relation between K3 fibrations admitting multiple elliptic fibrations and T-dual little strings with a Heterotic origin that we plan to explore further. From that perspective, the T-duality of 6d LSTs is clearly not a phenomenon of order two, involving just pairs of theories, instead it allows for rather large families of dual systems. This extends well-known features of 6d LSTs arising from M5 brane probes of $\mathbb C^2/\mathbb Z_k$ spaces \cite{Bastian:2018dfu}, as well as various recent results in \cite{DelZotto:2022ohj,Bhardwaj:2022ekc}.\footnote{\ It would be interesting to probe this effect further and see whether there are unbounded families of T-duals, adapting to the LST setup the arguments of \cite{Anderson:2017aux}.}

\medskip

This work is structured as follows: In Section~\ref{sec:reviewF} we introduce the basic ingredients to construct LSTs using F-theory emphasizing in particular the appearance of elliptic K3 fibrations. In Section~\ref{sec: HeteroticLSTs} we construct smooth elliptic threefolds systematically to derive pairs of T-dual LSTs. We probe the conjectures of \cite{DelZotto:2022ohj} with chosen examples dual to Heterotic instantons along all kinds of ALE spaces. These include singularities of all ADE types, with special emphasis on the case of exceptional singularities. In those cases, we focus on the examples that are the hardest to realise exploiting brane webs, namely those with global symmetries that are products of exceptional group factors, $E_6, E_7$ and $E_8$, which we study systematically. In Section~\ref{sec: MultiTDual} we start exploring more exotic families of LSTs to emphasise the power of the technique we have been employing: these include `discrete holonomy theories' following \cite{Aspinwall:1998xj}, as well as Multi-T-Dual LSTs with exotic fiber/base exchanges and, probably most interestingly, non-geometric Heterotic theories with maximal flavor ranks. In this last family of examples, we exhibit in particular a ``hexality'' system, which is completely determined thanks to the properties of the elliptic fibration structures of a special K3 with maximal Picard rank. 

\section{Review: Heterotic Little Strings \& Geometry}\label{sec:reviewF}
This section serves as a review of the geometric construction of LSTs using F-theory. As such we want to introduce the basic ingredients and set up the standard notation of the literature (see e.g \cite{Weigand:2018rez,Heckman:2018jxk} for a recent review on the geometry of F-theory). 
In the following section we first give a rough outline of the geometric engineering setup that give rise to LSTs, then review the M/F-theory duality in order to explain the T-duality in geometric terms. In practise, we employ the toric technique, which will be reviewed at a later stage in Section~\ref{ssec:ToricPrelim}, to obtain smooth geometries.  

Finally we give a short summary of K3 fibrations and the associated Kulikov degenerations of elliptic K3s, which are explained in more details in Appendix~\ref{app:Kulikov}.

\subsection{Review: Little Strings from F-theory}\label{sec:LST}

We turn to F-theory to go beyond the limitation of the perturbative string constructions. This allows us to construct the most complete collection of SCFTs \cite{Heckman:2015bfa} (for a review, see \cite{Heckman:2018jxk} and references therein) and LSTs \cite{Bhardwaj:2015oru} to date. This section will review some basic facts about the engineering of LSTs in F-theory; experts can safely skip it \cite{Bhardwaj:2015oru,Heckman:2018jxk}. An F-theory compactification to 6D is defined by the geometry of a CY threefold $X$, which is an elliptic fibration over a two-dimensional K\"ahler surface $B_2$ given as
\begin{align}
\begin{array}{rl} T^2 \rightarrow & X \\ & \downarrow \pi \\ & B_2 \end{array} \, .
\end{align}
This geometry can be viewed as a non-perturbative generalization of Type IIB strings by identifying the axio-dilaton $\tau$ with the complex structure of the torus fiber $T^2$ over any point of the physical compactified space $B_2$. This construction enables to consistently engineer configuration with non-local (p,q) 7-branes, which allow $\tau$ to attain values at strong couplings, such as exceptional symmetries. This approach is very powerful due to the ability to describe this fibration algebraically as a Weierstrass model of the form
\begin{align}
p= Y^2 + X^3 + f X Z^4 + g Z^6 \, ,
\end{align} 
where $X,Y,Z$ are the projective coordinates of a $\mathbb{P}^2_{2,3,1}$ ambient space and $s_0: \{ X,Y,Z\} = \{1,-1,0\}$ being the zero-section. Furthermore, $f$ and $g$ are sections given by powers of line bundles of the base anticannonical class, namely $\{ f,g  \} \in \{ \mathcal{O}(K^{-4}_{B_2}),\mathcal{O}(K^{-6}_{B_2})\}$. 
D7 brane stacks are located at codimension one components of the discriminant $\Delta=4f^3+27g^2$, where the elliptic fibre becomes singular. Since LSTs are decoupled from gravity, the base $B_2$ is non-compact. Therefore also the discriminant $\Delta$ admits compact and non-compact components that D7 branes can wrap. The world volume degrees of freedom residing on branes that wrap compact components yield gauge degrees of freedom, while those that wrap non-compact ones produce flavor symmetries.

This description exploits an elegant relationship between algebraic geometry and physics: Kodaira \cite{1023071969881} and Tate (see \cite{Bershadsky:1996nh,Katz:2011qp} for a direct connection to F-theory) have classified the types of singularities in terms of vanishing orders of $(f,g,\Delta)$ and additional monodromy data. This leads to an ABCDEFG type of classification that coincides with all possible compact lie algebras, which can appear as flavor factors $\mathfrak{g}_F$ or gauge factors $\mathfrak{g}$ on the respective D7 branes.
 
The geometric classification is realized through a de-singularization procedure of the fibre. In this process singular points on the elliptic curve are substituted by chains of compact fibral curves $f_i$ that are topologically $\mathbb{P}^1$s, whose intersection form $f_i \cdot f_j = - G_{i,j}$ coincides with the Cartan-matrix $G_{i,j}$ of the respective algebra. We perform many of such resolutions in depth in the sections that follow. Beyond the non-Abelian part of flavor and gauge algebras in F-theory, there is also finer data encoded in the Mordell-Weil group (MW) of the fibration $X$. This structure arises when the respective fibration admits extra sections $s_i$ with $i=1 \ldots r$ in addition to the zero-section $s_0$. These sections admit an addition law, that generates an Abelian group that is isomorphic to 
\begin{align}
MW(X) = \mathbb{Z}^r \times MW(X)_{Tor} \, .
\end{align}
Hence, the free part of the group has rank $r$ and generates an Abelian $U(1)^r$ group in F-theory (see \cite{Grimm:2010ez,Morrison:2012ei} and \cite{Cvetic:2018bni} for a recent review). In compact scenarios those Abelian symmetries are always gauged but in non-compact ones they become global flavor symmetries \cite{Lee:2018ihr,Apruzzi:2020eqi} at best.
Many of the cases we discuss in the following sections do have additional Abelian flavor factors.
For these symmetries to be present, they are required to be free of ABJ anomalies.

In the following we will leave a check for an ABJ anomaly free charge assignments of the hypermultiplets for future work but insist on their consistency by arguing via geometry and
flavor rank preservation under T-duality.  

Secondly, the Mordell-Weil torsion (sub-)group $MW(X)_{Tor}\sim \mathbb{Z}_n \times \mathbb{Z}_m$  gives rise to non-simply connected symmetries \cite{Aspinwall:1998xj,Mayrhofer:2014opa} by acting as a diagonal quotient on the (sub-)centers  $Z(G_F)$ and $Z(G)$ of flavor and gauge group respectively \cite{Dierigl:2020myk,Hubner:2022kxr}. This yields the total symmetry group
\begin{align}
G_T = \frac{ G_F \times G  }{ MW_{Tor}} \, .
\end{align}
The above quotient highlights in particular the presence of a gauged center one-form symmetry \cite{Gaiotto:2014kfa} and consequently a restricted set of matter and Wilson line operators\footnote{A finite MW group also results in the restriction of $SL(2,\mathbb{Z})$ monodromies to a congruence subgroup as e.g. argued in \cite{Hajouji:2019vxs}.} in 6D QFTs, regardless of whether they flow to SCFTs or LSTs, as demonstrated in recent literature, e.g. \cite{Dierigl:2020myk,Apruzzi:2020zot,Cvetic:2021sxm,Hubner:2022kxr}.

The non-critical strings and in particular the little strings in 6D arise from D3 branes that wrap compact curves $\Sigma$ in $B_2$ with a string tension determined by Vol$(\Sigma)$. The integral homology lattice $\Lambda = H_2 (B_2, \mathbb{Z})$ is further identified  with the (negative) string charge lattice and possesses a natural pairing 
\begin{align}
(\cdot , \cdot) : \quad  H_2 (B_2, \mathbb{Z}) \times H_2 (B_2, \mathbb{Z}) \rightarrow \mathbb{Z} \, .
\end{align}
For a certain choice of basis  $w^I \in H_2 (B_2, \mathbb{Z})$, we denote this intersection form as
\begin{align}
\eta^
{IJ} = -  w^I \cdot w^J \, . 
\end{align}
The rank $r$ of this pairing, determines the number of independent dynamical tensor multiplets in the 6D theory.
 
Due to a theorem by Arten, Grauert and Mumford (see reference in \cite{Heckman:2018jxk}), the intersection pairing $\eta^{IJ}$ must be positive definite for all compact curves to be contractable to a point. This implies that we can shrink the base to a point and eliminate all scales from the theory. The result is thus scale-free as required for an SCFT and the degrees of freedom are supported by the now tensionless BPS strings \cite{Seiberg:1996vs}.

Alternatively, if the intersection pairing is only positive \textit{semi-definite}, the intersection pairing $\eta^{IJ}$ does not have full rank $r$ but includes a zero eigenvalue with an eigencurve
\begin{align}
\label{eq:LScurve}
\Sigma^0 = \sum_{I=1}^{r+1} {l}_{LS,I} \cdot w^I \, .
\end{align}
$\Sigma^0$ can not be contracted to a point and thus all scales of the theory cannot be eliminated.  

The D3 brane wrapping $\Sigma^0$ therefore leads to a non-critical string with a finite string tension everywhere in the moduli space, the little string (LS). The volume of the respective curve Vol($\Sigma^0$) then sets the LS tension. Moreover, by reducing the $IIB$ four-form $C_4$ along $\Sigma^0$ we obtain a non-dynamical tensor multiplet that couples to the little string. This tensor field should be regarded as a background field that admits a continuous $\mathfrak{u}^{(1)}$-one-form symmetry. As $\Sigma^0$ is a linear combination of other curves $w^I$, this induces a non-trivial LS charge of all tensor multiplets under the LS $\mathfrak{u}^{(1)}$ symmetry. 

As of writing of this paper, there exist only two types of $\mathcal{N}=(1,0)$ 6D LSTs, distinguished by their birational base topologies \cite{Bhardwaj:2015oru}. These are obtained upon collapsing all one curves to smooth points and yield the two classes
\begin{align}
\label{eq:LSTbases}
\hat{B}_{LST,1} : \mathbb{P}^1 \times \mathbb{C} \, , \qquad \hat{B}_{LST,2}: (\mathbb{T}^2 \times \mathbb{C})/\Gamma \qquad \Gamma \in SU(2)\, .
\end{align} In this work we focus on theories of the class $B_{LST,1}$ as those are dual to heterotic LSTs.

The relevant data of  LSTs are packed  into the following type of quivers 
\begin{align}
\label{eq:quiver}
[G_{F_1}]   \, \, \overset{\mathfrak{g}_1}{n_1} \, \, \underset{[N_R=j]}{\overset{\mathfrak{g}_2}{n_2}} \ldots \underset{[G_{F_I}]}{\overset{\mathfrak{g}_I}{n_I}}  \,\,  \overset{\mathfrak{g}_N}{n_N} \,\,  [G_{F_{N_f}}]  \, ,
\end{align}
where the gauge algebra $\mathfrak{g}_I$ resides on the D7 brane that wraps a genus zero curve of  self-intersection $w^I \cdot w^I = -\eta^{II}= -n_I$. Neighboring curves intersect each other once, leading to bifundamental matter. A box in the picture highlights a non-compact flavor brane algebra $\mathfrak{g}_{F_i}$ that intersects the adjacent curve. We additionally depict extra $N_\mathbf{R}$ hypermultiplets in representations $\mathbf{R}$ of the respective gauge algebra, which are not originated from obvious brane intersections with some neighbouring curves. The matter of $\mathfrak{g}_I$ that couples to a tensor with charge $\eta^{II}$ is highly constrained by 6D gauge anomalies. In this work we will not discuss the details of this anomaly cancellation mechanism and its constrains explicitly but  refer to excellent reviews in the literature \cite{Taylor:2011wt,Johnson:2016qar,Weigand:2018rez}.
 
This quiver makes it straightforward to extract the intersection matrix $\eta^{I J}$ that determines its rank, LS curve $\Sigma^0$ and the corresponding LS charges $\vec{l}_{LS}$ from its nullspace.

A characteristic feature of all LSTs is the continuous 2-group symmetry sourced from the mixing of the $\mathfrak{u}_1^{(1)}$-one-form symmetry and the 6D Poincare-, SU(2)$_R$- and Flavor- zero-form symmetries \cite{Cordova:2020tij}. This mixing gives rise to the three structure constants, that can be geometrically computed as 
\begin{align}
\label{eq:2groupconstants}
\widehat{\kappa}_{\mathscr P} = -  \sum_{I=1}^{r+1} l_{LS,I} (\eta^{I I}-2) \, , \quad
\widehat{\kappa}_{\mathscr R}= \sum_{I=1}^{r+1} l_{LS,I}  h^\text{v}_{\mathfrak{g}_I}\, , \quad 
\kappa_{F_A }= - \sum_{I=1}^{r+1} l_{LS,I} \eta^{I A } \, ,
\end{align}
with  $h^\text{v}_{\mathfrak{g}_I}$  the dual Coxeter number of gauge algebra $\mathfrak{g}_I$ on the curve $w^I$. The $\kappa_{F_A }$ receives contributions when a compact curve $w^I$ is intersected by a non-compact one,
$w^A_{\text{nc}}$ carrying the respective flavor group $[G_{F_A}]$. The $\widehat{\kappa}_{\mathscr P}$  structure constant depends only on the base topology and is a birational equivalent. It can only take two values that are 
\begin{align}
\widehat{\kappa}_{\mathscr P} (\hat{B}_{LST,1}) = 2 \, , \quad \widehat{\kappa}_{\mathscr P} (\hat{B}_{LST,2}) = 0 \, .
\end{align}
These values have the proposed physical interpretation \cite{DelZotto:2020sop}: it counts the number of M9 branes in the theory.  
  
\subsection{Geometrization of T-duality}\label{sec:toricT}
LSTs can be related by T-duality similar to SUGRA theories: I.e. upon circle compactification and at a specific locus of the Coulomb branch (CB) and Wilson line (WL) background, two (or more) LSTs become identical.

For this duality to exist, a couple of key features must be shared among the two theories.  First, two T-dual LSTs must have the same Coulomb branch dimension and amounts of Wilson line parameters i.e. 
\begin{align}
\textbf{Dim(CB)} = T+\text{rk}(G) \, , \qquad \textbf{Dim(WL})= \text{rk}(G_F) \, .
\end{align}  
The aforementioned matching is evident from field theory, e.g. the tensor multiplets become vector multiplets in 5D with an additional scalar, that is the associated CB parameter. 
Second, the continuous 2-group symmetries provide an additional set of global symmetries necessary for the matching across T-duality \cite{DelZotto:2020sop}. Each of the relevant 2-group structure constant must match individually which places tight bounds on possible T-dual LSTs. The matching of $\widehat{\kappa}_{\mathscr P}$ for example implies that T-duality can not alter the birational base class, i.e. heterotic LSTs have to be mapped to heterotic LSTs.  

This heterotic-heterotic LST match may not be very surprising. However matching the various $\widehat{\kappa}_{\mathscr R}$ values is highly sensitive to tensor and gauge algebra structure and can therefore be used as a non-trivial check of two little string theories\footnote{Note that for pure $\fsu/\fsp$ gauge groups with LS charge $l_{LS,I}=1$, $\widehat{\kappa}_{\mathscr R}$ coincides with the CB dimension. Hence in such cases it does not yield new constrains.} being T-dual.

T-duality is naturally geometrized in terms of F/M-theory, where the LSTs are captured by the singular geometry of certain non-compact elliptic threefolds $X_i$. Compactifying those theories on a circle yields by definition a 5D theory that coincides with M-theory on the very same geometries $X_i$. In M-theory, also the F-theory elliptic fibre is physical and resolving the (non-)compact singularities corresponds to switching on non-trivial CB(WL) parameters, which breaks the theories to their maximal Cartan sub-algebras. If two theories are T-dual, they must become entirely identical for some values of those parameters. The M-theory duality therefore implies the $X_i$ to be \textit{birational equivalent} \cite{Aspinwall:1996vc}.

Put differently, we could equally well start from M-theory on a smooth non-compact threefold $X$ and explore its inequivalent torus fibrations, which by the virtue of M/F-theory lift to inequivalent 6D little string theories.   Such explorations have systematically been performed for compact threefolds that are CICYs \cite{Anderson:2017aux} and hypersurfaces in toric varieties \cite{Huang:2019pne}. This work focuses on toric hypersurfaces and the classification of toric hypersurface fibrations in \cite{Klevers:2014bqa} for non-compact threefolds.
 
 \subsection{Degenerate K3 fibrations and Heterotic LSTs}
 \label{ssec:K3Degens}
 In this part, we provide a brief outline of the structure of the geometries utilized in our heterotic LST constructions from the perspective of a degenerate K3.

 \paragraph{An elliptic K3 surface} In 8D, the F-theory dual of the Heterotic string is obtained from an elliptic K3 surface with the stable degeneration limit \cite{Friedman:1997yq,Donagi:1997aks} 
 \be
 \begin{gathered}\xymatrix{\mathbb T^2_f\ar[r]& K3\ar[d]^f\\& \mathbb P^1}\end{gathered} \qquad\to\qquad \begin{gathered} \xymatrix{\mathbb T^2_f\ar[r]& dP_9 \vee_{\mathbb T^2_H} dP_9
 \ar[d]^f\\& \mathbb P^1}\end{gathered}\ee
 
The eight-dimensional reduction of the $E_8 \times E_8$ Heterotic string on $T^2_H$ is equivalent to F-theory on such a K3, where the possible $E_8\times E_8$ bundles correspond to the possible Picard sublattices that can be realized within the homology lattice \cite{Friedman:1997yq} (see e.g. \cite{Harder_2015} for a recent review)
 \be
 \Lambda_{K3}= U^{\oplus 3} \oplus E_8^{\oplus 2} \cong \Pi_{3,19} ,
 \ee
 with $U$ being the hyperbolic plane lattice, $\Lambda_{K3}$ is isometric to the intersection form of the second cohomology group for the K3 surface in concern.
 
Let $S$ denote our K3 surface; the most crucial sublattice in this context is the Neron-Severi lattice $NS(S)$
 \begin{align}
 NS(S) := H^{1,1}(S) \cap H^2(K3, \mathbb{Z}) \, .
 \end{align}
The orthogonal complement to $NS(S)$ in $\Lambda_{K3}$ is denoted by the transcendental lattice $T_S$. While the transcendental lattice can be thought of as the complex structure deformations of a K3, the NS$(S)$ lattice is spanned by the divisors modulo rational equivalence. We are not interested in K3s with any NS lattice structure but those, that allow for an elliptic fibration. 

 In lattice theoretic terms, this fibration structure is displayed by the embedding of the hyperbolic lattice $U$ into $NS(S)$.  The K3 provides us with a manifest option of NS$(S)$, namely one can incorporate a hyperbolic factor $U$ directly by the orthogonal decomposition \cite{Braun:2013yya}
  \begin{equation}
	 NS(S) \cong U_{}\oplus W_{\text{frame}}\, .
 \end{equation}
 The class of the zero-section and the base are contained in $U$, while all
 other divisors not intersecting the zero-section are shrinkable and part of $W_{\text{frame}}$.   Those divisors produce the F/M-theory flavor group containing potential non-Abelian and Abelian algebra factors given by the Mordell-Weil group \cite{Morrison:1996pp,Grimm:2010ez,Morrison:2012ei}. According to the Tate classification \cite{Bershadsky:1996nh,Katz:2011qp}, we might potentially incorporate extra monodromies over the base $\mathbb{C}^1$, resulting in the possibility of non-simply laced flavor groups.  
The  associated intersection form of divisors spanning NS$(S)$ is embedded in the aforementioned frame lattice $W_{\text{frame}}$. Thus it allows us to read off the ADE type root lattice and the Mordell-Weil group directly. Therefore the most straightforward choice for a heterotic dual theory is to include two $E_8$ components as
 \begin{align}
 \label{eq:exampleK3}
 NS(S) = U \oplus (-E_8)^{\oplus 2} \,  .
 \end{align}
Note that when the NS$(S)$ lattice is relatively large, there may be more elliptic fibrations i.e., $U$ embeddings into the NS lattice. It is known \cite{Candelas:1996su} that the above choice of NS lattice admits another elliptic fibration, with an $\fso_{32}$ fibre and an order two MW group, which unifies $E_{8}^2/(Spin(32)/\mathbb{Z}_2)$ duality. In general, finding all torus fibrations for a given K3 requires classifying all $U$ embeddings into a single NS$(S)$ lattice, which is not an easy mathematical problem (see e.g. \cite{Braun:2013yya}). An important takeaway from \cite{Braun:2013yya} is that certain K3s with large Picard numbers ($\rho \geq 17$) can have more than two elliptic fibrations; in one case the number even reaches 63. Therefore, we should anticipate multi-T-dual heterotic LSTs found in Section~\ref{sec: MultiTDual} once we approach the 6D theory.
Note however that in 8D the heterotic/F-theory moduli spaces match perfectly. This moduli space is given by a point in the even and self-dual lattice II$_{2,18}$. This is exactly the $\lambda_{K3}$ modulo an elliptic fibration on the F-theory side, and the heterotic string on a 2D Narain torus. 
 
Note that the stable degeneration limit above allows to identify a heterotic torus at large volume in addition to some $E_8^2$ Wilson line configurations. However, there are also non-geometric loci where the heterotic torus is not at large volume,  that lead to gauge enhancements as we point out in Section~\ref{ssec:ExtremalK3}.

\paragraph{K3 fibrations} For the remainder of this study, we intend to explore the 6D LSTs over $B_{LST,1}$, i.e. bases that are birational to $\mathbb{P}^1 \times \mathbb{C}$. Consequently, we are fibering the elliptic K3s over a non-compact direction. Hence the non-compact threefold admits the following nested fibration structure
\begin{align}
 \begin{array}{cl}
 K3 \rightarrow &X \\
 &\downarrow \\
 &\mathbb{C}
 \end{array}
  \qquad \text{ with }\quad  \begin{array}{cl}
 T^2 \rightarrow &K3 \\
 &\downarrow \\ 
 &\mathbb{P}^1
 \end{array} \, ,
 \end{align}
which allows us to rephrase many features of an LST as properties of an elliptic K3 fibration and its degeneration.  Furthermore, the entire NS$(S)$ lattice of the K3 becomes the overlattice of the polarization lattice $\Lambda_S$ of the fibration. $\Lambda_S$ connects our K3 surface and the CY threefold $X$ as it is obtained from the intersection form of the non-compact divisors $D_i\in X$ restricted to the K3 surface.
 
\begin{figure}[ht!]
	 \begin{center}
	 \includegraphics[scale=0.5]{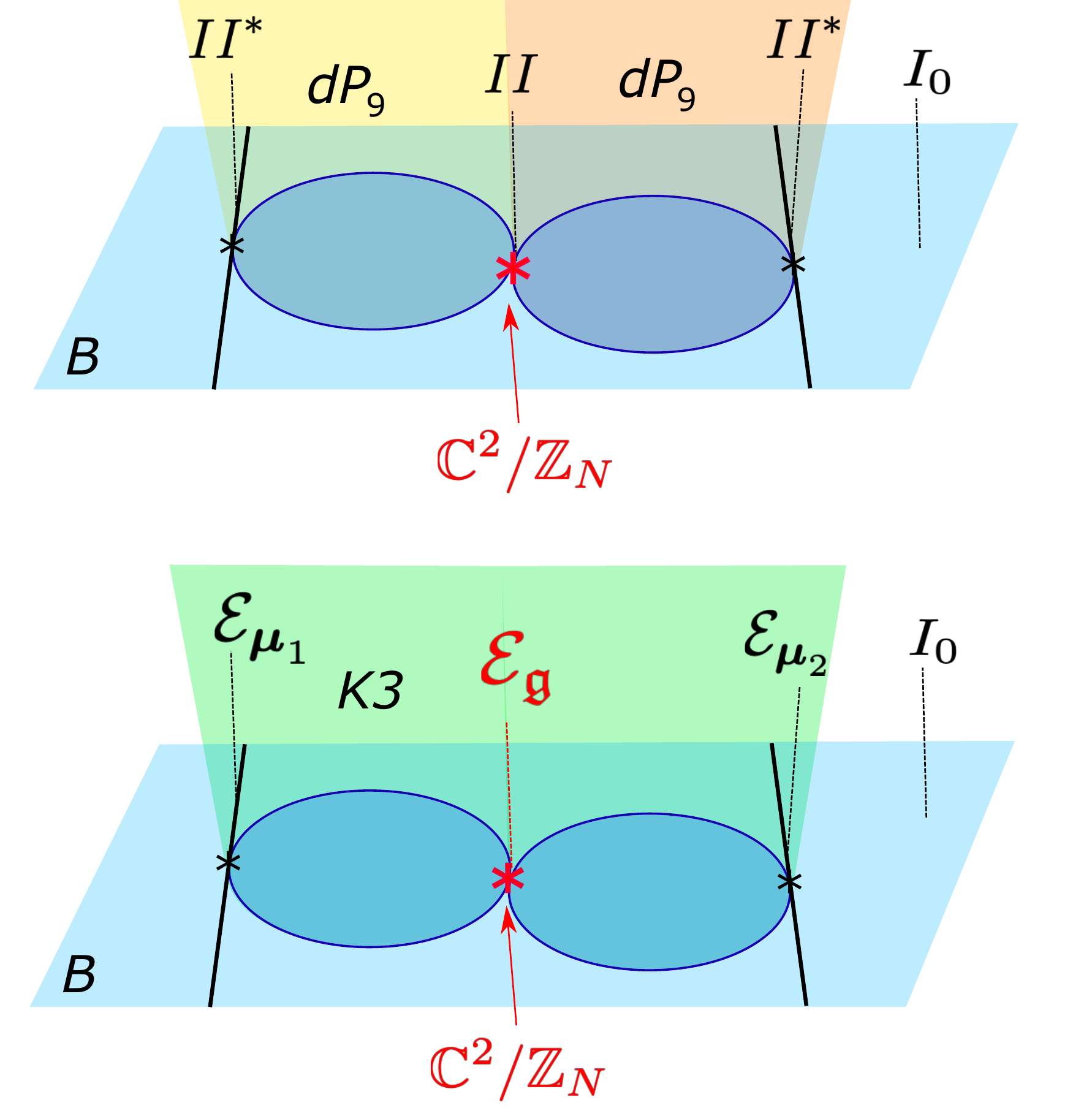}
	 \caption{\textsc{Top}: stable degeneration limit of a $K3$ with $E_8^{\oplus 2}$ Picard lattice in two copies of $dP_9$: Heterotic instantons are realized as a $\mathbb C^2/\mathbb Z_N$ singularity where the two 1 curves intersect.  \textsc{Bottom}: engineering of ALE Heterotic instantons in F-theory. The choice of $\boldsymbol{\mu}_a$ dictates the structure of the transcendental lattice of the K3 fibre, this gives rise to the two non-compact curves of Kodaira singularities $\mathcal E_{\boldsymbol{\mu}_a}$. In addition, the presence of the ALE background singularity in the Hořava-Witten setup is dualized by the presence of Kodaira type $\mathfrak g$ fiber whose locus is the $\mathbb C^2/\mathbb Z_N$ giving rise to the $N$ instantons as well as the two 1 curves in the base.}\label{fig:K3deg}
	 \end{center}
	  \end{figure}

On the other hand, a 7D gauge theory arises from a singularity $\mathbb C^2/\Gamma_{\mathfrak g}$ along the M9 brane in the M-theory description. The latter being a gauge symmetry lifted to F-theory as a Kodaira fibre $\fg$ at the origin. This however introduces additional compact curves $D_i$ at the origin of the K3 fibration and hence yields a degeneration of the K3 fibre. 
The choice of the two embeddings
 \be
 \boldsymbol{\mu}_a : \Gamma_\mathfrak{g} \to E_8
 \ee
 corresponds to tilting the transcendental lattice in the K3 fibre, thus impacting only the non-compact Kodaira fibers that encode the global symmetry of the associated theory. The Aspinwall-Morrison geometries correspond to the choices $\boldsymbol{\mu}_1=\boldsymbol{\mu}_2 = \text{id}_{E_8}$, leaving the flavor group unbroken. All other possible options of $\boldsymbol{\mu}_1$ and $\boldsymbol{\mu}_2$ in Figure ~\ref{fig:K3deg} can be achieved by suitably tilting the transcendental lattice, as we shall do for various choices in the next sections.
 \paragraph{K3 degeneration} In order to obtain a theory with non-trivial dynamics, we must add a couple of non-trivial compact curves to the fibration. We are therefore compelled to investigate degenerate K3s at this locus. In the mathematics literature, there exists a general theory of degenerate K3s in terms of Kulikov models.  A short review of the Kulikov model is provided in Appendix~\ref{app:Kulikov}. Roughly speaking, Kulikov degenerations concern smooth degenerated K3s around a neighborhood of the origin of the base $\mathbb{C}$. We denote the K3 fiber by $S_u$ with $u \in \mathbb{C}$ and split the discussion into two pieces. Those at a generic point $S_1=S_{u\neq 0}$ and the origin $S_{u=0}$, where the K3 fibration is allowed to degenerate. Note that the K3 fibre could become reducible at finite points over the base, the intersecting property of the splitting components classifies Kulikov models into three types. For instance, a type II model corresponds to the K3 fibre that decomposes into $N+1$ irreducible components intersecting in a chain with the ending components being rational elliptic surfaces. This is precisely the heterotic instanton illustration shown at the top of Figure~\ref{fig:K3deg} upon shrinking them to a $\mathbb{C}^2/\mathbb{Z}_N$ singularity.

The Kulikov classification, however, is very coarse. i.e., surface components with a higher multiplicity are allowed to be removed by birational base changes, which is forbidden in our construction as this would change the associated LST physics. Second, it is impossible to have non-trivial fibres over some ruled surfaces when intersecting with an $\mathfrak{g}$ type fibral singularity. Those are the degenerations obtained from the removal of non-minimal torus fibre singularities that characterize the LST.  Nonetheless, many of our K3 fibrations degenerate in a "Kulikov-like" way but with a substantially more intricate fibre structure. It would be highly interesting to extend the Kulikov model with e.g. the geometric properties of an elliptic threefold similar to \cite{Grassi1991OnMM,Grassi:2011hq} to mathematically classify heterotic LSTs as generalized Kulikov degenerations.
 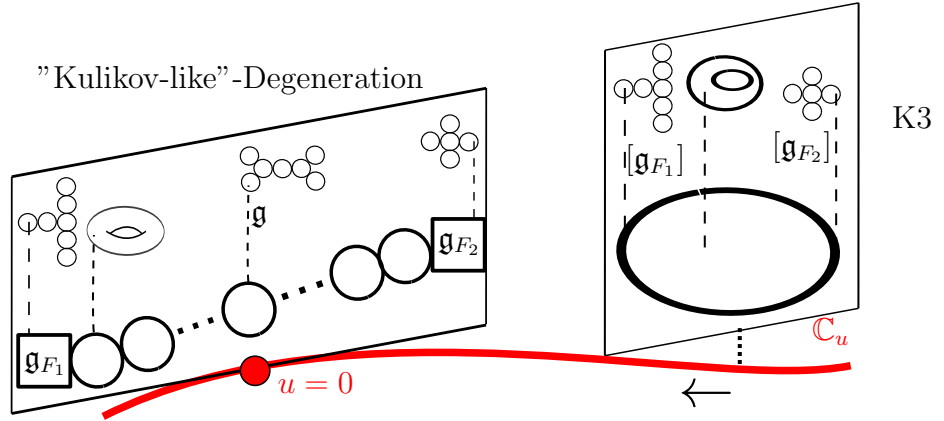
\begin{figure}[t!] 
	 \begin{picture}(0,150)
\put(100,20){\includegraphics[scale=1.0]{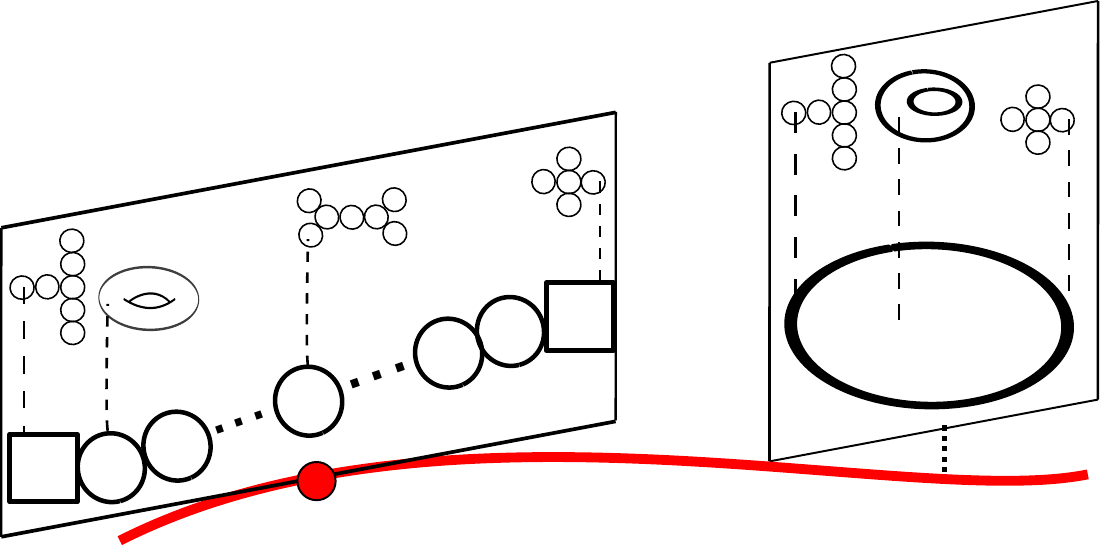}}
\put(430,130){K3}
\put(385,120){$[\mathfrak{g}_{F_2}]$}
\put(330,115){$[\mathfrak{g}_{F_1}]$}
\put(110,145){"Kulikov-like"-Degeneration}
\put(350,25){\LARGE $\leftarrow$}
\put(400,50){$\textcolor{red}{\mathbb{C}_u}$}
\put(200,30){$\textcolor{red}{u=0}$}
\put(190,95){$\mathfrak{g}  $}
\put(260,85){$\mathfrak{g}_{F_2}  $}
\put(105,40){$\mathfrak{g}_{F_1}  $}
\end{picture}
	 \caption{A little string theory as a "Kulikov-like" degeneration: Over a generic point in the base $\mathbb{C}_u$ there is an elliptic K3 whose frame lattice yields the flavor group. At the origin of $\mathbb{C}$, the K3 degenerates into compact surfaces with $\mathfrak{g}$ fibers that exhibits a fusion of $\mathcal{T}(\mathfrak{g},\mathfrak{g}_{F_i})$ conformal matter.}\label{fig:Kulikov}
 \end{figure}

Before proceeding to the constructions of such heterotic LSTs via toric geometry, we summarize the lessons gained from viewing them as degenerate K3 fibrations:  
 \begin{enumerate}
 \item The heterotic LSTs are described as elliptic K3 fibrations over $\mathbb{C}$, which degenerate in a Kulikov-like way at the origin.
 \item For a given K3 with a chosen torus-fibration, the flavor group is given by the frame lattice $W_{\text{frame}}$. The flavor group is contained in the even self-dual lattice II$_{2,18}$
 \item The maximal rank of the flavor group is 18.
 When the flavor group is purely non-Abelian, those are classified by extremal K3 surfaces. 
 \item As the K3 is compact, there can only be a finite set of heterotic LST flavor symmetries.\footnote{As argued in \cite{delZottoAppear2} upon removing all NS5 branes, we expect the theory to uplift to an 8D $\mathcal{N}=1$ SUGRA theory, given via F-theory on the same elliptic K3.} This is expected from the deep interplay with these degenerations and breaking patterns of $E_8\times E_8$ or $Spin(32)/\mathbb Z_2$, but captures also some slight generalisation arising from non-geometric heterotic backgrounds (see e.g. \cite{Font:2016odl,Braun:2013yya}).
 \item A chosen $NS(S)$ lattice of the K3 must include an elliptic fibration. T-dual LSTs are identified by other elliptic fibration structures of the same $NS(S)$ choice. 
 \end{enumerate}
 Hence for a fixed geometry, classifying T-dual LSTs is simply transferred to find inequivalent elliptic fibrations in the flavor K3 fibre. Such a classification is non-trivial but can be done via lattice theoretic techniques \cite{Braun:2013yya}. For large $NS(S)$ lattices, we should expect dozens of fibrations/LSTs.
 In the bulk of this work, we construct resolved threefolds with up to three of such fibrations. In Section~\ref{ssec:ExtremalK3} we discuss fibrations with extremal K3 surfaces that do not exhibit a toric resolution and propose six different LSTs that are all T-dual to each other.
 
\section{Geometrizing heterotic ALE instantons}\label{sec: HeteroticLSTs}
In the Hořava-Witten setup \cite{Horava:1995qa,Horava:1996ma}, the Heterotic instantons are realized by $N$ M5 branes parallel to the two M9s, each is dual to a copy of $E_8$ within the K3 NS lattice, in the stable degeneration limit, each $E_8$ is mapped in the $E_8$ sublattice of $dP_9$ \cite{Morrison:1996pp,Morrison:1996xf,Donagi:1997aks}. In IIB $N$ M5 branes emerge as a $\mathbb C^2/\mathbb Z_N$ singularity; hence to geometrically engineer the LST of $N$ Heterotic $E_8 \times E_8$ instantons within F-theory we start with the setup in the upper part of Figure~\ref{fig:K3deg}. Namely, we realize a $\mathbb C^2/\mathbb Z_N$ at the locus where the two -1 curves corresponding to the base $\mathbb P^1$'s of the $dP_9$ intersect in the stable degeneration limit.
 
 The T-duality for LSTs with $\widehat{\kappa}_{\mathscr P} = 2$ is often realized as a form of fiber/base duality. Note that $\widehat{\kappa}_{\mathscr P} = 2$ signals the presence of the M9 branes, it is anticipated that models corresponding to the theories $\widetilde{K}_N(\boldsymbol{\lambda};\mathfrak{g})$ would be realized by flipping the role of $\mathbb C^2/\mathbb Z_N$ and the singular Kodaira fiber $\mathcal E_{\mathfrak g}$. This remark is motivated by the examples we have discussed in Type I' in \cite{DelZotto:2022ohj}. There are, nevertheless, several (crucial) subtleties: first, the singularity $\mathbb C^2/\mathbb Z_N$ is part of a broader collection of curves with the structure of a $\mathcal I_{2N+1}/\mathbb Z_2$, where the $\mathbb Z_2$ acts by identifying the corresponding nodes along the diagonal
 \be
 \xymatrix{
 &&\bullet_1\ar@{-}[r]&\cdots&\bullet_{N}\ar@{-}[dr]\\
 \ar@{..}[rrrrrr]&\bullet_0 \ar@{-}[ur] \ar@{-}[dr]&&&&\bullet_{N+1}&\\
 &&\bullet_{2N+1}\ar@{-}[r]&\cdots&\bullet_{N+2}\ar@{-}[ur]
 }
 \ee
therefore we expect the emergence of an $\mathfrak{sp}_N$ algebra, which is accurate under the appearance of $Spin(32)/\mathbb Z_2$ instantons. Furthermore, this fiber will decorate a base containing a collection of intersecting curves, whose topology is dictated by the structure of the $\mathcal E_{\mathfrak g}$ fiber. Obviously, the intersection numbers of the associated curves must jump throughout the folding process (otherwise we would get $\widehat{\kappa}_{\mathscr{P}} = 0$ which cannot be the case for a T-dual pair).
  
 \medskip

\subsubsection*{Fusing SCFTs to LSTs}
    Before constructing the various LSTs geometrically we quickly recall the algorithm of how an $\mathcal{K}_N(\mu_1,\mu_2,\mathfrak{g})$ heterotic LST is obtained \cite{DelZotto:2022ohj}.  
    This algorithm exploits the fact  we can split up 
    an $\mathcal{K}_N(\boldsymbol{\mu}_1,\boldsymbol{\mu}_2;\mathfrak g)$ into three parts that are comprised of SCFTs, fused together, schematically given as  
\begin{align}
\mathcal  K_N(\boldsymbol{\mu}_1,\boldsymbol{\mu}_2;\mathfrak g) = \xymatrix{\mathcal T(\boldsymbol{\mu_1},\mathfrak g) \ar@{-}[r]^{\mathfrak {g}} &\mathcal T_{N-2}(\mathfrak g,\mathfrak g)  \ar@{-}[r]^{\mathfrak {g}}&\mathcal T(\boldsymbol{\mu_2},\mathfrak g) } \, .
\end{align}
Here $T(\boldsymbol{\mu_a},\mathfrak g)$ denotes minimal orbi-instanton theories attached to either of the two M9 brane sides classified by \cite{Frey:2018vpw} via fusion \cite{Heckman:2018pqx} (see also \cite{DelZotto:2014hpa,DelZotto:2018tcj}). The theories are specified by the singularity $\mathfrak{g}$ and the morphisms $\mu_a$ acting as
\begin{align}
\boldsymbol{\mu}_{a} \colon \pi_1(S^3/\Gamma_{\mathfrak g}) \simeq \Gamma_{\mathfrak g}\to E_8\,,
\end{align}
and breaking the $\fe_8$ into a commuting subgroup
\be
F^{(0)}_a \equiv \{ g \in E_8 \, | \, gh = hg, \forall h \in \boldsymbol{\mu}_a(\Gamma_{\mathfrak g})\} \qquad a=1,2\,. \\
\ee
 Finally the 
 $T_{N-2}(\mathfrak g,\mathfrak g)$ part is a minimal conformal matter theory associated to $N-2$ M5 branes probing an $\mathbb{C}^2/\Gamma_\mathfrak{g}$-type singularity. The fusion process can be viewed as gauging a common flavor group $\mathfrak{g}$ of two SCFTs that sits over an additional tensor multiplet with string charge $n$. Graphically this can be denoted by the quiver action:
\begin{align} 
\xymatrix{\cdots \, \, \overset{\mathfrak g_{i }  } \, \,  {n_{i } } \,\, \overset{\mathfrak g_{i+1}  }{n_{i+1} } \,\, \textcolor{red}{[\mathfrak g]} \ar@[red]@{-}[r]^{\textcolor{red}{\mathfrak g}}\,\,& \textcolor{red}{[\mathfrak g]}   \,\,  \overset{\mathfrak g_{j+1} }{n_{j+1}}\,\, \overset{\mathfrak g_{j } }{n_{j  }}  \cdots }\,\, \longrightarrow \,\,\cdots
 \overset{\mathfrak g_{i }  }{n_{i } } \,\, \overset{\mathfrak g_{i+1}  }{n_{i+1} } \, \, 
 \textcolor{red}{\overset{\mathfrak{g}}{n}}\, \, 
 \overset{\mathfrak g_{j+1} }{n_{j+1}}\,\, \overset{\mathfrak g_{j } }{n_{j  }}  \cdots  
\end{align}
The above structure is also very useful in the piece-wise construction of the associated smooth threefolds as well as the computation of LS charges:
I.e. we can take the various SCFT parts individually and stick the resolved pieces together, to obtain the full smooth threefold that constructs the LST. This procedure makes it straightforward to read off the second (or more) fibration and its fibre structure which yields T-dual LSTs. 
  
In some cases, we will actually be able to take "negative" numbers of M5-branes and to directly glue together the minimal orbi-instanton theories which yield exotic types of dual theories.

In order to resolve the respective threefolds, we make heavily use of toric  geometry, which will be introduced in the following. 

We will go through all ADE singularities but focus in particular details on the exceptional cases where a perturbative brane picture is lacking. 
 
We start with $A$-type singularities and their resolutions as warm up and introduce their toric resolutions. We then move to the exceptional cases where the additional conformal matter factors and various other technical details are discussed. 
For completeness we also briefly discuss the $D$-type cases. 
 
\subsection{Toric Preliminaries}
\label{ssec:ToricPrelim}
This section begins by reviewing a couple of facts about toric fibrations. We heavily rely on the construction of threefolds via polytopes 4D $\Delta$ which is standard for compact threefolds, which we adopt and sufficiently generalize for non-compact cases. For further details, see \cite{Bouchard:2003bu,Cox:2000vi,cox2011toric}. The geometric construction uses the following two steps: 
\begin{enumerate}
\item The non-compact threefold $X$ is built as the anti-canonical hypersurface in a non-compact smooth complex four-dimensional toric variety $\mathcal{A}$.  
\item The toric variety $\mathcal{A}$ is obtained from a Fan that is achieved via a regular fine star triangulation of the rays, which forms a four-dimensional \textit{semi-convex} polytope $\Delta$.
\end{enumerate} 
We employ the notion of a polytope by $\Delta$ and its dual $\Delta^*$ according to Batyrev \& Borisov \cite{Batyrev:1994pg} and make use of the results of \cite{Bouchard:2003bu}.

Consider a polytope $\Delta$ that is the convex hull of a finite set of points $v_p$ spanning a lattice $N \sim \mathbb{Z}^4$. Recall that every point $v_p \in \Delta$ is associated with a complex coordinate $x_p$ as a copy of $\mathbb{C}$, it corresponds to a toric divisor $D_{x_p}: x_p=0$. These divisors intersect whenever their associated vertices share a cone in the fan. Obtaining a desired triangulation of the fan is computationally very hard and scales with the number of points in $\Delta$. In practice we only need to require the existence of such a triangulation, which allows us to deduce the majority of the structures from the rays alone \footnote {Such triangulations are automated in computer algebra programs such as \textit{SAGE, CYTools} \cite{sagemath,Demirtas:2022hqf} but only for compact threefolds. We employ those programs as cross-checks for the compactified threefolds in the following.}. 

All the coordinates $x_p$ are subject to homogeneous scalings
\begin{align}
\label{eq:linequiv1}
(x_1, x_2 , x_3 \ldots x_n) \sim (\lambda^{k_1} x_1,\lambda^{k_2} x_2 , \lambda^{k_3}x_3 \ldots  \lambda^{k_n}x_n) \, ,
\end{align}
with $\lambda \in \mathbb{C}^*$ that relates to some torus action. Thus we can read off their linear relations from the vertices
\begin{align}
\label{eq:linequiv2}
\sum_p k_p v_p = 0 \, . 
\end{align}
Whenever there is a point $v$ that does
admit only a trivial linear relation with other vertices, it admits a trivial $\mathbb{C}^*$ action and hence is a non-compact direction in the associated toric variety. Moreover, we can derive linear equivalence relation among divisors using any lattice vector $m$ from the dual lattice $M$ and obtain
\begin{equation}\label{eq:DivEquiv}
\sum_i \langle v_p, m  \rangle D_{x_p} \sim 0 \, .
 \end{equation} 
 
To setup a toric fibration, the ambient variety $\mathcal{A}$ must admit a fibration structure, \cite{Klevers:2014bqa} from which the elliptic fibration on $X$ is inherited\footnote{In the compact case, such a sub-polytope guarantees the existence of a triangulation such that the ambient space admits a fibration by $F$ and thus a
 toric elliptic fibration 
\cite{Kreuzer:1997zg,Huang:2019pne} of the threefold modulo non-flat fibres. } In particular we require 
\begin{align}
\begin{array}{cl}
F \hookrightarrow & \mathcal{A} \\ 
& \downarrow \pi \\
& B_{2} 
\end{array} \, .
\end{align}
Here $B_2$ coincides with the non-compact LST base, while $F$ is a compact complex two-dimensional \textit{weak Fano} surface. The weak Fano property leads to the existence of an effective anti-canonical hypersurface, which is a CY one-fold and hence a torus. 

The above structure then implies that the anti-canonical hypersurface in $\mathcal{A}$ inherits the fibration structure and in particular becomes a torus-fibration over $B_2$. In most usages in the literature, the fibre ambient space is fixed to the Tate model, i.e. $F \sim \mathbb{P}^{2}_{1,2,3}$. We will also deal with different ambient spaces that are not necessarily of Tate-type. \footnote{In the toric context, there are sixteen toric ambient spaces as two-dimensional reflexive polytopes. Their fiber structure and F-theory physics have been extensively explored in \cite{Klevers:2014bqa}, and we adopt their notation in the following. }.

Recall that the reflexivity of a polytope is defined by the origin $0$ being the unique interior point in the convex hull spanned by the vertices of $\Delta$. Therefore, the polytope $\Delta$ in our consideration for the LST base is not reflexive as this would imply compactness. In our cases, the origin locates on a certain face of $\Delta$. The derivation of fibration and resolution structures does not require compactness.

To read off a toric fibration from a polytope $\Delta$, we require the following triangle form 
\begin{align}
\label{eq:fibrationPoly}
\Delta = \left( \begin{array}{cc}   F  &  (0) \\ T & B_2   \end{array}\right) \, ,
\end{align}
Here $(0)$ links to the origin of the base toric variety $B_2$ and represents the generic point. This triangular form implies a projection $\pi$ acting as
\begin{align}
\pi (\Delta) \rightarrow (B_2) \, .
\end{align} 
Over the generic point in $B_2$, we find the 2D toric variety is represented by the vertices of the 2D reflexive polytope $F$ \footnote{An important addition though is, that we also have a star triangulation of the rays which respects the toric morphism $\pi$, i.e. cones are mapped to cones. In \cite{Kreuzer:1997zg,Huang:2015sta,Huang:2019pne}, the argument for the existence of such a triangulation has been given, although the fibration might not always be flat. In Section~\ref{sec:non-flat} we will show how to deal with non-flatness.}.

This condition coincides with the toric fibration given in \cite{Bouchard:2003bu} inherited from the ambient space. We dualize the polytope $\Delta$ into its polar $\Delta^*$ following Batyrev
\begin{align}
\label{eq:Batyrev}
\Delta^* = \left\{  m \in M_R | \langle v, m  \rangle \geq -1 \text{ for all } v \in \Delta   \right\} \, .
\end{align}
If $\Delta$ is reflexive, also $\Delta^*$ would be. Again, this is not going to be the case in our study but we will continue with that assumption for a moment. The usage of the dual polytope $\Delta^*$ permits the dual points to play the role of monomials in the anti-canonical hypersurface defining equation. Hence the most generic CY hypersurface is written as
\begin{align}
\label{eq:hypersurface}
p= \sum_{m \in \Delta^*} \prod_{v_p \in \Delta} a_m x_p^{\langle   v_p , m    \rangle +1 } \, ,
\end{align}
which is a generic section of the anti-cannonical class of $\mathcal{A}$. For the non-reflexive polytopes we are considering, coordinates that are not restricted by $\mathbb{C}^*$ scalings appear as arbitrary power series in the defining equation which is equivalent to $\Delta^*$ being an infinitely extended prism.  

The toric hypersurface \eqref{eq:fibrationPoly} allows us to set up an elliptic fibration. Namely, over a generic point in the base, i.e. $(0)$ we assign all base coordinates that are not part of the slice $F $ to some generic value, such that the CY hypersurface equation $p$ reduces to a hypersurface in the 2D ambient space $F$, which is an elliptic curve. To establish an elliptic fibration, the structure of how a 2D reflexive polytope $F $ becomes a sub-polytope is crucial. Moreover, the base can be read off directly from a projection of the fan of the toric ambient space as
\begin{align}
\pi (\Delta) \rightarrow \Delta_{B_2} \, .
\end{align} The base $B_2$ must be birational to either $\mathbb{P}^2$ or Hirtzebruch surfaces $\mathbb{F}_n$ \cite{Grassi1991OnMM} for compactness.
 
Before moving to the total space, we consider the structure of the LST bases.  As reviewed in Section~\ref{sec:LST}, the little string base is featured by the non-compactness and the existence of a unique curve $\Sigma_0$ of self-intersection $\Sigma_0^2 = 0$.  In addition, there are only two distinct endpoint topologies after shrinking all possible curves: 
\begin{align}
B_{LST,1}: \mathbb{P}^1 \times \mathbb{C} \, \qquad B_{LST,2}: (\mathbb{T}^2 \times \mathbb{C})/\Gamma  \, \text{ with } \Gamma  \in SU(2)
\end{align}
In our toric setup,  these bases are depicted via 2D toric ambient spaces. However, there is no 2D toric ambient space isomorphic to $B_{LST,2}$. \footnote {To achieve this, our construction can be enlarged to CICY bases, similar as discussed in \cite{Anderson:2017aux}. A straightforward example would be the anti-cannonical hypersurface in $\mathbb{P}^2 \times \mathbb{C}$ by enlarging the ambient variety to a fivefold. }. Hence in the toric setup, we are left with LST bases of type $B_{LST,1}=\mathbb{P}^1\times \mathbb{C}$ as followed, up to $GL(2,\mathbb{Z})$ transformations
	  \begin{align}\label{fig:B2base}
	  \begin{tikzpicture}
		  \draw[gray, thick,->] (0,0) -> (0,1);
		  \draw[gray, thick,<->] (-1,0) -- (1,0);  
	  \filldraw[black] (0,0) circle (2pt) node[anchor=west]{};
		  \draw[black] (-1,0) circle (0pt) node[anchor=east]{$v_{w_0}$};
		  \draw[black] (1,0) circle (0pt) node[anchor=west]{$v_{w_1}$}; 
  \draw[black] (0,1.6) circle (0pt) node[anchor=north]{$v_{y_0}$}; 
  \end{tikzpicture}	 
\end{align}
It follows directly from Eqn.~\eqref{eq:linequiv1} and Eqn.~\eqref{eq:linequiv2} that $w_0,w_1$ are the two affine coordinates of a $\mathbb{P}^1$ and $y_0$ has no $\mathbb{C}^*$ scaling and hence just parameterize $\mathbb{C}$. The actually geometries we will consider, will have many more compact curves though and thus toric rays. As all heterotic LSTs have bases, that must be birational to $\mathbb{P}^1 \times \mathbb{C}$, since there does always exist a blow-down map, that shrinks the base down to the toric configuration as in \eqref{fig:B2base}. Note that the base itself is not longer convex since the two-dimensional cone spanned by $v_{w_0},v_{w_1}$ includes the origin as an interior point.

We can now start to investigate the general structure of the full polytope over such a base. Note that non-convexity of the base lifts to the full polytope $\Delta$ which can therefore not be reflexive either.  However, we can still dualize $\Delta$ to obtain the dual polytope  $\Delta^*$ using the Batyrev prescription   Eq.~\eqref{eq:Batyrev}. From this it is
easy to see, that a generic ray $v \in \Delta$ admits the form $v=(x,y,z,q)$ with $q \geq 0$.

Hence each dual vector  $m \in \Delta^*$ with coordinates $m=(x^*,y^*,z^*,q^*)$ admits a $q^*$ coordinate that is not bounded from above, resulting an infinite set of solutions.  Analogous the tops defined in \cite{Bouchard:2003bu},  $\Delta^*$ admits the structure of an infinitely long prism in the $q^*$ direction. 

Since $\Delta^*$ plays the role of the polytope providing monomials in the CY hypersurface, we conclude that $p$ in \eqref{eq:hypersurface} is an infinite power series in certain coordinates. This is evident from the fact that e.g. $y_0$ is simply a copy of $\mathbb{C}$ with trivial $\mathbb{C}^*$ scaling and thus can appear to any arbitrary degree in the CY hypersurface $p$. This fact is useful to keep in mind but in practice it suffices to truncate the polynomials at some suitable order. 

Before constructing some first LST configurations and analyzing their structure, we discuss some basic properties that can be readily inferred from the toric geometry.

\subsubsection*{Tops, Kac labels and LS charges from toric geometry}
The structure of toric geometry is powerful enough to construct and deduce many basic ingredients of LSTs, which we will employ throughout this work. 

As previously stated, all toric little string bases in our consideration are birational to $\mathbb{P}^1 \times \mathbb{C}^1$ as given in Figure~\ref{fig:B2base}. For some higher rank LST, we simply add extra vectors $v_{y_I}$  to the configuration with $\mathbb{Z}^2$ coordinates $v_{y_I}: (z_I ,q_I)$ such that $q_I>0$ for all $I$. In addition, each pair of vectors that span a convex cone must fulfill
\begin{align}
|\text{Det}( v_{y_I} , v_{y_J}) | = 1 \, .
\end{align}
to satisfy the smoothness condition of the base. The intersection form of base curves can then be easily computed using Eqn.~\eqref{eq:DivEquiv}. 

 All the respective divisors $D_{y_I}: y_I=0$ are then compact curves isomorphic to $\mathbb{P}^1$'s\footnote{We should keep in mind not to generate curves of negative self-intersection larger than 12, otherwise the crepant resolution of our elliptic threefold does not exist.}. The condition $q_I>0$ ensures the existence of a canonical equivalence relation, take the dual vector as $m=(0,1)$ and insert to \eqref{eq:linequiv2}, it gives us
\begin{align}
D_0 : \sum_I  q_I D_{y_I}  \, ,
\end{align}
and $D_0$ must be a principal divisor. Since each toric ray is also dual to a curve $w_I$, we can borrow the $q_I$ factor to rewrite the above principal divisor in terms of curves as
\begin{align}
 \Sigma_0 =  q_I w^I \, .
\end{align}
Here $\Sigma_0$ is precisely the LS curve and $q_I=l_{LS,I}$ are the LS charges of the tensors that wrap the respective curves $w_I$. Thus it is sufficient to know the $q_I$ coordinate of the toric rays along the LS direction to infer the LS charges.

On the other hand, the toric construction is also powerful enough to directly derive the rough ADE fibre structure of those rays without performing a full fledged intersection calculation.
ADE fibers on $\mathbb{P}^1$ can be constructed by a so called \textit{top} \cite{Bouchard:2003bu} technique, which refers to \textit{multiple} vertices $v_p \in \Delta$ projected onto the same base ray modulo certain multiplicity $q$. Consider a local configuration with $v_i \in \Delta$ with $(x_i,y_i,z_i,q_i), z_i=0, q_i>0$ which projects onto the base as
\begin{align}
\label{eq:divProject}
   \pi:(x_1, x_2 , x_3 \ldots x_n) \rightarrow   \hat{Y}: = \prod_{x_i} x_i^{ q_i  } \, .
\end{align}
Then $\hat{Y}$ is a divisor in the base, which is reducible in the full threefold. This  implies that there exists a group of divisors $D_{x_i}$ restricted onto  $\hat{Y}=0$ and hence the same codimension one locus in the base. However, they are also fibral curves $C_i$ which must intersect like an affine Dynkin diagram of some lie algebra $\mathfrak{g}$ according to the Kodaira classification. The multiplicity of such fibral curves is encoded in $\hat{Y}$, which are exactly given by $q_i$. Hence those values are identified with the Kac label $d_i$ of the affine Dynkin diagram \cite{Bouchard:2003bu}. Clearly, the same argument applies to a more generic vector with $v_i = (x_i,y_i,z_i,q_i)$ with $d_i = \text{gcd}({z_i, q_i})$ which restricts onto the \textit{primitive} base ray $b= (z_i, q_i)/d_i$. Note that the affine node always identifies such base ray as its Kac label $d_{aff}=1$.

Consequently, all other fibral base rays are a multiple $q_i$ of the primitive ray $v_{y_0}=(0,1)$.\footnote{Note that this holds only for vertices outside of faces of the top. Divisors within faces can also intersect a CY but only at codimension two, resulting in non-flat fibers \cite {Buchmuller:2017wpe,Dierigl:2018nlv,Apruzzi:2018nre,Apruzzi:2019vpe,Apruzzi:2019opn}. See Section~\ref{sec:non-flat} for explicit examples of this type.} In summary, we are able to deduce the rank and the Kac labels of some lie algebra from the data of the toric rays alone \footnote{In addition, we require the toric rays to be favorable generators of the K\"ahler cone. This is in general true but in Section~\ref{sssec:E8E8E8} we also discuss some examples where this fails to hold.}. From the toric structure, we can read off the number of fibral curves in the base as well as their Kac labels in a feasible way. This is great information, as it enables us to almost uniquely identify the fibre structure without the necessity to perform an intersection computation. E.g. a fibral curve of multiplicity six can only be contained in an $\hat{E}_8$ fibre as this is the only ADE algebra with this Kac label. Similarly for the elliptic fibration with a multiplicity four, we know it must be of $E_7$ type.
 
The geometry of fibral curves naturally connects to the geometry of T-dual bases, where the flips onto base curves lead to some other fibration (see \cite{Bhardwaj:2015oru} or \cite{Haghighat:2018dwe} for some recent work.). This ties the $\mathfrak{g}$ type singularity of certain gauging to the $\mathfrak{g}^{(1)}$ type of base topologies that we seek throughout this work. It is natural to expect the connection between the Kac labels of the affine $\mathfrak{g}^{(1)}$ fibres and the dual LS charges. This expectation however is almost right but obstructed by two observations: 
\begin{enumerate}
\item A base curve $C_i$ of some dual top may not be primitive, but rather a $d_i$ multiple. The corresponding dual LS charge $l_{LS,I}$ therefore corresponds to the Kac label $a_I$ of the $\mathfrak{g}^{(1})$ fibre which must be divided by the multiplicity $d_I$ leading to $l_{LS,I}=a_I/d_I$. On the other hand, the multiplicity $d_I$ is identified with the Kac label $a_I$ of some other fibral curve $w^I$, which for $a_I>1$ implies a D or E type of gauge algbera $\mathfrak{g}_I$  itself with Kac label $d_i>1$, such as a $D$ or $E$ type fibre.
\item In some cases it may happen, that several fibral curves of some $\mathfrak{g}$-type of singularity are mapped on top of each other upon the new base direction. This implies first, that this base curve must admit a non-trivial gauge algebra, but also that the topology of the dual base quiver is not $\mathfrak{g}^{(1)}$ type but reduced, often times by an automorphism of the affine quiver\footnote{For $\fe_6$ gaugings generically find an $\ff_4^{(1)}$ type of quiver in the dual (see Section~\ref{sssec:E6}).}. 
\end{enumerate}
The above conclusions hold for the main part of this paper, but we also find a couple of exotic cases where the picture  deviates slightly: E.g. when probing   a very small number of NS5 branes with an $\mathfrak{g}$-type   singularity (typically less than three), we find the dual base to miss some divisors to fully complete to an $\mathfrak{g}^{(1)}$ topology. Secondly, we also find some exotic cases, where the base topology resembles the affine folded topology of an $\mathfrak{g}$-singularity. I.e. in Section~\ref{sec: MultiTDual} we find cases of an $\fe_7$ singularity becomes an $\fe_6^{(2)}$ shaped base. Note again, that the base intersection form is still symmetric but the LS charges correspond to the Kac labels of an (twisted-)affine algebra. 
 
\subsubsection*{Warmup: $E_8^2 /(Spin(32)/\mathbb{Z}_2)$ Rank 0 T-dual LSTs} 
Now we are going to enhance the LST base by a $\mathbb{P}^2_{1,2,3}$ fiber polytope. The toric  semi-convex polytope and configuration matrix are summarized as follows
\begin{align}
\begin{array}{|cc|cc|}\hline
 {x_i} &$ Coordinate$ & \mathbb{C}^*_1 & \mathbb{C}^*_2 \\ \hline
X& (1,0,0,0) &   2 &0 \\ 
Y& (0,1,0,0) & 3 & 0 \\
Z& (-2,-3,0,0) & 1& -2  \\ \hline 
x_0& (-2,-3,1,0) &0&1  \\
x_1&(-2,-3,-1,0) &0&1  \\ \hline
y_0&(-2,-3,0,1)& 0 & 0 \\ \hline
p & - & 6 & 0   \\ \hline
\end{array}
\end{align}
where we have depicted two $\mathbb{C}^*$ actions. One refers to the $\mathbb{P}^2_{1,2,3}$ fiber weights and the other to those of the $\mathbb{P}^1$ base part. The weights of the anti-cannonical hypersurface $p=0$ of bidegree $(6,0)$ are given below, which yields the elliptic threefolds
\begin{align}
Y^2 = X^3 + a_1 XYZ + a_ 2  X^2 Z^2+ a_3 Y Z^3 + a_4 X Z^4 + a_6 Z^6 \, , 
\end{align}
In general, $a_i$ is a degree (2i) polynomial in the $x_0,x_1$ written e.g. as
\begin{align}
a_1 =  b_{1,1} x_0^2 + b_{1,2} x_1^2 + b_{1,3} x_0 x_1 \, ,
\end{align}
with $b_{i,j}$ being generic complex constants. As $y_0$ does not permit any $\mathbb{C}^*$ scaling, we could add monomials of arbitrary degree to the above equation.

Note that the above geometry admits not only an elliptic fibration over $\mathbb{P}^1 \times \mathbb{C} $ but also the structure of an elliptic K3 fibered over $\mathbb{C}$. This is directly evident from the facts explained before and noting that the rays of the coordinates $X,Y,Z,x_0,x_1$ span a reflexive 3D polytope fibered over $B_1 = \mathbb{C}$. Including this non-compact direction only leads to a reflexive half-polytope and an infinite prism for its dual.  
 
\begin{figure}[t!]
\begin{center}
\begin{picture}(0,170)
\put(-80,0){\includegraphics[scale=0.5]{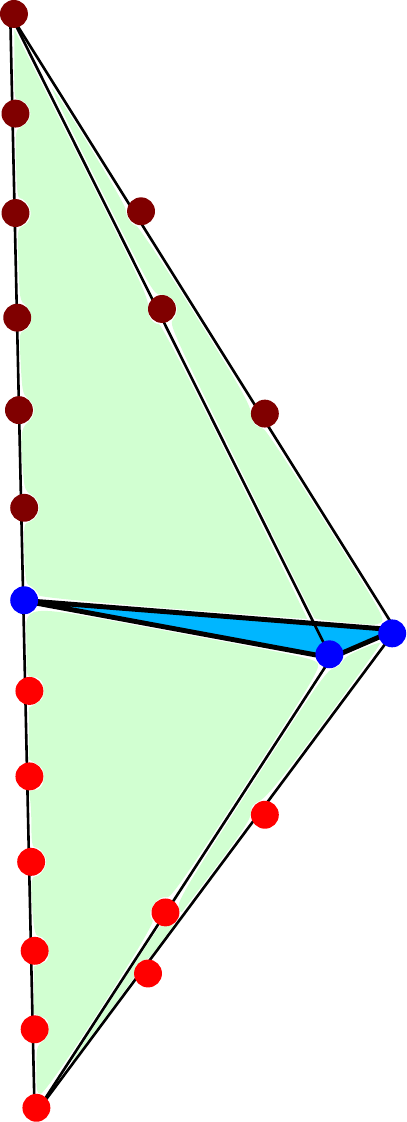}} 
\put(80,0){\includegraphics[scale=0.5]{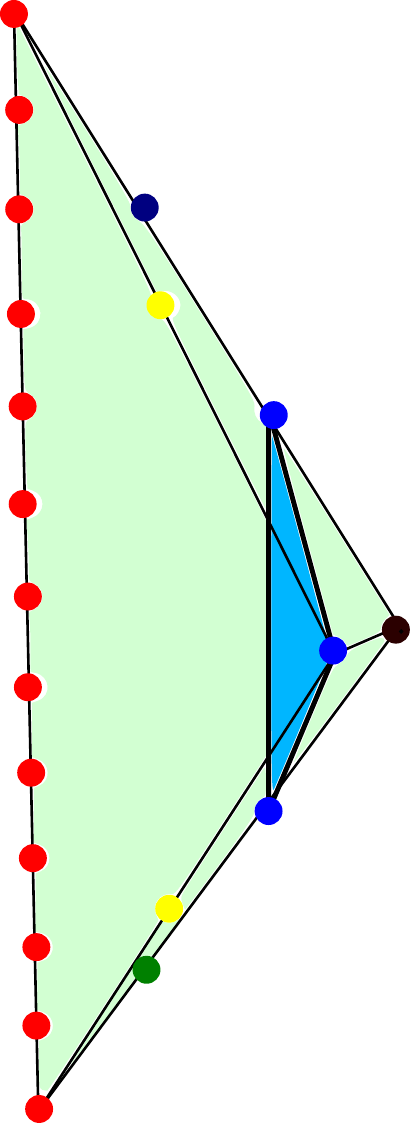}} 
\end{picture}
\caption{\label{fig:K3Dual}Two depictions of the same 3D reflexive polytope with a K3 hypersurface. The left graph displays a 2D reflexive sub-polytope of Tate-type and identifies the residual points as two affine $\fe_8$ fibers. On the right, we illustrate a 2D reflexive sub-polytope of $F_{13}$ type with an $\fso_{32}$ top in red, green and yellow points.}
\end{center}
\end{figure}

We proceed by dressing up the K3 fiber polytope. One classic example \cite{Candelas:1997pq} unifies the 8D SUGRA vacua of the heterotic $E_8\times E_8$ and that of  $Spin(32)/\mathbb{Z}_2$ string within a single M/F-theory description. This duality and its geometric description serve as the starting point to construct T-dual little string theories in lower dimensions. The rays of the reflexive 3D sub-polytope are depicted in Figure~\ref{fig:K3Dual} and the associated triangulation produces a toric variety whose toric hypersurface encodes a K3 with two explicit elliptic fibration structures.  

\begin{table}[t!]
\begin{center}
$\Delta_{K3}=$ 
\begin{tabular}{|c|c| }\hline 
\multicolumn{2}{|c|}{Generic Fiber}\\ \hline
Z & $(-2,-3,0 )$      \\
X & $(1,0,0 ) $  \\
Y & $(0,1,0 )$   \\ \hline
\end{tabular} ,
  \begin{tabular}{|c|c|}\hline
\multicolumn{2}{|c|}{E$_8 ^2$ tops} \\ \hline
$\alpha_{6,\pm}$  &$ (-2,-3,\pm 6  ) $ \\
$\alpha_{5,\pm}$ &$ (-2,-3,\pm 5  ) $  \\
$\alpha_{4,\pm}$ & $(-2,-3,\pm 4  ) $  \\
$\alpha_{3,\pm}$ & $(-2,-3,\pm 3   )$ \\
$\alpha_{2,\pm}$ &$ (-2,-3,\pm 2  )$ \\
$\alpha_{1,\pm}$ & $(-2,-3,\pm 1  )$ \\
$\alpha_{\hat{4},\pm}$ &$ (-1,-2,\pm 4  )$  \\
$\alpha_{\hat{2},\pm}$ & $(0,-1,\pm 2  )$  \\
$\alpha_{\hat{3},\pm}$ &$ (-1,-1,\pm 3  )$ \\ \hline 
\end{tabular} 
\caption{ \label{tab:E8E8k3}The toric rays, for the K3 polytope depicted in Figure~\ref{fig:K3Dual}. }
\end{center}
\end{table}

\textbf{The $\mathbf{E_8^2}$ fibration} is obtained by the projection onto the $z$ coordinates of the rays with coordinate $v=(x,y,z)$. The reflexive fiber polytope that lives over the generic base point $"(0)"$ is the $\mathbb{P}^2_{2,3,1}$ ambient space of the Tate/Weierstrass model $X,Y,Z$. The base $\mathbb{P}^1$ is given via the vertices $z=\pm 1$. Nine vertices project onto that vertex with the $z$ coordinate yields the $\fe_8$ Kac labels $d_i$. Note that also the intersections in Figure~\ref{fig:K3Dual} can be read off by noting that each ray has self intersection $-2$ and intersects an adjacent ray precisely once.

\textbf{The $\mathbf{Spin(32)/\mathbb{Z}_2}$ fibration} is obtained by the projection onto the $x$ coordinate of the $(x,y,z)$ rays.  The reflexive sub-polytope is not of $\mathbb{P}^{2}_{2,3,1}$ type anymore but $F_{13}$ according to the enumeration used in \cite{Klevers:2014bqa}. The 2D sub-polytope is spanned  by the vertices $Y$ and $\alpha_{\hat{2},\pm}$. The $\mathbb{P}^1$ base is given in a similar way as before. However over the base coordinate $"(+1)"$ given by  $X=0$ we find just a trivial top. On the other pole, we discover $17$ vertices sitting over the $"(-1)"$ side, with 12 of them having multiplicity referring to the Kac label two. This immediately indicates that this fiber must be of type  $I^*_{16}$ as the only option compatible with this combination of Kac labels. This claim can be verified by directly inspecting the intersection structure in the Figure~\ref{fig:K3Dual} on the right.
 
In actuality, the structure is slightly more intricate as the generic fiber admits two sections, given by $\alpha_{\hat{2},\pm}=0$ that generates a finite Mordell-Weil group of order two \cite{Aspinwall:1998xj,Mayrhofer:2014opa}. Again, their intersection structure is feasibly read off from Figure~\ref{fig:K3Dual}, with one section intersecting the affine node and the other one touching the spinor/co-spinor root. This suffices to write down the torsion Shioda map, which implements the restriction of the weight lattice  \cite{Aspinwall:1998xj,Mayrhofer:2014opa}.\footnote{Also see \cite{GarciaEtxebarria:2019caf,Hajouji:2019vxs,Apruzzi:2020zot,Cvetic:2021sxm,Hubner:2022kxr} for a recent discussion in terms of restricted monodromies.} To be fully explicit, we discuss the hypersurface equation in details that is given as
{\footnotesize \begin{align}  
\label{eq:E8E8Tate}
p=& \alpha_{\hat{3},-} \alpha_{\hat{3},+} Y^2 +
\alpha_{\hat{2},-}^2 \alpha_{\hat{2},+}^2 \alpha_{\hat{4},-} \alpha_{\hat{4},+} X^3  \nonumber \\  
&+\alpha_{1,-} \alpha_{1,+} \alpha_{2,-} \alpha_{2,+} \alpha_{3,-} \alpha_{3,+} \alpha_{4,-} \alpha_{4,+} \
\alpha_{5,-} \alpha_{5,+} \alpha_{6,-} \alpha_{6,+} \alpha_{\hat{2},-} \alpha_{\hat{2},+} \alpha_{\hat{3},-} \alpha_{\hat{3},+}  
\alpha_{\hat{4},-} \alpha_{\hat{4},+} X Y Z \hat{a}_1 \nonumber \\   
&+\alpha_{1,-}^2 \alpha_{1,+}^2 \alpha_{2,-}^2 \alpha_{2,+}^2 \alpha_{3,-}^2 \alpha_{3,+}^2  
\alpha_{4,-}^2 \alpha_{4,+}^2 \alpha_{5,-}^2 \alpha_{5,+}^2 \alpha_{6,-}^2 \alpha_{6,+}^2  
\alpha_{\hat{2},-}^2 \alpha_{\hat{2},+}^2 \alpha_{\hat{3},-} \alpha_{\hat{3},+} \alpha_{\hat{4},-}^2 \alpha_{\hat{4},+}^2 X^2 Z^2\hat{a}_2 \nonumber \\  
&+   \alpha_{1,-}^3 \alpha_{1,+}^3 \alpha_{2,-}^3 \alpha_{2,+}^3 \alpha_{3,-}^3 \alpha_{3,+}^3  
\alpha_{4,-}^3 \alpha_{4,+}^3 \alpha_{5,-}^3 \alpha_{5,+}^3 \alpha_{6,-}^3 \alpha_{6,+}^3 \alpha_{\hat{2},-} \
\alpha_{\hat{2},+} \alpha_{\hat{3},-}^2 \alpha_{\hat{3},+}^2 \alpha_{\hat{4},-}^2 \alpha_{\hat{4},+}^2 Y Z^3  \hat{a}_3 \nonumber \\ 
& +  \alpha_{1,-}^4 \alpha_{1,+}^4 \alpha_{2,-}^4 \alpha_{2,+}^4 \alpha_{3,-}^4 \alpha_{3,+}^4 \
\alpha_{4,-}^4 \alpha_{4,+}^4 \alpha_{5,-}^4 \alpha_{5,+}^4 \alpha_{6,-}^4 \alpha_{6,+}^4 \
\alpha_{\hat{2},-}^2 \alpha_{\hat{2},+}^2 \alpha_{\hat{3},-}^2 \alpha_{\hat{3},+}^2 \alpha_{\hat{4},-}^3 \alpha_{\hat{4},+}^3 X  
Z^4 \hat{a}_4 \nonumber \\
&+ \alpha_{1,-}^5 \alpha_{1,+}^5 \alpha_{2,-}^4 \alpha_{2,+}^4 \alpha_{3,-}^3 \alpha_{3,+}^3 \
\alpha_{4,-}^2 \alpha_{4,+}^2 \alpha_{5,-} \alpha_{5,+} Z^6 \hat{a}_6 \, .
\end{align}}
We have factorized the above equation exactly in such a way, that the Tate-type fibration is evident. Here the $\hat{a}_i$s are some generic constants. This however is not true for $\hat{a}_6$ which is a polynomial of degree two in the \textit{appropriate} base coordinates $\hat{Y}_+,\hat{Y}_-$ as  
\begin{align}
\hat{a}_6 = \hat{Y}_+^2 b_1 + \hat{Y}_-^2 b_2 +\hat{Y}_+ \hat{Y}_- b_3 \, ,
\end{align}
with $b_i$ some generic complex constants. $\hat{Y}_{\pm}$ are defined in Eqn.~\eqref{eq:divProject} that given as
\begin{align}
\label{eq:divisor}
\hat{Y_\pm}= (\alpha_{1,\pm}) (\alpha_{2,\pm}^2) (\alpha_{3,\pm}^3) (\alpha_{4,\pm}^4) (\alpha_{5,\pm}^5) \
(\alpha_{6,\pm}^6) (\alpha_{\hat{2},\pm}^2) (\alpha_{\hat{3},\pm}^3) (\alpha_{\hat{4},\pm}^4)  \, .
\end{align}
The  $\hat{Y_\pm}$ are the base divisors projecting onto the two poles of the base $\mathbb{P}^1$ and are themselves $\mathbb{P}^1$ fibrations. They pullback to the nine fibral curves of the affine $\fe_8$ with powers of the Dynkin multiplicities/Kac labels as given in Figure~\ref{fig:K3Dual}.

One can double check consistency by going to the singular model. To achieve this, we shrink down all nodes except fibral components intersected by the zero-section $Z=0$, which is $\alpha_{1,\pm}$. The vanishing order of the respective Tate coefficients is then given as
\begin{align}
\text{ord}_{\text{van}}[a_1,a_2,a_3,a_4,a_6] = [1,2,3,4,5] \, ,
\end{align}  
and hence compatible with the Tate-classification of a type $II^*$ singularity, i.e. an $\fe_8$.  

The same applies to the $Spin(32)/\mathbb{Z}_2$ model. For that we rewrite Eqn~\eqref{eq:E8E8Tate} in a different factorized form. First, we rename the generic fiber coordinates $Y$ and $\alpha_{\hat{2},\pm}$ in a more suggestive way as
\begin{align}
Y \rightarrow z,  \qquad \alpha_{\hat{2},+} \rightarrow x\qquad \alpha_{\hat{2},-} \rightarrow y\, .
\end{align}
We choose $x=0$ as the zero-section and shrink down all but the fibral components that intersect it. These are further renamed as the $\mathbb{P}^1$ coordinates $w_i$, namely $\alpha_{\hat{4},+}\rightarrow w_0$  and $X\rightarrow w_1$. For simplicity, we blow down all other fibral curves by setting them to $1$.  Making all these substitutions and inserting $\hat{a}_6$ yields the hypersurface equation as
\begin{align}
\label{eq:hyperSO32}
p=& z^2+  x^2 y^2 w_0 w_1^3  +  x y z w_0 w_1 a_1 + x^2 y^2 w_0^2 w_1^2 a_2 + xyz w_0^2 a_3 +x^2 y^2 w_0^3 w_1 a_4 \nonumber \\ &+  (x^4 w_0^8 b_1 + y^4 b_2 +x^2 y^2 w_0^4 b_3) \,.
\end{align}
As expected we find the fiber to be $F_{13}$ type, according to \cite{Klevers:2014bqa}, with the general form
\begin{align}
\label{eq:F13fibre}
p=x^4 s_1 + x^2 y^2 s_2 + y^4 s_3 + x y z s_6 + z^2 s_9 \, .
\end{align}
Here the $s_i$ are analogous to the Tate-coefficients, that admit the following map into Weierstrass form
\begin{align}
\label{eq:f13toTate}
\begin{split}
f=&\frac{1}{48} (-s_6^4 + 8 s_2 s_6^2 s_9 - 16 (s_2^2 - 3 s_1 s_3) s_9^2) \, , \\ 
g=&\frac{1}{864} (s_6^2 - 4 s_2 s_9) (s_6^4 - 8 s_2 s_6^2 s_9 + 
   8 (2 s_2^2 - 9 s_1 s_3) s_9^2), 
\end{split}
\end{align}
The above equation is going to become relevant when we consider other types of models with additional fibers. Comparing coefficients with the generic form in Eqn.~\eqref{eq:hyperSO32}, we obtain the following replacements
\begin{align}
 s_1 \rightarrow w_0^8 b_1\, ,\qquad  s_2\rightarrow w_0 d_3\, ,\qquad  s_3 \rightarrow b_2\, ,\qquad  s_6 \rightarrow w_0 d_1\, ,\qquad s_9 \rightarrow 1\,.
\end{align}
The codimension one singularity over $w_0=0$ yields the vanishing orders $\text{van}_{\text{ord}}(f,g,\Delta ) = (2,3,18)$ which is exactly the desired $I_{16}^*$ type of fiber\footnote{In addition we find that the Weierstrass model is simply a specialization of the most general order two MW group model, given in \cite{Aspinwall:1998xj}.}.

With the structure of the K3 fibration and its T-duality being manifest, we simply need to complete the K3  fiber over $\mathbb{C} $ by embedding the K3 reflexive polytope $\Delta_{K3}$ into a toric fourfold ambient space. We essentially combine the generic LST base of the previous section with the constructed $E_8^2/(Spin(32)/\mathbb{Z}_2$) polytope. The resulting polytope is then given as
\begin{align}
\Delta = \left( \begin{array}{c} (\Delta_{K3}, 0,0) \\ 
 (-2,-3,0,1)   \end{array}\right) \, .
\end{align} 
It is important to note that neither the $E_8^2$ nor the $Spin(32)/\mathbb{Z}_2$ are gauge symmetries anymore but only flavor symmetries as they live over non-compact divisors. 

In conclusion, this simple setup yields a pair of T-dual LST theories 
\begin{align}
[\fe_8 ] - (0)-[\fe_8] \quad \leftarrow \text{ T-dual}  \rightarrow \quad  [\fso_{32}] - (0) \, .
\end{align}
At next we generalize the setup further by decorating the base with more  divisors, such that this action respects the two fibrations.

\subsubsection*{Toric Realization of higher rank theories}
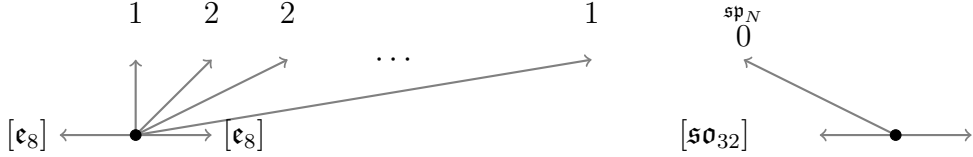
\begin{figure}[ht!]
	  \centering
	  \begin{tikzpicture}
		  \draw[gray, thick, ->] (0,0) -> (0,1);
		  \draw[gray, thick,<->] (-1,0) -- (1,0);
		  \draw[gray, thick,->] (0,0) -> (1,1);
		  \draw[gray, thick,->] (0,0) -> (2,1);
		  \draw[black] (3,1) circle (0pt) node[anchor=west]{$\cdots$};
		  \draw[gray, thick,->] (0,0) -> (6,1);

		  \filldraw[black] (0,0) circle (2pt) node[anchor=west]{};
		  \draw[black] (-1,0) circle (0pt) node[anchor=east]{$[\fe_8]$};
		  \draw[black] (1,0) circle (0pt) node[anchor=west]{$[\fe_8]$};
		  \draw[black] (0,1.9) circle (0pt) node[anchor=north]{${1}$};
	  \draw[black] (1,1.9) circle (0pt) node[anchor=north]{${2}$};
   
		  \draw[black] (2,1.9) circle (0pt) node[anchor=north]{${2}$};
	  \draw[black] (6,1.9) circle (0pt) node[anchor=north]{${1}$};

	\draw[gray, thick,<->] (9,0) -- (11,0);
		\draw[gray, thick,->] (10,0) -> (8,1);

		\filldraw[black] (10,0) circle (2pt) node[anchor=west]{};
		\draw[black] (7,0) circle (0pt) node[anchor=west]{$[\fso_{32}]$};
		\draw[black] (8,1.9) circle (0pt) node[anchor=north]{$\overset{\mathfrak{sp}_N}{0}$};

  \end{tikzpicture}	
  \caption{\label{fig:E8E8SP0}Toric diagram of $N$ instantons $E_8$ on the left, its T-dual  $Spin(32)/\mathbb{Z}_2$ theory with $\fsp_N$ gauge group is given on the right.}
  \end{figure}

Now we decorate the rank $0$ LST configuration with additional rays, which on the $E_8\times E_8$ fibration yield $N$ small $E_8$ instantons. This is achieved by including $N$ additional rays between $E_8$ factors with $v_{y_i}=(-2,-3,i,1), i=1 \ldots N$. For clarity we have omitted the two $E_8$ factors, which should be added according to the Table~\ref{tab:E8E8k3in4fold}.  
\begin{table}[t!]
\begin{center}
$ 
\begin{array}{|cc|c|cccccc|}\hline
 x_i  &$ vertex$ & \mathbb{C}^*_l &\mathbb{C}^*_0 & \mathbb{C}^*_1 & \mathbb{C}^*_2&  \ldots &\mathbb{C}^*_{N-1}  & \hat{\mathbb{C}^*_{N}}  \\ \hline 
Z& (-2,-3,0,0) &-2 &  -1& 0 &  0&\ldots & 0 &-1                \\ \hline 
\alpha_1^+& (-2,-3,1,0)&1 &0&0 &0& \ldots & 0 &1 \\
\alpha_1^-&(-2,-3,-1,0)&1 &1&0 &0&\ldots & 0 & 0 \\ \hline
y_0&(-2,-3,0,1)& 0&-1 & 1 &0&\ldots&0&0 \\  
y_1&(-2,-3,1,1)& 0&1 & -2&1 & \ldots & 0 &0\\  
y_2&(-2,-3,2,1)&0& 0 & 1&-2 & \ldots & 0 &0 \\ 
\vdots & \vdots& 0 &\vdots  & \vdots  &\vdots  &\vdots  &\vdots  & \vdots  \\
y_{N-1}&(-2,-3,N-1,1) &0&    0 & 0 & 0 & \ldots & -2 & 1  \\ 
y_N &(-2,-3,N,1) & 0&0 & 0 &0&\ldots & 1& -1 \\ \hline  
\end{array} $
\end{center}
\caption{\label{tab:E8E8k3in4fold} The toric rays of the ambient fourfold, including the zero-section $Z$ and their $\mathbb{C}^*$ relations. One of the $\mathbb{C}^*$ relations is linear dependent. }
\end{table}
Here the $\mathbb{C}^*_i$ scalings are written in a redundant way to read off intersections easier, with $N$ independent $\mathbb{C}^*$ scalings between the $N+1$ rays $y_i$. The little string curve $\Sigma_0 = l_{LS,I} w^I$ in the $E_8$ picture determines the following LS charge 
\begin{align}
\vec{l}_{LS}= (  \underbrace{1,1,\ldots,1,\ldots,1,1}_{\times N+1}   ) \, .
\end{align}
The T-dual configuration is again obtained by projecting onto the $(x,q)$ plane, where all toric rays map down to the same ray with $(-2,1)$ coordinates. This is the same $0$ curve described in the warmup example before but now we include $\mathfrak{sp}_N$ non-trivial gauge algebra as depicted in Figure~\ref{fig:E8E8SP0}. 

Note that we find no gauge algebra but only $N+1$ tensors on the $E_8$ side, with E-strings contributed as "matter content" in the 6D anomaly. On the $Spin(32)/\mathbb{Z}_2$ description there is only one single $0$ LST tensor with the $\mathfrak{sp}_N$ gauge algebra on top of it. It is known that both configurations have an additional unbroken $SU(2)_L$ flavor symmetry, which is not realized as any toric rays in our setup. Nevertheless, the T-dual picture allows us to track the locus of that $SU(2)_L$ divisor. 

We start on the $Spin(32)/\mathbb{Z}_2$ side first, such that every irrelevant elements are blown down in order to only keep the compact curves resolved. Then the hypersurface becomes
\begin{align}
p=d_2 w_0 x^2 y^2 + d_3 y^4 \left(\prod_{i=1}^{N} y^i_{N-i}\right) + 
 d_1 w_0^8 x^4 \left(\prod_{i=1}^N y_i^i \right)  + d_6 w_0 x y z + z^2
\end{align}
whereas $d_1, d_2,d_3$ and $d_6$ are polynomials in the $\mathbb{P}^1$ coordinates $w_0, w_1$ as well as the $\mathfrak{sp}_N$  zero curve $\hat{Y}=0$ given by
\begin{align}
\hat{Y}= \prod^N_{i=0} y_i \, .
\end{align}
where $\hat{Y}=0$ determines exactly the little string curve in the T-dual $E_8$ picture. We denote $y_N$ as the affine node of the $\mathfrak{sp}_N$, which intersects $x=0$, and another fibral divisor $y_0=0$ that intersects the torsion section $y=0$. Note that $d_i$ are additional polynomials in the $w_i$ and $\hat{Y}$. As there is no $\mathbb{C}^*$ scaling acting on the $\hat{Y}$, the polynomial is of infinite order. In practice, we truncate it to simply a constant such that $[d_1]=[d_3]=1$, $[d_2]=3$ and $[d_6]=1$. 6D anomaly cancellation requires the $\fsp_N$ to have $16$ Fundamentals and one antisymmetric representation hypermultiplet. The former is supported at intersection points along the $Spin(32)/\mathbb{Z}_2$ flavor group, which can be read off from Figure~\ref{fig:K3Dual} \footnote{The bi-fundamentals are only $\frac12$ hypermultiplets.}. However, the antisymmetric is not directly evident as it is non-localized, due to the fact that the anti-symmetrics of $\fsp_N$ originate from a $\mathbb{Z}_2$ monodromy effect that folds the $\fsu_{2N}$ to $\fsp_N$ in the fibre.
The respective group theory decomposition is traced from the adjoint of $\fsu_{2N}$, which splits into an adjoint and the antisymmetric of $\fsp_N$. The folding originate from the non-trivial intersection of the $\mathbb{Z}_2$ ramification divisor $D_L$ and the curve that carries the $\fsu_{2N}$, which interchanges the two classes of nodes. 
  
This locus is actually positioned at codimension two component of the discriminant where the resolved fiber becomes further reducible. The finding task becomes more convenient after shrinking all fibral curves but the affine node $f_N$ by setting them to $1$ and mapping into the Weierstrass form via Eqn~\eqref{eq:f13toTate}, which yields 
\begin{align}
\begin{split}
f=&-\frac{1}{48} w_0^2 (16 d_2^2 - 8 d_2 d_6^2 w_0 + d_6^4 w_0^2 - 48 w_0^6 y_N^N) \, ,\\
g=&-\frac{1}{864} w_0^3 (4 d_2 - d_6^2 w_0) (16 d_2^2 - 8 d_2 d_6^2 w_0 + d_6^4 w_0^2 - 
   72 w_0^6 y_N^N) \, ,\\
\Delta=&\frac{1}{16} w_0^{18} y_N^{
 2 N} (-(-4 d_2 + d_6^2 w_0)^2 + 64 w_0^6 y_N^N) \, .
\end{split}
\end{align}
Indeed we find the expected $I_{16}^*$ fiber over $w_0=0$ as well as the $\mathfrak{sp}_N$ over $y_N=0$. The bifundamental matter arises at their collision and the ramification locus is given by
 \begin{align}
D_L= -4 d_2 + d_6^2 w_0=0  \, .
\end{align}
Naively this locus looks like a further enhancement as the discriminant seems to have an enhanced vanishing order but it is actually the aforementioned monodromy divisor.

By construction, $D_L$ intersects the $\hat{Y}=0$ divisor and thus its fibral components $y_i=0$ for $i=1\ldots N-1$. 
It is noteworthy that the fibre degenerates into two copies of $\mathbb{P}^1$ over such points. E.g. in the fully resolve fibration we find
\begin{align}
p_{D_L=f_i=0} = \frac14 (d_6 w_0 x y + 2 z)^2 \qquad \text{ for } i = 1 \ldots N-1.
\end{align}
The same type of factorization does not occur for the $y_0=0$ and $y_N=0$ fibral curves i.e. the affine and opposite $\fsp_N$ fibral curves.

This demonstrates that the fibral divisors $y_i=0$ with $i=1\ldots N-1$ precisely map to the (negative) self-intersection $2$ curves in the T-dual of the $E_8^2$ fibration, while $y_0=0$ and $y_N=0$ are the two $1$ curves. As argued before, the single divisor $D_L=0$ intersects all the $y_i=0$ divisors,  leading to a subtle degeneration of the respective curves observed in the T-dual picture. 

This suggests that $D_L$ plays the role of the $SU(2)_L$ flavor divisor in the $E_8$ picture and explains why there is only a single $SU(2)_L$ flavor group, as opposed to a copy to each of the  $(N-1) \times $ $2$ curves.  

We end with various field theory checks for T-dual pairs. I.e. the 5D Coulomb branch must be exactly $N$-dimensional with a rank 16+1 flavor group. In addition, the matched 2-group structure constants are given by
   \begin{equation}
   \begin{split}
   \widehat{\kappa}_{\mathscr R}=h_g^\vee~=N+1~, \qquad \hat \kappa _{\mathscr P}=2~.
   \end{split}
   \end{equation} 

\subsubsection*{T-duality breakdown by SCFT limit}
In the following, we intend to use the same techniques to construct many more non-trivial T-dual little string theories. One might then wonder if the same T-duality procedure is possible for QFTs that flow to SCFTs instead of LSTs. This is of course a contradiction, as SCFTs have a unique stress energy tensor and no intrinsic scale along which we could perform T-dualization. However it is instructive to reach the same conclusion from the toric geometry point of view.
 
 One can obtain an SCFT from an LST simply by decompactifying any compact curve in the LST configuration\cite{Bhardwaj:2015oru}. For instance, let us decompactify the whole configuration to a rank $n$ E-string theory in the $E_8$ picture. We do so by 
 removing all base divisors right next to the curve $y_n$. From the perspective of toric geometry, we simply remove all rays $(x,y,z,q)$ with $z>n$. This procedure removes $N-n$ vertices in addition to  one $\fe_8$ flavor factor and changes the base as to the $n$-th blow-up of $\mathbb{C}^2$ which is an SCFT configuration. Now we note however, that the T-dual fibration got removed by the this decompactification: To complete the second fibration we require the vertices $\alpha_{\hat{2},\pm }: (0,-1,\pm2, 0)$ to complete the $F_{13}$ polytope.
 Those vertices however, were part of the second $\fe_8$ flavor top that we just removed. 
 Hence when degenerating the LST to an SCFT configuration, the disappearance of one $\fe_8$ flavor top forbids us to preserve the second fibration structure.
A similar argument holds on the $\fso_{32}$ side: For instance, decompactifying the base $\mathbb{P}^1$ into $\mathbb{C}^2$ by removing the vertex $w_1$: $(1,0,0,0)$. This ray corresponds to the $\mathbb{P}^2_{1,2,3}$ coordinate $X$,  that is necessary for the Tate-model description in the $\fe_8^2$ dual model. As anticipated by the field theory, we also find that T-duality here to be incompatible with an SCFT.

\subsection{A-Type Singularities}
Equipped with the toric machinery used throughout this work, we will explore various ADE singularities and their T-duals. For the $A$ and $D$ type cases, we first stick to trivial flavor holonomies but also discuss some special choices in Section~\ref{sec:DiscreteHolonomy}. Some of those cases for the generic number of heterotic instantons $M$ have been discussed in \cite{Aspinwall:1997ye,Font:2016odl}, which we will repeat here plus special cases for $M=1$ that correspond to the fractionalization of the pure heterotic string (see \cite{delZottoAppear2}) and their T-duals.

Below, we begin with the configuration of $E_8\times E_8$ heterotic string probing the $\mathbb C^2 / \mathbb Z_3$ singularity. The tensor branch geometry is given via the quiver
\begin{align}
    \lbrack \fe_8 \rbrack \, \,  1\, \,  2 \, \, \overset{\fsu_2 }{2}    \, \,    \overset{\fsu_3 }{\underset{[N_F=2]}{2}}    \, \,    \overset{\fsu_2}{2}\, \,  2\, \,  1\, \,  \lbrack \fe_8 \rbrack \, ,
\end{align}
with the toric data given in Table~\ref{topE8}.
\begin{table}[ht!]\label{topE8}
	\begin{center}
	\begin{tabular}{|c|c|c| }
	\hline
$x_i$ & vertices   & $\fe_8$-frame group  \\
	\hline
$y_0 $&	$(-2,-3,0,1) $& $\emptyset$    \\
	\hline
$y_1$	& $(-2,-3,1,1)$ & $\emptyset$     \\
	\hline
$y_{2,i}$& 	$(-2,-3,2,1), (-1,-2,2,1)$ & $SU(2)$ \\
	 \hline
$y_{3,i}$&  $(-2,-3,3,1) , (-1,-1,3,1),(-1,-2,3,1)$ & $SU(3)$   \\
	 \hline
	$y_{4,i}$& $ (-2,-3,4,1) , (-1,-2,4,1)$ & $SU(2)$ \\
	 \hline
	$y_{5 }$& $ (-2,-3,5,1)$ & $\emptyset$     \\
	 \hline
	$y_{6 }$&  $(-2,-3,6,1)$ & $\emptyset$   \\
	 \hline
	\end{tabular}
	\caption{Toric data for the $SU(3)\times SU(2)\times SU(2)$ gauge group in the $E_8\times E_8$ picture. } 
 \end{center}
\end{table}
The T-dual description is obtained by projecting the rays onto the $(x,q)$ plane. The tensor branch consists of six ${Spin}(32)/\mathbb{Z}_2$ instantons probing an $A_3$ singularity expressed by the quiver below  
\begin{equation} 
	\lbrack \fso_{32} \rbrack  \, \, \overset{\mathfrak{sp}_6}{1}  \, \,  \overset{\mathfrak{su}_4}{1}   \, \, \lbrack N_A=1\rbrack 
\end{equation} 

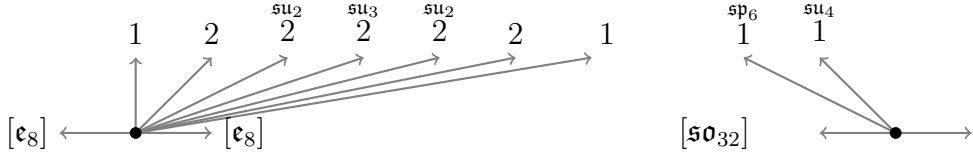
\begin{figure}[ht!]\label{fig:E8E8SU3}
\begin{center}
	\begin{tikzpicture}
		\draw[gray, thick,->] (0,0) -> (0,1);
		\draw[gray, thick,<->] (-1,0) -- (1,0);
		\draw[gray, thick,->] (0,0) -> (1,1);
		\draw[gray, thick,->] (0,0) -> (2,1 );
		\draw[gray, thick,->] (0,0) -> (3,1);
		\draw[gray, thick,->] (0,0) -> (4,1);
		\draw[gray, thick,->] (0,0) -> (5,1);
		\draw[gray, thick,->] (0,0) -> (6,1);

		\filldraw[black] (0,0) circle (2pt) node[anchor=west]{};
		\draw[black] (-1,0) circle (0pt) node[anchor=east]{$[\fe_8]$};
		\draw[black] (1,0) circle (0pt) node[anchor=west]{$[\fe_8]$};
		\draw[black] (1,1.6) circle (0pt) node[anchor=north]{$2$  };
		\draw[black] (2,1.9) circle (0pt) node[anchor=north]{$\overset{\mathfrak{su}_2}{2}$  };
		\draw[black] (0,1.6) circle (0pt) node[anchor=north]{$1$  };

		\draw[black] (3,1.9) circle (0pt) node[anchor=north]{$\overset{\mathfrak{su}_3}{2}$  };
		\draw[black] (4,1.9) circle (0pt) node[anchor=north]{$\overset{\mathfrak{su}_2}{2}$  };

		\draw[black] (5,1.6) circle (0pt) node[anchor=north]{$2$  };

		\draw[black] (6.2,1.6) circle (0pt) node[anchor=north]{$1$  };

	     \draw[gray, thick,<->] (9,0) -- (11,0);
		\draw[gray, thick,->] (10,0) -> (8,1);
		\draw[gray, thick,->] (10,0) -> (9,1);

		\filldraw[black] (10,0) circle (2pt) node[anchor=west]{};
		\draw[black] (7,0) circle (0pt) node[anchor=west]{$[\fso_{32}]$};

            \draw[black] (8,1.9) circle (0pt) node[anchor=north]{$\overset{\mathfrak{sp}_6}{1}$  };
		\draw[black] (9,1.9) circle (0pt) node[anchor=north]{$\overset{\mathfrak{su}_4}{1}$  }; 
\end{tikzpicture}	
\caption{Toric diagram that corresponds to the $E_8\times E_8$ flavor symmetry with the $SU(3)\times SU(2)\times SU(2)$ gauge group on the left and its T-dual on the right.}
\end{center}
\end{figure}
Having constructed two T-dual theories we can compute and confirm matched 2-group structure constants as shown below:
\begin{equation}\label{eq:2groupE8}
	\begin{split}
 \widehat{\kappa}_{\mathscr R}=11~, \qquad\;\;\;\;  \hat \kappa _{\mathscr P}=2~.
  \end{split}
\end{equation}
The above configuration is easily generalized to arbitrary $\fsu_N$ distinguished by two different types, e.g. $\fsu_{2N}$ and $\fsu_{2N+1}$. Note that the relevant superconformal matter is
\begin{align}
\mathcal{T}(\fe_8, \fsu_N):\qquad    [\fe_8]\, \, 1 \, \, 2 \, \, \overset{\fsu_2}{2}\, \, \overset{\fsu_3}{2} \ldots  \overset{\fsu_{N-1}}{2}\, \, [\fsu_N] \, .
\end{align}
Upon inserting the respective superconformal matter we will always end up with $\fsu_N$ on a $2$ curve. Hence, we can never have an $\fsp_N$ gauging for those type of curves as it is not consistent with anomalies (e.g. see \cite{Johnson:2016qar}). 
The tensor branch on the $\fe_8$ side then becomes
\begin{align}
     \lbrack \fe_8 \rbrack \overset{\fsu_{N}}{0}   \lbrack \fe_8 \rbrack \qquad   \rightarrow       \lbrack \fe_8 \rbrack \, \,  1\, \,  2 \, \, \overset{\fsu_2 }{2} \ldots  \overset{\fsu_{N-1}}{2}    \, \,    \overset{\fsu_N }{\underset{[N_F=2]}{2}}    \, \,            \overset{\fsu_{N-1}}{2} \, \,      \ldots   \overset{\fsu_2}{2}\, \,  2\, \,  1\, \,  \lbrack \fe_8 \rbrack \, .
\end{align}
We highlight that the $\fsu_N$ gauging requires two more flavors in order to be consistent with anomalies. The LS charges 
are universally one in this type of case. 

The T-dual theory for the $\fsu_{2N}$ guaging is given as
\begin{align}
\lbrack \fso_{32} \rbrack \, \,   \overset{\fsp_{4N }}{1  }  \, \,   \  \underbrace{\overset{\fsu_{8(N-1) }}{2   } \ldots   \overset{\fsu_{8(N-1- k) }}{2   } \ldots    \overset{\fsu_{8 }}{2   }}_{(N-1) \times } \, \,   \ 1 \, .
\end{align}
similarly for the $\fsu_{2N+1}$ gauging the chain is given as
\begin{align}
\lbrack \fso_{32} \rbrack \, \,   \overset{\fsp_{4N+2 }}{1  }  \, \,   \  \underbrace{\overset{\fsu_{8N-4 }}{2   } \ldots   \overset{\fsu_{8(N- k)-4 }}{2   } \ldots    \overset{\fsu_{12 }}{2   }}_{(N-1) \times } \, \,   \ \overset{\fsu_{4 }}{\underset{[N_A=1]}{1}   }  \, .
\end{align}
In all theories above, we find that the dual quiver in the $\fso$ side forms  the shape of an affine $\fsp_N^{(1)}$ type when starting with some $\fsu_{2N}$ gauging. 

The above theories have special properties as additional localized flavors group in the $\fe_8^2$ LST as well as the different structure carried by the final curve in its T-dual. However, they often go away when considering gauging $M$ curves given by the quiver \footnote{When setting $N=2$, this quiver coincides with the $\mathcal K_N(1+1,1+1; \mathfrak{su}_2)$ model presented in Table 1 of  \cite{DelZotto:2022ohj} after applying the notation $M=N+1$. Meanwhile, the $N=3$ setup coincides with the $\mathcal K_N(1+1+1,1+1+1;\mathfrak{su}_3)$ model presented in Eqn. (4.41) of the same literature when setting the notation $M=N+1$.}
\begin{align}
 \lbrack \fe_8 \rbrack \underbrace{ \overset{\fsu_{N}}{1}  \overset{\fsu_{N}}{2} \ldots   \overset{\fsu_{N}}{2}     \overset{\fsu_{N}}{1 }}_{\times M}   \lbrack \fe_8 \rbrack \, ,
\end{align}
which admits a similar tensor branch as before with extra $\fsu_N$ plateau of length $M$. Thus for $\fsu_{2N}^M$ gaugings we have the associated T-dual description\footnote{When setting $N=1$, this quiver coincides with the $\tilde{\mathcal K}_{\tilde{N}}(0,16; \mathfrak{su}_2)$ model presented in Eqn. (4.20) of \cite{DelZotto:2022ohj} after replacing the notation $M=N+1$.}
\begin{align} 
	\lbrack \fso_{32} \rbrack \, \,   \overset{\mathfrak{sp}_{4N+M-1}}{1} \, \,   \underbrace{\overset{\mathfrak{su}_{8N-10 +2M}}{2}\ldots   \overset{\mathfrak{su}_{8(N-k)+2M-2}}{2} \ldots  \, \,    \overset{\mathfrak{su}_{6+ 2M}}{2}  \, \, \overset{\mathfrak{sp}_{M-1}}{1}}_{\times N}\, ,
	\end{align}
and for $\fsu_{2N+1}$ we have
\begin{align} 
\lbrack \fso_{32} \rbrack \, \,   \overset{\mathfrak{sp}_{4N+M+1}}{1} \, \,   \underbrace{\overset{\mathfrak{su}_{8N+2M-6}}{2}\ldots   \overset{\mathfrak{su}_{8(N-k)+2M+2}}{2} \ldots  \, \,    \overset{\mathfrak{su}_{2M+10}}{2} \, \, \overset{\mathfrak{su}_{2M+2}}{1}}_{\times N}\,.
\end{align}
The CB dimension and the $\widehat{\kappa}_{\mathscr R}$ can be matched and given below as
\begin{align}
	\begin{array}{|c|c|c|c|c|}\hline
	& [\fe_8]-\fsu_{2N}^M-[\fe_8] \, & \,  \text{T-dual} & [\fe_8]-\fsu_{2N+1}^M-[\fe_8] \, & \, \text{T-dual}  \\
	\hline
	\text{Dim(CB)} & \multicolumn{2}{|c|}{4N^2+2NM-2N+1} & \multicolumn{2}{|c|}{4N^2+2NM+2N+M+1} \\
	\hline
\widehat{\kappa}_{\mathscr R}  & \multicolumn{2}{|c|}{4N^2+2NM-2N+2} & \multicolumn{2}{|c|}{4N^2+2NM+2N+M+2} 
 \\ \hline
	\end{array} 
\end{align}

\subsection{E-Type Singularities}\label{sec:non-flat}
With the above strategy, we are now equipped to decorate the $E_n $ flavor symmetries and LSTs with exceptional gauge groups. I.e. we want to set up LST chains of the form
\begin{align}
[\fe_m]\,\,\underbrace{\overset{\mathfrak{e}_n}{1}\,\, \overset{\mathfrak{e}_n}{2} \ldots \overset{\mathfrak{e}_n}{2} \,\,\overset{\mathfrak{e}_n}{1}}_{\times M}\,\,[ \fe_o ]\, ,
\end{align}
for $m,n,o=8,7,6$ that refers to the $\fe$ side and compute their T-duals.
The results are summarized in Table~\ref{table:T-duals}.  Note that the above geometry requires the insertion of sufficient superconformal matter factors, which can be found in \cite{DelZotto:2014hpa} between the various $\fe_n$ factors, they require additional blow-ups in base and fibre directions. 

All of the geometries are constructed to explicitly admit a second fibration, similar to the simple examples in the preceding sections. The geometry proves the T-duality via the match of computational results as the CB dimension, the flavor rank and the 2-group structure constants. 
We illustrate our approach in several examples and discuss their geometric properties. This involves in particular the appearance of non-toric flavor group factors on the $Spin(32)/\mathbb{Z}_2$ side, as well as the non-flatness of both fibrations. I.e. we will find that at codimension two locus, the fibre dimension jumps and becomes a surface.  Such effects are well understood by now \cite{Buchmuller:2017wpe,Dierigl:2018nlv,Apruzzi:2018nre,Apruzzi:2019vpe,Apruzzi:2019opn} and they generally highlight the presence of superconformal matter. Moreover, it is expected that the flat resolution over a birational base is part of the extended K\"ahler cone of the threefold, with the same Hodge numbers. Hence we can treat the non-flat fibers as base and fibral divisors of the flat resolution, which allows us to extract sufficient information for the tensor branches of the LST that we are constructing in the following.

\subsubsection{$ [\fe_8]- \mathfrak{e}_8^M-[\fe_8]$ LSTs}
\label{sssec:E8E8E8}
We start by appending the $\fe_8^2$ type of configuration with rank $M$ heterotic instanton and then gauging this chain of curves with an $\fe_8$ symmetry on each of the tensors
\begin{align}
[\fe_8 ]\,\,\underbrace{ \overset{\mathfrak{e}_8}{1} \, \,\overset{\mathfrak{e}_8}{2} \ldots \overset{\mathfrak{e}_8}{2}\, \,\overset{\mathfrak{e}_8}{1}}_{\times M}\,\,[\fe_8] \, .
\end{align}
The mutual $\fe_8$ intersections do not lead to a smooth elliptic fibration or a perturbative LST tensor branch, but rather the insertion of $\mathcal{T}(\fe_8,\fe_8)$ conformal matter at each intersection, which has been systematically determined in \cite{DelZotto:2014hpa}, we repeat them here modulo the hypermultiplet content as
\begin{align}
\label{eq:e8e8SCM}
	[\fe_8]\, \, 1 \, \, 2 \, \, \overset{\fsp_1}{2}\, \, \overset{\fg_2}{3}\, \, 1\, \,   \overset{\ff_{4}}{5}\,\, 1 \, \,  \overset{\fg_2}{3}  \, \,  \overset{\fsp_1}{2}\, \,  2\,\, 1\,\, [\fe_8] \, . 
\end{align}
By doing so, we enhance the $1$ and $2$ negative self-intersections to $11$ and $12$ respectively. In the special scenario when $M=1$, the $0$ curve is enhanced to a $10$ curve. However an $\fe_8$ requires a $12$ curve from the anomaly to be fully consistent.  Blowing up E-string as additional $(4,6,12)$ points over the $10$ and $11$ curves leads to a 12 curves as desired in \cite{Morrison:2012np}, which results in the chain
\begin{align}
\label{eq:e8chain}
\begin{array}{ccccclc}
 & 1 & &&&1 & \\
\lbrack \fe_8\rbrack   &    \overset{\mathfrak{e}_8}{2} &\overset{\mathfrak{e}_8}{2} &\ldots &\overset{\mathfrak{e}_8}{2}& \overset{\mathfrak{e}_8}{2}&  \lbrack \fe_8 \rbrack \, .
\end{array}
\end{align}  
Now we employ toric geometry to engineer the above setup in a fully resolved manner. Note that only linear chains can be resolved, thus we are obstructed by the additional $1$ curves at the beginning and the end $\fe_8$'s. However, the fibration structure of the resolved threefold can still be utilized but in a non-flat way. To show how this works we consider a very concrete example for $M=1$ first. The relevant toric rays are given in Table~\ref{tab:E8E8E8}.
\begin{table}[ht!]
\begin{center}  
\begin{tabular}{|c|c| }\hline 
\multicolumn{2}{|c|}{Tate-Fiber}\\ \hline
Z & (-2,-3,0,0 )      \\
X & (1,0,0,0 )   \\
Y & (0,1,0,0 )   \\ \hline 
\multicolumn{2}{|c|}{$\fe_8^2$ Flavor} \\ \hline
$\alpha_{6,\pm}$  &$ (-2,-3,\pm 6 ,0 ) $ \\
$\alpha_{5,\pm}$ &$ (-2,-3,\pm 5 ,0) $  \\
$\alpha_{4,\pm}$ & $(-2,-3,\pm 4 ,0 ) $  \\
$\alpha_{3,\pm}$ & $(-2,-3,\pm 3 ,0  )$ \\
$\alpha_{2,\pm}$ &$ (-2,-3,\pm 2 ,0 )$ \\
$\alpha_{1,\pm}$ & $(-2,-3,\pm 1,0  )$ \\
$\alpha_{\hat{4},\pm}$ &$ (-1,-2,\pm 4,0  )$  \\
$\alpha_{\hat{2},\pm}$ & $(0,-1,\pm 2 ,0 )$  \\
$\alpha_{\hat{3},\pm}$ &$ (-1,-1,\pm 3 ,0 )$ \\ \hline 
\end{tabular} 
\begin{tabular}{|c|c| }\hline 
\multicolumn{2}{|c|}{$\fe_8$ gauge factor } \\ \hline
$f_{6}$  &$ (-2,-3,0 ,6 ) $ \\
$f_{5 }$ &$ (-2,-3,0,5) $  \\
$f_{4 }$ & $(-2,-3,0 ,4 ) $  \\
$f_{3 }$ & $(-2,-3,0 ,3  )$ \\
$f_{2 }$ &$ (-2,-3,0 ,2 )$ \\
$f_{1 }$ & $(-2,-3,0,1  )$ \\
$f_{\hat{4} }$ &$ (-1,-2,0,4  )$  \\
$f_{\hat{2} }$ & $(0,-1,0 ,2 )$  \\
$f_{\hat{3} }$ &$ (-1,-1,0 ,3 )$ \\ \hline
$f_{nf}$ &$ (0,0,0,1 )$ \\ \hline 
\end{tabular}

\begin{tabular}{|c|l| }\hline 
\multicolumn{2}{|c|}{$\mathcal{T}(\fe_8,  \fe_8)$ Conformal Matter } \\ \hline
$s_{\pm 1}$  &$ (-2,-3, \pm 1 ,5 )  $ \\
$s_{\pm2 }$ &$ (-2,-3,\pm1,4) $  \\
$s_{\pm3,i }$ & $(-2,-3,\pm1 ,3 ),(-1,-2,\pm1,3) $  \\
$s_{\pm4, j}$ & $(-2,-3,\pm1 ,2  ),(-1,-1,\pm1,2) ,(-2,-3,\pm2,4) $ \\
$s_{\pm5 }$ &$ (-2,-3,\pm2 ,3 )$ \\
$s_{\pm6,k }$ & $(-2,-3,\pm1,1  ),(0,-1,\pm1,1),(-1,-2,\pm2,2),(-2,-3,\pm2,2),(-2,-3,\pm3,3)$ \\
$s_{\pm7 }$ &$ (-2,-3,\pm3,2  ) $  \\
$s_{\pm 8,j }$ & $(-2,-3,\pm2 ,1 ),(-1,-1,\pm2,1) ,(-2,-3,\pm4,2)$  \\
$s_{\pm 9,i }$ &$ (-2,-3,\pm3 ,1 ),(-1,-2,\pm3,1)$ \\ 
$s_{\pm10 }$ &$ (-2,-3,\pm4,1 )$ \\  
$s_{\pm11 }$ &$ (-2,-3,\pm5,1 )$ \\ \hline  
\end{tabular}   
\caption{\label{tab:E8E8E8} Toric vertices that resolve the $\fe_8 -\fe_8 -\fe_8 $ little string configuration.
The $f_i$ are the $\fe_8$ fibral divisors and the $s_{\pm j,k}$ resolve the superconformal matter. 
 }
\end{center} 
 \end{table}

This configuration meets all necessary requirements for the fibration to exist and it is also evident from the rays, that they lead to $11$ base rays on the left and right of gauge $\fe_8$ on the $10$ curve given by the affine node $f_1$. The curve configuration from the  projection to the base as well as the gauge algebra factors are pictured in Figure~\ref{fig:E8E8Toric} on the left.
\begin{figure}[ht!]
	\centering
	\begin{tikzpicture}
\draw[gray, thick,<->] (-1,0) -- (1,0);
		\draw[gray, thick,->] (0,0) -> (1,5);
		\draw[gray, thick,<->] (0,0) -- (1,4);
		\draw[gray, thick,->] (0,0) -> (1,3);
		\draw[gray, thick,->] (0,0) -> (1,2);
             \draw[gray, thick,->] (0,0) -> (2,3);
		\draw[gray, thick,->] (0,0) -> (1,1);
		\draw[gray, thick,->] (0,0) -> (2,1);
		\draw[gray, thick,->] (0,0) -> (3,2);
		\draw[gray, thick,->] (0,0) -> (3,1);
       	\draw[gray, thick,->] (0,0) -> (4,1); 
\draw[gray, thick,->] (0,0) -> (5,1); 
\draw[gray, thick,->] (0,0) -> (0,1); 
		\draw[gray, thick,->] (0,0) -> (-1,5);
		\draw[gray, thick,<->] (0,0) -- (-1,4);
		\draw[gray, thick,->] (0,0) -> (-1,3);
		\draw[gray, thick,->] (0,0) -> (-1,2);
             \draw[gray, thick,->] (0,0) -> (-2,3);
		\draw[gray, thick,->] (0,0) -> (-1,1);
		\draw[gray, thick,->] (0,0) -> (-2,1);
		\draw[gray, thick,->] (0,0) -> (-3,2);
		\draw[gray, thick,->] (0,0) -> (-3,1);
       	\draw[gray, thick,->] (0,0) -> (-4,1); 
\draw[gray, thick,->] (0,0) -> (-5,1);

	 	\draw[black] (-1.5,0) circle (0pt) node[anchor=east]{$|\fe_8|$};
		 \draw[black] (1.5,0) circle (0pt) node[anchor=west]{$|\fe_8|$};
		 \draw[black] (-5.2,1.5) circle (0pt) node[anchor=north]{$1$ };

 \draw[black] (-4.2,1.5) circle (0pt) node[anchor=north]{$2$ }; 
\draw[black] (-3.2,1.7) circle (0pt) node[anchor=north]{$\overset{\mathfrak{sp}_1}{2}$ };
\draw[black] (-2.2,1.6) circle (0pt) node[anchor=north]{$\overset{\mathfrak{g}_2}{3}$ };

\draw[black] (-3.2,2.5) circle (0pt) node[anchor=north]{$1$ };
\draw[black] (-1.1,1.7) circle (0pt) node[anchor=north]{$\overset{\mathfrak{f}_4}{5}$ };
\draw[black] (-2.2,3.5) circle (0pt) node[anchor=north]{$1$ };

\draw[black] (-1,2.8) circle (0pt) node[anchor=north]{$\overset{\mathfrak{g}_2}{3}$ }; 
\draw[black] (-1.2,3.7) circle (0pt) node[anchor=north]{$\overset{\mathfrak{sp}_1}{2}$ };
\draw[black] (-1.2,4.5) circle (0pt) node[anchor=north]{$2$ };
\draw[black] (-1.2,5.5) circle (0pt) node[anchor=north]{$1$ };
\draw[black] (0,2 ) circle (0pt) node[anchor=north]{$\overset{\mathfrak{e}_8}{10}$ };

		  \draw[black] ( 5,1.5) circle (0pt) node[anchor=north]{$1$ };

 \draw[black] (4.2,1.5) circle (0pt) node[anchor=north]{$2$ }; 
\draw[black] (3.2,1.7) circle (0pt) node[anchor=north]{$\overset{\mathfrak{sp}_1}{2}$ };
\draw[black] (2.2,1.6) circle (0pt) node[anchor=north]{$\overset{\mathfrak{g}_2}{3}$ };

\draw[black] (3.2,2.5) circle (0pt) node[anchor=north]{$1$ };
\draw[black] (1.15,1.7) circle (0pt) node[anchor=north]{$\overset{\mathfrak{f}_4}{5}$ };
\draw[black] (2.2,3.5) circle (0pt) node[anchor=north]{$1$ };

\draw[black] (1.1 ,2.7) circle (0pt) node[anchor=north]{$\overset{\mathfrak{g}_2}{3}$ }; 
\draw[black] (1.2,3.7) circle (0pt) node[anchor=north]{$\overset{\mathfrak{sp}_1}{2}$ };
\draw[black] (1.2,4.5) circle (0pt) node[anchor=north]{$2$ };
\draw[black] (1.2,5.5) circle (0pt) node[anchor=north]{$1$ };

\draw[gray, thick,<->] (8,0) -- (10,0);
		\draw[gray, thick,->] (9,0) -> (7,1);
\draw[gray, thick,->] (9,0) -> (7,1);
\draw[gray, thick,->] (9,0) -> (8,1);
\draw[gray, thick,->] (9,0) -> (8,2);
	\draw[gray, thick,->] (9,0) -> (7,3);
\draw[gray, thick,->] (9,0) -> (9,1);
\draw[gray, thick,->] (9,0) -> (7,5);
\draw[gray, thick,->] (9,0) -> (8,4);
\draw[gray, thick,->] (9,0) -> (8,3);

		\draw[black] (6.8,1.7) circle (0pt) node[anchor=north]{$\overset{\mathfrak{sp}_{10}}{1}$};

\draw[black] (7.8,1.7) circle (0pt) node[anchor=north]{$\overset{\mathfrak{so}_{24}}{4}$};
\draw[black] (6.8,3.7) circle (0pt) node[anchor=north]{$\overset{\mathfrak{sp}_{6}}{1}$};
\draw[black] (7.8,2.7) circle (0pt) node[anchor=north]{$\overset{\mathfrak{so}_{16}}{4}$};
\draw[black] (6.9,5.7) circle (0pt) node[anchor=north]{$\overset{\mathfrak{sp}_{2}}{1}$};
\draw[black] (7.9,3.7) circle (0pt) node[anchor=north]{$\overset{\mathfrak{so}_{7}}{3}$};
\draw[black] (8,4.5) circle (0pt) node[anchor=north]{$1$};
\draw[black] (9,1.7) circle (0pt) node[anchor=north]{$\overset{\mathfrak{so}_{8}}{4}$};
\draw[black] (7.5,0.2) circle (0pt) node[anchor=north]{$|\fso_{32}|$};

\end{tikzpicture}	
\caption{\label{fig:E8E8Toric}Toric rays for two T-dual LST configurations
with self-intersection and gauge algebra decorations
: On the left is a $E_8\times E_8$ flavor symmetry with $\fe_8$ gauge group on a $10$ curve, with two non-flat fibers.
On the right we show the T-dual toric base of the $Spin(32)/\mathbb{Z}_2$ side.}
\end{figure}

In order to check the non-flat fiber resolution, we need to study the hypersurface. For simplicity, we blow down all but the affine $\fe_8$ fibral divisor $f_1=0$ and the non flat fiber surface $f_{nf}=0$. The local Tate-model around the $10$ curve is then written as
\begin{align}
p=f_{nf} X^3 +f_{nf} Y^2 &+ a_1 X Y Z f_1  f_{nf}  + a_2 X^2 Z^2 f_1^2  f_{nf} \nonumber \\&+ a_3 Y Z^3 f_1^3 f_{nf}  + a_4 X Z^4 f_1^4 f_{nf}  +a_6  Z^6 f_1^5.
\end{align} 
$f_{nf}=0$ restricts down to the same base locus as $f_1=0$, however it misses the CY hypersurface in codimension one. It restricts onto the hypersurface when the $a_6=0$ locus intersects the $10$ curve at codimension two. We can check by blowing down $f_{nf}$ and find that over $f_1 = a_6=0$ there is a $(1,2,3,4,6)$ point in terms of Tate-coefficient. Note that $a_6$ is a polynomial of degree two in local coordinates $s_{1,\pm}$, which can easily be seen by noting that this is locally a $\mathbb{P}^1 \times \mathbb{C}^1$ base, with $f_1$ and $f_{nf}$ living over $\mathbb{C}^1$ and $a_6$ must therefore be of degree 2 in the $\mathbb{P}^1$ coordinates. Thus there are two intersection points leading to two solutions of the non-flat fiber \footnote{This single toric ambient divisor $f_{nf}=0$ contributes to \textit{two} Kahler deformations on the CY hypersurface, those configurations have been called \textit{ non-K\"ahler favorable} \cite{Anderson:2017aux}.}. 

To fully remove this locus we need to blow-up the $10$ curve at the non-toric intersection loci $f_1 = a_6 =0$. Keeping in mind that this base blow-up exists we can still infer the flat resolution though. The actual base intersection of the flat fibration is therefore given by 
\begin{align} \arraycolsep=2.5pt\def\arraystretch{1}
\begin{array}{cccccccccccccccccccccccccc}
&&&&&&&&&&&&1&&&&&&&&&&&&& \\
\lbrack \fe_8\rbrack  & 1 &2 &\overset{\mathfrak{sp}_{1}}{2} &\overset{\mathfrak{g}_{2}}{3}& 1& \overset{\mathfrak{f}_{4}}{5}& 1 &\overset{\mathfrak{g}_{2}}{3}& \overset{\mathfrak{sp}_{1}}{2}&  2 & 1&  \overset{\mathfrak{e}_{8}}{12} &1&2&\overset{\mathfrak{sp}_{1}}{2}&\overset{\mathfrak{g}_{2}}{3}& 1& \overset{\mathfrak{f}_{4}}{5}& 1 &\overset{\mathfrak{g}_{2}}{3}& \overset{\mathfrak{sp}_{1}}{2}&   2 & 1&\lbrack \fe_8\rbrack  \\
&&&&&&&&&&&&1&&&&&&&&&&&&& 
\end{array}
\end{align}
We proceed by computing the little string charges of the above configuration given as
\begin{align}
	\vec{l}_{LS} = (1, 1, 1, 1, 2, 1, 3, 2, 3, 4, 5, \begin{array}{c}1 \\ 1, \\ 1 \\  \end{array}  5, 4, 3, 2, 3, 1, 2, 1, 1, 1, 1)\, \, .
\end{align}

The T-dual configuration is obtained by projecting toric rays in Table~\ref{tab:E8E8E8} onto the $(x,q)$ surface as depicted in Figure~\ref{fig:E8E8Toric} on the right. This leads to an $F_{13}$ type fiber with a $\mathbb{Z}_2$ MW group as a non-trivial global flavor group of the LST. There are two tricks to deduce the correct gauge algebra factors. We can either blow down all but the affine node per base ray and determine the singularity type in the Weierstrass model, or we use the fact that each collection of rays projecting onto the same base ray is a \textit{top} \cite{Candelas:1996su,Bouchard:2003bu} and derive the local fibral intersection structure under the assumption that a desired triangulation exists. In practice, the number of toric rays and their multiplicities are usually sufficient to determine the ADE type. The Weierstrass model can then be applied to double check whether a configuration is (non-)split inferred from the gauge anomalies and the base curve structures. For instance, consider the following four rays 
\begin{align}
s_{+,6,2}:(0,-1,-1,1)\, , \, s_{-,6,2}: (0,-1,1,1)\, ,\,  f_{nf}: (0,0,0,1) \, ,\,  f_{\hat{2}}:(0,-1,0,2) \,,
\end{align}
that project onto the $y_0=(0,1)$ base ray, which is a $4$ curve. The fibral divisor $f_{\hat{2}}$ is $2 y_0$ with Dynkin multiplicity two, while the rest have Dynkin multiplicity one, which naively suggests an $\fso_7$ gauge algebra. However, this is in contradiction with the gauge anomaly, as this is a $4$ curve in the base. Note that $f_{nf}=0$ contains two divisors of  Dynkin multiplicity one as it was non-K\"ahler favorable and intersects the CY hypersurface twice, thus the actual gauge algebra is $\fso_8$ and consistent with the anomaly. This fact can be double checked in an explicit triangulation by computing the intersection form of the resolved structure and the singular Weierstrsass model.

To summarize, we obtain the following T-dual chain and LS charge
\begin{align}
\lbrack \fso_{32} \rbrack \,\, \overset{\mathfrak{sp}_{10}}{1}\,\,\overset{\mathfrak{so}_{24}}{4}\,\,\overset{\mathfrak{sp}_{6}}{1}\,\,\overset{\mathfrak{so}_{16}}{4}\,\, \overset{\mathfrak{sp}_{2}}{1}\,\,\overset{\mathfrak{so}_{7}}{3}\,\, 1\,\, \overset{\mathfrak{so}_{8}}{4}\,,  \qquad 
\vec{l}_{LS} = ( 1, 1, 3, 2, 5, 3, 4, 1)\, .
\end{align}
The T-duality of the above two LSTs can be verified by the matched data below 
\begin{align}
	\begin{array}{|c|c|c|}\hline
	&\lbrack  \fe_8 \rbrack -\fe_8- \lbrack \fe_8 \rbrack & \text{T-dual} \\
	\hline
 \text{Dim(CB)} & \multicolumn{2}{|c|}{ 52 } 
 \\ \hline
	 \widehat{\kappa}_{\mathscr R}&\multicolumn{2}{|c|}{ 122  }  \\
	\hline
	\end{array} 
\end{align} 
  
\subsubsection*{Generalization to arbitrary M}
Having discussed the case of a single $0$ curve gauged by an $\fe_8$ and how to resolve it, we can generalize the above construction by additional M$\times \fe_8$ ray collections\footnote{Note that for low $M$ we can double check the proposed ray collection, by compactifying the toric variety by adding the  $(-2,-3,0,-1)$ ray and confirm reflexivity of the respective polytope.}
given as\footnote{This coincides with $[\fe_8]-\fe_8^N-[\fe_8]$ quiver presented in Table 7 of \cite{DelZotto:2022ohj} when replacing the notation $M\rightarrow N+1$.}
\begin{align}
\label{eq:tab:e8i}
 \begin{tabular}{|c|c| }\hline 
\multicolumn{2}{|c|}{$\fe_8^m$ gauge factor } \\ \hline
$f_{6,n}$  &$ (-2,-3, 6m ,6 ) $ \\
$f_{5,n }$ &$ (-2,-3,5m,5) $  \\
$f_{4,n }$ & $(-2,-3,4m ,4 ) $  \\
$f_{3,n }$ & $(-2,-3,3m ,3  )$ \\
$f_{2,n }$ &$ (-2,-3,2m ,2 )$ \\
$f_{1,n }$ & $(-2,-3,m,1  )$ \\
$f_{\hat{4},n }$ &$ (-1,-2,4m,4  )$  \\
$f_{\hat{2},n }$ & $(0,-1,2m ,2 )$  \\
$f_{\hat{3},n }$ &$ (-1,-1,3m ,3 )$ \\ \hline
$f_{nf,0}$ &$ (0,0,0,1 )$ \\ \hline 
$f_{nf,N}$ &$ (0,0,M,1 )$ \\ \hline 
\end{tabular} \qquad   \text{ for } m= 1 \ldots M \, ,
\end{align}
This yields a resolved chain of $M\times$ $\fe_8$'s as in \eqref{eq:e8chain}, with two non-flat fiber divisors $f_{nf,0}, f_{nf,M}$ that work in a similar way as for the $M=1$ case. 

Introducing all the mutually intersecting $\fe_8$ factors, also requires us to incorporate their superconformal matter to obtain a fully consistent tensor branch geometry. This enhances the self-intersection of every compact $\fe_8$ curve by 10 and obtain a chain as given in \eqref{eq:e8e8SCM}. From such a quiver we can then easily compute the LS charge as  
\begin{align}
\vec{l}_{LS} = (& 1, 1, 1, 1, 2, 1, 3, 2, 3, 4, 5,  \begin{array}{c}\, 1 \\ 1 \\ \\ \end{array} , 
6, 5, 4, 3, 5, 2, 5, 3, 4, 5, 6  ,1 \ldots \times M-4 \ldots  \nonumber \\ \ldots & 1, 6, 5, 4, 3, 5, 2, 5, 3, 4, 5, 6,
\begin{array}{c}1 \\ 1 \\ \\ \end{array}  , 5, 4, 3, 2, 3, 1, 2, 1, 1, 1, 1 ) \, .
\end{align} 
The respective rays that resolve the full threefold are given in Table~\ref{tab:ConfMatter}. Note that the conformal matter between the  $\fe_8$ flavor factors and the $\fe_8^1$ and $\fe_8^M$ gauge factors are slightly differently although they yield the same base and fibral blow-ups. This is clear, as their fibral divisors have different LS charges that are represented by the $q-th$ component of the 4d ray as discussed in Section~\ref{ssec:ToricPrelim}. This in fact allows us to easily deduce the full resolution.
\begin{table}[ht!]
{\scriptsize
\begin{center}
\begin{tabular}{|c|l| }\hline 
\multicolumn{2}{|c|}{ \normalsize{$\mathcal{T}(\fe_8  \fe_8^1)$ Conformal Matter }} \\ \hline
$s_{-1}$  &$ (-2,-3, - 5 ,1 )  $ \\
$s_{-2 }$ &$ (-2,-3,-4 ,1) $  \\
$s_{-3,i }$ & $(-2,-3,-3 ,1 ),(-1,-2,-3,1) $  \\
$s_{-4, j}$ & $(-2,-3,-2 ,1  ),(-1,-1,-2,1) ,(-2,-3,-4,2) $ \\
$s_{-5 }$ &$ (-2,-3,-3 ,2 )$ \\
$s_{-6,k }$ & $(-2,-3,-1,1  ),(0,-1,-1,1),(-1,-2,-2,2),(-2,-3,-2,2),(-2,-3,-3,3)$ \\
$s_{-7 }$ &$ (-2,-3,-2,3  ) $  \\
$s_{- 8,j }$ & $(-2,-3,-1 ,2 ),(-1,-1,-1,2) ,(-2,-3,-2,4)$  \\
$s_{- 9,i }$ &$ (-2,-3,-1 ,3 ),(-1,-2,-1,3)$ \\ 
$s_{-10 }$ &$ (-2,-3,-1,4 )$ \\  
$s_{-11 }$ &$ (-2,-3,-1,5 )$ \\ \hline  
\end{tabular} \\
\begin{tabular}{|c|l| }\hline 
\multicolumn{2}{|c|}{\normalsize{$\mathcal{T}(\fe_8^M \fe_8)$ Conformal Matter} } \\ \hline
$s_{+1}$  &$ (-2,-3, 5M-4 ,5 )  $ \\
$s_{+2 }$ &$ (-2,-3,4M-3,4) $  \\
$s_{+3,i }$ & $(-2,-3,3M-2 ,3 ),(-1,-2,3M-2,3) $  \\
$s_{+4, j}$ & $(-2,-3, 2M-1 ,2  ),(-1,-1,2M-1,2) ,(-2,-3,4M-2,4) $ \\
$s_{+5 }$ &$ (-2,-3,3M-1 ,3 )$ \\
$s_{+6,k }$ & $(-2,-3,M ,1  ),(0,-1,M,1),(-1,-2,2M,2),(-2,-3,2M,2),(-2,-3,3M,3)$ \\
$s_{+7 }$ &$ (-2,-3,2M+1,2  ) $  \\
$s_{+8,j }$ & $(-2,-3,M+1 ,1 ),(-1,-1,M+1,1) ,(-2,-3,2M+2,2)$  \\
$s_{+9,i }$ &$ (-2,-3,M+2 ,1 ),(-1,-2,M+2,1)$ \\ 
$s_{+10 }$ &$ (-2,-3,M+3,1 )$ \\  
$s_{+11 }$ &$ (-2,-3,M+4,1 )$ \\ \hline  
\end{tabular}

\begin{tabular}{|c|l| }\hline 
\multicolumn{2}{|c|}{\normalsize{$\mathcal{T}(\fe_8^m \fe_8^{m+1})$ Conformal Matter for all $m=0\ldots M-1$ if $M>1$ }} \\ \hline
$s_{1,m}$  &$ (-2,-3, 1+6m ,6 )  $ \\
$s_{2,m }$ &$ (-2,-3,1+5m,5) $  \\
$s_{3,m,i }$ & $(-2,-3,1+4m ,4 ),(-1,-2,1+4m,4) $  \\
$s_{4,m, j}$ & $(-2,-3, 1+3m ,3  ),(-1,-1,1+3m,3) ,(-2,-3,2+6m,6) $ \\
$s_{5,m }$ &$ (-2,-3,2+5m ,5 )$ \\
$s_{6,m,k }$ & $(-2,-3,1+2m ,2  ),(0,-1,1+2m ,2),(-1,-2,2+4m ,4),(-2,-3,2+4m ,4),(-2,-3,3+6m ,6)$ \\
$s_{7,m }$ &$ (-2,-3,3+5m,5  ) $  \\
$s_{8,m,j }$ & $(-2,-3,2+3m ,3 ),(-1,-1,2+3m ,3) ,(-2,-3,4+6m ,6)$  \\
$s_{9,m,i }$ &$ (-2,-3,3+4m ,4 ),(-1,-2,3+4m ,4)$ \\ 
$s_{10,m }$ &$ (-2,-3,4+5m ,5)$ \\  
$s_{11,m}$ &$ (-2,-3,5+6m ,6 )$ \\ \hline  
\end{tabular} 
\end{center}
}
\caption{\label{tab:ConfMatter}Depiction of the toric rays that resolve the $M+1$ $\mathcal{T}(\fe_8,\fe_8)$ conformal matter between the $M\times $ $\fe_8$ gauge factors and the two $\fe_8$ flavor factors.}
\end{table} 

 \subsubsection*{The T-dual LST}
The T-dual theory is again obtained by the new base projection on the first and fourth component which yields the same base configuration as given in Figure~\ref{fig:E8E8Toric}. However now we have larger tops over each base ray and thus a higher rank gauge group. Unlike in the $E_8$ frame, the number of tensors in this T-dual frame is fixed to $8+1^*$. The additional $1^*$ tensor appears only for $M>2$ and is again realized through a non-flat fiber which must be properly resolved.

The general fiber structure over each base ray follows the same algorithm as discussed before. E.g. over the base ray $(-1,1)$ we have
\begin{align*}
\begin{array}{ll}
 s_{-3,2}: (-1,-2,-3,1) \, ,  &   s_{-4,2}:(-1,-1,-2,1)\, , \\  s_{+8,2}:(-1,-1,M+1,1)\, ,  &  s_{+9,2}:(-1,-2,M+2,1) \, .
\end{array}
\end{align*}
In addition there are the following rays that restrict with multiplicity two
\begin{align*}
\begin{array}{lll}
 s_{-4,3}: (-2,-3,-4,2) &   s_{-5}:(-2,-3,-3,2)  & s_{-6,4}:(-2,-3,-2,2)      \\    s_{-8,1 }: (-2,-3,-1,2)
&   s_{+4,1} : (-2,-3,2M-1,2)  & s_{+6,4}:(-2,-3,2M,2) \\  s_{+7}:(-2,-3,2M+1,2) &  s_{8,3 }: (-2,-3,2M+2,2)   \\
 f_{2,m} :(-2,-3,2m,2)&  s_{6,\hat{m},1}:(-2,-3,1+2\hat{m},2) 
\end{array}
\end{align*}
where $m$ ranges from 1 to $ M$ and $\hat{m}$ ranges between 0 and  $M-1$.  Hence we have four fibral curves of  Kac label one and $8+M+(M-1)=2M+7$ rays of  Kac label two. Only $\fso_{2k}$ type of (affine) Dynkin diagrams are of those type and from the multiplicity two roots we can easily deduce its rank as $\fso_{4M+20}$. 

We can proceed similarly for the other rays, but the base ray $(-1,3)$ needs a bit more care: From \eqref{eq:tab:e8i} and Table~\ref{tab:ConfMatter} we deduce the top rays again. Lets first consider those of multiplicity two, i.e. the ones with $(-2,-,-,6)$ which are concretely given by
\begin{align*}
&f_{6,m}\quad\text{ for }\quad m=1 \ldots M    \\ &  s_{1,m}\, ,  s_{4,m,3}\, ,  s_{6,m,5}\, ,  s_{8,m,3}\, ,  s_{11,m} \quad  \text{with } m=0 \ldots M-1    \quad\text{for }\quad M>1 \, ,
\end{align*}
which in total yield $6M-5$ rays. These rays are the middle roots of Dynkin multiplicity two in some $\fso$ type and thus count the rank to be $12M-4$. However, this base curve is just a $3$ curve, which is not consistent with the 6D gauge anomalies. Now let us count the rays of multiplicity one, for which we expect to find four (or three) just as before. Instead we find the following collection: 
\begin{align*}
&f_{\hat{3},m} \quad \text{ with }m=1 \ldots M \quad  \text{ and } \quad s_{+3,2}\, ,  s_{-9,3} \, , \\ & s_{4,m,2}\, ,  s_{8,m,2}  \quad \text{ with } m=0 \ldots M-1  \quad \text{for } M>1 \, .
\end{align*}
This totals $3M$ rays instead of the four  we would expect. It turns out though that again $3M-4$ of those rays are actually non-flat fibers. When $M>1$, these non-flat fibers can be removed by a single base blow-up and results in a $1$ curve with an  $\fsp_{3M-5}$ gauge algebra over it. This blowup enhances the base $3$ curve to a $4$ curve and thus becomes consistent with gauge anomalies. We sum up the final tensor branch configuration as\footnote{This coincides with the dual theory of the $[\fe_8]-\fe_8^N-[\fe_8]$ quiver presented in Table 7 of \cite{DelZotto:2022ohj} when replacing the notation $M\rightarrow N+1$.}
\begin{align}
\lbrack \fso_{32} \rbrack  \overset{\mathfrak{sp}_{M+9}}{1}\overset{\mathfrak{so}_{4M+20}}{4} \,\,
  \overset{\mathfrak{sp}_{3M+3}}{1} \,\,
 \overset{\mathfrak{so}_{8M+8}}{4}  \,\,
 \overset{\mathfrak{sp}_{5M-3}}{1}  \,\,
  \begin{array}{c}      
 \overset{\mathfrak{sp}_{3M-5}}{1^*}      \\     \overset{\mathfrak{so}_{12M-4}}{4^*}  \\ \\ \\ \end{array}  \,\,
 \overset{\mathfrak{sp}_{4M-4}}{1}    \,\,
 \overset{\mathfrak{so}_{4M+4}}{4}  
\, .
\end{align}
Here the $*$ indicates that the presence of the $1^*$ curve is only for $M>1$ and that the $4^*$ is a $3$ curve otherwise. Therefore, the LS charges are given as
\begin{align}
\label{eq:E8TdLST}
\vec{l}_{LS} = ( 1, 1, 3, 2, 5, \begin{array}{c}3^* \\ 3, \\ \\ \end{array} 4, 1) \, .
\end{align}
Also here the $3^*$ is absent, as we saw for $M=1$, but the other LS charges do not change. Finally lets comment on the global structure of the flavor group: Each flavor and gauge group factor contributes to the naive total centre as
\begin{align}
 Z(G_f ;\prod_i^9 G_i)  = (\mathbb{Z}_2^2;\mathbb{Z}_2,\mathbb{Z}_2^2,\mathbb{Z}_2,\mathbb{Z}_2^2,\mathbb{Z}_2,\mathbb{Z}_2,\mathbb{Z}_2^2 ) \, ,
\end{align}
Each individual factor is broken by bi-fundamental matter but a diagonal combination can survive. Indeed, due to the presence of a
 $\mathbb{Z}_2$ MW group such a diagonal group is present and in fact gauged \cite{Dierigl:2018nlv,Hubner:2022kxr}. 

The geometry proves the T-duality of the two theories in terms of the matched CB branch dimension and the 2-group structure coefficient:
\begin{align}
	\begin{array}{|c|c|c|}\hline
	& \lbrack \fe_8 \rbrack -\fe_8^M-\lbrack \fe_8 \rbrack  & \text{T-dual} \\
	\hline
 \text{Dim(CB)} & \multicolumn{2}{|c|}{ 22+30M } 
 \\ \hline
	 \widehat{\kappa}_{\mathscr R}&\multicolumn{2}{|c|}{ 120M+2  }  \\
	\hline
	\end{array} 
\end{align}    

\subsubsection*{T-dual base from the fiber}
A remarkable observation of our T-dual models is, that its base resembles the exact intersection picture of an affine $E_8$ Dynkin diagram. This follows directly from the geometry as discussed in the section before. We observe that the toric rays of the affine $E_8$ gauge group become the base rays, when projecting onto the T-dual base. The $E_8$ Kac labels $d_i$ become the new LS charges
\begin{align}
	l_{LS,i} = d_i/\hat{d_i}\, .
\end{align}
Note that $\hat{d}_i=GCD(z_i,w_i) $ of the ray $v_i=(x_i,w_i)$ which we have to divide by to obtain the primitve ray. Note that $\hat{d}_i$ on the other hand yields the Kac label of the respective fibral curve that sits over the ray. Hence if $\hat{d}_i>1$, we know that there must be a non-trivial fibre which is at least of $D$ or $E$ type. 

  We depict the fibral intersection of the later but factor out the multiplicity factor $\hat{d}_i$
\begin{align}
 \begin{array}{cccccccc}
&&&&&(f_{\hat{3}})  && \\  
(f_1) &  2 (f_2)   &(f_3)  &2 (f_4)  &(f_5)  &2 (f_6)  & (f_{\hat{4}})  &2 (f_{\hat{2}}) \, ,
\end{array}
\end{align}
after dividing by the multiplicity factor $\hat{d}_i$, we find the T-dual LS charges of \eqref{eq:E8TdLST}. As every second vertex admits a multiplicity of two, it is clear that we must have an ortho-symplectic quiver, already from this qualitative picture.

\subsubsection{$ [\fe'_7] -  \mathfrak{e}_7^M  - [\fe'_7]$ LSTs }\label{ssec:e7e7Me7}
We continue by the toric construction a similar family of LST theories but with $E'_7$\footnote{We employ the notation $E'_s$ used in \cite{DelZotto:2022ohj} to clarify specific type of 6d SCFTs of type $\mathcal{T}(\mu,E_8)$ that corresponds to the toric construction in this work.} flavor and gauge groups, the set of rays are given below,
\begin{table}[ht!]
\begin{center}
\begin{tabular}{cc}
\begin{tabular}{|c|c| }\hline  
Z & (-2,-3,0,0)     \\
X & (1,0,0,0) \\
Y & (0,1,0,0) \\ \hline
$\alpha_{4,\pm} $ & $(-2,-3,\pm 4 ,0 ) $  \\
$\alpha_{3,\pm} $ & $(-2,-3,\pm 3 ,0  )$ \\
$\alpha_{2,\pm} $ &$ (-2,-3,\pm 2 ,0 )$ \\
$\alpha_{1,\pm} $ & $(-2,-3,\pm 1 ,0 )$  \\
$\alpha_{\hat{3},\pm} $ &$ (-1,-2,\pm 3 ,0 )$  \\
$\alpha_{\hat{2},\pm} $ & $(0,-1,\pm 2 ,0 )$  \\
$\alpha_{\hat{1},\pm} $ &$ (0,0,\pm 1 ,0 )$ \\ 
$\alpha_{\tilde{2},\pm}$&$ (-1,-1,-2,0)$ \\
\hline  
\end{tabular}
& 
\begin{tabular}{|c|c |}\hline 
\multicolumn{2}{|c|}{$\fe_7^m$ gauge factor, $m=1\ldots M$ } \\ \hline
$f_{4,m} $ & $(-2,-3,  4 m ,4 ) $  \\
$f_{3,m} $ & $(-2,-3,  3m ,3  )$ \\
$f_{2,m} $ &$ (-2,-3,  2m ,2 )$ \\
$f_{1,m} $ & $(-2,-3,  m ,1 )$  \\
$f_{\hat{4},m}  $ &$ (-1,-2,  4m ,4 )$  \\
$f_{\hat{2},m} $ & $(0,-1,  2m  ,2 )$  \\
$f_{\hat{3},m} $ &$ (-1,-1,  3m ,3 )$ \\
$f_{\hat{1},m} $ &$ (0,0,  m ,1 )$ \\
$f_{\tilde{2},m} $ &$ (-1,-1,  2m ,2 )$ \\
 \hline  

\end{tabular} 
\end{tabular}
\end{center}
\caption{\label{tab:E7E7E7fibre}The toric rays that resolve an $(E'_7)^2$ flavor group and $M\times \fe_7$'s.}
\end{table}
whereas the $\alpha_{i,\pm}$ are the non-compact $\fe'_7$ flavor divisors and the $f_{i,m}$ the $M\times 8$ compact $\fe_7$ gauge divisors. The affine node of each $\fe_7$ is given by $\alpha_{1,\pm}$, $f_{1,m}$. Thus the little string chain looks then like\footnote{This coincides with $[\fe'_7]-\fe_7^N-[\fe'_7]$ quiver presented in Table 6 of  \cite{DelZotto:2022ohj} when replacing the notation $M\rightarrow N+1$.} 
\begin{align}
[\fe'_7 ]\, \, \,  \underbrace{ \overset{\mathfrak{e}_{7}}{1} \, \,   \overset{\mathfrak{e}_{7}}{2} \ldots \overset{\mathfrak{e}_{7}}{2} \, \, \overset{\mathfrak{e}_{7}}{1} }_{\times M}\, \, [\fe'_7] \, .
\end{align}
As the intersection of two $\fe_7$ divisors leads to a non-minimal singularity in the elliptic fiber we shall resolve this via the inclusion of superconformal matter given as \cite{DelZotto:2014hpa}
\begin{align}
\label{eq:E7E7SCM}
	\begin{array}{ccccccc  }
		& &   \fsp_1 & \fso_7 &  \fsp_1 & &  \\
	\lbrack \fe_7]&	1 & 2  &   3 &  2&1 & [\fe_7] \, 
	\end{array}.
\end{align}
The toric rays for the full resolved space are depicted in Table~\ref{tab:e7e7ConfMatter},
\begin{table}[t!]
\begin{center}
\begin{tabular}{|c|l| }\hline 
\multicolumn{2}{|c|}{$\mathcal{T}(\fe'_7, \fe_7^1)$ Conformal Matter } \\ \hline
$s_{-1}$  &$ (-2,-3, - 3 ,1 )  $ \\
$s_{-2 }$ &$ (-2,-3,-2 ,1),(-1,-2,-2,1) $  \\ 
$s_{-3,k }$ & $(-2,-3,-1,1  ),(-2,-3,-2,2),(0,-1,-1,1),(-1,-1,-1,1) $ \\ 
$s_{-4 }$ &$ (-2,-3,-1,2 ),(-1,-2,-1,2)$ \\  
$s_{-5 }$ &$ (-2,-3,-1,3 )$ \\ \hline  
\end{tabular} \\
\begin{tabular}{|c|l| }\hline 
\multicolumn{2}{|c|}{$\mathcal{T}(\fe_7^M, \fe'_7)$ Conformal Matter } \\ \hline
$s_{+1}$  &$ (-2,-3, 3M- 2 ,3 )  $ \\
$s_{+2 }$ &$ (-2,-3,2M-1 ,2),(-1,-2,2M-1,2) $  \\ 
$s_{+3,k }$ & $(-2,-3,M ,1  ),(-2,-3,2M ,2),(0,-1,M ,1),(-1,-1,M ,1),$ \\ 
$s_{+4 }$ &$ (-2,-3,M+1,1 ),(-1,-2,M+1,1)$ \\  
$s_{+5 }$ &$ (-2,-3,M+2,1 )$ \\ \hline   
\end{tabular}

\begin{tabular}{|c|l| }\hline 
\multicolumn{2}{|c|}{$\mathcal{T}(\fe_7^m \fe_7^{m+1})$ Conformal Matter for all $m=0\ldots M-1$ if $M>1$ } \\ \hline
$s_{1,m}$  &$ (-2,-3, 4m+1 ,4 )  $ \\
$s_{2,m }$ &$ (-2,-3,3m+1 ,3),(-1,-2,3m+1,3) $  \\ 
$s_{3,m,k }$ & $(-2,-3,2m+1,2  ),(-2,-3,4m+2,4),(0,-1,2m+1,2),(-1,-1,2m+1,2)$ \\ 
$s_{4,m }$ &$ (-2,-3,3m+2,3 ),(-1,-2,3m+2,3)$ \\  
$s_{5,m }$ &$ (-2,-3,4m+3,4 )$ \\ \hline  
\end{tabular} 
\end{center}
\caption{\label{tab:e7e7ConfMatter}Depiction of the toric rays that resolve the $(M+1)\times$ $\mathcal{T}(\fe_7,\fe_7)$ conformal matter between the $M\times$ $\fe_7$ gauge factors and the two $\fe'_7$ flavor factors. }
\end{table}
whereas the inclusion of the superconformal matter \ref{eq:E7E7SCM} increases the (negative) self-intersection by 6. Hence, the first and the last $\fe_7$ in the chain\footnote{For the special case $M=1$ we actually have two $\mathbf{56}$-half hypermultiplets.} admit a $\mathbf{56}$ half-hypermultiplet.  It is straightforward to read off the LS charges of the tensors, which are given as
\begin{align}
	\vec{l}_{LS} = (1, 1, 1,2  ,3, \underbrace{1,
4,3,2,3,4,  1, \ldots ,1, 4,3,2,3,4,
1}_{\times M} , 3,2,1,1,1) \, .
\end{align}

\subsubsection*{The T-Dual LST}
The T-dual configuration are extracted from a second 2D reflexive sub-polytope given by the $(0,y,z,0)$ projection. This is again an $F_{13}$ type and hence admits a $\mathbb{Z}_2$ MW group, but with two more non-compact toric rays $\alpha_{\hat{1},\pm}$ that were not present in the $E_8$ T-dual. This highlights the presence of two additional non-toric $\fsu_2$ flavor branes, which can be seen by explicitly writing the model in the $F_{13}$ way after renaming the coordinates as
\begin{align}
Y\rightarrow z\,,\quad  \alpha_{\hat{2},+} \rightarrow x\, , \quad  \alpha_{\hat{2},-} \rightarrow y \, , \quad  \alpha_{\hat{1},+}\rightarrow e_1\, , \quad \alpha_{\hat{1},-}\rightarrow e_2 \, , 
\end{align} 
such that over a generic base point, the elliptic fiber equation becomes
\begin{align}
p=s_1 e_1^2 x^4 + s_3 e_2^2 y^4 + s_6 e_1  e_2 xyz + s_2  e_2  e_1 x^2 y^2 + 
  e_1 e_2 z^2 \, .
\end{align}
As opposed to the $E_8$ example, $s_1$ and $s_3$ are no longer constants but polynomials in the coordinates $w_0: (-1,0)$ and $w_1: (1,0)$. It is then evident that over $s_1=0$ or $s_3=0$, the fiber becomes of $I_2^{s}$ type, i.e. an $\fsu_2$ flavor algebra resolved by $e_1, e_2$\footnote{The same structure can also be deduced from the compact K3, which is fibered over $\mathbb{P}$.}.

The toric rays of the T-dual base are depicted in Figure~\ref{fig:E7E7ToricDual} with two $3$ curves that host non-flat fibers when $M>1$ \footnote{For $M=1$ the final $3$ curve admits an $\fso_7$ gauge algebra only.}.
\begin{figure}[ht!]
	\centering
	\begin{tikzpicture} 

\draw[gray, thick,<->] (8,0) -- (10,0);
		\draw[gray, thick,->] (9,0) -> (7,1);
\draw[gray, thick,->] (9,0) -> (7,1);
\draw[gray, thick,->] (9,0) -> (8,1);
\draw[gray, thick,->] (9,0) -> (8,2);
	\draw[gray, thick,->] (9,0) -> (7,3);
\draw[gray, thick,->] (9,0) -> (9,1);

\draw[gray, thick,->] (9,0) -> (8,3);

		\draw[black] (6.7,1.5) circle (0pt) node[anchor=north]{$  \overset{\mathfrak{sp}_{M+5}}{1}    $};

\draw[black] (7.5,1.7) circle (0pt) node[anchor=north]{$   \overset{\mathfrak{so}_{4M+12}}{4}    $};
\draw[black] (6.8,3.9) circle (0pt) node[anchor=north]{$   \overset{\mathfrak{sp}_{3M-1}}{1}          $};
\draw[black] (7.9,2.7) circle (0pt) node[anchor=north]{$  \overset{\mathfrak{so}_{8M}}{3^*}                  $};
\draw[black] (8 ,3.9) circle (0pt) node[anchor=north]{$  \overset{\mathfrak{sp}_{3M-3}}{1}                  $};
\draw[black] (9.1,1.9) circle (0pt) node[anchor=north]{$        \overset{\mathfrak{so}_{4M+4}}{3^*}       $};
\draw[black] (7,0.4) circle (0pt) node[anchor=north]{$[\fso_{24}]$};

\end{tikzpicture}	
\caption{\label{fig:E7E7ToricDual}The toric base of the $\fe_7^M$ T-dual model. Curves with a $*$ admit additional non-flat fibers whose resolution goes beyond the toric base. The two additional $\fsu_2$ flavor factors live over a non-toric locus and are therefore not depicted here. }
\end{figure} 
The combination of the $SO(24)$ flavor group plus the aforementioned $\fsu_2$ flavor factors match the rank 14 flavor algebra of the $E_7$ side.  Furthermore, both $s_1$ and $s_3$ polynomials over the locus $y_0=0$ intersect the $3^*_{\fso_{4M+4}}$ curve in a single point, thus we expect a bifundamental representation as being consistent with the expected anomalies. Moreover, when $M>2$, resolving the non-flat fibers introduces another $1$ curve with an $\fsp_{M-3}$ and enhances the $3$ curve to a $4$ curve.
In the following we summarize the full flat quiver as\footnote{This is denoted by the dual theory of the $[\fe'_7]-\fe_7^N-[\fe'_7]$ quiver presented in Table 6 of \cite{DelZotto:2022ohj} after replacing the notation $M\rightarrow N+1$.} 
\begin{align}
\begin{array}{r}
{ \overset{\mathfrak{\fsp}_{2M-4}}{1^*}}     \, \, \, 
\hspace{8 mm}   {\overset{\mathfrak{\fsp}_{M-3}}{1^*}} \, \, \, \,  \, \quad  \\ 
\lbrack \fso_{24} \rbrack  \, {\overset{\mathfrak{\fsp}_{M+5}}{1}} \, {\overset{\fso_{4M+12}}{4}} \,  {\overset{\mathfrak{\fsp}_{3M-1}}{1}} \,\,  {\overset{\fso_{8M}}{4^*}}\, \, {\overset{\mathfrak{\fsp}_{3M-3}}{1}} \,   {\overset{\fso_{4M+4}}{4^*}} \lbrack \fsu_2^2 \rbrack 
\end{array} 
\end{align} 
The LS charges are then straightforward to compute and given as
\begin{align} 
\begin{array}{rrrrrl} 
         &&&\, \, \, \,2^* &&  1^* \\ 
\vec{l}_{LS}= (1, &  1,& 3,&  2,\,\,& 3, &  1^*)
\end{array}  
\end{align} 
Note that the base configuration is an affine $\fe_7$ Dynkin diagram coming from the $\fe_7$ fibers of the T-dual gauge groups. Similar as the $E_8$ example, we project onto the base and factor out the multiplicity factor $\hat{d}_i$ and get 
\begin{align}
\begin{array}{rrrrrr}
         &&&(f_{\tilde{2}}) &&  (f_{\tilde{1}})\, \,  \\ 
 (f_1) & 2 (f_2)& (f_3)&  2(f_4) &   (f_{\hat{3}}) &   2(f_{\hat{2}}) \, ,
\end{array}
\end{align} 
which coincides with the LS charges for $l_{LS,I} = d_I / \hat{d}_I $\, . 

We close this section by two remarks: First, the LST configuration found here is non-simply connected due to the MW torsion group. I.e. a $\mathbb{Z}_2$ center gauging acts diagonally on flavor and gauge group factors\footnote{Also see \cite{Dierigl:2020myk,Hubner:2022kxr} for a recent discussion.}; Second, for $M>1$, the base two-cycles intersect exactly in the form of an affine $\fe_7$ Dynkin diagram, modulo self-intersections. The 2-groups and CB dimension can be matched and are given below as  
\begin{align}
	\begin{array}{|c|c|c|}\hline
	& [\fe'_7]-\fe_7^M-[\fe'_7] & \text{T-dual} \\
	\hline
\text{Dim(CB)}  & \multicolumn{2}{|c|}{ 9+18M } 
 \\ \hline
 \widehat{\kappa}_{\mathscr R}  &\multicolumn{2}{|c|}{ 48M   }  \\
	\hline
	\end{array} 
\end{align}     

\subsubsection{$ [\fe_6] -  \mathfrak{e}_6^M -[\fe_6]$ LSTs }
\label{sssec:E6}
In this section we construct an $E_6$ type of theory using the same algorithm as before. This case however admits a substantially changed T-dual ambient space where the MW group is not finite anymore and highlights the presence of an Abelian flavor group. We first start with the $E_6$ theory by employing the following type of resolutions
\begin{center}
\begin{tabular}{cc}
\begin{tabular}{|c|c| }\hline  
Z & (-2,-3,0,0)     \\
X & (1,0,0,0) \\
Y & (0,1,0,0) \\  \hline
$\alpha_{3,\pm} $ & $(-2,-3,\pm 3 ,0  )$ \\
$\alpha_{2,\pm} $ &$ (-2,-3,\pm 2 ,0 )$ \\
$\alpha_{1,\pm} $ & $(-2,-3,\pm 1 ,0 )$  \\ 
$\alpha_{\hat{2},\pm} $ & $(-1,-1,\pm 2 ,0 )$  \\
$\alpha_{\hat{1},\pm} $ &$ (0,0,\pm 1 ,0 )$ \\    
$\alpha_{\tilde{2},\pm} $ & $(-1,-2,\pm 2 ,0 )$  \\
$\alpha_{\tilde{1},\pm} $ &$ (0,-1,\pm 1 ,0 )$ \\ \hline  
\end{tabular}
& 
\begin{tabular}{|c|c |}\hline 
\multicolumn{2}{|c|}{$\fe_6^m$ gauge factor, $m=1\ldots M$ } \\ \hline
$f_{3,n} $ & $(-2,-3,  3m ,3  )$ \\
$f_{2,n} $ &$ (-2,-3, 2m ,2 )$ \\
$f_{1,n} $ & $(-2,-3,  m ,1 )$  \\ 
$f_{\hat{2},n} $ & $(-1,-1,2m ,2 )$  \\
$f_{\hat{1},n} $ &$ (0,0, m ,1 )$ \\    
$f_{\tilde{2},n} $ & $(-1,-2, 2m ,2 )$  \\
$f_{\tilde{1},n} $ &$ (0,-1,  m ,1 )$ \\ \hline  
\end{tabular} 
\end{tabular}
\end{center} 
When projected onto the (primitive) base rays, we obtain a chain as of\footnote{This coincides with the $\mathcal K_N(E_6 \times U(1),E_6 \times U(1);\mathfrak g = \mathfrak e_6)$ quiver presented in Sec 6.1 \cite{DelZotto:2022ohj} when replacing the notation $M\rightarrow N+1$.}
\begin{align}
[\fe_6]\, \,    \underbrace{\overset{\mathfrak{e}_{6}}{1} \, \, \overset{\mathfrak{e}_{6}}{2} \ldots \overset{\mathfrak{e}_{6}}{2}\, \, \overset{\mathfrak{e}_{6}}{1}}_{\times M}|\, \,  [\fe_6]\, ,
\end{align}
which needs to be supplemented by the superconformal matter given by
\begin{align}
\label{eq:E6E6SCM}
	\begin{array}{ccccc   }
		& &   \fsu_3 &  &   \\
	\lbrack \fe_6]&	1 & 3  &   1 &  [\fe_6 \rbrack   
	\end{array} \, .
\end{align}
Thus we need to add $3M$ base blowups for $M\times$ $\fe_6$ gauge algebra factors with two more fibral divisors. These superconformal matter configurations are resolved via the toric rays as given in Table~\ref{tab:e6e6ConfMatter}.
\begin{table}[ht!]
\centering
\begin{tabular}{|c|l| }\hline 
\multicolumn{2}{|c|}{$\mathcal{T}(\fe_6, \fe_6^1)$ Conformal Matter } \\ \hline
$s_{-1}$  &$ (-2,-3, - 2 ,1 )  $ \\ 
$s_{-2,k }$ & $(-2,-3,-1,1  ),(-1,-2,-1,1),(-1,-1,-1,1)  $ \\  
$s_{-3 }$ &$ (-2,-3,-1,2 )$ \\ \hline  
\end{tabular} \\
\begin{tabular}{|c|l| }\hline 
\multicolumn{2}{|c|}{$\mathcal{T}(\fe_6^M, \fe_6)$ Conformal Matter  } \\ \hline
$s_{+1}$  &$ (-2,-3, 2M- 1 ,2 )  $ \\ 
$s_{+2,k }$ & $(-2,-3,M,1  ),(-1,-2,M,1),(-1,-1,M,1)  $ \\  
$s_{+3 }$ &$ (-2,-3,M+1,1 )$ \\ \hline   
\end{tabular}

\begin{tabular}{|c|l| }\hline 
\multicolumn{2}{|c|}{$\mathcal{T}(\fe_6^m, \fe_6^{m+1})$ Conformal Matter for all $m=0\ldots M-1$ if $M>1$ } \\ \hline
$s_{1,m}$  &$ (-2,-3, 3m+1 ,3 )  $ \\ 
$s_{2,m,k }$ & $(-2,-3,2m+1,2  ),(-1,-2,2m+1,2),(-1,-1,2m+1,2)  $ \\  
$s_{3,m }$ &$ (-2,-3,3m+2,3 )$ \\ \hline   
\end{tabular} 
\caption{\label{tab:e6e6ConfMatter}Depiction of the toric rays that resolve the $(M+1)\times$ $\mathcal{T}(\fe_6,\fe_6)$ conformal matter between the $M\times$ $\fe_6$ gauge and the two $E_6\times U(1)$ flavor factors. }
\end{table}
There are no charged hypermultiplets in the theory except the $\mathbf{27}$'s of the first and last $\fe_6$.\footnote{For $M=1$ there are again two $\mathbf{27}$ plets on the single $\fe_6$ gauge factor.} In 6D LST, each naive $E_6$ enhances to the real flavor group of type $E_6\times U(1)$ according to the fusion theory described in our companion work\cite{DelZotto:2022ohj}.
 
It is straightforward to read off the LS charges of the tensors, which are given as
\begin{align}
	\vec{l}_{LS} = (1, 1, 2 ,\underbrace{1,
 3,2,3,   1, \ldots ,1, 3,2,3,
1}_{\times M} ,  2, 1,1) \, .
\end{align} 

\subsubsection*{The T-Dual LST}
We identify the second fibration by the 2D reflexive sub-polytope given via vertices of the form $(0,y,z,0)$ with the toric rays
\begin{align}
Y: (0,1,0,0)\, ,  \quad \alpha_{\hat{1},\pm}: (0,0,\pm1,0)\, , \quad\alpha_{\tilde{1},\pm}:(0,-1,\pm1,0)\, .
\end{align}
This is a fiber type $F_9$ according to the enumeration used in \cite{Klevers:2014bqa}. Once we rename the above coordinates to
\begin{align}Y \rightarrow w, \quad  \alpha_{1,+} \rightarrow e_1, \quad \alpha_{1,-} \rightarrow v, \quad \alpha_{\tilde{1},-} \rightarrow e_3, \quad \alpha_{\tilde{1},+} \rightarrow u \, .
\end{align}  Then the generic fiber equation becomes
\begin{align}
p_{F9}=e_1^2 s_1 u^3 + e_1 s_2 u^2 v + s_3 u v^2 + e_1^2 s_5 u^2 w + e_1 s_6 u v w + 
 s_7 v^2 w + e_1 v w^2 \, ,
\end{align} 
which can be viewed as a restricted cubic equation in $\mathbb{P}^2_{u,v,w}$ where the $s_i$ are some polynomials in the base. There are three sections $u=0, e_1=0, v=0$ that generate a rank two free MW group, which in the F-theory lift becomes a $U(1)^2$ gauge group for compact bases \cite{Klevers:2014bqa} and a flavor group in the non-compact limit \cite{Lee:2018ihr,Apruzzi:2020eqi}.

\begin{figure}[ht!]
	\centering
	\begin{tikzpicture} 

\draw[gray, thick,<->] (8,0) -- (10,0);
		\draw[gray, thick,->] (9,0) -> (7,1);
\draw[gray, thick,->] (9,0) -> (7,1);
\draw[gray, thick,->] (9,0) -> (8,1);
\draw[gray, thick,->] (9,0) -> (8,2);
	\draw[gray, thick,->] (9,0) -> (7,3);
\draw[gray, thick,->] (9,0) -> (9,1);

		\draw[black] (6.7,1.5) circle (0pt) node[anchor=north]{$   \overset{\mathfrak{sp}_{M+3}}{1}                $};

\draw[black] (7.5,1.7) circle (0pt) node[anchor=north]{$  \overset{\mathfrak{so}_{4M+8}}{4}          $};
\draw[black] (7,3.9) circle (0pt) node[anchor=north]{$       \overset{\mathfrak{sp}_{3M-3}}{1}   $};
\draw[black] (8.1,2.9) circle (0pt) node[anchor=north]{$     \overset{\mathfrak{su}_{4M-2}}{2}     $};

\draw[black] (9 ,1.8) circle (0pt) node[anchor=north]{$  \overset{\mathfrak{su}_{2M}}{2}              $};
\draw[black] (7,0.4) circle (0pt) node[anchor=north]{$[\fso_{20}]$};

\end{tikzpicture}	
\caption{\label{fig:E6E6ToricDual}The toric base of the $\fe_6^M$ T-dual model. }
\end{figure}
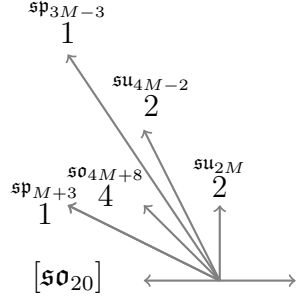

The gauge algebra factors of the T-dual theory can then be deduced just following the general strategy as before with no non-flat fibers for any $M$. A novelty in these models are the $\fsu_k$ gauge factors present over $2$ curves in the chain. The toric base and their toric gauge and flavor factors are summarized in Figure~\ref{fig:E6E6ToricDual}. 

When investigating the toric fibral divisors $f_i$ for some $M$ in the $E_6$ frame, with $i$ being their Kac label as fibral curves, we project onto the base coordinates as
\begin{align}\label{eq:E6Tdualcoord}
f_1 \rightarrow y_4\, , \quad f_2 \rightarrow y_3 \, , \quad f_3 \rightarrow  y_2\, , \quad  \left\{ f_{\hat{2}}, f_{\tilde{2}} \right\} \rightarrow y_1\, , \quad    \left\{ f_{\hat{1}}, f_{\tilde{1}} \right\} \rightarrow y_0\, .
\end{align}
This identification is akin to an $\fe_6 \rightarrow \ff_4$ folding. The LS charges coincide with the Kac labels of affine $\ff_4$ whereas the LS charge of $f_2$ had to be divided by two, due to the $\fso$ gauge algebra factor. 

In the T-dual frame, $v=0$ and $e_1=0$ become purely Abelian flavor group generators, these $U(1)$ flavor factors contribute to the total flavor rank matching and are also important for the $\fsu_{4M-2}$ and $\fsu_{2M}$ gauge factors over the last two $2$ curves respectively. Both of them require two more fundamentals to be consistent with anomalies. 

We can deduce those factors by mapping our $F_9$ elliptic curve into a singular Weierstrass model and derive the matter from the discriminant. For the simplest case when $M=3$, we blow down all fibral divisors and the MW generators $e_1$ and $e_3$ but the affine node, and arrive at the singular restricted cubic 
\begin{align}
p=& \hat{s}_2 u^2 v y_0 y_3 + \hat{s}_5 u^2 w y_0 y_1^4 y_2^6 y_3^4 + \hat{s}_6 u v w w_0 y_3 y_4 + 
 \hat{s}_1 u^3 y_0^2 y_1^4 y_2^7 y_3^5 y_4 \nonumber \\ &+ \hat{s}_7 v^2 w w_0^4 y_0 y_4^6 + 
 \hat{s}_3 u v^2 w_0^4 y_0^2 y_2 y_3 y_4^7 + v  w^2  w_0 y_1   \, ,
\end{align}
In the equation above, we have named the compact base divisors $y_i$ $i=0\ldots 4$ according to the rays in Figure~\ref{fig:E6E6ToricDual} in counterclockwise direction, starting at at the "12" position. 
We can map this configuration into the singular Weierstrass model via the methods layed out in \cite{Klevers:2014bqa,Buchmuller:2017wpe} which then yields
\begin{equation}
  \resizebox{0.92\textwidth}{!}{%
$\begin{aligned}
f=&\frac{1}{48} w_0^2 y_3^2 [-w_0^2 y_3^2 y_4^4 (\hat{s}_6^2 - 
      4 \hat{s}_5 \hat{s}_7 w_0^2 y_0^2 y_1^4 y_2^6 y_3^2 y_4^4)^2 + 
   8 w_0 y_0 y_1 y_3 y_4^2 (  \hat{s}_2 (\hat{s}_6^2 + 2 \hat{s}_5 \hat{s}_7 w_0^2 y_0^2 y_1^4 y_2^6 y_3^2 y_4^4) \\ &   \qquad \, \, \quad 
    -3 \hat{s}_6 (\hat{s}_3 \hat{s}_5 + 
         \hat{s}_1 \hat{s}_7) w_0^3 y_0^2 y_1^4 y_2^7 y_3^3 y_4^6)   + 
   16 y_0^2 y_1^2 (-\hat{s}_2^2 + 3 \hat{s}_1 \hat{s}_3 w_0^4 y_0^2 y_1^4 y_2^8 y_3^4 y_4^8) ] \, , 
\\ 
g=&-\frac{1}{864}  w_0^3 y_3^3 [w_0^3 y_3^3 y_4^6 (\hat{s}_6^2 - 
      4 \hat{s}_5 \hat{s}_7 w_0^2 y_0^2 y_1^4 y_2^6 y_3^2 y_4^4)^3 - 
   12 w_0^2 y_0 y_1 y_3^2 y_4^4 (\hat{s}_6^2 - 
      4 \hat{s}_5 \hat{s}_7 w_0^2 y_0^2 y_1^4 y_2^6 y_3^2 y_4^4)  \\ &\qquad \quad \qquad \times (-3 \hat{s}_6 (\hat{s}_3 \hat{s}_5 + 
         \hat{s}_1 \hat{s}_7) w_0^3 y_0^2 y_1^4 y_2^7 y_3^3 y_4^6 + 
      \hat{s}_2 (\hat{s}_6^2 + 2 \hat{s}_5 \hat{s}_7 w_0^2 y_0^2 y_1^4 y_2^6 y_3^2 y_4^4)) \\   &\qquad \quad \qquad   
      + 2 \hat{s}_2^2 (\hat{s}_6^2 + 2 \hat{s}_5 \hat{s}_7 w_0^2 y_0^2 y_1^4 y_2^6 y_3^2 y_4^4) +  32 \hat{s}_2 y_0^3 y_1^3 (-2 \hat{s}_2^2 + 
      9 \hat{s}_1 \hat{s}_3 w_0^4 y_0^2 y_1^4 y_2^8 y_3^4 y_4^8)   \\ 
      & \qquad \quad \qquad    + 
         3 w_0^4 y_0^2 y_1^4 y_2^8 y_3^4 y_4^8 (-\hat{s}_1 \hat{s}_3 \hat{s}_6^2 + (3 \hat{s}_3^2 \hat{s}_5^2 -  2 \hat{s}_1 \hat{s}_3 \hat{s}_5 \hat{s}_7 + 
            3 \hat{s}_1^2 \hat{s}_7^2) w_0^2 y_0^2 y_1^4 y_2^6 y_3^2 y_4^4)) 
         \\ &  \qquad \quad\qquad    + 
   24 w_0 y_0^2 y_1^2 y_3 y_4^2 (-6 \hat{s}_2 \hat{s}_6 (\hat{s}_3 \hat{s}_5 + 
         \hat{s}_1 \hat{s}_7) w_0^3 y_0^2 y_1^4 y_2^7 y_3^3 y_4^6] \, .
\end{aligned}$%
}
\end{equation}
We can then obtain the discriminant 
\begin{align}
\Delta=   w_0^{12}  y_0^6 y_1^{10} y_2^{12} y_3^{12} y_4^{12}  \hat{\Delta} \, .
\end{align}
From this discriminant we can again, read off the singularities to double check flavor and gauge group factors, that we inferred from the toric rays.  The matter can be deduced from the codimension two loci where the vanishing order of the discriminant further enhances. In particular for the $\fsu_{10}$ and $\fsu_6$ factor over $y_1$ and $y_0$ we will further factorize the residual discriminant, which reveals the two reducible loci
\begin{align}
Q_A= \hat{s}_2 \hat{s}_5 - \hat{s}_1\hat{s}_6 w_0 y_2 y_3 y_4^2\, , \qquad Q_B= \hat{s}_2 \hat{s}_7 - \hat{s}_3 \hat{s}_6 w_0 y_2 y_3 y_4^2 \, .
\end{align}
Indeed we find the discriminant to enhance as
\begin{align}
\begin{array}{l}
\text{van}_{\text{ord}}(f,g,\Delta)_{|y_0=0}:(0,0,6) \xrightarrow{Q_{A/B}=0}(0,0,7)   \\  \text{van}_{\text{ord}}(f,g,\Delta)_{|y_1=0}:(0,0,10) \xrightarrow{Q_{A/B}=0}(0,0,11) \end{array} \, ,
\end{align} 
which is expected from additional fundamentals using the Kac-Vafa method. In particular $\hat{s}_2$ is a linear polynomial over the vanishing of $y_0$ and $y_1$ respectively while the other are constants. Hence $[y_{1/2}]\cdot [Q_{A,B}]=1$ and therefore we find exactly one fundamental over each vanishing locus, consistent with anomalies. The chain is given by
\begin{equation}
    [\fso_{20}] \, {\overset{\mathfrak{\fsp}_{M+3}}{1}} \,  {\overset{\fso_{4M+8}}{4}} \,  {\overset{\mathfrak{\fsp}_{3M-3}}{1}} \, \underset{[\fsu_{2}]} {\overset{\mathfrak{\fsu}_{4M-2}}{2}} \, {\overset{\mathfrak{\fsu}_{2M}}{2}} \, [\fsu_{2}]
\end{equation}
It is straightforward to read off the LS charges of the tensors above Figure~\ref{fig:E6E6ToricDual}
\begin{align}
\vec{l}_{LS}=(1, 1, 3, 2, 1) \, ,
\end{align}
which coincide with the second component of the $y_i$ vectors in \eqref{eq:E6Tdualcoord}. 
Finally we confirm the match of Coulomb branches and 2-group structure constants.
\begin{align}
	\begin{array}{|c|c|c|}\hline
	& [\fe_6]-\fe_6^M-[\fe_6] & \text{T-dual} \\
	\hline
 \text{Dim(CB)} & \multicolumn{2}{|c|}{ 4+12M } 
 \\ \hline
 \widehat{\kappa}_{\mathscr R} &\multicolumn{2}{|c|}{ 24M    }  \\
	\hline
	\end{array} 
\end{align}

\subsubsection{$ [\fe_8] -  \mathfrak{e}_6^M -[\fe_7 ]$ LSTs }
Finally we consider another LST configuration, where the two $E_n$ flavor factors as well as the $\fe_m^M$ gauging is different.  We exemplify this with an infinite family of type 
\begin{align}
[\fe_8]  \,  \, \, \underbrace{ \overset{\mathfrak{e}_{6}}{1}        \, \,            \overset{\mathfrak{e}_{6}}{2}     \ldots \overset{\mathfrak{e}_{6}}{2}  \, \,\overset{\mathfrak{e}_{6}}{1}       }_{\times M}\, \, \,   [\fe_7 ]\, .
\end{align}
As usual, we need to include the $\mathcal{T}(\fe_6,\fe_6)$, $\mathcal{T}(\fe_8,\fe_6)$   and $\mathcal{T}(\fe_6,\fe_7)$ conformal matter. We repeat the later at the tensor branch as
\begin{equation}
  \resizebox{0.89\textwidth}{!}{%
$\begin{aligned}
 \label{eq:E6E7SCM}
	\begin{array}{cccccccccccccccccc }
		& & & \fsp_1 & \fg_2 & & \ff_4&  & \fsu_3&  \\
	\lbrack \fe_8\rbrack &	1 & 2  & 2 & 3 & 1 & 5 & 1 & 3 & 1 &  \lbrack \fe_6\rbrack  \, ,
	\end{array} \,  \quad
	\begin{array}{ccccccc  }
		& &   \fsp_1 & \fso_7 &  \fsp_1 &&  \\
	\lbrack \fe_6\rbrack &	1 & 2  &   3 &  2&1 & \lbrack \fe_7\rbrack   
	\end{array} \, .
\end{aligned}$%
}
\end{equation}
The toric rays that implement the resolution is given in Table~\ref{tab:e8e6e7ConfMatter}. 
\begin{table}[ht!]
\centering
\begin{tabular}{|c|l| }\hline 
\multicolumn{2}{|c|}{$\mathcal{T}(\mathfrak{e}_8, \mathfrak{e}_6^1)$ Conformal Matter } \\ \hline
$s_{-1}$  &$ (-2,-3,-5,1 )  $ \\ 
$s_{-2}$  &$ (-2,-3,-4,1 )  $ \\ 
$s_{-3,k }$ & $(-2,-3,-3,1 ),(-1,-2,-3,1)  $ \\  
$s_{-4,k}$  &$ (-2,-3,-2,1 ) ,(-1-2,-3,-4,2),(-1,-1,-2,1) $ \\ 
$s_{-5}$  &$ (-2,-3,-3,2 )  $ \\ 
$s_{-6,k}$  &$ (-2,-3,-1,1 ) ,(0,-1,-1,1),(-2,-3,-2,2) ,(-1,-2,-2,2)  ,(-2,-3,-3,3)    $ \\ 
$s_{-7 }$ &$ (-2,-3,-2,3)$ \\    
$s_{-8,k}$  &$ (-2,-3,-1,2 ),(-1,-2,-1,2),(-1,-1,-1,2)  $ \\ 
$s_{-9}$  &$ (-2,-3,-1,3 )  $ \\ \hline  
\end{tabular} \\
\begin{tabular}{|c|l| }\hline 
\multicolumn{2}{|c|}{$\mathcal{T}(\mathfrak{e}_6^M, \mathfrak{e}_7)$ Conformal Matter  } \\ \hline
$s_{+1}$  &$ (-2,-3, 3M- 2 ,3 )  $ \\
$s_{+2 }$ &$ (-2,-3,2M-1 ,2),(-1,-2,2M-1,2) $  \\ 
$s_{+3,k }$ & $(-2,-3,M ,1  ),(-2,-3,2M ,2),(0,-1,M ,1),(-1,-1,M ,1),$ \\ 
$s_{+4 }$ &$ (-2,-3,M+1,1 ),(-1,-2,M+1,1)$ \\  
$s_{+5 }$ &$ (-2,-3,M+2,1 )$ \\ \hline   
\end{tabular} 
\caption{\label{tab:e8e6e7ConfMatter}Toric rays that resolve the  $\mathcal{T}(\fe_8,\mathfrak{e}_6^1)$ and $\mathcal{T}(\mathfrak{e}_6^M , \fe_7)$ conformal matter.}
\end{table}
The resolution of the $E_8$ and $E_7$ flavor groups can be taken from the sections before, just as the $\fe_6^M$ gauge and conformal matter factors. Note that for every $M$, the $\mathfrak{e}_6$ factors sit over $6$ curves and hence, have no matter. The LS charges are
\begin{align}
\vec{l}_{LS} = (1, 1, 1 , 1,2,1,3,2,3, \underbrace{1,
 3,2,3,   1, \ldots ,1, 3,2,3,
1}_{\times M} ,  3,2,1,1,1) \, .
\end{align}

\subsubsection*{The T-dual LST}
We now collect the rays in the second projection and deduce the gauge group and matter factors. The relevant toric rays for the fibre ambient space are given as
\begin{align}
z: (0,1,0,0)\, ,\quad  x: (0,-1,2,0)\, ,\quad y: (0,-1,-2,0)\, ,\quad e_1 : (0,0,1,0) \, ,
\end{align}
which is again of $F_{13}$ type with a $\mathbb{Z}_2$ MW group. As opposed to the $\fe_7^M$ case, there is only a single non-toric $\fsu_2$ factor appearing, which is consistent with anomalies and the full flavor rank matching. The toric diagram of the base is exactly the same as in the $\mathfrak{e}_6^M$ type given in Figure~\ref{fig:E6E6ToricDual} however with some enhanced flavor and gauge group factors. The chain is given as
\begin{align}
[\fso_{28}] \,     \overset{\mathfrak{sp}_{M+7}}{1}  \, \,     \overset{\mathfrak{so}_{4M+16}}{4}   \, \, 
 \overset{\mathfrak{sp}_{3N+1}}{1}\, \, 
 \overset{\mathfrak{su}_{4N+2}}{2}\, \, 
 \overset{\mathfrak{su}_{2N+2}}{2}\, \,  
 [\fsu_2] \, ,
\end{align}
and thus also admits the very same LS charges given as
\begin{align}
\vec{l}_{LS}=(1, 1, 3, 2, 1) \, .
\end{align}
The 2-groups data and CB dimension can be matched and are given below as  
\begin{align}
	\begin{array}{|c|c|c|}\hline
	& \lbrack  \fe_8\rbrack -\fe_6^M-\lbrack \fe_7 \rbrack & \text{T-dual} \\
	\hline
\text{Dim(CB)}  & \multicolumn{2}{|c|}{ 22+12M } 
 \\ \hline
 \widehat{\kappa}_{\mathscr R} &\multicolumn{2}{|c|}{ 24M+34   }  \\
	\hline
	\end{array} 
\end{align}

\subsubsection{The $[\fe_8]-\ff_4-\fe_6^{M-2}-\ff_4-[\fe_8]$ LSTs}
Next we discuss an unbroken $\fe_8$ flavor group but probing an $\fe_6$ singularity. However we write the quiver little bit differently to match this quiver to also access  some exotic lower rank theories. The starting quiver is of type
\begin{align}
[ \fe_8]    \, \,     \underbrace{ \overset{\mathfrak{f}_{4}}{1}        \, \,            \overset{\mathfrak{f}_{6}}{2}     \ldots \overset{\mathfrak{f}_{6}}{2}  \, \, \overset{\mathfrak{f}_{4}}{1}       }_{\times M}  \, \,     [\fe_8 ]\, .
\end{align}
The tensor branch of $\mathcal{T}(\fe_8,\ff_4)$ and $\mathcal{T}(\ff_4,\ff_4)$ conformal matter are given as
\begin{align}
	\begin{array}{ccccccccccccccc  }
		& &  & \fsp_1 & \fg_2 &    \\
	\lbrack \fe_8]&	1 & 2  & 2 & 3 & 1 &  [\ff_4 \rbrack \,\, ,
	\end{array}   \quad  \begin{array}{ccccc   }
		& &   \fsu_3 & &   \\
	\lbrack \ff_4]&	1 & 3  &   1 &  [\ff_4 \rbrack \, .
	\end{array} 
\end{align}
Note that here one might have wondered, if one could have replaced the $\fe_6$  gauge factors on the $2$ curves in the configuration above with $\ff_4$ factors.
This however is not possible, as the conformal matter will automatically enhance these curves to $6$ curves and thus also the gauge algebra to $\fe_6$ due to anomalies. This way of writing however, also leads to the possibility to start with $M=1$ or $M=2$ configurations for which the $\mathcal{T}(1,\fe_6)$ orbi-instanton theories in the fusion process start to overlap. I.e. we obtain the two exotic cases. For $M=1$ and $M=2$ we therefore have 
\begin{align}
\label{eq:E8E8F4Fusion}
\lbrack \fe_8\rbrack  \, \, 1\, \, 2 \, \,   \overset{\fsp_1}{2} \, \,   \overset{\fg_2}{3 } \, \,      1 \, \,       \overset{\ff_4}{\underset{[N_F=1]}{4} }        \, \, 1 \, \,   \overset{\fg_2}{3 } \, \,   \overset{\fsp_1}{2}\, \,   2   \, \,    1        \, \,  \lbrack \fe_8\rbrack \, \, , \quad 
\lbrack \fe_8\rbrack \, \,   1\, \,  2 \, \, \overset{\fsp_1}{2}     \, \,      \overset{\fg_2}{3 }\, \,   1  \, \,     \overset{\ff_4}{5}    
\, \, 1   \, \,     \overset{\fsu_3}{3 }     \, \, 1 
\, \,     \overset{\ff_4}{5}     
           \, \, 1 \, \,   \overset{\fg_2}{3 } \, \,   \overset{\fsp_1}{2}\, \,   2   \, \,    1        \, \,  \lbrack \fe_8\rbrack \,  
\end{align} 
Their respective T-duals are then given as 
\begin{align} 
\label{eq:SO32F4Duals}
\lbrack \fso_{32} \rbrack \, \,  \overset{\mathfrak{sp}_{8}}{1} \, \,   \overset{\mathfrak{so}_{16}}{4}\, \,   1 \, \,2\, \,  2\, , \, \text{ and }\qquad
\lbrack \fso_{32} \rbrack \, \,  \overset{\mathfrak{sp}_{9}}{1} \, \,   \overset{\mathfrak{so}_{20}}{4}\, \,   \overset{\mathfrak{sp}_{3}}{1} \, \,\overset{\mathfrak{su}_{4}}{2}\, \,  \overset{\mathfrak{su}_{2}}{2}\, , \, \,   
\end{align} 
respectively. The LS charges are again simply 
\begin{align}
	\vec{l}_{LS} =(1,1,3,2,1) \, .
\end{align}
This allows to explicitly match the 2-groups and CB dimension that are given as
\begin{align}
(\text{Dim}(\text{CB}),  \widehat{\kappa}_{\mathscr R})= (20, 29) \, , \qquad \text{ and } \qquad (\text{Dim}(\text{CB}), \widehat{\kappa}_{\mathscr R})= (30, 48) \, .
\end{align}
Note that the $\ff_4$ flavor factors for $M>1$ have no hypermultiplets. For $M>2$ we need to include $\fe_6$ rays and their conformal matter. The resolution thus follows by first taking   
the $(M-2) \times$ $\fe_6$ toric rays are given as 
\begin{center}
\begin{tabular}{cc}
\begin{tabular}{|c|c| }  \hline 
\multicolumn{2}{|c|}{$\ff_4$ gauge factors, $m=1, M$ } \\ \hline
$f_{3,m} $ & $(-2,-3,  3m ,3  )$ \\
$f_{2,m} $ &$ (-2,-3, 2m ,2 )$ \\
$f_{1,m} $ & $(-2,-3,  m ,1 )$  \\  
$f_{\tilde{2},m} $ & $(-1,-2, 2m ,2 )$  \\
$f_{\tilde{1},m} $ &$ (0,-1,  m ,1 )$ \\ \hline  
\end{tabular}
& 
\begin{tabular}{|c|c |}\hline 
\multicolumn{2}{|c|}{$\fe_6^m$ gauge factor, $m=2\ldots M-1$ } \\ \hline
$f_{3,m} $ & $(-2,-3,  3m ,3  )$ \\
$f_{2,m} $ &$ (-2,-3, 2m ,2 )$ \\
$f_{1,m} $ & $(-2,-3,  m ,1 )$  \\ 
$f_{\hat{2},m} $ & $(-1,-1,2m ,2 )$  \\
$f_{\hat{1},m} $ &$ (0,0, m ,1 )$ \\    
$f_{\tilde{2},m} $ & $(-1,-2, 2m ,2 )$  \\
$f_{\tilde{1},m} $ &$ (0,-1,  m ,1 )$ \\ \hline  
\end{tabular} 
\end{tabular}
\end{center} 
where the toric rays of $E_8$ flavor group and $\mathbb{P}^2_{1,2,3}$ fibre coordinates from Table~\ref{tab:E8E8k3in4fold} should be added to the full Fan. The superconformal matter is then simply given by the toric rays in Table~\ref{tab:e8e6f4ConfMatter}. The LS charges can readily be read off from the toric rays as
\begin{align}
	\vec{l}_{LS} = (1, 1, 1 , 1,2, \underbrace{1,
 3,2,3,   1, \ldots ,1, 3,2,3,
1}_{\times M} ,   2,1,1,1,1) \, .
\end{align}
\begin{table}[ht!]
\centering
\begin{tabular}{|c|l| }\hline 
\multicolumn{2}{|c|}{$\mathcal{T}(\fe_8 ,\ff_4)$ Conformal Matter } \\ \hline
$s_{-1}$  &$ (-2,-3,-5,1 )  $ \\ 
$s_{-2}$  &$ (-2,-3,-4,1 )  $ \\ 
$s_{-3,k }$ & $(-2,-3,-3,1 ),(-1,-2,-3,1)  $ \\  
$s_{-4,k}$  &$ (-2,-3,-2,1 ) ,(-1-2,-3,-4,2),(-1,-1,-2,1) $ \\ 
$s_{-5}$  &$ (-2,-3,-3,2 )  $ \\  \hline  
\end{tabular} \\
\begin{tabular}{|c|l| }\hline 
\multicolumn{2}{|c|}{$\mathcal{T}(\ff_4 , \fe_8)$ Conformal Matter} \\ \hline 
$s_{+1}$  &$ (-2,-3,2M-12 ) $ \\ 
$s_{+2,k}$  &$ (-2,-3,M ,1 ) ,(-1-2,-3,2M ,2),(-1,-1,M,1) $ \\ 
$s_{+3,k }$ & $(-2,-3,M+1,1 ),(-1,-2,M+1,1)  $ \\  
$s_{+4}$  &$ (-2,-3,M+2,1 ) $\\
$s_{+5}$  &$ (-2,-3,M+3,1 )$\\   \hline 
\end{tabular} 
\begin{tabular}{|c|l| }\hline 
\multicolumn{2}{|c|}{$\mathcal{T}(\ff_4/\fe_6^m ,\fe_6^{m+1}/\ff_4)$ Conformal Matter for all $m=0\ldots M-1$ if $M>1$ } \\ \hline
$s_{1,m}$  &$ (-2,-3, 3m+1 ,3 )  $ \\ 
$s_{2,m,k }$ & $(-2,-3,2m+1,2  ),(-1,-2,2m+1,2),(-1,-1,2m+1,2)  $ \\  
$s_{3,m }$ &$ (-2,-3,3m+2,3 )$ \\ \hline   
\end{tabular}   
\caption{\label{tab:e8e6f4ConfMatter}Depiction of the toric rays that resolve the $\mathcal{T}(\fe_8,\ff_4)$ conformal matter and those in between and $\mathcal{T}(\ff_4, \ff_4), \mathcal{T}(\ff_4, \fe_6),\mathcal{T}(\ff_6, \ff_6)$ conformal matter.}
\end{table} 

Next we are switching to the T-dual theory. The base quiver then becomes
\begin{align}
[\fso_{32}] \,     \overset{\mathfrak{sp}_{M+7}}{1}  \, \,     \overset{\mathfrak{so}_{4M+12}}{4}   \, \, 
 \overset{\mathfrak{sp}_{3M-3}}{1}\, \, 
 \overset{\mathfrak{su}_{4M-4}}{2}\, \, 
 \overset{\mathfrak{su}_{2M-2}}{2}  \, .
\end{align}
 with the usual LS charges. 
We close by the matching of the Coulomb branch dimension and the 2-group structure constants for large $M>2$, which is given by 
\begin{align}
	\begin{array}{|c|c|c|}\hline
	& \lbrack \fe_8\rbrack - \mathfrak{f}_4-\mathfrak{e}_6^{M-2} -\ff_4 -\lbrack \fe_8\rbrack & \text{T-dual} \\
	\hline
 \text{Dim(CB)} & \multicolumn{2}{|c|}{ 6+12M } 
 \\ \hline
	\widehat{\kappa}_{\mathscr R} &\multicolumn{2}{|c|}{24M+2}  \\
	\hline
	\end{array} 
\end{align}

\subsection{D-Type Singularities}
The D-type singularities work similarly as the exceptional ones, but are a bit simpler. In the following we omit the details of the exact toric resolution and give just the results of the two T-dual pairs which coincide with those given in \cite{Aspinwall:1996vc} modulo some low rank cases. We close by adding some novel cases with broken flavor groups.

In the following it is beneficial to discuss the $\fso_{4N+2}$ and $\fso_{4N}$ type of singularities separately. We start with the cases with very low number of NS5 branes that is the $M=1$ cases as these oftentimes have unexpected T-duals. Let us also discuss the particular low case gauge group case of an $\fso_8$ type of singularity.
The full tensor branch an LS charge is then given as  
\begin{align}
\lbrack \fe_8 \rbrack       \, \, 1 \, \,  2   \, \,   \overset{\fsu_{ 2}}{2}  \, \,    \overset{\fg_{ 2}}{3}  \, \,   1   \, \,         \overset{\fso_{ 8}}{4}   \, \,  1   \, \,    \overset{\fg_{ 2}}{3}     \, \,    \overset{\fsu_{ 2}}{2}  \, \,    2  \, \, 1     \, \,  \lbrack \fe_8 \rbrack \, , \qquad  \vec{l}_{LS} = (1,1,1,1,2,1,2,1,1,1,1) \, ,
\end{align} 
with the T-dual is given by
\begin{align}
\begin{array}{cccccc}
&   1&  & &  1 &   \\
\lbrack \fso_{32} \rbrack    \, \,   \overset{\fsp_{ 8}}{1  }& \overset{\fso_{ 16}}{4  }   &    1   \, \,  , \qquad  & \vec{l}_{LS} = ( 1, & 1, & 1)     \, ,    \\
&   1&  & & 1 & 
\end{array}
\end{align}
which is exactly of $\fso_8$ affine type. When proceeding to arbitrary $M$ we obtain the same type of quiver with an additional $\mathcal{T}(\fso_8,\fso_8)$ conformal matter interpreted as an E-string. Thus the generalized quiver looks like\footnote{This coincides with the $\mathcal K_N(E_8,E_8;\mathfrak{so}_8)$ quiver presented in Eqn.(5.7) of \cite{DelZotto:2022ohj} when setting the notation $M=N+1$.}
\begin{align}
\lbrack \fe_8 \rbrack       \underbrace{ \overset{\fso_{ 8}}{1}   \overset{\fso_{ 8}}{2} \ldots     \overset{\fso_{ 8}}{2}   \overset{\fso_{ 8}}{1}}_{\times M} \lbrack \fe_8 \rbrack \, , 
\end{align} 
and we obtain the T-dual\footnote{This quiver coincides with the model presented in Eqn. (5.3) of \cite{DelZotto:2022ohj} after replacing the notation $M=N+1$.}
\begin{align}
\begin{array}{ccccc}
&   \overset{\fsp_{ M-1}}{1  }&   \\
\lbrack \fso_{32} \rbrack    \, \,      \overset{\fsp_{ 8+M-1}}{1  }& \overset{\fso_{ 16+4M-4}}{4  }&   \overset{\fsp_{ M-1}}{1  }            \\
 & \overset{\fsp_{ M-1}}{1  } & 
\end{array} \, .
\end{align}
The above theory retains the full $\mathbb{S}_3$ symmetry of $\fso_8$ as a symmetry of the tensors, which is a highly special feature of $\fso_8$ alone and no longer hold for more general $\fso_{4N+8}$ type.

We generalize this quiver to $\fso_{4N+8}$ and include a chain containing $M$ copies of $\fso_{4N+8}$
\begin{align}
\lbrack \fe_8\rbrack    \, \,  \ldots    \, \,   \overset{\fso_{4N+7 }}{4} \, \,  \overset{\fsp_{2N }}{ \underset{[N_F=1]}{1}} \, \,    \underbrace{\overset{\fso_{4N+8 }}{4} \, \,   \overset{\fsp_{2N }}{1} \, \,  \ldots   \overset{\fso_{4N+8 }}{4} \ldots  \, \,    \overset{\fsp_{2N }}{1} \overset{\fso_{4N+8 }}{4}}_{\times 2M-1 }  \, \,  \overset{\fsp_{2N }}{ \underset{[N_F=1]}{1}}   \, \,  \overset{\fso_{4N+7 }}{4}       \ldots   \, \,     \lbrack \fe_8\rbrack \, . 
\end{align} 
whereas the $\fso$ group has no additional flavors. The LS charge is then given as
\begin{align}
	\vec{l}_{LS}= (1,1,1,1, \ldots 1,2, \underbrace{ 1,2, \ldots 1  \ldots   2 ,  1}_{\times 2M-1} ,  2 , 1 \ldots  , 1,1,1,1) \, ,
\end{align} 
and there is the T-dual LST
\begin{align}
	{\scriptsize
	\begin{array}{cc cl}
	&   \overset{\fsp_{4N+M-1}}{1  }& \qquad \qquad \qquad \qquad \qquad \qquad \qquad \qquad  \qquad \qquad \qquad \qquad \qquad \qquad\qquad \, \, \overset{\fsp_{M-1}}{1} & \\
	\lbrack \fso_{32} \rbrack     \,      \overset{\fsp_{4N+M+7}}{1  }&   \overset{\fso_{16N+4M+12}}{4  }&   \underbrace{  \overset{\fsp_{8N+2M-2 }}{1  } \,        \overset{\fso_{16N +4M-4}}{4  } \ldots     \overset{\fsp_{8(N-k)+2M-2 }}{1  }      \,  \overset{\fso_{16(N-k)+4M-4 }}{4  }  \ldots               \overset{\fsp_{ 2M+6 }}{1  } \,         \overset{\fso_{4M+12}}{4  }}_{\times 2N}&        \overset{\fsp_{M-1}}{1  }              \\
	\end{array}}
\end{align} 
Note that the T-dual LST admits a $\mathbb{Z}_2$ symmetry for the last two spinor legs and admits the exact $\fso_{4N+8}^{(1)}$ shape with LS charges
\begin{align}
\begin{array}{clclc}
& 1& \qquad \qquad \qquad \qquad   1& \\
\vec{l}_{LS}=(1, & 1, &   \underbrace{2,  \, \, 1, \ldots    2,    \, \,  1,  \ldots      2,  \, \,    1}_{\times 2N},  &   1       )   \, .   
\end{array}
\end{align}
The 2-groups and CB dimension can be matched and are given below as  
\begin{align}
	\begin{array}{|c|c|c|}\hline
	& [\fe_8]-\fso_{4N+8}^M-[\fe_8] \, &  \text{T-dual}  \\
	\hline
\text{Dim(CB)} & \multicolumn{2}{|c|}{8N^2+4NM+22N+6M+14}  
 \\ \hline
	\widehat{\kappa}_{\mathscr R} &\multicolumn2{|c|}{16N^2+8NM+32N+8M+18} \\
	\hline
	\end{array} 
\end{align}

Next we consider the $\fso_{4N+6}$ type of gauging, whose T-dual is slightly different to that of the $\fso_{4N+8}$. On the $E_8^2$ side we start with a quiver in the form of
\begin{align}
\lbrack \fe_8 \rbrack    \underbrace{\overset{\fso_{4N+6 }}{1}      \ldots   \overset{\fso_{4N+6 }}{2}   \ldots   \overset{\fso_{4N+6 }}{1}}_{\times M }   \, \,  \lbrack \fe_8 \rbrack \, .
\end{align}
The superconformal matter and the LS charges are the same as before. However, the T-dual differs slightly on the last node as 
\begin{align}
	{\footnotesize
\begin{array}{cl }
&   \, \overset{\fsp_{4N+M-3}}{1  }  \\
\lbrack \fso_{32} \rbrack    \, \,      \overset{\fsp_{4N+M+5}}{1  } &  \underbrace{ \overset{\fso_{16N+4M + 4}}{4  }  \,\,     \overset{\fsp_{8N+2M -6 }}{1  }  \ldots   \overset{\fso_{16(N-k)+4M+4 }}{4  }  \, \,   \overset{\fsp_{8(N-k)+2M-6 }}{1  }  \ldots               \overset{\fso_{4M+20 }}{4  }  \, \,     \overset{\fsp_{2M+2 }}{1  }}_{2N \times }  \, \,  \overset{\fsu_{2M+2 }}{2  }        \, ,
\end{array}}
\end{align}
with the LS charge
\begin{align}
\begin{array}{clclc}
& 1 & &   & \\
\vec{l}_{LS}=(1, & 1, &   2,   \ldots    1,    \, \,  2,  \ldots      1,   & 2,  &   1       )   \, .   
\end{array}
\end{align}
The quiver ends on an $\fso_{4M+20}$ node with a single neighboring $2$ curve, such that the T-dual looks more like an $\fso_{4N+5}^{(1)}$ type of quiver. The 2-groups and CB dimension can be matched and are given below as  
\begin{align}
	\begin{array}{|c|c|c|}\hline
	& [\fe_8]-\fso_{4N+6}^M-[\fe_8] \, &  \text{T-dual}  \\
	\hline
\text{Dim(CB)} & \multicolumn{2}{|c|}{8N^2+4NM+14N+4M+5}  
 \\ \hline
	\widehat{\kappa}_{\mathscr R} &\multicolumn{2}{|c|}{16N^2+8NM+16N+4M+6} \\
	\hline
	\end{array} 
\end{align} 

An interesting case are very low rank cases i.e. when taking $\fso_8$ but for very low values of $M$. The resulting quivers we can construct are given as
\begin{align}
\lbrack \fe_8\rbrack \, \,   1\, \,  2 \, \,      \overset{\fsp_1}{2 }     \, \,      \overset{\fg_2}{ \underset{[ N_F=2]}{2 } }  \, \,   \overset{\fsp_1}{2}\, \,   2   \, \,    1        \, \,  \lbrack \fe_8\rbrack \, ,  
\quad \text{ and } \quad 
\lbrack \fe_8\rbrack \, \,   1\, \,  2  \, \,      \overset{\fsp_1}{2 }     \, \,      \overset{\fg_2}{3  }   \, \,1   \, \,      \overset{\fg_2}{ 3 }  \, \,   \overset{\fsp_1}{2}\, \,   2   \, \,    1        \, \,  \lbrack \fe_8\rbrack \, ,
\end{align}
and their T-duals
\begin{align} 
\lbrack \fso_{32} \rbrack \, \,  \overset{\mathfrak{sp}_{6}}{1} \, \,   \overset{\mathfrak{so}_{7}}{\underset{\lbrack N_F=2 \rbrack}{1}}\,    \, , 
 \quad \text{ and } \qquad
\lbrack \fso_{32} \rbrack \, \,  \overset{\mathfrak{sp}_{7}}{1} \, \,   \overset{\mathfrak{so}_{12}}{\underset{\lbrack N_S=\frac12 \rbrack}{1}}\,    \, ,  
\end{align}
 with Coulomb branch and 2-group structure constants
\begin{align}
(\text{Dim}(\text{CB}) ,\widehat{\kappa}_{\mathscr R})= (10, 12) \, , \qquad \text{ and } \qquad (\text{Dim}(\text{CB}), \widehat{\kappa}_{\mathscr R})= (14,18) \, .
\end{align}
.

In a similar way, we can also take the two $\mathcal{T}(\fe_8, \fso_{2n})$ conformal matter (e.g. see \cite{DelZotto:2014hpa} but fuse  at some $\fso_{2n-1}$ factor that appears at the chain. This yields quivers as in 
\begin{align}
\lbrack \fe_8\rbrack   \overset{\fso_{4N+5 }}{1} \, \,  \overset{\fso_{4N+5 }}{2}  \ldots \overset{\fso_{4N+5 }}{2}  \, \, \overset{\fso_{4N+5 }}{2} \ldots \overset{\fso_{4N+5 }}{1}     \lbrack \fe_8\rbrack   \, ,
\end{align}
and the analogous chain for $\fso_{4N+7}$ we are forced to put in $\mathcal{T}(\fso_{4N+5}, \fso_{4N+5})$ conformal matter blocks, that are given as
\begin{align}
\lbrack \fso_{4N+5} \rbrack \overset{\fsp_{2N-1 }}{1}  \lbrack  \fso_{4N+5} \rbrack\, , \qquad \lbrack \fso_{4N+7}  \rbrack   \overset{\fsp_{2N  }}{1} \lbrack  \fso_{4N+7} \rbrack 
\end{align}
From the conformal matter point of view it is evident, that we can not start with a configuration that had an $\fso_{4N+6\pm1}$ factor on a $2$ curve: Whenever we tried to do that, we obtained a chain like 
\begin{align*}
\ldots \overset{\fsp_{2N-1}}{1}  \,\,  \overset{\fso_{4N+5 }}{4}  \,\, \overset{\fsp_{2N-1 }}{1} \ldots  \, , \qquad \text{ and } \qquad \ldots \overset{\fsp_{2N }}{1}  \,\, \overset{\fso_{4N+7 }}{4}  \,\, \overset{\fsp_{2N  }}{1} \, ,
\end{align*}
which is not consistent with anomalies. Instead the $\fso$ algebras enhance again to  $\fso_{4M+6}$ and $\fso_{4N+8}$ singularities
which brings us back to the $\mathfrak{g}=\fso_{4M+7 \pm 1}$ and $M=1$ cases that we discussed before. Thus there are only two new exotic families for each type of singularity. The first two are given as
\begin{align} 
\lbrack \fe_8\rbrack     \, \,   1 \, \,  2   \, \,   \overset{\fsu_{ 2}}{2}  \, \,    \overset{\fg_{ 2}}{3}  \, \,    \ldots   \overset{\fso_{4N+3}}{4 } \, \,      \overset{\fsp_{2N-2}}{1 }  \, \,    \overset{\fso_{4N+5}}{\underset{[N_F=1]}{4}} \, \,        \overset{\fsp_{2N-2}}{1 }  \, \,    \overset{\fso_{4N+3}}{4 }              \, \,    \ldots  \, \,    \overset{\fg_{ 2}}{3}     \, \,    \overset{\fsu_{ 2}}{2}  \, \,    2  \, \, 1    \, \,  \lbrack \fe_8\rbrack \, .
\end{align} and second two as
\begin{align} 
\lbrack \fe_8 \rbrack   \, \,   1 \, \,  2   \, \,   \overset{\fsu_{ 2}}{2}  \, \,    \overset{\fg_{ 2}}{3}  \, \,     \ldots   \overset{\fso_{4N+5}}{4 } \, \,      \overset{\fsp_{2N-1}}{1 }  \, \,    \overset{\fso_{4N+7}}{\underset{[N_F=1]}{4}} \, \,        \overset{\fsp_{2N-1}}{1 }  \, \,    \overset{\fso_{4N+5}}{4 }              \, \,    \ldots   \, \,    \overset{\fg_{ 2}}{3}     \, \,    \overset{\fsu_{ 2}}{2}  \, \,    2  \, \, 1    \, \,  \lbrack \fe_8 \rbrack \, .
\end{align}
Their LS charges of both theories is in the form of
\begin{align}
	\vec{l}_{LS} = (1,1,1,1, \ldots 1,2, 1, 2 ,  1 \ldots  , 1,1,1,1)
\end{align} 
Their associated T-duals  are given by
\begin{align}
\begin{array}{cl  }
&   \overset{\fsp_{4N-4}}{1  }  \\
\lbrack  \fso_{32} \rbrack    \, \,      \overset{\fsp_{4N+4}}{1  }&  \underbrace{ \,\, \overset{\fso_{16N }}{4  }  \,\,    \overset{\fsp_{ 8N-8}}{1  }    \,\,   \overset{\fso_{16N-16 }}{4  } \ldots     \overset{\fsp_{8N-8}}{1  }    \ldots               \overset{\fsp_{8 }}{1  }    \,\,    \overset{\fso_{16 }}{4  }    1}_{2N \times }  \, \,   2      \, ,      
\end{array}
\end{align}
with the LS charges 
\begin{align}
\begin{array}{cl }
& 1 \\
\vec{l}_{LS}=(1, &  \underbrace{ 1,    2,  \, \, 1, \ldots    2,    \, \,  1,  \ldots      1,    2  }_{2N \times},      1     )   \, .   
\end{array}
\end{align}
and 
\begin{align}
\begin{array}{cl  }
&  \, \, \overset{\fsp_{4N-2}}{1  }  \\
\lbrack \fso_{32} \rbrack    \, \,      \overset{\fsp_{4N+6}}{1  }&   \,\, \overset{\fso_{16N+8 }}{4  }  \,\,  \underbrace{  \overset{\fsp_{ 8N-4  }}{1  }    \,\,   \overset{\fso_{16N-8}}{4  } \ldots     \overset{\fsp_{8(N-k)-4}}{1  }   \,\, \overset{\fso_{16(N-k)-8 }}{4  } \ldots               \overset{\fsp_{12 }}{1  }    \,\,    \overset{\fso_{24 }}{4  }      \, \,           \overset{\fsp_{4 }}{1  }       \overset{\fso_{7 }}{ \underset{[N_F=1]}{2}}       }_{2N \times }  \, ,
\end{array}
\end{align}
with respective LS charges 
\begin{align}
\begin{array}{cl }
& 1 \\
\vec{l}_{LS}=(1, & 1,  \, \, \underbrace{   2,  \, \, 1, \ldots    2,    \, \,  1,  \ldots      2,  \, \,    1 }_{2N \times}     )   \, .   
\end{array}
\end{align} 
The 2-groups and CB dimension can be matched and are given below as  
\begin{align}
	\begin{array}{|c|c|c|c|c|}\hline
	& [\fe_8]-\fso_{4N+5}-[\fe_8] \, & \,  \text{T-dual} & [\fe_8]-\fso_{4N+7}-[\fe_8] \, & \,  \text{T-dual}  \\
	\hline
\text{Dim(CB)}  & \multicolumn{2}{|c|}{8N^2+10N+2} & \multicolumn{2}{|c|}{8N^2+18N+10} 
 \\ \hline
	\widehat{\kappa}_{\mathscr R} &\multicolumn{2}{|c|}{16N^2+8N+3 } & \multicolumn{2}{|c|}{16N^2+24N+11}   \\
	\hline
	\end{array} 
\end{align}   
For the $M=2$, the chain in the $E_8\times E_8$ side are simply obtained by inserting one copy of $\fso_{4N+5}$ or $\fso_{4N+7}$ and resolved by the SCM. Their corresponding T-duals are given by
 \begin{align}
\begin{array}{cl }
&   \overset{\fsp_{4N-3}}{1  }  \\
\lbrack \fso_{32} \rbrack    \, \,      \overset{\fsp_{4N+5}}{1  } &  \underbrace{ \overset{\fso_{16N+4}}{4  }  \,\,     \overset{\fsp_{8N-6}}{1  } \ldots   \overset{\fso_{16(N-k)+4 }}{4  }\, \,   \overset{\fsp_{8(N-k)-6 }}{1  }     \ldots               \overset{\fso_{20 }}{4  }  \, \,     \overset{\fsp_{2 }}{1  }}_{2N \times }  \, \,   \overset{\fsu_{2}}{2  }       \, ,   
\end{array}
\end{align}
and 
\begin{align}
\begin{array}{cl }
&   \overset{\fsp_{4N-1}}{1  }  \\
\lbrack \fso_{32} \rbrack    \, \,      \overset{\fsp_{4N+7}}{1  } &  \overset{\fso_{16N+12}}{4  }  \,\,    \underbrace{  \overset{\fsp_{8N-2}}{1  }   \, \,  \overset{\fso_{16N-4 }}{4  }   \ldots   \overset{\fsp_{8(N-k)-2 }}{1  } \, \, \overset{\fso_{16(N-k)-4 }}{4  }     \ldots               \overset{\fsp_{6 }}{1  }  \, \,        \overset{\fso_{12 }}{   \underset{ \lbrack  N_S=1 \rbrack }{2}  }             }_{2N \times }     \, .
\end{array}
\end{align} 
The 2-groups and CB dimension can be matched and are given below as  
\begin{align}
	\begin{array}{|c|c|c|c|c|}\hline
	& [\fe_8]-\fso_{4N+5}^2-[\fe_8] \, & \,  \text{T-dual} & [\fe_8]-\fso_{4N+7}^2-[\fe_8] \, & \,  \text{T-dual}  \\
	\hline
\text{Dim(CB)}  & \multicolumn{2}{|c|}{8N^2+14N+5} & \multicolumn{2}{|c|}{8N^2+22N+14} 
 \\ \hline
	\widehat{\kappa}_{\mathscr R} &\multicolumn{2}{|c|}{16N^2+16N+6 } & \multicolumn{2}{|c|}{16N^2+32N+18}   \\
	\hline
	\end{array} 
\end{align}   

\subsubsection*{Example with Broken Flavor group}
We close the D-type singularities, by also presenting an example with broken flavor symmetry. In the following we focus on a simple example, with only a single broken $\fe_8$ flavor factor. More  exotic cases are presented in Section~\ref{sec: MultiTDual}. 
We use $\mathfrak{g}=\mathfrak{so}_{8+4N}$ as the starting singularity and break one $E_8$ to an $E_7 \times \fsu_2 $ flavor group. The LST is then given as the following chain
\begin{align}  
    [\fe_8] \, \, \underbrace{ \textcolor{red}{\overset{\fso_{8+4N}}{1}} \, \, 2 \ldots 2 \, \, \textcolor{red}{ \underset{[\fsu_2]}{\overset{\fso_{8+4N}}{1}}} }_{M\times } \, \, [\fe_7]
\end{align}
for $M>0$. We recall the conformal matter theories $\mathcal{T}(\fe_8, \fso_{8+4N})$ in the following as
\begin{align}
    [\fe_8] \, 1\,  2\,  \overset{\fsu_2}{2}\,  \overset{\fg_2}{3}\,  1\, \overset{\fso_9}{4}\, \overset{\fsp_1}{1} \, \overset{\fso_{11}}{4} \, \overset{\fsp_2}{1}   \ldots   \overset{\fso_{7+2k}}{4}\,  \overset{\fsp_k}{1}\ldots   \overset{\fso_{7+4N}}{4} \, \overset{\fsp_{2N}}{1}     [\fso_{8+4N}] \, , 
\end{align}
 The 
  $\mathcal{T}( (\fso_{8+4N},\fe_7)/\mathbb{Z}_2)$   conformal matter that we fuse on the right of the LS chain are given in \cite{Dierigl:2020myk}   as
\begin{align}
\lbrack  \fso_{8+4N}  \rbrack \overset{\fsp_{2N-1}}{1}  \overset{\fso_{8+4(N-1)}}{1}   \overset{\fsp_{2(N-1)-1}}{1}              \ldots \overset{\fsp_{2(N-k)-1}}{1} \, \,  \overset{\fso_{8+4(N-k)}}{1}        \ldots    \overset{\fso_{16}}{4} \, \,  \overset{\fsp_3}{1}\, \,   \overset{\fso_{12}}{4} \, \,  \overset{\fsp_1}{1} \, \,  \overset{\fso_7}{3} \, \,   \overset{\fsu_2}{2}  \, \,  1 \, \,  \lbrack  \fe_7 \rbrack
\end{align} 
and for $\mathcal{T}(  \fso_{8+4N},\fso_{8+4N}  )$ there is the tensor branch
\begin{align}
     \lbrack  \fso_{8+4N}  \rbrack \overset{\fsp_{2N}}{1}  \lbrack  \fso_{8+4N}   \rbrack\, .
\end{align}
Fusing together all of the above contributions  we obtain the total LS charges
\begin{align}
   \vec{l}_{LS} = (1,1,1,1,2,1,2,1,2, \ldots \textcolor{red}{1}, 2,  1 \, \ldots 1,2, \textcolor{red}{1},2,1,2,1,\ldots,1,2,1,2, 1,1,1 ) \, .
\end{align}

The T-dual LST admits an $(SO(28) \times SO(4) )/\mathbb{Z}_2$ Flavor group. The dual LS quiver is given as
\begin{align}{\footnotesize
\begin{array}{rcl}
\overset{\fsp_{ 3N+M-2}}{1}\, \,   &&\overset{\fsp_{M-2}}{1} \\ 
    \lbrack \fso_{28} \rbrack \, \,  \overset{\fsp_{3N+M+5}}{1} \, \,  \overset{\fso_{12N+4M+8}}{4} &   \overset{\fsp_{6N+2M-3}}{1} \, \, \ldots  \overset{\fso_{12(N-k)+4M+8}}{4} \, \,   \overset{\fsp_{ 6(N-k)+2M-3}}{1} \ldots     \overset{\fsp_{2M+3}}{1}\,\, & \overset{\fso_{8+4M}}{4} \, \, \overset{\fsp_{M-1}}{1} \, \,  \lbrack  \fso_4  \rbrack 
    \end{array}}
\end{align}
The LS charges are simply those that resemble the $\fso_{8+4N}$ singularity as before
\begin{align}
\begin{array}{c ll }
& 1  \qquad \qquad \qquad \qquad \, \,   1 \\
\vec{l}_{LS}=(1, & 1,  \, \,    2,  \, \,\underbrace{ 1, \ldots    2,    \, \,  1,  \ldots      2,    1}_{2N \times}  ,      1     )   \, .   
\end{array}
\end{align}

The 2-groups and CB dimension can be matched and are given below as  
\begin{align}
	\begin{array}{|c|c|c|}\hline
	& [\fe_8]-\fso_{4N+8}^M-[\fe_7 \times \fsu_2 ]  \, &  \text{T-dual}  \\
	\hline
\text{Dim(CB)} & \multicolumn{2}{|c|}{6N^2+4NM+15N+6M+8}  
 \\ \hline
	\widehat{\kappa}_{\mathscr R} &\multicolumn2{|c|}{12N^2+8NM+20N+8M+10} \\
	\hline
	\end{array} 
\end{align}
Breaking the $E_8$ flavor group with a discrete holonomy as above, allows to construct theories that still exhibit the full flavor rank. Such theories exhibit some more interesting exotic properties that we discuss in the next section.

\section{Exotic Little String Theories} \label{sec: MultiTDual}
In this section we discuss some theories, that are also engineered via toric geometry but go beyond the fusion algorithm discussed before, which yields exotic types of little strings. 

 First we start by engineering elliptic K3s, that admit different fibre type structures that generally have more than just fibrations and thus also more than just two T-duals LSTs. 
   Indeed those general fibre constructions are key for another class of exotic theories, where the flavor holonomies $\mu_i$ admit a discrete piece which breaks the $E_8$ group to some non-simply connected maximal subgroups. Those theories have in general multiple T-duals and in particular bases, that seem to be related to the original singularity by an twisted (affine) folding. 

Finally, we give an outlook on the generality of our proposals, by abandoning the toric framework to discuss six different LSTs with a rank 18 flavor group that we propose to be all T-dual to each other. 
 
\subsection{Discrete Holonomy Theories}\label{sec:DiscreteHolonomy}

In this section, we explore the geometric models with a non-trivial global structure. Notably, we will study the discrete holonomy LSTs that are systematically constructed in \cite{delZottoAppear2} and focus here on the $\mathbb{Z}_2$ and $\mathbb{Z}_3$ cases. They contain a beautiful toric realization of the fibre type $F_{16}$ and $F_{13}$ respectively, according to the enumeration in \cite{Klevers:2014bqa}. As discussed in more detail in \cite{Aspinwall:1998xj}, those theories are characterized by breaking the $E_8$ flavor factors in a maximal rank preserving manner via a discrete holonomy $\mu_i = \mathbb{Z}_n$. These factors correspond to the exclusion of a node in the $E_8$ affine Cartan matrix with Kac label $n$. The resulting flavor group then admits a non-simply connected structure that acts diagonally on the gauge group.
From the geometric perspective, such models can be achieved by having a non-trivial but finite Mordell-Weil group, which occurs naturally in the $Spin(32)/\mathbb{Z}_2$ type of theories that contain an $\mathbb{Z}_2$ MW group in general. Now, we directly break the $E_8$ flavor group such that the discrete holonomies yield a finite MW group. Although more restricted than the most general flavor deformation, those theories are still very rich. For the sake of brevity, we focus on the rank preserving case and leave a more general classification for future work.
 
\subsubsection{A Multi-T-Dual LST}
We start by realizing the flavor branes with a single $0$ curve and then keep decorating with more compact divisors. We begin with the elliptic K3, which engineers the flavor group, and then the compact curves that engineer the dynamical degrees of freedom. The $SO(16)^2/\mathbb{Z}_2$ flavor group is induced by the rays given in Table~\ref{tab:So16m2rays}.
\begin{table}[t!]
\begin{center}
\begin{tabular}{|c|c| }\hline  
Z & (1,0,0,0)     \\
X & (-1,2,0,0) \\
Y & (-1,-2,0,0) \\  \hline
$\beta_{1,\pm} $ & $(-1,2,\pm 2 ,0  )$ \\
$\beta_{2,\pm} $ &$ (-1,-2,\pm 2 ,0 )$ \\ 
$\beta_{3,\pm} $ & $(-1,-1,\pm 2 ,0 )$  \\ 
$\beta_{4,\pm} $ & $(-1,0,\pm 2 ,0 )$  \\
$\beta_{5,\pm} $ &$ (-1,1,\pm 2 ,0 )$ \\    
$\alpha_{1,\pm} $ & $(-1,-2,\pm 1 ,0 )$  \\ 
$\alpha_{2,\pm} $ & $(-1,2,\pm 1 ,0 )$  \\
$\alpha_{3,\pm} $ &$ (0,-1,\pm 1 ,0 )$ \\  
$\alpha_{4,\pm} $ &$ (0,1,\pm 1 ,0 )$  \\ \hline  
$y_0$ & $(-1,-2,0,1)$ \\ \hline
\end{tabular}
\caption{\label{tab:So16m2rays}The toric rays the yield an $F_{13}$ type of fibre \cite{Klevers:2014bqa} and the resolution divisors of two  $\fso_{16}$ flavor branes. }
\end{center} 
\end{table}
For each toric fibration, it is necessary to identify a reflexive 2D sub-polytope that resides over the generic point of the residual base.  For a generic vector $v=(x,y,z,q)$, we can generate such a 2D reflexive polytope via:
\begin{enumerate}
\item \textbf{Fibration 1}: With base in the $(z_i,q_i)$ plane and fibre spanned by $Z,X,Y$, with $F_{13}$ generic ambient space, containing a $\mathbb{Z}_2$ MW group and $SO(16)^2/\mathbb{Z}_2$ flavor group.
\item \textbf{Fibration 2}: With base in the $(y_i,q_i)$ plane and fibre spanned by $Z,\beta_{4,\pm}$, with $F_{13}$ generic ambient space, containing a $\mathbb{Z}_2$ MW group and $SO(16)^2/\mathbb{Z}_2$ flavor group.
\item \textbf{Fibration 3}: With base in the $(x_i,q_i)$ plane and fibre spanned by $\alpha_{3,\pm},\alpha_{4,\pm}$, that is an $F_{15}$ fibre type and an $SU(16)/\mathbb{Z}_2 \times U(1)$ flavor group. 
\end{enumerate} Now we decorate the graph with more compact divisors via the following set of vertices
\begin{align}
y_i : (-1,-2,n,1) \, , \qquad \text{ for } i =1 \ldots M \, .
\end{align}
This results to three different T-dual frames with the following little string theories:
\begin{enumerate}
\item A trivial gauge theory but $M$ additional tensors and hence a chain 
\begin{align}
[\fso_{16}  ]  \, \,   \underbrace{1 \, \,    2 \ldots  2   \, \,  1}_{\times M+1}  \, \,     [\fso_{16} ] \, .
\end{align} 
\item An single tensor theory with gauge group as
\begin{align}
[\fso_{16} ]  \, \,  \overset{\mathfrak{sp}_{M }}{0}      [\fso_{16}  ] \, ,
\end{align} 
with bi-fundamentals of $\fsp_M$ as well as one more antisymmetric representation.
\item An single tensor theory with gauge group as
\begin{align}
[\fsu_{16}]  \, \,  \overset{\mathfrak{su}_{M+1 }}{0}  \, ,
\end{align} 
with bi-fundamental matter as well as two more antisymmetric representations.
\end{enumerate}
These three different theories all have uniform LS charge $1$ and thus their 2-group structure constant is simply the already matched Coulomb branch\footnote{For those models one might actually expect a flavor enhancement of $\fso_{16}^2$ to $E_8^2$ and $\fso_{32}$ respectively.}.

We can further decorate the above configuration in some other non-trivial way by adding the following set of vertices
\begin{align}
y_i : (-1,-1,n,1) \, , \qquad \text{ for } i =1 \ldots M \, .
\end{align}
This decoration affects each LST differently, given as
\begin{enumerate}
\item An $\fsu_2^{M+1}$ given by the following quiver
\begin{align}
[\mathfrak{so}_{16} ]  \, \,   \underbrace{\overset{\mathfrak{su}_{2 }}{1}  \, \,    \overset{\mathfrak{su}_{2 }}{2} \ldots   \overset{\mathfrak{su}_{2 }}{2}   \, \,    \overset{\mathfrak{su}_{2 }}{1}}_{\times M+1}  \, \,     [\mathfrak{so}_{16}  ] \, .
\end{align} The matter consists of bifundamentals only. Note that the diagonal $\mathbb{Z}_2$ gauging allows only for bifundamental matter. 
\item A two tensor theory with
\begin{align}
[ \mathfrak{so}_{16} ]  \, \,  \overset{\mathfrak{sp}_{M }}{1}   \, \,  \overset{\mathfrak{sp}_{M }}{1}\, \,    [\mathfrak{so}_{16}   ]  \, .
\end{align} 
Again, with a bifundamental modding as before. 
\item  A single tensor theory with 
\begin{align}
[\mathfrak{su}_{16}   \times  \mathfrak{u}_1  ]  \, \,  \overset{\mathfrak{su}_{2M+1 }}{ \underset{[N_A = 2 ]}{  0}}   \, ,
\end{align} 
Again, we have only bifundamentals and some antisymmetrics at the last curve.  
\end{enumerate}
The above three cases summarize that: First we find an expected T-dual where the affine fibres of the gauging enter the base, but a third fibration that has not appeared in the sections before. Instead of the two expected curves, we find just one as if the affine $\fsu_2^{(1)}$ base quiver has likely been further reduced to a $\fsu_2^{(2)}$. This is indeed the correct interpretation in the more generalized examples presented in the next sections.
 
\subsubsection{$\mathbb{Z}_2$ Discrete Holonomy LSTs}
We start by the flavor breaking given below, which is also considered in \cite{Dierigl:2018nlv,Aspinwall:1998xj}
\begin{align}
\mathbb{Z}_2: \fe_8 \rightarrow \{ \fe_7 \times \fsu_2  , \fso_{16} \} \, .
\end{align}
In the case before, we already studied the $SO(16)^2$ flavor group breaking, which lead to various new T-dual theories. Here we  mainly focus the $(E_7 \times SU(2))^2/\mathbb{Z}_2$ flavor group that produces more exotic T-duals. We start of with $\fe_7$ gaugings leading to an LST\footnote{This coincides with the $\mathcal K_N(E_7 \times SU(2),E_7 \times SU(2);\mathfrak g = \mathfrak e_7)$ quiver described in Sec 6.2 \cite{DelZotto:2022ohj} when replacing the notation $M\rightarrow N+1$.}
\begin{align}  
\begin{array}{c } 
{\overset{ \lbrack\mathfrak{su}_2\rbrack}{1}}\, \, \,  \qquad \qquad \qquad \qquad \quad \qquad \qquad {\overset{ \lbrack\mathfrak{su}_2\rbrack}{1}}\\ 
\lbrack \fe_7\rbrack  \,\,  1 \,  {\overset{\mathfrak{su}_2}{2}}\,\, {\overset{\mathfrak{so}_7}{3}} \,\, {\overset{\mathfrak{su}_2}{2}} \, \, 1\, \,  \underbrace{\overset{\mathfrak{e}_7}{8} \,\,  1 \,  {\overset{\mathfrak{su}_2}{2}}\,\, {\overset{\mathfrak{so}_7}{3}} \, \,  {\overset{\mathfrak{su}_2}{2}} \, \, 1 \, \,   \overset{\mathfrak{e}_7}{8}  \ldots \overset{\mathfrak{e}_7}{8}  \,\,  1 \,  {\overset{\mathfrak{su}_2}{2}}\,\, {\overset{\mathfrak{so}_7}{3}} \,\, {\overset{\mathfrak{su}_2}{2}} \, \, 1\, \,  \overset{\mathfrak{e}_7}{8}}_{\times M} \,\,  1 \,  {\overset{\mathfrak{su}_2}{2}}\,\, {\overset{\mathfrak{so}_7}{3}}  {\overset{\mathfrak{su}_2}{2}} \, \, 1 \, \,   
 \lbrack \fe_7\rbrack    \, .
\end{array}
\end{align}
It is straightforward to read off the LS charges of the tensors, which are given as
\begin{align}
\begin{array}{c}
\qquad 1\, \,  \, \,\, \quad \qquad \qquad \qquad \quad \qquad \qquad 1\\
\vec{l}_{LS}=(1, 1, 1, 2, 3, \underbrace{1, 4, 3, 2, 3, 4, 1 \ldots 1, 4, 3, 2, 3, 4, 1}_{\times M}, 3, 2, 1, 1, 1  )   \, .   
\end{array}
\end{align}
This is distinct from a simple $\fe_8 \rightarrow \fe_7$ breaking due to the absence of $\fe_7$ matter multiplets, implemented by the global $\mathbb{Z}_2$ gauging in the center of gauge and flavor groups. The resolution of the model is done via an $F_{13}$ type of ambient space summarized in Table~\ref{tab:e7e7ConfMatter}. This choice admits two more inequivalent toric fibrations both also with an $F_{13}$ type of fibre that has a $\mathbb{Z}_2$ MW group. The first one is given as 
\begin{align}
\begin{array}{c} 
  {\overset{\mathfrak{sp}_{2M-4}}{1^*}} \, \, \, \,  \\ 
\lbrack \fso_8\rbrack    \, \,  {\overset{\mathfrak{sp}_{M-1}}{1}} \, \,  {\overset{\mathfrak{so}_{4M+4}}{4}} \, \, {\overset{\mathfrak{sp}_{3M-3}}{1}} \, \,   {\overset{\mathfrak{so}_{8M^*}}{4^*}} \, \,    {\overset{\mathfrak{sp}_{3M-1}}{1}} \, \,    {\overset{\mathfrak{so}_{4M+12}}{4}} \, \,   {\overset{\mathfrak{sp}_{M+5}}{1}} \, \,   \lbrack \fso_{24}\rbrack  \, ,
\end{array}
\end{align}
with the LS charges given by
\begin{align}
\begin{array}{c}
\qquad 2^*\\
\vec{l}_{LS}=(1,1,3,2^*,3,1,1 )   \, .   
\end{array}
\end{align}

\begin{figure}[t!]
\begin{picture}(0,60)
\put(30,0){\includegraphics[scale=0.5]{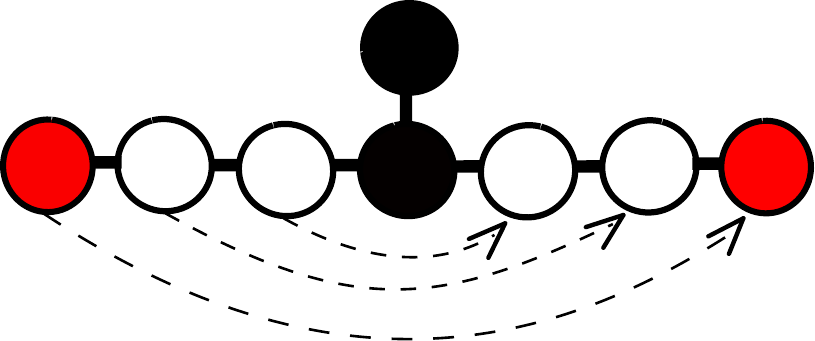}} 
\put(290,15){\includegraphics[scale=0.5]{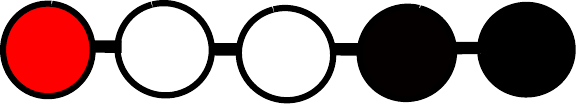}}
\put(200,25){\Large $\xrightarrow{\text{T-dual}}$}
\end{picture}
\caption{\label{fig:E7Folding} Depiction of an $\fe_{7}^{(1)}$ fibre that becomes a $\fe_{6}^{(2)}$ type of base upon T-duality. The nodes intersected by the two torsion sections are marked in red and are mapped onto each other taking all other nodes but the middle ones, marked in black, along. The right picture shows the shape of the resulting base quiver. }
\end{figure}

Note that for $M=1$ the $*$ curve is just a $\overset{\fso_7}{3}$ and the attached $1^*$ is missing. We find the chain to have the full $\fe_7^{(1)}$ topology. The third fibration has the quiver given as
\begin{align}   
\lbrack \mathfrak{u}_{16} \rbrack    \, \,  {\overset{\mathfrak{su}_{2M+10}}{2}} \, \,  {\overset{\mathfrak{su}_{4M+4}}{2}} \, \, {\overset{\mathfrak{su}_{6M-2}}{2}} \, \,   {\overset{\mathfrak{sp}_{4M-4}}{1}} \, \,   {\overset{\mathfrak{so}_{4M+4}}{4^*}} \, ,
\end{align}
with the LS charges given by
\begin{align}
	\vec{l}_{LS} = (1,2,3,4,1^*) \, .
\end{align}
That is those of an $\fe_6^{(2)}$ modulo the expected division by two due to the  $\fso_{4M+4}$ gauge algebra. Hence, we find the twisted affine folded type of the $\fe_7^{(1)}$ base, see Figure~\ref{fig:E7Folding} for a depiction. The 2-groups and CB dimension and matched are given below as  
\begin{align}
\text{Dim}(\text{CB}) = 18M+11  \,, \quad \widehat{\kappa}_{\mathscr R}= 48M+2\, .
\end{align}

\begin{figure}[t!]
\begin{picture}(0,60)
\put(20,0){\includegraphics[scale=0.5]{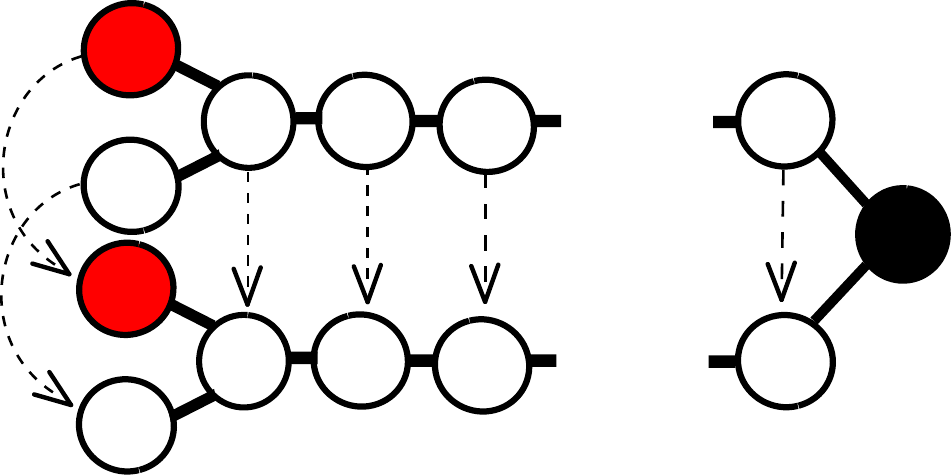}} 
\put(290,15){\includegraphics[scale=0.5]{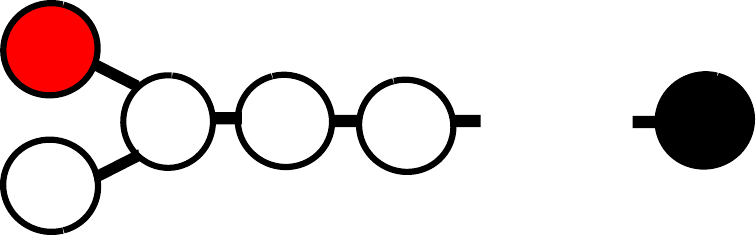}} 
\put(200,25){\Large $\xrightarrow{\text{T-dual}}$}
\end{picture}
\caption{\label{fig:So4nFolding} Depiction of an $\fso_{4N+8}^{(1)}$ fibre that becomes $\fsu_{3N+2}^{(2)}$ type of base upon T-duality. The nodes intersected by the two torsion sections are marked in red and are mapped onto each other taking all other nodes but the middle one, marked in black along. The right picture shows the shape of the resulting base quiver. }
\end{figure}

Such type of foldings occur for various other choices of gaugings with the same flavor group. E.g. take the $M\times$ $\fso_{4N+8}$ gauging which yields  
\begin{align}
 \lbrack \fe_7\rbrack   \, \, \underbrace{ {\overset{\mathfrak{so}_{4N+8}}{\underset{[\fsu_2]}{1}}} \, \,  {\overset{\mathfrak{so}_{4N+8}}{2}}   \ldots {\overset{\mathfrak{so}_{4N+8}}{2}} \, \,     {\overset{\mathfrak{so}_{4N+8}}{\underset{[\fsu_2]}{1}}} }_{M\times}  \, \,           \lbrack \fe_7\rbrack     
\end{align}
upon the inclusion of the sufficient conformal matter
\begin{align}
\lbrack \fso_{4N+8}   \rbrack  \, \,    \overset{\fsp_{2N}}{1}    \, \,   \lbrack \fso_{4N+8}   \rbrack \,      
\end{align} 
 Note that the conformal matter of such theories differs slightly due to the $\mathbb{Z}_2$ gauging. E.g. for $M=1$ it is given as
\begin{equation}
  \resizebox{0.9\textwidth}{!}{%
$\begin{aligned} 
\lbrack \fe_7\rbrack  \,\, 1 \, \, {\overset{\mathfrak{su}_{2}}{2}}  \, \, {\overset{\mathfrak{so}_{7}}{3}} \, \, {\overset{\mathfrak{sp}_{1}}{1}}    
  \ldots {\overset{\mathfrak{sp}_{2N-3}}{1}}\,\,   {\overset{\mathfrak{so}_{4N+4}}{4}}  \, \,
 {\overset{\mathfrak{sp}_{2N-1}}{1}}\, \,  
 {\overset{\mathfrak{so}_{4N+8}}{ \underset{\lbrack \fsu_2^2 \rbrack }{4}}}\, \,  {\overset{\mathfrak{sp}_{2N-1}}{1}}  {\overset{\mathfrak{so}_{4N+4}}{4}} \, \,  {\overset{\mathfrak{sp}_{2N-3}}{1}} ,  \ldots
{\overset{\mathfrak{sp}_{1}}{1}} \, \, {\overset{\mathfrak{so}_{7}}{3}} \, \,   {\overset{\mathfrak{su}_{2}}{2}}  \, \, 1  \, \,   \lbrack \fe_7\rbrack  \, .
\end{aligned}$%
}
\end{equation}
There are again two T-dual theories. The second fibration admits the quiver 
\begin{equation}
  \resizebox{0.9\textwidth}{!}{%
$\begin{aligned} 
& \,\, \overset{\fsp_{2N+ M-3}}{1 }\qquad \qquad\qquad\quad   \qquad \quad \qquad \quad \,\,\,\,\,\, \quad \qquad \overset{\fsp_{M-3}}{1^* } \\ 
\lbrack \fso_{24}\rbrack \, \,  \overset{\fsp_{2N+M+3}}{1 } & \, \, \overset{\fso_{8N+4M+4}}{4 } \, \, \underbrace{ \ldots \, \, \overset{\fsp_{4(N-k)+2M-4}}{1 } \, \, \overset{\fso_{8(N-k)+4M-4 }}{4 } \, \, \ldots
\overset{\fsp_{2M}}{1 } \, \,  \overset{\fso_{4M+4}}{4^* } }_{2 N\times } \, \,  \overset{\fsp_{M-1}}{1 } \, \,  
\, \,  \lbrack \fso_{8}\rbrack \, .
\end{aligned}$%
}
\end{equation}
Here we find that the dual base has the $\fso_{4N+8}^{(1)}$ shape, whereas the last node is completed for $M>2$. The LS charges of the tensors are given as
\begin{align}
\begin{array}{c}
 \, \, \,  \,\,\, \, \quad 1\, \,  \, \,\, \, \quad \qquad \qquad 1^*\\
\vec{l}_{LS}=(1, 1, \underbrace{\ldots 2, 1, \ldots 2, 1^*,}_{2 N\times } 1  )   \, .   
\end{array}
\end{align}
For the third fibration on the other hand we have the following quiver, 
\begin{align} 
\begin{array}{c cl}
 \lbrack \fsu_{16} \times \mathfrak{u}_1 \rbrack & \overset{\fsu_{4N+2M+6}}{2}&  \\ 
   & \overset{\fsu_{8N+4M-4}}{2}&  \, \, \underbrace{\overset{\fsu_{8N+4M-12}}{2}\, \, \overset{\fsu_{8N+4M-20}}{2} \ldots \overset{\fsu_{4M+4}}{2} \, \, \overset{\fsp_{2M-2}}{1} }_{N \times} \\
& \overset{\fsu_{4N+2M-2}}{2} & 
\end{array} 
\end{align}
With the LS charges given as
\begin{align}
\begin{array}{cc}
 1 \quad  \, \, \, \, \quad\\
 \vec{l}_{LS} =(\,2, 2, 2, \ldots 2, 2 )   \, .  \\
 1\quad \, \, \, \,   \quad\\
\end{array}
\end{align}
Here the $\fso_{4 N+8}^{(1)}$ is folded by a $\mathbb{Z}_2$ to an $\fsu_{N+3}^{(2)}$(see Figure~\ref{fig:So4nFolding} for a depiction). From the $\fso_{4N+8}^{(1)}$ point of view, only the middle node is fixed under the involution, which can also be seen from the fact that it hosts an $\fsp$ type of gauge algebra. The 2-groups and CB dimension can be matched and are given below as  
\begin{align}
	\begin{array}{|c|c|c|}\hline
	& [\fe_7]-\fso_{4N+8}^M-[\fe_7] & \text{T-duals}  \\
	\hline
\text{Dim(CB)} & \multicolumn{2}{|c|}{4N^2+4NM+8N+6M+2}  
 \\ \hline
	\widehat{\kappa}_{\mathscr R} &\multicolumn{2}{|c|}{8N^2+8NM+8N+8M+2} \\
	\hline
	\end{array} 
\end{align}

These are features of the flavor group from the respective $K3$ in use, hence we also observe them in the last possible gauging, i.e. of $\fsu_{2N}$ type. For simplicity we again look at the fractionalization of the pure heterotic string to obtain the generalized form
\begin{align} 
\lbrack \fe_7\rbrack\, \,  \overset{ \fsu_{2N} }{0} \, \,   \lbrack \fe_7\rbrack\, , \quad \rightarrow \quad  
\begin{array}{c } 
\lbrack \fsu_2\rbrack   \\ 
\lbrack \fe_7\rbrack  \,\,  1 \,  {\overset{\mathfrak{su}_2}{2}}\,\,   {\overset{\mathfrak{su}_4}{2}}\ldots 
 {\overset{\mathfrak{su}_{2N-2}}{2}}\, \,  
 {\overset{\mathfrak{su}_{2N}}{2}}\, \,  {\overset{\mathfrak{su}_{2N-2}}{2}}\ldots
 {\overset{\mathfrak{su}_4}{2}} \, \,  {\overset{\mathfrak{su}_2}{2}}\, \, 1 \, \,    \lbrack \fe_7\rbrack \, .   \\
\lbrack \fsu_2\rbrack    
\end{array}
\end{align}
The first T-dual is given as, 
\begin{align} 
\lbrack \fso_{24}\rbrack \, \,  \overset{\fsp_{2N}}{1 }  \underbrace{  \, \, \overset{\fsu_{4N-4}}{2 }   \ldots \overset{\fsu_{4(N-n)-4}}{2 }  \ldots \overset{\fsu_{4 }}{2 } \, \, 1}_{\times N} \, \,  
\, \,  \lbrack \fso_{8}\rbrack \, ,
\end{align} 
which looks like a quiver of $\fsp_{N}^{(1)}$ type associated to a regular reduction of the $\fsu_{2N}$ Cartan matrix.
The third T-dual quiver is given as
\begin{align} 
\begin{array}{c }
 \lbrack \fsu_{16}\rbrack \\ 
\underbrace{1 \, \, \overset{\fsu_{8}}{2} \ldots \overset{\fsu_{4N-8}}{2} \,\, \overset{\fsu_{4N}}{2 }\,\,  \overset{\fsu_{4N-8}}{2} \ldots \overset{\fsu_{8}}{2}\, \,  1 }_{\times (N+1) \text{ for N even}} 
\end{array} \,, \,\,
\begin{array}{c }
 \lbrack \fsu_{16}\rbrack \\ 
\underbrace{\overset{\fsu_{4}}{\underset{N_A=1}{1}} \, \, \overset{\fsu_{12}}{2} \ldots \overset{\fsu_{4N-8}}{2} \,\, \overset{\fsu_{4N}}{2 }\,\,  \overset{\fsu_{4N-8}}{2} \ldots \overset{\fsu_{12}}{2 }\, \,  \overset{\fsu_{4}}{\underset{N_A=1}{1}}}_{\times N \text{ for N odd}} 
\end{array} 
\end{align}
Notably we find that the LS charges are universally one  in all models and all nodes are of $\fsu_k$ gauge algebra type which suggests a folding which leaves none of the $\fsu_{2N}^{(1)}$ nodes fixed. Instead it identifies the affine node with the $N$-th node in the $\fsu_{2N}^{(1)}$ Cartan matrix. The dimension of the Coulomn branch and $\hat{\kappa}_R$ are given by $\text{Dim}(\text{CB})+1=\hat{\kappa}_R=2N^2+2$.

\subsubsection{$\mathbb{Z}_3$ Discrete Holonomy LSTs}\label{Sec:Z3Discrete}
We continue with $\mathbb{Z}_3$ discrete holonomy theories, which admit additional exotic features as well as self-T-duality. The flavor breaking of the two $\fe_8$s are given by  
\begin{align}
\fe_8 \rightarrow \{ \fe_6 \times \fsu_3, \fsu_9  \} \, ,
\end{align}
where only $\fsu_{3N}$ and $\fe_6$ singularities can lead to this type of breaking. These theories have similar properties as the ones before as well as self T-dual cases presented in the next section. An interesting example to start with are those specified with the quivers  
\begin{align}     
\begin{array}{l }  
\qquad  \qquad  \, \,  {\overset{\lbrack \fsu_3\rbrack}{1}}   \\
\lbrack \fe_6\rbrack \, \, 1  \, \, {\overset{\mathfrak{su}_{3}}{3}}  \, \,   1  \, \, \underbrace{{\overset{\mathfrak{e}_{6}}{6}}  \, \, 1  \, \, {\overset{\mathfrak{su}_{3}}{3}}  \, \,   1  \, \, {\overset{\mathfrak{e}_{6}}{6}} \ldots {\overset{\mathfrak{e}_{6}}{6}}}_{M\times}   \,\,  1 \, \, {\overset{\mathfrak{su}_3}{2}} \,\, {\overset{\mathfrak{su}_6}{2}}  \,  \,  \lbrack \fsu_{9}\rbrack   
\end{array}
\end{align}
It is straightforward to read off the LS charges of the tensors, which are given as
\begin{align}
	\begin{array}{l}
		\qquad  \qquad  \qquad \, \, \, \, \,  1\\
\vec{l}_{LS}=(1, \, \, 1, \, \,  2, \, \,  \underbrace{ 1, \, \, 3,\, \,2,\, \,3,\, \,1,   \, \,  \ldots      1,}_{\times M} \, \, 3,\, \,2,\, \,1  )   \, .   
	\end{array}
	\end{align}	
where we have included in total $M\times $ $\fe_6$ factors. This theory can be resolved via the toric rays given in Table~\ref{tab:Z3MultiTdual}, which admits again two more inequivalent fibrations.
\begin{table}[ht!]
\centering
\begin{tabular}{ccc}
\begin{tabular}{|c|c| }\hline  
\multicolumn{2}{|c|}{$F_{16}$ fibre } \\ \hline
Z & (-1,-1,0,0)     \\
X & (-1,2,0,0) \\
Y & (2,-1,0,0) \\   
$e_1$& (1,0,0,0) \\   
$e_2$& (0,1,0,0) \\  \hline
\end{tabular}
&
 \begin{tabular}{|c|c| }\hline  
\multicolumn{2}{|c|}{$\fe_6$ flavor} \\ \hline
$\alpha_{3}$&(-1,-1,-3,0) \\ 
$\hat{\alpha_{1}}$&(-1,-1,-1,0)  \\ 
$\hat{\alpha_{2}}$&(-1,-1,-2,0) \\ 
$\alpha_{2}$&(-1,0,-2,0) \\ 
$\tilde{\alpha_{1}}$&(-1,1,-1,0)  \\ 
$\tilde{\alpha_{2}}$&(0,-1,-2,0) \\
$\alpha_{1}$&(1,-1,-1,0)  
 \\ \hline 
\end{tabular}
&
 \begin{tabular}{|c|c| }\hline  
\multicolumn{2}{|c|}{$\fsu_9$ flavor} \\ \hline
$\beta_{0}$& (-1,-1,1,0) \\
$\beta_{1}$&(-1,0,1,0) \\
$\beta_{2}$&(-1,1,1,0) \\
$\beta_{3}$& (-1,2,1,0)  \\
$\beta_{4}$&(0,1,1,0)  \\
$\beta_{5}$&(1,0,1,0) \\
$\beta_{6} $& (2,-1,1,0)\\ 
$\beta_{7}$&(1,-1,1,0) \\
$\beta_{8}$&(0,-1,1,0) \\ \hline 
\end{tabular}

\end{tabular}
\begin{tabular}{cc}
\begin{tabular}{|c|l|} \hline
\multicolumn{2}{|c|}{$\fe_6^i  $  $i=1 \ldots M$ } \\ \hline
$\alpha_{3,i}$&(-1,-1,3i,3) \\ 
$\hat{\alpha_{1,i}}$&(-1,-1,i,1)  \\ 
$\hat{\alpha_{2,i}}$&(-1,-1,2i,2) \\ 
$\alpha_{2,i}$&(-1,0, 2i,2) \\ 
$\tilde{\alpha_{1,i}}$&(-1,1,i,1)  \\ 
$\tilde{\alpha_{2,i}}$&(0,-1,2i,2) \\
$\alpha_{1,i}$&(1,-1,-1,0)  \\ \hline
$f_{0}$&(0,0,0,1)  \\ \hline
\end{tabular}
&
\begin{tabular}{|c|l|} \hline
\multicolumn{2}{|c|}{$\mathcal{T}(\fe_6,\fe_6^1) $ conformal matter } \\ \hline
$s_{-1}$ &(-1,-1,-2,1)  \\
$s_{-2,k}$&(-1,-1,-1,1) , (-1,0,-1,1), (0,-1,-1,1) \\
$s_{-3}$&(-1,-1,-1,2)  \\ \hline   \multicolumn{2}{|c|}{ } \\ \hline
\multicolumn{2}{|c|}{$\mathcal{T}(\fe_6^{i},\fe_6^{i+1}), i= 1 \ldots M-1$  conformal matter } \\ \hline
$s_{3,i}$&(-1,-1,3i-2,3)  \\
$s_{2,i,k}$&(-1,-1,2i-1,2) , (-1,0,2i-1,2), (0,-1,2i-1,2) \\
$\hat{s}_{3,i,k}$&(-1,-1,3i-1,3)  \\ \hline
\end{tabular}
\end{tabular}
{\small
\begin{tabular}{|c|l|} 
\hline
\multicolumn{2}{|c|}{$\mathcal{T}(\fe_6^M,\fsu_9) $ conformal matter } \\ \hline
$s_{+3}$&(-1,-1,3M+1,3)  \\
$s_{+2,k}$&(-1,-1,2M+1,2) , (-1,0,2M,2), (0,-1,2M,2) \\
$s_{+1,k}$ & (-1,-1,M+1,1) ,(-1,0,M+1,1),(-1,1,M+1,1) ,(0,-1,M+1,1), (0,0,M+1,1) ,(1,-1,M+1,1)  \\ \hline 
\end{tabular}
}
 \caption{\label{tab:Z3MultiTdual}The toric resolution of $(E_6 \times SU(3) \times SU(9))/\mathbb{Z}_3$ probing $M\times$ $\fe_6$ singularity. }
\end{table}
Here, all rays are toric and flat except $f_0$.  Both other fibrations however, admit the very same flavor group and in fact also the same type of fibres, i.e. they are   T-self-dual. The fibre type is that of $F_{11}$ and admits an additional $U(1)$ MW group generator which mixes with the various gauge group factors. The quiver itself admits the shape of an $\ff_4^{(1)}$ similar to all other $\fe_6$ T-duals and is given as
\begin{align}
\lbrack \mathfrak{u}_{11} \rbrack \, \,  \overset{\fsu_{6+2M}}{2} \, \, \overset{\fsu_{1+4M}}{2} \, \, \overset{\fsp_{3M-2}}{1}  \, \,  \overset{\fso_{6+4M}}{4}  \, \,  \overset{\fsp_{M}}{1}  \, \,  \lbrack \fso_{10} \rbrack \, ,
\end{align}  
with the LS charge
\begin{align}
\begin{array}{clclc}
\vec{l}_{LS}=(1, & 2, &   3,  &1,  &   1       )   \, .   
\end{array}
\end{align}
Note that this duality is not a simple identification but is really due to a non-trivial symmetry of moduli space of the M-theory compactification. I.e. when compactifying the second T-dual, we obtain the same 5D theory only after a non-trivial action in the CB moduli space. Finally we verify the matching of CB and 2-group structure constants.
\begin{align}
	\begin{array}{|c|c|c|}\hline
	& [\fe_6]-\fe_6^M-[\fsu_9] & \text{T-dual} \\
	\hline
 \text{Dim(CB)} & \multicolumn{2}{|c|}{12M +10} 
 \\ \hline
	\widehat{\kappa}_{\mathscr R} &\multicolumn{2}{|c|}{ 24M+10  }  \\
	\hline
	\end{array} 
\end{align}  
 
Similarly we can consider a theory that have two $E_6$ flavor factors  given via the quiver\footnote{This coincides with the $\mathcal K_N(E_6 \times SU(3),E_6 \times SU(3);\mathfrak g = \mathfrak e_6)$ quiver presented in Sec 6.1 \cite{DelZotto:2022ohj} when replacing the notation $M\rightarrow N+1$.}
\begin{align}     
\begin{array}{c }  
  { \overset{\lbrack \fsu_3\rbrack}{1}}  \, \, \,\,\,\,\qquad \quad   \qquad  \quad  \qquad   \qquad {\overset{\lbrack \fsu_3\rbrack}{1}}   \\
\lbrack \fe_6\rbrack \, \, 1  \, \, {\overset{\mathfrak{su}_{3}}{3}}  \, \,   1  \, \, \underbrace{{\overset{\mathfrak{e}_{6}}{6}}  \, \, 1  \, \, {\overset{\mathfrak{su}_{3}}{3}}  \, \,   1  \, \, {\overset{\mathfrak{e}_{6}}{6}} \ldots {\overset{\mathfrak{e}_{6}}{6}} \ldots {\overset{\mathfrak{e}_{6}}{6}}\, \ 1  \, \, {\overset{\mathfrak{su}_{3}}{3}}  \, \,   1  \, \,{\overset{\mathfrak{e}_{6}}{6}}}_{M\times}  \, \, 1  \, \, {\overset{\mathfrak{su}_{3}}{3}}  \, \,   1  \, \, \lbrack \fe_{6}\rbrack   \,.
\end{array}
\end{align}
It is straightforward to read off the LS charges of the tensors, which is given as
\begin{align}
	\begin{array}{l}
	\qquad  \qquad  \, \,\,\, \,	 \qquad 1 \qquad  \qquad  \qquad \qquad \qquad \qquad \qquad \quad  1\\
	\vec{l}_{LS}=(1, \, \, 1, \, \,  2, \, \,  \underbrace{ 1, \, \, 3,\, \,2,\, \,3,\, \,1,   \, \, \ldots 1, \,\,  \ldots 1, \,\, 3,\,\, 2,\,\, 3,\,\, 1, }_{\times M} \, \, 2,\, \,1,\, \,1  )   \, .   
	\end{array}
\end{align}		
The resolution process is fully analogous and can be obtained, when exchanging the $SU(9)$ flavor group for $E_6$ as given in the table together with its matching conformal matter. 
The resulting theory admits again two more identical fibration with an $F_{13}$ fibre type which hosts an $U(1) \times \mathbb{Z}_2$ MW group. The resulting quiver is given as 
\begin{align} 
\lbrack \mathfrak{u}_{12}\rbrack \, \,  \overset{\fsu_{2M+6}}{2} \, \, \overset{\fsu_{4M}}{2} \, \, \overset{\fsp_{3M-3}}{1}  \, \,  \overset{\fso_{4+4M}}{4}  \, \,  \overset{\fsp_{M-1}}{1}  \, \,  \lbrack \fso_{8} \rbrack \, , \quad \vec{l}_{LS}=(1, \, \, 2, \, \,  3,  \, \,1, \, \,  1    )   \, .  
\end{align} 
 Note that the theory is very similar to the one before but with a slightly modifed flavor group. The 2-groups and CB dimension can be matched and are given below as  
 \begin{align}
	\begin{array}{|c|c|c|}\hline
	& [\fe_6]-\fe_6^M-[\fe_6] &  \text{T-dual}  \\
	\hline
\text{Dim(CB)} & \multicolumn{2}{|c|}{12M+6}  
 \\ \hline
	\widehat{\kappa}_{\mathscr R} &\multicolumn{2}{|c|}{24M+2} \\
	\hline
	\end{array} 
\end{align}

Similarly, we can consider the T-duals of the $SU(9)^2/\mathbb{Z}_3$ type of flavor groups probing $\fe_6^M$ singularities. Their tensor branch is just as in the example above but with the quiver
\begin{align}     
\begin{array}{l }   
\lbrack \fsu_9\rbrack \, \,  {\overset{\mathfrak{su}_6}{2}}  \, \, {\overset{\mathfrak{su}_{3}}{2}}  \, \,   1  \, \, \underbrace{{\overset{\mathfrak{e}_{6}}{6}}  \, \, 1  \, \, {\overset{\mathfrak{su}_{3}}{3}}  \, \,   1  \, \, {\overset{\mathfrak{e}_{6}}{6}} \ldots {\overset{\mathfrak{e}_{6}}{6}}}_{\times M}   \,\,  1 \, \, {\overset{\mathfrak{su}_3}{2}} \,\, {\overset{\mathfrak{su}_6}{2}}  \,  \,  \lbrack \fsu_{9}\rbrack     \, .   
\end{array}
\end{align}
It is straightforward to read off the LS charges of the tensors, which is given as
\begin{align}
N_I = (1, \, \, 2, \, \,3,  \, \,\underbrace{1, \, \,
 3, \, \,2, \, \,3, \, \,   1, \, \, \ldots  \,\,
1,}_{\times M}  \, \,  3, \, \, 2, \, \,1) \, ,
\end{align} 

The toric resolution is realized by the combination of the $\fsu_9$ flavor rays $\beta_i$ and the $\mathcal{T}(\fsu_9,\fe_6)$ conformal matter, then reflect them to the $(x,y,-1,0)$ side. There are two self-dual theories with the fiber being the $F_{15}$ type, which admits a free and a torsional MW generator. Therefore the total quiver is given as
\begin{align}
\lbrack \mathfrak{u}_{10} \rbrack \, \,  \overset{\fsu_{6+2M}}{2} \, \, \overset{\fsu_{2+4M}}{2} \, \, \overset{\fsp_{3M-1}}{1}  \, \,  \overset{\fso_{8+4M}}{4}  \, \,  \overset{\fsp_{1+M}}{1}  \, \,  \lbrack \fso_{12} \rbrack  \, , \quad \vec{l}_{LS}=(1, \, \, 2, \, \,  3,  \, \,1, \, \,  1    )   \, .
\end{align}
The 2-groups and CB dimension can be matched and are given below as  
 \begin{align}
	\begin{array}{|c|c|c|}\hline
	& [\fsu_9]-\fe_6^M-[\fsu_9]  &   \text{T-dual}  \\
	\hline
\text{Dim(CB)}  & \multicolumn{2}{|c|}{12M+14}  
 \\ \hline
	\widehat{\kappa}_{\mathscr R} &\multicolumn{2}{|c|}{24M+22} \\
	\hline
	\end{array} 
\end{align}

\begin{figure}[t!]
\begin{picture}(0,100)
\put(270,45){\includegraphics[scale=0.5]{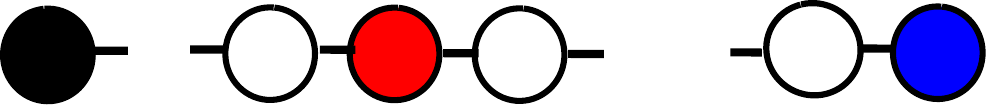}}
\put(30,00){\includegraphics[scale=0.5]{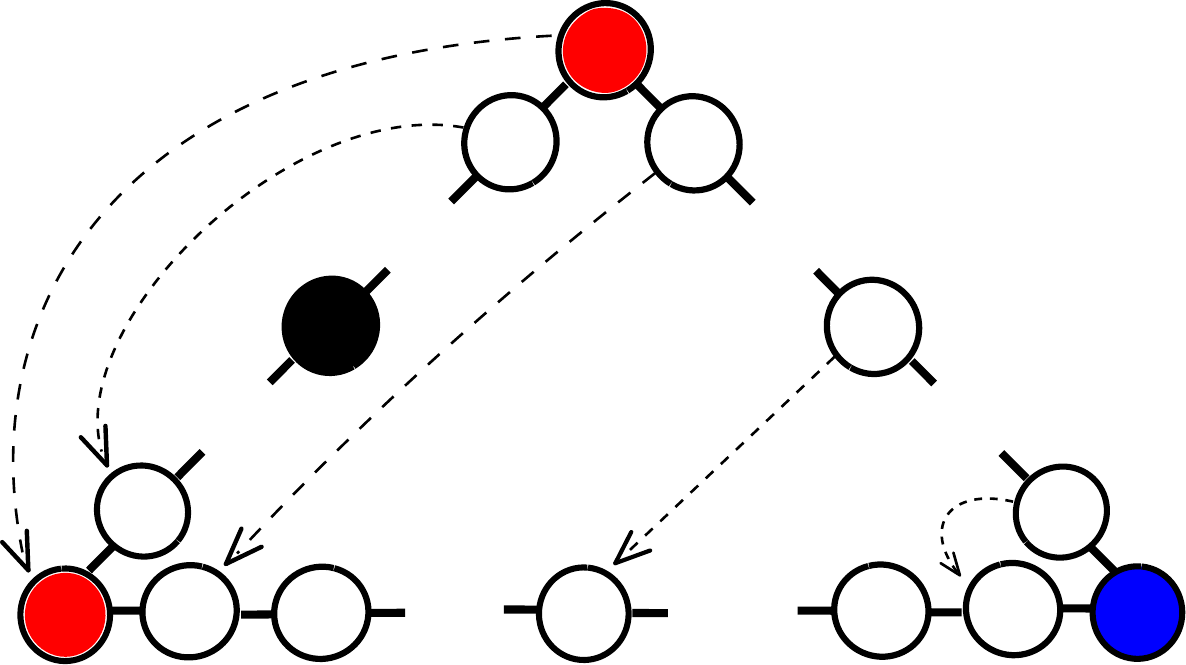}}
\put(210,45){\Large $\xrightarrow{\text{T-dual}}$}
\end{picture}
\caption{\label{fig:SU3nFolding} Depiction of the identification of an $\fsu_{3N}^{(1)}$ fibre under T-duality. The red nodes are mapped on top of each other, carrying the other ones along, leading to the base quiver on the right. From red to blue, there are $N$ nodes, while there are N/2 from red to black, for odd $N$. }
\end{figure}

At next we consider $\fsu_{ 3N}$ gaugings which surprisingly will be more exotic than the $\fe_6$ cases. For simplicity we start with the $(E_6\times SU(3))^2/\mathbb{Z}_3$ flavor group and no additional NS5 brane i.e. M=1, for which we get \begin{align}
\lbrack \fe_6\rbrack\  \, 1 \, \,   {\overset{\mathfrak{su}_{3}}{2}}  \ldots  {\overset{\mathfrak{su}_{3n-3}}{2}} \, \,         {\overset{\mathfrak{su}_{3n}}{ \underset{\lbrack \fsu_3^2 \rbrack\  }{2}}} \ldots   {\overset{\mathfrak{su}_{3n-3}}{2}}  \, \,                     {\overset{\mathfrak{su}_{3}}{2}}     \, \,  1 \, \,              \lbrack \fe_6\rbrack\, \, .
\end{align}
as discussed in \cite{delZottoAppear2}. As the Flavor group is purely determined by the K3, we already know that it is the same as in the models before, but the quiver now looks like  
\begin{align} 
\begin{array}{c }
 \lbrack \mathfrak{u}_{12}\rbrack \qquad \\ 
 \underbrace{1 \, \, \overset{\fsu_{8}}{2} \ldots \overset{\fsu_{4N-8}}{2}}_{\times N/2 } \,\, \overset{\fsu_{4N}}{2 } \,\, \underbrace{\overset{\fsu_{4N-4}}{2} \ldots \overset{\fsu_{4}}{2}\, \,  1   }_{\times (N) \text{ for N even}}\, \, \lbrack  \fso_{8}\rbrack \, ,
\end{array} 
\begin{array}{c }
\lbrack \mathfrak{u}_{12}  \rbrack  \\ 
 \underbrace{\overset{\fsu_{4}}{\underset{[N_A=1]}{1}}     \, \, \overset{\fsu_{12}}{2} \ldots \overset{\fsu_{4N-8}}{2}}_{\times (N-1)/2} \,\, \overset{\fsu_{4N}}{2 } \,\, \underbrace{\overset{\fsu_{4N-4}}{2} \ldots \overset{\fsu_{4}}{2}\, \,  1   }_{\times (N) \text{ for N odd}} \, \, \lbrack \fso_{8}\rbrack \, .
\end{array} 
\end{align}
All LS charges in the above theories are one, the dimension of the Coulomb branch and $\hat{\kappa}_R$ differ by one, namely Dim(CB)+1$=\hat{\kappa}_R=3N^2+2$.
The above result is very surprising as it looks almost like an $\fsp_{2N}^{(1)}$ type  of quiver. In terms of the geometry of the $\fsu_{3N}^{(1)}$ fibral curves, we can deduce this action directly in the T-dual geometry as depicted in Figure~\ref{fig:SU3nFolding}. 
The affine Dynkin diagram shapes exactly as an equilateral triangle with the corners nodes intersected by the $\mathbb{Z}_3$ torsion sections each. Under T-duality, two red nodes are mapped on top of each other, while the blue one stays invariant. Besides, there can be another node invariant under the action, colored in black, in case that $N$ is odd, which is why we had to distinguish those two cases.

\subsection{Quotient base of the heterotic string}
In this section, we want to consider a small class of heterotic LST bases, that is not (directly) related to those of some N NS5 branes supsended between the two $E_8$ walls. 
As we have seen, those configurations can be related to two $1$ curves in between a crepant $\mathbb{Z}_N$ singularity. 

Instead, here we consider the geometry of no NS5 brane, that is M=1 but take a quotient of the configuration 
 \footnote{For more details on those types of bases and a connection to 2-form gauge symmetries, see \cite{Braun:2021sex}}, given as
\begin{align}
    B = (\mathbb{P}^1 \times \mathbb{C}) / \Gamma_M \, .
\end{align}
The quotient $\Gamma_M$ is only crepant for $M=2$ and it can be implemented as a refinement of the 2D lattice lattice  in which the toric rays live
\begin{align}
w_0: (-1,0) \, ,\quad  w_1: (1,0)\, ,  \quad y_0: (1,0)  \rightarrow \hat{y}_M:(1,M) \, .
\end{align}
This action implements the discrete action on
the coordinates, which yields the singular points 
\begin{align}
   \{ \gamma   w_0, \gamma^{ 1} \hat{y}_M\} \qquad \text{ and} \qquad \{ \gamma \hat{y}_M , \gamma^{- 1} w_1\} \, ,
\end{align}
with $\gamma$ being an $M$-th root of unity.
The first singularity is in general a non-crepant singularity, while the second one is. The resolved base quiver  is then given as 
\begin{align}
    M \, \, 1 \, \, \underbrace{2 \, \, 2 \ldots 2}_{\times M-1} \, , \qquad \vec{l}_{LS} = (1, M, M-1, M-2, \ldots 1 )\, .
\end{align}
Torically, we can resolve these singularities by adding the rays that split the singular cones, which yields 
\begin{align}
    w_0: (-1,0) , y_0:(0,1) \, , \hat{y}: (1,M) \,, y_{m}: (1,m) \, , w_1: (1,0) \, ,\text{ for } m=1 \ldots M-1 \, .
\end{align}
This type of bases are limited to the families $M=2 \ldots 12$ in order to allow for a crepant resolvable threefold. Again, the LS charges are given by the second coordinate of the toric vectors and are given as
\begin{align}
    \vec{l}_{LS} = (1,M,M-1, M-2, \ldots ,1 )\, .
\end{align}
From this it is directly evident that the 2-group structure constant is $\widehat{\kappa}_{\mathscr P}$ and hence we have a heterotic LST. This is also clear by noting, the we can blow-down the $1$ curve consecutively again and obtain a base of the type  $\mathbb{P}^1 \times \mathbb{C}$ consistent with the 2-group structure constant.

The presence of the single non-Higgsable cluster  on the left of the chain yields additional singular fibres i.e. gauge groups in the F-theory context \cite{Morrison:2012ei,Morrison:2012np}\, ,
such as $\fsu_3, \fso_8, \ff_5, \fe_6, \fe_7$ for $M=2 \ldots 7$. Some care has to be taken when $M=9$ as this leads to $12-M$ additional $(4,6,12)$ points that can be removed via a $1$ curve blow-up adding to the CB dimension and the 2-group structure constant. In  Table~\ref{tab:Purequot} we have listed all Coulomb branch dimensions and 2-group structure constants of the respective models.
\begin{table}[t!]
    \centering
    \begin{tabular}{|c|c|c|c|c|c|c|c|c|c|c|c|}\hline
    M& 2 & 3 & 4 & 5 & 6 & 7 & 8 & 9 & 10 & 11 & 12 \\ \hline
    CB & 2 & 5 & 8 & 10 & 12 & 14 & 15 & 20& 20 & 20 & 20 \\
 $\widehat{\kappa}_R$ &   3 &   9 & 16 & 24 & 33 & 46 & 54 & 78 &   87 & 97 & 108  \\ \hline
    \end{tabular}
    \caption{Coulomb branches and 2-group structure constants of $(\mathbb{P}^1 \times \mathbb{C})/\Gamma_M$ type of LSTs.}
    \label{tab:Purequot}
\end{table}
 In general, we can add some non-trivial flavor groups and hence also gauge group factors. A couple of simple options where we just decorate by some $\fsu_k$ factors are given as 
 \begin{align}
     \overset{\fsu_3}{3} \, \, 1 \, \,  \overset{\fsu_6}{2}\, \, \overset{\fsu_{12}}{2}\, \, [\mathfrak{u}_{18}] \, , \qquad \{ \text{Dim(CB)}, \kappa_R \} = \{  21,30\}
 \end{align}
 or 
 \begin{align}
        \overset{\fe_6}{6} \, \, 1 \, \,  \overset{\fsu_3}{2}\, \, \overset{\fsu_{6}}{2}
        \, \, \overset{\fsu_{9}}{2}
        \, \, \overset{\fsu_{12}}{2}
        \, \, \overset{\fsu_{15}}{2} 
        \, \, [\mathfrak{u}_{18}] \, , \qquad\{ \text{Dim(CB)}, \kappa_R \} = \{   52,123\}
 \end{align}
 and similarly
 \begin{align}
       \overset{\fe_7}{8} \, \, 1 \, \,  \overset{\fsu_2}{2}\, \, \overset{\fsu_{4}}{2}
        \, \, \overset{\fsu_{6}}{2}
        \, \, \overset{\fsu_{8}}{2}\, \, \overset{\fsu_{10}}{2}
        \, \, \overset{\fsu_{12}}{2} 
        \, \, \overset{\fsu_{14}}{2}  
        \, \, [\mathfrak{u}_{16}] \, , \qquad\{ \text{Dim(CB)}, \kappa_R \} = \{  64,194\}
 \end{align}
 The large rank of the flavor group and the absence of an obvious $\fe_8^2$ HW picture makes it hard to explicitly construct T-dual LSTs\footnote{ The toric construction of the last model does not admit any other fibrations and hence T-dual models. A different phase in the K\"ahler cone however, might admit another fibration.}.  
 
Moreover, the high rank of the flavor symmetry does not allow us to engineer the first two models torically at all. This is due to the reason, that toric geometry confines us to K3s with a rank 17 flavor group at most.

In cases of large flavor groups like those, we might have more general arguments at our disposal to deduce possible T-dual fibrations.

\subsection{Extremal Flavor Rank and T-Hexality}
\label{ssec:ExtremalK3}
We conclude this paper by giving an example that goes beyond the toric construction and illustrates the validity of our more general proposal. 
Here we present Heterotic LSTs that have a maximal flavor group from the K3 perspective. 
In all these examples the Poincar\'e 2-group structure constant $\kappa_{\mathscr{P}}$ equals two, which signals the presence of two M9 branes in the M-theory dual. However, the flavor symmetry here enhances to groups with rank 18. The origin of such a rank 2 enhancement is harder to trace in the Horava-Witten picture \cite{Horava:1995qa,Horava:1996ma}. A possible geometric origin for such an enhancement can be indeed recognised: by construction transverse directions to branes often give rise to global symmetries along their worldvolumes. For M5 branes in M-theory, we indeed have an $SO(5)$ transverse isometry which is interpreted in terms of the 6d (2,0) R-symmetry. In presence of M9 branes such $SO(5)$ is broken to $SO(4) \simeq SU(2) \times SU(2)$ where the first factor is the 6d (1,0) R-symmetry, while the second factor is enhancing the global symmetry to $SU(2) \times E_8 \times E_8$. In the non-geometric Heterotic theories we are constructing in this section we see an enhancement of rank two. We believe this is related to the extra dimensions of F-theory, which give a further transverse direction which justifies the further rank enhancement. In this sense the Heterotic/F-theory duality building on K3 is giving a slight generalisation of the HW picture.

As usual, our starting points will be the elliptic K3s fibres that encode the flavor group of the LST configuration once we add additional singularities. 
The underlying K3s discussed here differ drastically from those considered in the bulk of this work: I.e., any of the 3145 toric Kreutzer-Skarke K3s can at most have an NS lattice of rank 19 and hence a frame lattice of rank 17.\footnote{Since all toric hypersurfaces are geometric, they must admit at least one K\"ahler class inherited from the ambient space. On the other hand, all toric polytopes are closed under the mirror symmetry. As there exists at least one K\"ahler modulus, there must also be at least one dual complex structure modulus. For K3s, this implies that the NS lattice can never fill out the full 20-dimensional second cohomology.}. However, there exist plenty of elliptic K3s with maximal NS lattice as classified in \cite{Miranda1986,2005math......5140S}. To get a better understanding of the underlying geometry, it is beneficial to view the flavor configuration inherited from the K3 from an 8D SUGRA perspective \cite{Font:2016odl,Font:2017cmx,Font:2017cya}.  

Those cases admit a Weierstrass description as well as a heterotic dual in terms of a (Narain-) torus, simply due to the perfect match of 8D F-theory and heterotic string vacua. Due to a large amount of symmetries, these lack a geometric description in the heterotic string perspective (see \cite{delZottoAppear2} for more details)
E.g., in the following, we consider the heterotic string with a $E_8^2 \times SU(3)$ flavor group. In this case, the 8D heterotic string is compactified on a torus with fixed complex structure $\tau = e^{-2\pi i/3}$ at the self-dual radii, which enhances the $\fsu_2^2$ symmetry to $\fsu_3$. This theory maps to F-theory on an extremal rational elliptic surface, with the very same type of fibres.
\begin{figure}[t!]
\begin{center}
    \begin{picture}(0,130)
    \put(-70,10){\includegraphics[scale=0.4 ]{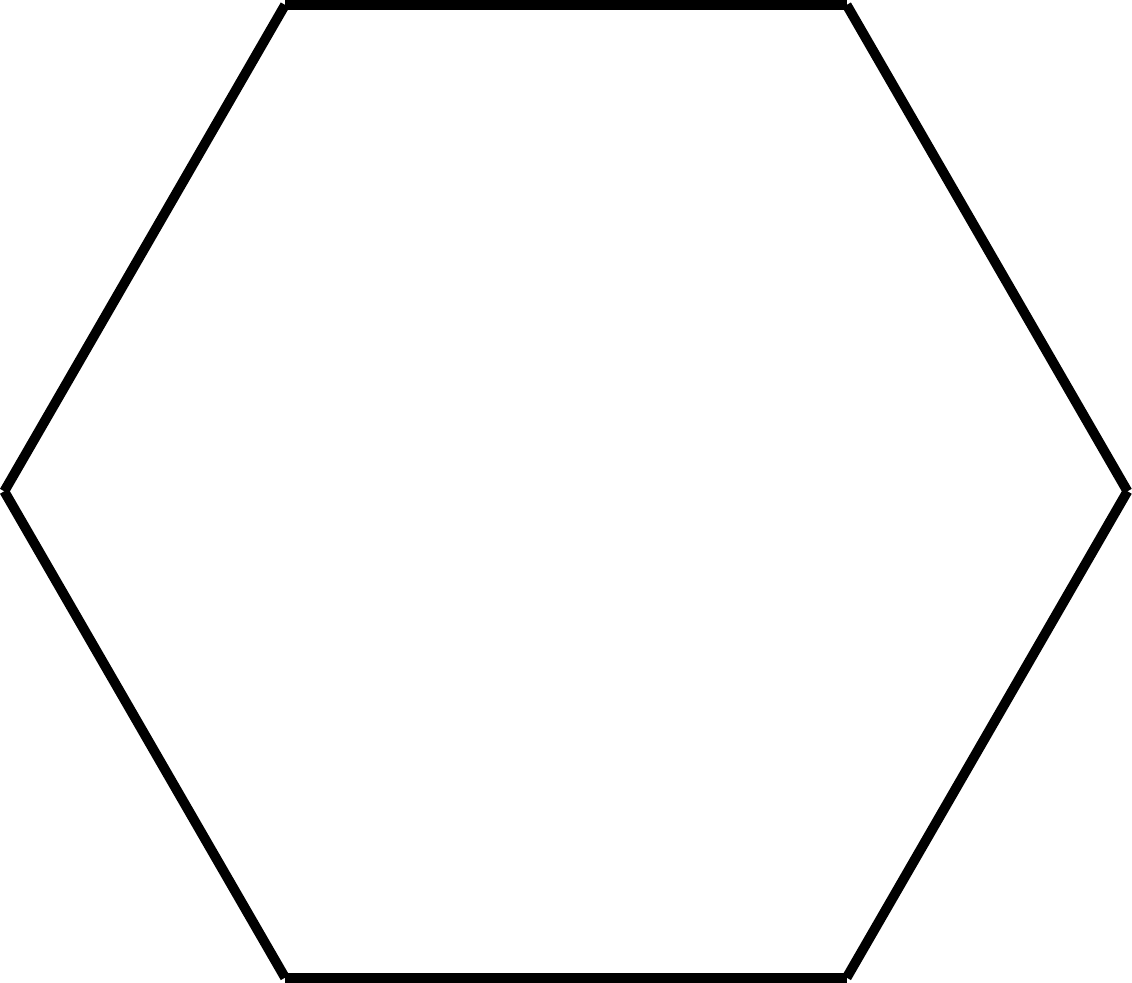}}
    \put(-105,115){$E_8^2 \times SU(3)$}
    \put(35,115){$ \text{Spin}(32)/ \mathbb{Z}_2 \times SU(3)$}
    \put(-115,65){$E_6^3/\mathbb{Z}_3$}
     \put(70,65){$U(18)/\mathbb{Z}_3$}
     \put(-165,10){$(\text{Spin}(14)\times SU(12))/\mathbb{Z}_4$}
     \put(35,10){$(\text{Spin}(20)\times E_7)/\mathbb{Z}_2 \times U(1)$}
     \put(-35,63){\framebox{T-Hexality}}
    \end{picture}
\end{center}
\caption{\label{fig:K3Hexagon}
The six different flavor groups, obtained from different elliptic fibration structure of the same K3 \cite{Braun:2013yya}. Those map to non-geometric heterotic string backgrounds and yield the universal flavor groups of six different 6D LSTs.
}
\end{figure}
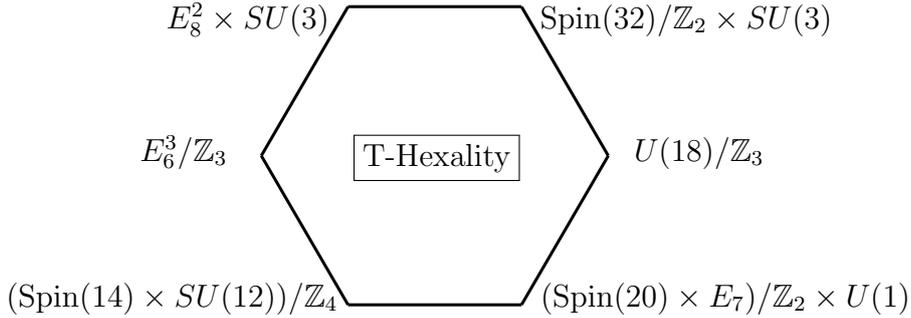

As pointed out in Section~\ref{ssec:K3Degens}, we expect T-duality of LSTs to be inherited from multiple fibration structures of the underlying non-compact elliptic K3 fibre. Hence, the problem of determining inequivalent elliptic fibrations reduces to classify elliptic fibrations of the K3. This is not an easy task in general, but lattice theoretic techniques of K3 can substantially help to do so. For some cases, such classifications have been done  \cite{Braun:2013yya}, and it includes the $E_8^2 \times SU(3)$ case discussed before as one of six different elliptic fibrations in the same K3. This Hexality of dual theories is given in  Figure~\ref{fig:K3Hexagon}.

Note that we have reversed the logic that is usually applied in order to construct LSTs or SCFTs: We start by knowing the flavor group but without knowing the actual tensor branch and gauge group of the full configuration. In fact, there might be multiple configurations with the same flavor group, as we have seen for the other $\fe_8^2$ cases before, but with different tensor branches\footnote{Note that the 2-group symmetries lead in general to an additional set of global symmetries to distinguish those theories.}. Finding a working configuration, however, is highly non-trivial due to the rigidity of the extremal K3, which makes an ADE classification with arbitrary NS5 brane numbers impossible. Instead, we should expect only very few configurations, if at all.

By conjecture, though, we expect that any consistent LST configuration with a fixed flavor group, must admit dual theories for each of the other flavor groups and matching Coulomb branch as well as 2-group structure constants.
Then the best starting point is the pure heterotic string i.e. simply a 0 curve probed by some singularity, as this gives us the most flexibility \cite{delZottoAppear2}. 
 
To give evidence for our proposal, we   construct six different LSTs based on the flavor groups as given in Figure~\ref{fig:K3Hexagon} and the following Coulomb branch and 2-group data
\begin{align}
\text{Dim(CB)} = 21 \, , \quad \widehat{\kappa}_{\mathscr R} = 30 \, .
 \end{align}
 This gives very strong evidence that all configurations are T-hexal to each other. Note that all but one theory are discrete holonomy theories as well as that $\text{Dim(CB)} \neq \widehat{\kappa}_{\mathscr R}$ from which follows that we can not have $l_{LS,I}=1$ with type $I_m$ fibres. The restricted structure of the K3 fibre and its elliptic curve will help us to deduce all of them correctly.
 
 \paragraph{1. $\mathbf{E_8^2 \times SU(3)}$ Flavor: } This configuration
 is very close to those we discussed in the bulk of this work but with an enhanced $\fsu_3$ flavor group which coincides with the very same self-dual $\fsu_3$ torus that the heterotic string is compactified on. Since this is the torus that maps to the F-theory torus, we find that the complex structure must be fixed to the very same elliptic point. This rules out singularities such as $\fe_7$. 
 Note that we have constructed already a quite close cousin of his theory with $\text{dim}(CB) = 20 \, ,\widehat{\kappa}_{\mathscr R} = 29 $ and an $E_8^2$ flavor group. It 
 is the $M=1$ configuration given by
 \eqref{eq:E8E8F4Fusion}.
This configuration admits enough room, to add an have an flavor factor, which then yields  
\begin{align}  
\begin{array}{ccc}
 \lbrack \fsu_3\rbrack  \\
 1 \\ 
\lbrack \fe_8\rbrack \, \, 1   \, \,2 \, \,   {\overset{\mathfrak{sp}_1}{2}}  \, \,   {\overset{\mathfrak{g}_2}{3}} \, \, 1 \, \,   \overset{\ff_4}{5}   \, \, 1 \, \,    {\overset{\mathfrak{g}_2}{3}}  \, \,   {\overset{\mathfrak{sp}_1}{2}}  \, \,  2 \,  \,  1 \,  \lbrack \fe_8\rbrack 
\end{array} \, ,  \qquad 
\begin{array}{c }
 \quad \, \, \, 1 \\ 
 \vec{l}_{LS}= (  1,1,1,1,2,1,2,1,1,1,1,1)
\end{array} \, 
\end{align} 
\paragraph{2. $\mathbf{ (SO(32)/\mathbb{Z}_2 \times SU(3)}$ Flavor:}In this case we simply T-dualize the known configuration of the one before.
This duality is simply inherited from 10D and thus, we still expect simply the heterotic string to be on an $\fsu_3$ self-dual torus that the $\fso_{32}$ heterotic string is compactified on. Moreover, we already know the T-dual of configuration \eqref{eq:E8E8F4Fusion} which is \eqref{eq:SO32F4Duals}, which only needs the addition of an $\fsu_3$ flavor group. The only possible location is  at the final $2$ curve, which yields the configuration  
 \begin{align}
\lbrack \fso_{32}\rbrack     \, \,     {\overset{\mathfrak{sp}_8}{1 }}  \, \,   {\overset{\mathfrak{so }_{16}}{4}}  \, \,  1  \, \,    2 \, \,  \underset{[N_F=1]}{\overset{\mathfrak{su}_2}{2}}   \, \,         \lbrack \fsu_{3 }\rbrack \, , \qquad \vec{l}_{LS}=(1, 1, 3, 2, 1) \, .
\end{align}
Notably, the inserted conformal mater that starts from the $\fso_{16}$ curve looks like that of  $\mathcal{T}(\fe_8,\fsu_3)$. Note that  the Mordell-Weil $\mathbb{Z}_2$ modding does affect this sub part of the quiver, which would be inconsistent either with the odd $\fsu_3$ center and the single fundamental of the $\fsu_2$ gauge group.

\paragraph{3. $\mathbf{E_6^3/\mathbb{Z}_3}$ Flavor: } This singular K3 surface admits actually a very nice construction as the Jacobian of the global toroidal quotient  $T^2\times T^2/\mathbb{Z}_3$  \cite{Dasgupta:1996ij,Hayashi:2019fsa,Kohl:2021rxy}. For this orbifold to be consistent, the complex structures of both ambient tori must again be fixed to $\tau = e^{2 \pi i /3}$ which is relevant for the possible F-theory singularities we can add to the LST. From the 8D heterotic
string perspective we obtain a naive $\fsu_3$ self-dual lattice torus in 8D which again would enhance the group to an $\fsu_3$. However this time we have a fully non-geometric compactification where also both $E_8^2$ are broken to $(E_6 \times SU(3))/\mathbb{Z}_3$ by a discrete holonomy. Then the three $\fsu_3$ factors recombine to a third $E_6$. The presence of the MW torsion and the fact that the complex structure is fixed to an elliptic point allows for $\fsu_{3n}$ and $\fe_6$ singularities only. We choose the later which yields the configuration 
\begin{align}  
\begin{array}{c}
 \lbrack \fe_6\rbrack  \\
 1 \\
 {\overset{\mathfrak{su}_3}{3}} \\
 1 \\
 \lbrack   \fe_6\rbrack \, \, 1 \,\, {\overset{\mathfrak{su}_3}{3}}\,  \,  1 \, {\overset{\mathfrak{e}_6}{6}} \, \, 1 \,\, {\overset{\mathfrak{su}_3}{3}}\,  \,  1 \,  \lbrack \fe_6\rbrack 
\end{array}
 \begin{array}{c} 
\quad \qquad \qquad  1 \\
\quad  \qquad \qquad 1 \\
\quad \qquad \qquad 2 \\
\, , \qquad  \vec{l}_{LS}= ( 1 ,\,\, 1,\,  \,  2, \, 1, \, \, 2, \,\, 1,\,  \,  1 \, )
\end{array} 
\end{align}
Note that we admit a full $\mathbb{S}_3$ permutation symmetry of the three tensor legs.

\paragraph{4. $\mathbf{ (SO(14)\times SU(12))/\mathbb{Z}_4)  }$ Flavor:}
For this case, we find a $\mathbb{Z}_4$ MW group which again restricts the admissible singularities to $\fso_{6+4n}$ and $\fsu_{4n}$. Moreover due to the $\fe_6$ singularity appearing in the models before and the  $\fso_n \times \fsu_m$ flavor group here \cite{DelZotto:2022ohj} one might expect a base of type $\ff_4^{(1)}$ similar to what we found in Section~\ref{Sec:Z3Discrete}. Indeed, when choosing an $\fso_{10}$ singularity and insert the right type of conformal matter with $\mathbb{Z}_4$ global structure \cite{Dierigl:2020myk} we obtain 
\begin{align}
\lbrack \fso_{14}\rbrack     \, \,     {\overset{\mathfrak{sp}_2}{1 }}  \, \,   {\overset{\mathfrak{so }_{10}}{4}}  \, \,  1  \, \,    {\overset{\mathfrak{su}_4}{2}} \, \,   {\overset{\mathfrak{su}_8}{2}}  \, \,            \lbrack \fsu_{12}\rbrack \, ,  \qquad \vec{l}_{LS}=(1, 1, 3, 2, 1) \, .
\end{align}
 
 \paragraph{5. $\mathbf{  U(18)/\mathbb{Z}_3   }$ Flavor:} For this case, we again have restricted monodromies due to the $\mathbb{Z}_3$ MW group. I.e. when gauging with a $\fsu_3$ type of singularity one obtains the chain
\begin{align}
   \, \,   {\overset{\mathfrak{su }_{3}}{3}}  \, \,  1  \, \,    {\overset{\mathfrak{su}_6}{2}} \, \,   {\overset{\mathfrak{su}_{12}}{2}}  \, \,            \lbrack \mathfrak{u}_{18}   \rbrack  \, ,  \qquad \,  \vec{l}_{LS}=(1, 3, 2, 1) \, .
\end{align}
Note here, that we require the $\mathfrak{u}_1$ flavor to complete the $\fe_8$ flavor symmetry of the E-string curve, when being attached to the $\fsu_3$ and $\fsu_6$ neighbours.
\paragraph{6. $\mathbf{ (SO(20)\times E_7)/\mathbb{Z}_2 \times U(1)  }$  Flavor:}
The final configuration is similar as before, but with an $\mathbb{Z}_2$ MW group as well as an overall $\mathfrak{u}_1$ part  
\cite{Dierigl:2020myk} we obtain 
\begin{align}
\lbrack \fso_{20}\rbrack     \, \,     {\overset{\mathfrak{sp}_4}{1 }}  \, \,   {\overset{\mathfrak{so }_{12}}{4}}  \, \,  1  \, \,    {\overset{\mathfrak{su}_2}{2}} \, \,   {\overset{\mathfrak{so}_7}{3}}  \, \, 
{\overset{\mathfrak{su}_2}{2}} \, \,  
1 \, \, 
\lbrack \fe_{7}\times \mathfrak{u}_1 \rbrack   \, ,   \qquad \vec{l}_{LS}=(1, 1, 3, 2,1,1, 1) \, .
\end{align}
 This concludes all six models.
 
At this point, we have not directly proven the full birational equivalence of the six underlying threefolds.
However there is very strong evidence for it, as the 
six different fibration structure are all part of the same K3 and in particular the Coulomb branch and 2-group structure constants match.   We also note that it is highly non-trivial for those six models to exist at all: On the one hand this is because the K3 fibre in each case is rigid and thus also the elliptic fibre in it, which fixes itself to specific values. On top of that, we often times find finite MW groups that constrain the global structure considerably, such that only a handfull of singularities are possible over the compact curves. 

  E.g. for the $\mathbb{Z}_3$ or $\mathbb{Z}_4$ model, there is almost no other choice but the ones we proposed.

Also adding any additional NS5 brane before adding some singularity would essentially not be possible as they lead to non-crepant singularities. This makes LSTs based on such extremal flavor groups an interesting and concise class of models to classify. We hope to return to this question in future work.
 
\section{Conclusion and Outlook}
This work scratches just the surface of the vast geometrical interplay which underlies little string theories and their networks of T-dualities.  Here we have taken a  deep dive into the  geometric engineering of the Heteorotic theories using F-theory on non-compact elliptic threefolds, giving a complementary analysis to the one of our previous paper, where field theoretical methods were exploited to predict novel T-duals from the match of the 2-groups  \cite{DelZotto:2022ohj}. Here we probe such conjectures exploiting geometric techniques confirming the field theoretical predictions. Our technique is to exploit  hypersurfaces in non-compact toric varieties, which admit multiple elliptic fibration structures, inherited from the ambient space. This leads us to propose a couple of universal features purely from geometry: First we find that all threefolds must underlie a nested elliptic K3 fibration structure over a complex line. The K3 structure makes the Heterotic/F-theory duality manifest and encodes the flavor group which bounds the full 6D flavor group to be of rank 18 at most. The elliptic K3 fibre may split into multiple compact surfaces, akin to a finer generalisation of Kulikov degenerations, which corresponds to the location of the LST, where more interesting degrees of freedom are localised.

In the examples we consider here T-duality is geometrically realized, when the threefold admits multiple inequivalent fibration structures. In all the examples considered in this work this structure directly descends from a property of the K3 fiber, thus akin to an eight dimensional SUGRA origin for the Heterotic LST T-dualities. 
\medskip

We use those general considerations to construct a plethora of LSTs using toric geometry techniques. Concretely, we start by considering NS5 branes that probe some ADE singularity and some flat connections that break the $E_8^2$ flavor group and map it to the corresponding $Spin(32)/\mathbb{Z}_2$ T-dual, focusing mainly on exceptional groups.  
The toric framework we use, has the advantage that it makes many physics properties very simple to read off
: First, different elliptic fibrations are realized 
via inequivalent 2D sub-polytopes. This relates
fibral curves in one fibration, to base curves in another. This clarifies the fibre-base duality flip geometrically, where the base topology resembles the affine $\mathfrak{g}^{(1)}$ topology, of the respective gauge group in the dual configuration. This perspective also allows to directly infer the little string charges in terms of the Kac label of the respective singularity. Second, we also explore many  exotic little string theories that yield interesting new behavior under T-duality:  We engineer non-simply connected flavor groups via a discrete flat connections that break the $\fe_8$s  in a rank preserving way \cite{Aspinwall:1998xj}. Those models generically admit up to three distinct T-dual models with  bases that  resemble the structure of a twisted affine diagram $\mathfrak{g}^{(n)}$. The structure of the gauge groups we find in this context is of course compatible with the analysis of real versus complex or quaternionic representations according to the MacKay correspondence \cite{Douglas:1996sw,Blum:1997fw,Blum:1997mm,Intriligator:1997dh}. 

Finally we consider cases that are neither confined to $\mathfrak{e}^8$ flavor holonomies, nor to toric geometry: We discuss LSTs with the maximal possible rank 18 flavor groups which we argue must map to non-geometric heterotic compactifications  \cite{Font:2016odl,Font:2017cya} in 8D. The F-theory geometry is based on extremal elliptic K3 surfaces, which due to their rigidity is quite remarkable to exist at all. In fact we are able to construct six different LSTs with maximal Flavor ranks that are closed under a T-hexality,
which is fixed solely by analysing the inequivalent elliptic structures for their fixed K3 fibre, building on \cite{Braun:2013yya}. This is probably the most interesting result in this paper, and we plan to explore this connection further to develop a better dictionary between Heterotic LSTs in geometric and non-geometric cases and elliptic/genus-one structures of K3 manifolds. This will give a very efficient way of generating very large families of T-dual models.
  
\medskip
  
While this work puts first steps towards a complete understanding of heterotic LSTs, we leave various future research directions: First, a more complete exploration of all other types of flavor groups and gaugings for the heterotic LSTs, potentially with the help of machine learning techniques would be desirable. Secondly, we have only considered heterotic LSTs but not the cases corresponding to non-Heterotic LSTs, given by systems without M9 branes (i.e. with 
$\hat{\kappa}_p=0$), that required the extension to a CICY base in terms of our toric framework. From this perspective it would also be extremely interesting to try to construct consistent geometries with $\hat{\kappa}_p=1$, corresponding to little CHL-like strings \cite{Chaudhuri:1995fk}.
Third, we have so far left out the possibility for twisted compactifications, following the methods used in \cite{Braun:2014oya,Bhardwaj:2019fzv} and \cite{DelZotto:2020sop,AndersonToAppear,Anderson:2018heq,Anderson:2019kmx,Bhardwaj:2022ekc} we expect the latter are related to genus-one fibered threefolds in F-theory, and it would be extremely fruitful to explore their interplay with genus-one fibered K3s, for the case of Heterotic models.

\medskip
 
 Finally it would be very interesting to connect heterotic LSTs directly with the mathematical classification of degenerate K3 surfaces: As pointed out, the reducible K3 fibrations we find share many features with Kulikov degenerations. Unfortunately the Kulikov classification itself is too coarse for our purposes as it allows for quadratic base changes and no higher multiplicity surfaces. \footnote{See \cite{Lee:2021qkx} for a generalization that includes specifically elliptic K3s and some more general degeneration's.} Classifying heterotic LSTs should then be equivalent to the classification of such generalized elliptic Kulikov degenerations, with a close interplay with the classification of Heteoritc bundles on ALE generalising the beautiful results by Friedman-Morgan-Witten to this more general scenario \cite{Friedman:1997ih,Friedman:1997yq}.

\section*{Acknowledgements}
We thank Andreas Braun, Jonathan Heckmann, Joseph Minahan, Thorsten Schimannek and Timo Weigand for discussions. The work of MDZ, PK and ML has received funding from the
European Research Council (ERC) under the European Union’s Horizon 2020 research
and innovation programme (grant agreement No. 851931). MDZ also acknowledges
support from the Simons Foundation Grant \# 888984 (Simons Collaboration on Global
Categorical Symmetries). PK has also received funding from  the NSF CAREER grant PHY-1848089. PK also thanks the Cornell group for hospitality during the completion of this work.
 
\begin{table}
	\centering
	\resizebox{0.95\textwidth}{!}{%
	\begin{tabular}{c|c|c|c|c}
	\hline
$\mathfrak{g}$ & $F(\boldsymbol{\mu_1},\boldsymbol{\mu_2})$ & $\widehat{\kappa}_{\mathscr R}$ &T-dual theory description & Dim(CB) \\
	\hline
	\hline
&&&& \\ 	
$\fe_6^M$ & $E_8\times E_8$ & $24M+50$ 
& $[\fso_{32}] \, {\overset{\mathfrak{\fsp}_{M+9}}{1}} \, {\overset{\fso_{4M+20}}{4}}\,  {\overset{\mathfrak{\fsp}_{3M+3}}{1}} \, {\overset{\mathfrak{\fsu}_{4M+4}}{2}} \, {\overset{\mathfrak{\fsu}_{2M+2}}{2}} \,  $ & $12M+30$\\ 
&&&& \\ 
& $E_8\times E_7$ & $24M+34$ 
& $[\fso_{28}] \, {\overset{\mathfrak{\fsp}_{M+7}}{1}} \, {\overset{\fso_{4M+16}}{4}}\,  {\overset{\mathfrak{\fsp}_{3M+1}}{1}} \, {\overset{\mathfrak{\fsu}_{4M+2}}{2}} \, {\overset{\mathfrak{\fsu}_{2M+2}}{2}} \,[\fsu_2] \,  $ & $12M+22$\\ 
&&&& \\ 
& $E_8\times E_6\times U(1)$ & $24M+25$ 
& $[\mathfrak{u}_1 \times \fso_{26}] \, {\overset{\mathfrak{\fsp}_{M+6}}{1}} \,  {\overset{\fso_{4M+14}}{4}} \, {\overset{\mathfrak{\fsp}_{3M}}{1}} \,  \underset{[N_F=1]} {\overset{\fsu_{4M+1 }}{2}}  \, \,  \underset{[N_F=1]} {\overset{\fsu_{2M+1 }}{2}} \,  $ \, & $12M+17$\\ 
	&&&& \\
& $E_7\times E_7 $ & $24M+18$ 
& $[\fso_{24}] \, {\overset{\mathfrak{\fsp}_{M+5}}{1}} \, {\overset{\fso_{4M+12}}{4}} \,  {\overset{\mathfrak{\fsp}_{3M-1}}{1}} \,  {\overset{\fsu_{4M}}{2}}\, \overset{\fsu_{2M+2}}{2} [\fsu_2^2] \,  $ & $12M+14$\\ 
&&&& \\
& $E_7\times E_6\times U(1)$ & $24M+9$ 
& $[\fso_{22}] \, {\overset{\mathfrak{\fsp}_{M+4}}{1}} \,  {\overset{\fso_{4M+10}}{4}} \, {\overset{\mathfrak{\fsp}_{3M-2}}{1}} \,  \underset{[N_F=1]}  {\overset{\mathfrak{\fsu}_{4M-1}}{2}} \,  \underset{[N_F=1]}  {\overset{\mathfrak{\fsu}_{2M+1}}{2}} \, [\mathfrak{u}_{2}]\, $ \, & $12M+9$\\ 
&&&& \\
& $(E_6\times U(1))^2$ & $24M$ 
& $[\fso_{20}] \, {\overset{\mathfrak{\fsp}_{M+3}}{1}} \,  {\overset{\fso_{4M+8}}{4}} \,  {\overset{\mathfrak{\fsp}_{3M-3}}{1}} \, \underset{[\fsu_{2}]} {\overset{\mathfrak{\fsu}_{4M-2}}{2}} \, {\overset{\mathfrak{\fsu}_{2M}}{2}} \, [\fsu_{2}]  $ \, & $12M+4$\\ 
&&&& \\
& $E_6\times E_6$ & $24M+42$ 
& $ \lbrack \fso_{20}\rbrack \overset{\fsp_{5+M}}{1}\, \, \overset{\fso_{16+4M}}{4}\, \, \underset{[N_F=2]}{\overset{\fsp_{3+3M}}{1}}\, \, \overset{\fsu_{4+4M}}{2}\, \, \overset{\fsu_{2+2M}}{2}  $ \, & $12M+24$\\ 
&&&& \\
& $(E_6\times SU(9)\times SU(3))/\mathbb{Z}_3$ & $24M+10$ 
& $[\mathfrak{u}_{11}] \, {\overset{\mathfrak{\fsu}_{2M+6}}{2}} \,  {\overset{\fso_{4M+1}}{2}} \,  {\overset{\mathfrak{\fsp}_{3M-2}}{1}} \, {\overset{\mathfrak{\fso}_{4M+6}}{4}} \, {\overset{\mathfrak{\fsp}_{M}}{1}} \, [\fso_{10}]  $ \, & $12M+10$\\ 
&&&& \\
& $(SU(9)\times SU(9))/\mathbb{Z}_3$ & $24M+22$ 
& $[\mathfrak{u}_{10}] \, {\overset{\mathfrak{\fsu}_{2M+6}}{2}} \,  {\overset{\fso_{4M+2}}{2}} \,  {\overset{\mathfrak{\fsp}_{3M-1}}{1}} \, {\overset{\mathfrak{\fso}_{4M+8}}{4}} \, {\overset{\mathfrak{\fsp}_{M+1}}{1}} \, [\fso_{12}]  $ \, & $12M+14$ \\ 
&&&& \\
& $(E_6\times E_6\times SU(3)^2)/\mathbb{Z}_3$ & $24M+2$ 
& $[\fso_{8}] \, {\overset{\mathfrak{\fsp}_{M-1}}{1}} \,  {\overset{\fso_{4M+4}}{4}} \,  {\overset{\mathfrak{\fsp}_{3M-3}}{1}} \, {\overset{\mathfrak{\fsu}_{4M}}{2}} \, {\overset{\mathfrak{\fsu}_{2M+6}}{2}} \, [\mathfrak{u}_{12}]  $ \, & $12M+6$\\ 
&&&& \\
\hline
&&&& \\
$\fe_7^M$ & $E_8\times E_8$ & $48M+50$ 
&\hspace{40 mm} $ {\overset{\mathfrak{\fsp}_{2M-2}}{1^*}} \,  $  
\hspace{8.5 mm} $ {\overset{\mathfrak{\fsp}_{M-3}}{1^*}} \,  $   & $18M+29$\\ 
&&& $[\fso_{32}] \, {\overset{\mathfrak{\fsp}_{M+9}}{1}} \, {\overset{\fso_{4M+20}}{4}}\,  {\overset{\mathfrak{\fsp}_{3M+3}}{1}} \,   {\overset{\fso_{8M+8}}{4^*}} \, {\overset{\mathfrak{\fsp}_{3M-1}}{1}} \,  {\overset{\fso_{4M+4}}{4^*}} \,  $& \\ 
&&&& \\ 
& $E_8\times E_7'$ & $48M+25$ 
&\hspace{33 mm} $ {\overset{\mathfrak{\fsp}_{2M-3}}{1^*}} \,  $  
\hspace{8.5 mm} $ {\overset{\mathfrak{\fsp}_{M-3}}{1^*}} \,  $   & $18M+19$\\ 
&&&$[\fso_{28}] \,{\overset{\mathfrak{\fsp}_{M+7}}{1}} \,  {\overset{\fso_{4M+16}}{4}} \, {\overset{\mathfrak{\fsp}_{3M+1}}{1}} \,  {\overset{\mathfrak{\fso}_{8M+4}}{4^*}} \, {\overset{\mathfrak{\fsp}_{3M-2}}{1}} \,  {\overset{\mathfrak{\fso}_{4M+4}}{4^*}} \, [\fsu_{2}] \,  $& \\ 
&&&& \\ 	
& $E_8\times E_6$ & $48M+25$ 
&	\hspace{50 mm} $ {\overset{\mathfrak{\fsp}_{2M-3}}{1^*}} \,  $  
\hspace{8 mm} $ {\overset{\mathfrak{\fsp}_{M-3}}{1^*}} \,  $    & $18M+19$\\ 
&&&$[\mathfrak{u}_1\times \fso_{26}] \, 
\underset{[N_F=1]}  {\overset{\mathfrak{\fsp}_{M+7}}{1}} \,  {\overset{\fso_{4M+16}}{4}} \, {\overset{\mathfrak{\fsp}_{3M+1}}{1}} \,  {\overset{\mathfrak{\fso}_{8M+4}}{4^*}} \, {\overset{\mathfrak{\fsp}_{3M-2}}{1}} \, \underset{[N_F=1]}  {\overset{\mathfrak{\fso}_{4M+4}}{4^*}} \, $ & \\ 
&&&& \\ 	
& $E_7'\times E_7'$ & $48M$ 
&\hspace{31.5 mm} $ {\overset{\mathfrak{\fsp}_{2M-4}}{1^*}} \,  $  
\hspace{7.5 mm} $ {\overset{\mathfrak{\fsp}_{M-3}}{1^*}} \,  $   & $18M+9$\\ 
&&& $[\fso_{24}] \, {\overset{\mathfrak{\fsp}_{M+5}}{1}} \, {\overset{\fso_{4M+12}}{4}} \,  {\overset{\mathfrak{\fsp}_{3M-1}}{1}} \,  {\overset{\fso_{8M}}{4^*}}\, {\overset{\mathfrak{\fsp}_{3M-3}}{1}} \, \overset{\fso_{4M+4}}{4^*} \, [\fsu_{2}^2] $& \\ 
&&&& \\ 	
& $E_7'\times E_6$ & $48M$ 
&	\hspace{31 mm} $ {\overset{\mathfrak{\fsp}_{2M-4}}{1^*}} \,  $  
\hspace{7 mm} $ {\overset{\mathfrak{\fsp}_{M-3}}{1^*}} \,  $    & $18M+9$\\ 
&&&$[\fso_{22}] \, \underset{[N_F=1]} {\overset{\mathfrak{\fsp}_{M+5}}{1}} \,  {\overset{\fso_{4M+12}}{4}} \, {\overset{\mathfrak{\fsp}_{3M-1}}{1}} \,  {\overset{\mathfrak{\fso}_{8M}}{4^*}} \, {\overset{\mathfrak{\fsp}_{3M-3}}{1}} \,  {\overset{\mathfrak{\fso}_{4M+4}}{4^*}} \, [\mathfrak{u}_{2}]\, $ & \\ 
&&&& \\
 & $(E_7\times E_7\times SU(2))/\mathbb{Z}_2$ & $48M+2$ 
&	$[\fsu_{16}\times \mathfrak{u}_1] \, {\overset{\mathfrak{\fsu}_{2M+10}}{2}} \, {\overset{\fsu_{4M+4}}{2}} \, {\overset{\mathfrak{\fsu}_{6M-2}}{2}} \, {\overset{\mathfrak{\fsp}_{4M-4}}{1}} \,{\overset{\mathfrak{\fso}_{4M+4}}{4}} \,   $  & $18M+11$\\ 
&&&& \\
&&&	\hspace{-4 mm} $ {\overset{\mathfrak{\fsp}_{2M-4}}{1^*}} \,  $    & \\ 
&&& $[\fso_{8}] \, {\overset{\mathfrak{\fsp}_{M-1}}{1}} \, {\overset{\fso_{4M+4}}{4}}\,  {\overset{\mathfrak{\fso}_{3M-3}}{4}} \,  {\overset{\fso_{8M^*}}{4^*}}\, {\overset{\mathfrak{\fsp}_{3M-1}}{1}} \,   {\overset{\fso_{4M+12}}{4}} \, {\overset{\mathfrak{\fsp}_{M+5}}{1}} \, [\fso_{24}]\,$ & \\ 
&&&& \\ 
\hline
\end{tabular}
	}
	\caption{T-duality theories verified so far via the matching of 2-group stucture constants, dimension of the Coulomb branch Dim(CB) and the rank of the flavor groups. Note that for $M=1$, the $(-1)$ curve highlighted with $*$ is absent.}
	\label{table:T-duals}
\end{table}

\begin{table}
	\centering
	\resizebox{1\textwidth}{!}{%
	\begin{tabular}{c|c|c|c|c}
	\hline
$\mathfrak{g}$ & $F(\boldsymbol{\mu_1},\boldsymbol{\mu_2})$ & $\widehat{\kappa}_{\mathscr R}$ &T-dual theory description & Dim(CB) \\
	\hline
	\hline
&&&& \\
$\fe_7^M$ & $E_6\times E_6$ & $48M$ 
&\hspace{48 mm} $ {\overset{\mathfrak{\fsp}_{2M-4}}{1^*}} \,  $  
\hspace{7 mm} $ {\overset{\mathfrak{\fsp}_{M-3}}{1^*}} \,  $    & $18M+9$\\ 
&&& $[\mathfrak{u}_1^2\times \fso_{20}] \, {\overset{\mathfrak{\fsp}_{M+5}}{1}} \,  {\overset{\fso_{4M+12}}{4}} \, {\overset{\mathfrak{\fsp}_{3M-1}}{1}} \,  {\overset{\mathfrak{\fso}_{8M}}{4^*}} \, {\overset{\mathfrak{\fsp}_{3M-3}}{1}} \, \underset{[N_F=1]}  {\overset{\mathfrak{\fso}_{4M+4}}{4^*}} \, $& \\ 
&&&& \\
\hline
$\fe_8^M$ & $E_8\times E_8$ & $120M+2$ 
&	\hspace{40.5 mm} $ {\overset{\mathfrak{\fsp}_{3M-5}}{1^*}} \,  $    & $30M+22$\\ 
&&& $[\fso_{32}] \, {\overset{\mathfrak{\fsp}_{M+9}}{1}} \, {\overset{\fso_{4M+20}}{4}}\,  {\overset{\mathfrak{\fsp}_{3M+3}}{1}} \,  {\overset{\fso_{8M+8}}{4}}\, {\overset{\mathfrak{\fsp}_{5M-3}}{1}} \,   {\overset{\fso_{12M-4}}{4}} \, {\overset{\mathfrak{\fsp}_{4M-4}}{1}} \,  {\overset{\fso_{4M+4}}{4}}\,$& \\ 
&&&& \\
& $E_7\times E_7$ & $120M$ 
&	\hspace{41 mm} $ {\overset{\mathfrak{\fsp}_{3M-5}}{1^*}} \,  $    & $30M+20$\\ 
&&& $[\fso_{24}]  \, {\overset{\mathfrak{\fsp}_{M+7}}{1}} \, \underset{[\fsu_2^{2}]} {\overset{\fso_{4M+20}}{4}} \,  {\overset{\mathfrak{\fsp}_{3M+3}}{1}} \,  {\overset{\fso_{8M+8}}{4}}\, {\overset{\mathfrak{\fsp}_{5M-3}}{1}} \, {\overset{\mathfrak{\fso}_{12M-4}}{4^*}} \,  {\overset{\mathfrak{\fsp}_{4M-4}}{1}} \, {\overset{\mathfrak{\fso}_{4M+4}}{4}} $& \\ 
&&&& \\
& $E_8\times E_7$ & $120M+1$ 
&	\hspace{41 mm} $ {\overset{\mathfrak{\fsp}_{3M-5}}{1^*}} \,  $    & $30M+21$\\ 
&&& $[\fso_{28}]  \, {\overset{\mathfrak{\fsp}_{M+8}}{1}} \, \underset{[\fsu_2]} {\overset{\fso_{4M+20}}{4}} \,  {\overset{\mathfrak{\fsp}_{3M+3}}{1}} \,  {\overset{\fso_{8M+8}}{4}}\, {\overset{\mathfrak{\fsp}_{5M-3}}{1}} \, {\overset{\mathfrak{\fso}_{12M-4}}{4^*}} \,  {\overset{\mathfrak{\fsp}_{4M-4}}{1}} \, {\overset{\mathfrak{\fso}_{4M+4}}{4}} $& \\ 
&&&& \\
& $E_8\times E_6$ & $120M-2$ 
&	\hspace{50 mm} $ {\overset{\mathfrak{\fsp}_{3M-5}}{1^*}} \,  $    & $30M+19$\\ 
&&& $[\mathfrak{u}_1\times \fso_{26}]  \, {\overset{\mathfrak{\fsp}_{M+7}}{1}} \,  {\overset{\fso_{4M+18}}{4}} \, \underset{[N_F=1]} {\overset{\mathfrak{\fsp}_{3M+3}}{1}} \,  {\overset{\fso_{8M+8}}{4}}\, {\overset{\mathfrak{\fsp}_{5M-3}}{1}} \, {\overset{\mathfrak{\fso}_{12M-4}}{4^*}} \,  {\overset{\mathfrak{\fsp}_{4M-4}}{1}} \, {\overset{\mathfrak{\fso}_{4M+4}}{4}} $& \\ 
&&&& \\
& $E_7\times E_6$ & $120M-3$ 
&	\hspace{41 mm} $ {\overset{\mathfrak{\fsp}_{3M-5}}{1^*}} \,  $    & $30M+18$\\ 
&&& $[\fso_{22}]  \, {\overset{\mathfrak{\fsp}_{M+6}}{1}} \, \underset{[\mathfrak{u}_2]}  {\overset{\fso_{4M+18}}{4}} \, 
\underset{[N_F=1]}  
{\overset{\mathfrak{\fsp}_{3M+3}}{1}} \,  {\overset{\fso_{8M+8}}{4}}\, {\overset{\mathfrak{\fsp}_{5M-3}}{1}} \, {\overset{\mathfrak{\fso}_{12M-4}}{4^*}} \,  {\overset{\mathfrak{\fsp}_{4M-4}}{1}} \, {\overset{\mathfrak{\fso}_{4M+4}}{4}} $& \\ 
&&&& \\
& $E_6\times E_6$ & $120M-6$ 
&	\hspace{50 mm} $ {\overset{\mathfrak{\fsp}_{3M-5}}{1^*}} \,  $    & $30M+16$\\ 
&&& $[\mathfrak{u}_1^2\times \fso_{20}]  \, {\overset{\mathfrak{\fsp}_{M+5}}{1}} \,  {\overset{\fso_{4M+16}}{4}} \, \underset{[N_F=2]} {\overset{\mathfrak{\fsp}_{3M+3}}{1}} \,  {\overset{\fso_{8M+8}}{4}}\, {\overset{\mathfrak{\fsp}_{5M-3}}{1}} \, {\overset{\mathfrak{\fso}_{12M-4}}{4^*}} \,  {\overset{\mathfrak{\fsp}_{4M-4}}{1}} \, {\overset{\mathfrak{\fso}_{4M+4}}{4}} $& \\ 
&&&& \\
\hline
&&&& \\
$\fsu_4^M$ & $E_8\times E_6$ & $4M+11$ 
&	$[\mathfrak{u}_1\times \fso_{26}] \, {\overset{\mathfrak{\fsp}_{M+5}}{1}} \, \underset{[\fsu_2]} {\overset{\fsu_{2M+5}}{2}} \, \underset{[N_F=1]} {\overset{\mathfrak{\fsp}_{M-1}}{1}} \,  $  & $4M+10$\\ 
&&&& \\ 
\hline
&&&& \\
$\fg_2$ & $E_8\times E_8$ & $12$ 
&	$[\fso_{32}] \, {\overset{\mathfrak{\fsp}_{6}}{1}} \, \underset{[N_S=2]} {\overset{\fso_{7}}{1}} \,   $ & $10$\\ 
&&&& \\ 
\hline
&&&& \\
$\fg_2^2$ & $E_8\times E_8$ & $18$ 
&	$[\fso_{32}] \, {\overset{\mathfrak{\fsp}_{7}}{1}} \, \underset{[N_S=3]} {\overset{\fso_{12}}{1}} \,   $ & $14$\\ 
&&&& \\ 
\hline
&&&& \\
$\ff_4\times \fe_6^{M-2}\times \ff_4$ & $E_8\times E_8$ & $24M+2$ 
&	$[\fso_{32}] \, {\overset{\mathfrak{\fsp}_{M+7}}{1}} \,  {\overset{\fso_{4M+12}}{4}} \, {\overset{\mathfrak{\fsp}_{3M-3}}{1}} \,   {\overset{\mathfrak{\fsu}_{4M-4}}{2}} \, {\overset{\mathfrak{\fsu}_{2M-2}}{2}} $  & $12M+6$\\ 
&&&& \\ 
\hline
&&&& \\
$\ff_4\times \fe_7^{M}\times \ff_4$ & $E_7\times E_7$ & $48M+48$ 
&	\hspace{11.5 mm} $ {\overset{\mathfrak{\fsp}_{2M-2}}{1^*}} \,  $    & $18M+27$\\ 
&&& $    \lbrack \fso_{24}\rbrack \overset{\fsp_{7+M}}{1}\, \, \underset{[\fso_4]}{\overset{\fso_{20+4M}}{4}}  \, \,  \overset{\fsp_{3+3M}}{1} \, \, \overset{\fso_{8+8M}}{4}\, \, \overset{\fsp_{3M-1 }}{1}  \, \, \overset{\fso_{4+4M}}{4} \, \, \overset{\fsp_{M-3 }}{1^*} $ &\\ 
&&&& \\ 
&&&& \\
& $E_6\times E_6$ & $48M+42$ 
&	\hspace{11.5 mm} $ {\overset{\mathfrak{\fsp}_{2M-2}}{1^*}} \,  $ 
& $18M+23$\\ 
&&& 
$\lbrack \fso_{20}\rbrack \overset{\fsp_{5+M}}{1}\, \, \overset{\fso_{16+4M}}{4}   \, \,  \underset{ [ N_F=2] }{\overset{\fsp_{3+3M}}{1}} \, \, \overset{\fso_{8+8M}}{4}\, \, \overset{\fsp_{3M-1 }}{1}  \, \, \overset{\fso_{4+4M}}{4^*} \, \, \overset{\fsp_{M-3 }}{1^*}$    &\\ 
&&&& \\ 
\hline					
\end{tabular}
	}
	\caption{Continuation of \ref{table:T-duals}} 
\end{table}

\begin{landscape}
\begin{table}
	\centering
	\resizebox{1.2\textwidth}{!}{%
	\begin{tabular}{c|c|c|c|c}
	\hline
$\mathfrak{g}$ & $F(\boldsymbol{\mu_1},\boldsymbol{\mu_2})$ & $\widehat{\kappa}_{\mathscr R}$ &T-dual theory description & Dim(CB) \\
\hline
\hline	
&&&& \\
$\ff_4\times \fe_6^{M}\times \ff_4$ & $E_7\times E_7$ & $24M+48$ 
&	$ [\fso_{24}]\, \,  \overset{\fsp_{7+M}}{1} \, \,  \underset{[\fso_4]}{\overset{\fso_{20+4M}}{4}} \, \,  \overset{\fsp_{3+3M}}{1}  \overset{\fsu_{4+4M}}{2}  \overset{\fsu_{2+2M}}{2}$   & $12M+28$\\ 
&&&& \\ 
\hline
&&&& \\
$\fe_6\times \fe_8^{M}\times \fe_6$ & $E_6'\times E'_6$ & $120M+90$ 
&	\hspace{48 mm} $ \overset{[N_F=2]}{\overset{\mathfrak{\fsp}_{3M-1}}{1  }}  \,  $    & $30M+38$\\ 
&&& 
$\lbrack \fso_{20}\rbrack  \,  \,{\overset{\mathfrak{\fsp}_{5+M}}{1}} \,\,    \overset{\fso_{16+4M}}{4}  \, \,    \overset{\mathfrak{\fsp}_{3+3M  }}{1}  \, \,  {\overset{\mathfrak{\fso}_{8M+12}}{4}} \, {\overset{\mathfrak{\fsp}_{5M+1}}{1}}  \,\,  {\overset{\mathfrak{\fso}_{12M+8}}{4 }} \, {\overset{\mathfrak{\fsp}_{4M }}{1}}\,   \,  \overset{\mathfrak{\fso}_{4M+8}}{4} $ &\\
&&&& \\ 
\hline
&&&& \\
$\fsu_{2N}^M$ & $E_8\times E_8$ & $4N^2 + 2N M - 2N + 2$ 
&	$[\fso_{32}] \, {\overset{\mathfrak{\fsp}_{4N+M-1}}{1}} \,  \underbrace{{\overset{\fsu_{8N+2M-10}}{2}} \, \ldots   \, {\overset{\fsu_{8(N-k)+2M-2}}{2}} \,   {\overset{\mathfrak{\fsu}_{2M+6}}{2}} \, {\overset{\mathfrak{\fsp}_{M-1}}{1}}}_{\times N} $  & $4N^2 + 2N M - 2N + 1$\\ 
&&&& \\ 		
\hline	
&&&& \\
$\fsu_{2N+1}^M$ & $E_8\times E_8$ & $4N^2 + 2N M + 2N +M + 2$ 
&	$[\fso_{32}] \, {\overset{\mathfrak{\fsp}_{4N+M+1}}{1}} \,  \underbrace{{\overset{\fsu_{8N+2M-6}}{2}} \, \ldots   \, {\overset{\fsu_{8(N-k)+2M+2}}{2}} \,   {\overset{\mathfrak{\fsu}_{2M+10}}{2}} \, {\overset{\mathfrak{\fsu}_{2M+2}}{1}}}_{\times N} $  & $4N^2 + 2N M + M + 2N + 1$\\ 
&&&& \\ 	
\hline
&&&& \\
$\fso_{4N+8}^M$ & $E_8\times E_8$ &$16N^2 + 8N M + 32N +8M+ 18$
&	\hspace{14 mm} $ {\overset{\mathfrak{\fsp}_{4N + M-1}}{1}} \,  $  \hspace{76 mm} $ {\overset{\mathfrak{\fsp}_{M-1}}{1}} \,  $   & $8N^2 + 4N M + 22N +6M+ 14$ \\ 
&&&$\lbrack \fso_{32} \rbrack    \, \,      \overset{\fsp_{4N+M+7}}{1  }  \overset{\fso_{16N+4M+12}}{4  }  \underbrace{ \overset{\fsp_{8N+2M-2 }}{1  } \, \,   \ldots     \overset{\fsp_{8N-8k+2M-2 }}{1  }     \, \,  \overset{\fso_{16N-16k+4M-4 }}{4  }  \ldots   \, \,   \overset{\fso_{4M+12}}{4  } }_{\times 2N}    \, \,     \overset{\fsp_{M-1}}{1  }  $& \\ 
&&&& \\
&&&& \\
& $E_8\times E_7 \times SU(2)$ &$12N^2 + 8N M + 20N +8M+ 10$
&	\hspace{4 mm} $ {\overset{\mathfrak{\fsp}_{3N + M-2}}{1}} \,  $  \hspace{74.5 mm} $ {\overset{\mathfrak{\fsp}_{M-2}}{1}} \,  $   & $6N^2 + 4N M + 15N +6M+ 8$ \\ 
&&&$\lbrack \fso_{28} \rbrack    \, \,      \overset{\fsp_{3N+M+5}}{1  }  \overset{\fso_{12N+4M+8}}{4  }  \underbrace{ \overset{\fsp_{6N+2M-3 }}{1  } \, \,   \ldots     \overset{\fsp_{6N-6k+2M-3 }}{1  }     \, \,  \overset{\fso_{12N-12k+4M-4 }}{4  }  \ldots   \, \,   \overset{\fso_{4M+8}}{4  } }_{\times 2N}    \, \,     \overset{\fsp_{M-1}}{1  }\,\, \lbrack \fso_{4} \rbrack  $& \\ 
&&&& \\
\hline	
&&&& \\
$\fso_{4N+6}^M$ & $E_8\times E_8$ &$16N^2 + 8N M + 16N +4M+ 6$ 
&	\hspace{-55 mm} $ {\overset{\mathfrak{\fsp}_{4N + M-3}}{1}} \,  $   &$8N^2 + 4N M + 14N +4M+ 5$  \\ 
&&&	$\lbrack \fso_{32} \rbrack    \, \,      \overset{\fsp_{4N+M+5}}{1  }  \underbrace{ \overset{\fso_{16N+4M+4}}{4  }   \ldots     \overset{\fsp_{8N-8k+2M-6 }}{1  }     \, \,  \overset{\fso_{16N-16k+4M-12 }}{4  }  \ldots   \overset{\fsp_{2M+2}}{1  }  }_{\times 2N}        \overset{\fsu_{2M-2}}{2}  $     & \\ 
&&&& \\ 
\hline	
&&&& \\
$\fso_{4N+5}$ & $E_8\times E_8$ &  $16N^2 + 8N +3$
&	\hspace{-43 mm} $ {\overset{\mathfrak{\fsp}_{4N -4}}{1}} \,  $   &  $8N^2 +  10N + 2$ \\ 
&&&$\lbrack  \fso_{32} \rbrack    \, \,      \overset{\fsp_{4N+4}}{1  } \underbrace{ \,\, \overset{\fso_{16N }}{4  }  \,\,    \overset{\fsp_{ 8N-8}}{1  }    \,\,   \overset{\fso_{16N-16 }}{4  } \ldots     \overset{\fsp_{8N-8}}{1  }    \ldots               \overset{\fsp_{8 }}{1  }    \,\,    \overset{\fso_{16 }}{4  }    1}_{2N \times }  \, \,   2     $     & \\ 
&&&& \\ 
\hline
$\fso_{4N+7}$ & $E_8\times E_8$ & $16N^2 + 24N +11$
&	\hspace{-73 mm} $ {\overset{\mathfrak{\fsp}_{4N -2}}{1}} \,  $   & $8N^2 +  18N + 10$\\
&&&$\lbrack \fso_{32} \rbrack    \, \,      \overset{\fsp_{4N+6}}{1  }   \,\, \overset{\fso_{16N+8 }}{4  }  \,\,  \underbrace{  \overset{\fsp_{ 8N-4  }}{1  }    \,\,   \overset{\fso_{16N-8}}{4  } \ldots     \overset{\fsp_{8(N-k)-4}}{1  }    \overset{\fso_{16(N-k)-8 }}{4  } \ldots               \overset{\fsp_{12 }}{1  }    \,\,    \overset{\fso_{24 }}{4  }      \, \,           \overset{\fsp_{4 }}{1  }       \overset{\fso_{7 }}{ \underset{[N_F=2]}{2}}       }_{2N \times }        $     & \\ 
&&&& \\ 
\hline	
\end{tabular}
	}
\caption{Continuation of \ref{table:T-duals}} 
\end{table}
\end{landscape}

\begin{landscape}
	\begin{table}
		\centering
		\resizebox{1.35\textwidth}{!}{%
		\begin{tabular}{c|c|c|c|c}
		\hline
	$\mathfrak{g}$ & $F(\boldsymbol{\mu_1},\boldsymbol{\mu_2})$ & $\widehat{\kappa}_{\mathscr R}$  &T-dual theory descriptions & Dim(CB) \\
		\hline
		\hline
&&&& \\
$\fso_{4N+5}^2$ & $E_8\times E_8$ & $16N^2 + 16N +6$
&	\hspace{-53 mm} $ {\overset{\mathfrak{\fsp}_{4N -3}}{1}} \,  $   &  $8N^2 +  14N + 5$  \\
&&&	$\lbrack \fso_{32} \rbrack    \, \,      \overset{\fsp_{4N+5}}{1  }   \underbrace{ \overset{\fso_{16N+4}}{4  }  \,\,     \overset{\fsp_{8N-6}}{1  } \ldots   \overset{\fso_{16(N-k)+4 }}{4  }\, \,   \overset{\fsp_{8(N-k)-6 }}{1  }     \ldots               \overset{\fso_{20 }}{4  }  \, \,     \overset{\fsp_{2 }}{1  }}_{2N \times }  \, \,   \overset{\fsu_{2}}{2  }  $     & \\ 
&&&& \\ 
\hline	
&&&& \\
$\fso_{4N+7}^2$ & $E_8\times E_8$ & $16N^2 + 32N +18$
&	\hspace{-65 mm} $ {\overset{\mathfrak{\fsp}_{4N -3}}{1}} \,$   & $8N^2 +  22N + 14$  \\
&&&		$\lbrack \fso_{32} \rbrack    \, \,      \overset{\fsp_{4N+7}}{1  }  \overset{\fso_{16N+12}}{4  }  \,\,    \underbrace{  \overset{\fsp_{8N-2}}{1  }   \, \,  \overset{\fso_{16N-4 }}{4  }   \ldots   \overset{\fsp_{8(N-k)-2 }}{1  } \, \, \overset{\fso_{16(N-k)-4 }}{4  }     \ldots               \overset{\fsp_{6 }}{1  }  \, \,        \overset{\fso_{12 }}{   \underset{ \lbrack  N_S=2 \rbrack }{2}  }             }_{2N \times }  $     & \\ 
&&&& \\ 
\hline
&&&& \\
$\fso_{4N+8}^M$ & $(E_7\times E_7\times SU(2))/\mathbb{Z}_2$ &   $8N^2 + 8NM+8N+8M +2$
&	\hspace{3.5 mm} $ {\overset{\mathfrak{\fsp}_{2N+M -3}}{1}} \,  $ \hspace{96 mm} $ {\overset{\mathfrak{\fsp}_{M -3}}{1^*}} \,$   &$4N^2+4NM+8N+6M+2$ \\
&&&		$\lbrack \fso_{24}\rbrack \, \,  \overset{\fsp_{2N+M+3}}{1 }  \, \, \overset{\fso_{8N+4M+4}}{4 } \, \, \underbrace{ \overset{\fso_{8N+4M-4 }}{4 } \, \, \overset{\fsp_{4N+2M-8}}{1 } \ldots    \overset{\fso_{8N+4M-8k+4 }}{4 } \, \, \overset{\fsp_{4N+2M-4-4k}}{1 }   \ldots
\overset{\fsp_{2M}}{1 } \, \,  \overset{\fso_{4M+4}}{4^* } }_{2 N\times } \, \,  \overset{\fsp_{M-1}}{1 } \, \,  
\, \,  \lbrack \fso_{8}\rbrack    $  & \\ 
&&&& \\
&&&	\hspace{-47 mm} $ {\overset{\mathfrak{\fsu}_{4N+2M+6}}{2}} \,  $    & \\ 
&&& $[\fsu_{16}\times \mathfrak{u}_1] \, \, \overset{\fsu_{8N+4M-4}}{2}  \, \, \underbrace{\overset{\fsu_{8N+4M-12}}{2}\, \, \overset{\fsu_{8N+4M-20}}{2} \ldots \overset{\fsu_{4M+4}}{2} \, \, \overset{\fsp_{2M-2}}{1} }_{N \times}$ & \\ 
&&&	\hspace{-47 mm} $ {\overset{\mathfrak{\fsu}_{4N+2M-2}}{2}} \,  $    & \\ 
&&&& \\ 
\hline
&&&& \\
$\fsu_{2N}$ & $(E_7\times E_7\times SU(2))/\mathbb{Z}_2$ & $2N^2 + 2$ 
& $ \lbrack \fso_{24}\rbrack \, \,  \overset{\fsp_{2N}}{1 }  \underbrace{  \, \, \overset{\fsu_{4N-4}}{2 }   \ldots \overset{\fsu_{4(N-n)-4}}{2 }  \ldots \overset{\fsu_{4 }}{2 } \, \, 1}_{N\times} \, \,  
\, \,  \lbrack \fso_{8}\rbrack $  & $2N^2  + 1$\\
&&&& \\ 	
&&&\hspace{-20 mm} $ [{\mathfrak{\fsu}_{16}}] $ \hspace{46 mm}  $ [{\mathfrak{\fsu}_{16}}] $ \\ 
&&&	$\underbrace{1 \, \, \overset{\fsu_{8}}{2} \ldots \overset{\fsu_{4N-8}}{2} \,\, \overset{\fsu_{4N}}{2 }\,\,  \overset{\fsu_{4N-8}}{2} \ldots \overset{\fsu_{8}}{2}\, \,  1 }_{\times (N+1) \text{ for N even}}  \,\, , \,\, \underbrace{\overset{\fsu_{4}}{\underset{[N_A=1]}{1}} \, \, \overset{\fsu_{12}}{2} \ldots \overset{\fsu_{4N-8}}{2} \,\, \overset{\fsu_{4N}}{2 }\,\,  \overset{\fsu_{4N-8}}{2} \ldots \overset{\fsu_{12}}{2 }\, \,  \overset{\fsu_{4}}{\underset{[N_A=1]}{1}}}_{\times N \text{ for N odd}}  $  & \\ 
&&&& \\ 	
\hline
&&&& \\
$\fsu_{3N}$ & $(E_6\times SU(3))^2/\mathbb{Z}_3$ & $3N^2 + 2$ 
&  \hspace{-20 mm} $ [{\mathfrak{\fsu}_{12}}\times \mathfrak{u}_1] $ \hspace{52 mm} $ [{\mathfrak{\fsu}_{12}}\times \mathfrak{u}_1] $   & $3N^2  + 1$\\
&&&	$\underbrace{1 \, \, \overset{\fsu_{8}}{2} \ldots \overset{\fsu_{4N-8}}{2}}_{N/2 \times} \,\, \overset{\fsu_{4N}}{2 } \underbrace{\overset{\fsu_{4N-4}}{2} \ldots \overset{\fsu_{4}}{2}\, \,  1 \, \,  \lbrack \fso_{8}\rbrack  }_{\times (N) \text{ for N even}}   $ \,\, , \,\, $\underbrace{\overset{\fsu_{4}}{\underset{[N_A=1]}{1}}     \, \, \overset{\fsu_{12}}{2} \ldots \overset{\fsu_{4N-8}}{2}}_{(N-1)/2 \times} \,\, \overset{\fsu_{4N}}{2 } \underbrace{\overset{\fsu_{4N-4}}{2} \ldots \overset{\fsu_{4}}{2}\, \,  1 \, \,  \lbrack \fso_{8}\rbrack  }_{\times (N) \text{ for N odd}}   $  & \\ 
&&&& \\ 	
\hline
&&&& \\
$\fe_7-\fe_8^M-\fe_7$ & $E_7\times E_7$ & $120M+144$
&	\hspace{37 mm} $ {\overset{\mathfrak{\fsp}_{3M -1}}{1}} \,$   & $30M+54$  \\
&&&		$\lbrack \fso_{24}\rbrack  \,  \,{\overset{\mathfrak{\fsp}_{7+M}}{1}} \,\,    \overset{\fso_{20+4M}}{4}  \, \,    \overset{\mathfrak{\fsp}_{5+3M  }}{1}  \, \,  {\overset{\mathfrak{\fso}_{8M+16}}{4}} \, {\overset{\mathfrak{\fsp}_{5M+3}}{1}}  \,\,  {\overset{\mathfrak{\fso}_{12M+12}}{4 }} \, {\overset{\mathfrak{\fsp}_{4M+2}}{1}}\,   \, {\overset{\mathfrak{\fso}_{4M+12}}{4}}[\fso_4] $   & \\ 
&&&& \\ 
\hline
	\end{tabular}
		}
		\caption{Continuation of \ref{table:T-duals}} 
	\end{table}
\end{landscape}

\appendix

\section{Kulikov model - A quick review}\label{app:Kulikov}
We present a quick recap of its definition \cite{Kulikov_1977,Braun:2016sks,Lee:2021qkx} in Appendix \ref{app:Kulikov}.
\begin{defi}
  Let $\pi: X\rightarrow \text{Disc}$ be a family of surfaces over the unit Disc$:= \{u\in \mathbb{C}|\lvert u \rvert < 1\}$, and $\pi$ is a morphism of a complex manifold $X$, dim$X=3$ such that all fibers $S_u:=\pi^{-1}(u)$ for $u\in \text{Disc}\backslash \{0\}$ are non-singular K3 surfaces. The central fibre $S_0 := \pi^{-1}(0)$ is degenerate.
\end{defi}
These K3 degenerations concern smooth degenerated K3s around a neighborhood of the origin of the base $\mathbb{C}^1$. 
A K3 degeneration is semi-stable if the following two conditions are satisfied:  
\begin{itemize}
  \item The central fiber $S_0=V_0\cup V_1\cup \ldots \cup V_n$ is a divisor of normal crossings.
  \item All $V_i$ are smooth surfaces with multiplicity one.
\end{itemize}

Every degeneration then becomes an inequivalent semi-stable degeneration upon a possible base change $u = s^N$ \cite{kempf2006toroidal} and a chain of birational transformations. Finally, in addition if the total space has $K_X = 0$, then degeneration is a Kulikov model, just as our total space.

Kulikov degenerations are generally classified into three different types via an monodromy matrix $T=exp(N)$ that acts on the second cohomology  $H^2(S_u; \mathbb{Z})$ for $u\neq 0$ around the point of degeneration $u = 0$. The three types are distinguished by the order of the nilpotency of the matrix $N$ as 
\begin{itemize}
  \item \textbf{Type I}: $S_0=V_0$  is a irreducible nonsingular K3 surface with $N=0$.
  \item \textbf{Type II}: $S_0=V_0\cup V_1\cup \ldots \cup V_n$ consists of $M+1>1$ irreducible components intersecting in a chain with $V_0$ and $V_M$ are rational surfaces and $V_1, \cdots , V_{n-1}$ have a minimal model that are elliptic ruled surfaces. The double curves $C_{i,j}=V_i \cap V_j$ are non-empty elliptic curves if and only if $\lvert j- i\rvert = 1$; $N \neq 0$, but $N^2 = 0$.
  \item \textbf{Type III}: $S_0=V_0\cup V_1\cup \ldots \cup V_n$, where all $V_i$ are rational surfaces; The double curves $C_{i,j}$ are rational and form a cycle on each of the surfaces $V_i$; $N^2 \neq 0$, but $N^3 = 0$.
\end{itemize}
The number of irreducible components is common to by a pair of Type II degenerations of K3 surfaces that are birational transformations of one another. The rational surfaces $V_0$ and $V_n$ of one degeneration however may not be isomorphic to each other. 

Note that type II degenerations directly correspond to heterotic LSTs with M NS5 branes, as we discuss in Section~\ref{sec: HeteroticLSTs}. The Kulikov degenerations are in many regards to coarse as they allow for base changes and reducible surfaces of multiplicity one only. As the base is physical, we can not allow for such changes. Moreover, \cite{Braun:2016sks} proposed \textit{non-Kulikov} degenerations (see also \cite{Lee:2021qkx}). Without the base changes, such degenerations can have fractional monodromy actions \cite{Braun:2016sks} that correspond to fractional branes, similar to the perspective in \cite{DelZotto:2014hpa}.
 
\section{Details of all the models in Table \ref{table:T-duals}} 
In this appendix we discuss the T-dual LST pairs that are summarized in Table~\ref{table:T-duals} in more detail.
Here we give more details in particular on the $\fe_n$ models, discuss their general curve structure and number of tensors. For all models we explicitly summarize the LS charges, requires to compute the 2-group structure constants and comment on their global structure.  

\addtocontents{toc}{\protect\setcounter{tocdepth}{1}}
\subsection{The $[\fe_8]-\fe_7^M-[\fe_8]$ LST } 
We construct the $E_8$ theory by the following LST chain of curves
\begin{align}
[\fe_8] \,\,   \underbrace{  \overset{\fe_{7}}{1 }    \,\, \overset{\fe_{7}}{2 }\ldots \overset{\fe_{7}}{2 } \ldots \overset{\fe_{7}}{2 } \, \, \overset{\fe_{7}}{1 } }_{\times M} \,\,  [\fe_8]\, ,
\end{align}
where the $\fe_7$ resolution divisors are given in Table~\ref{tab:ConfMatterE7NE8}. In addition we need to complete the quiver with the superconformal matter theories, which is simply that of $\mathcal{T}(\fe_8,\fe_8)$ but with one tensor less on the $\fe_7$ end. From this data we can compute the number of tensors and 5D Coulom branch as  
\begin{align}
\{ \text{T}, \text{Dim(CB)}\}(\mathcal{T}(\fe_8, \fe_7))=\{ 10,20\} \, ,  \quad \{ \text{T},\text{Dim(CB)} \}(\mathcal{T}(\fe_7,\fe_7))=\{ 5,10 \} \, ,  
\end{align}
and then in total
\begin{equation}
\begin{aligned}
\{ \text{T} ,  \text{Dim(CB)}\}( M\cdot \fe_7 + 2\cdot \mathcal{T}(\fe_8,\fe_7) + (M-1)\cdot \mathcal{T}(\fe_7,\fe_7)  -1\}=\{ 14+6M,29+18M\}\, . 
\end{aligned}
\end{equation} 
The LS charges of the curves, upon inserting the superconformal matter are given as
\begin{align}
	\vec{l}_{LS}=(1, 1, 1, 1, 2, 1, 3, 2, 3, 4, \underbrace{1,   4, 3, 2, 3, 4,  \ldots   1 \ldots  , 4, 3, 2, 3, 4 ,1}_{\times M} 4, 3, 2, 3, 1, 2, 1, 1, 1, 1) \, .
\end{align}
In order to resolve the full elliptic threefold, we can simply use the $\fe_8$ resolutions from Table~\ref{tab:ConfMatterE7NE8}, the $\fe_7^M$'s and $\mathcal{T}(\fe_7^m, \fe_7^{m+1})$ conformal matter.
The $\mathcal{T}(\fe_8, \fe_7^1)$ and $\mathcal{T}(\fe_7^{M}, \fe_8)$ need a slightly different resolution, which can be inferred from those of $\fe_8$ by dropping the one additional tensor ray. For convenience we depict those rays again in   Table~\ref{tab:ConfMatterE7NE8}.
\begin{table}[t!]
\centering
\begin{tabular}{|c|l| }\hline 
\multicolumn{2}{|c|}{$\mathcal{T}(\fe_8, \fe_7^1)$ Conformal Matter } \\ \hline
$s_{-1}$  &$ (-2,-3, - 5 ,1 )  $ \\
$s_{-2 }$ &$ (-2,-3,-4 ,1) $  \\
$s_{-3,i }$ & $(-2,-3,-3 ,1 ),(-1,-2,-3,1) $  \\
$s_{-4, j}$ & $(-2,-3,-2 ,1  ),(-1,-1,-2,1) ,(-2,-3,-4,2) $ \\
$s_{-5 }$ &$ (-2,-3,-3 ,2 )$ \\
$s_{-6,k }$ & $(-2,-3,-1,1  ),(0,-1,-1,1),(-1,-2,-2,2),(-2,-3,-2,2),(-2,-3,-3,3)$ \\
$s_{-7 }$ &$ (-2,-3,-2,3  ) $  \\
$s_{- 8,j }$ & $(-2,-3,-1 ,2 ),(-1,-1,-1,2) ,(-2,-3,-2,4)$  \\
$s_{- 9,i }$ &$ (-2,-3,-1 ,3 ),(-1,-2,-1,3)$ \\ 
$s_{-10 }$ &$ (-2,-3,-1,4 )$ \\  \hline  
\end{tabular} \\
\begin{tabular}{|c|l| }\hline 
\multicolumn{2}{|c|}{$\mathcal{T}(\fe_7^M, \fe_8)$ Conformal Matter } \\ \hline 
$s_{+1 }$ &$ (-2,-3,4M-3,4) $  \\
$s_{+2,i }$ & $(-2,-3,3M-2 ,3 ),(-1,-2,3M-2,3) $  \\
$s_{+3, j}$ & $(-2,-3, 2M-1 ,2  ),(-1,-1,2M-1,2) ,(-2,-3,4M-2,4) $ \\
$s_{+4 }$ &$ (-2,-3,3M-1 ,3 )$ \\
$s_{+5,k }$ & $(-2,-3,M ,1  ),(0,-1,M,1),(-1,-2,2M,2),(-2,-3,2M,2),(-2,-3,3M,3)$ \\
$s_{+6 }$ &$ (-2,-3,2M+1,2  ) $  \\
$s_{+7,j }$ & $(-2,-3,M+1 ,1 ),(-1,-1,M+1,1) ,(-2,-3,2M+2,2)$  \\
$s_{+8,i }$ &$ (-2,-3,M+2 ,1 ),(-1,-2,M+2,1)$ \\ 
$s_{+9 }$ &$ (-2,-3,M+3,1 )$ \\  
$s_{+10 }$ &$ (-2,-3,M+4,1 )$ \\ \hline  
\end{tabular}
\caption{\label{tab:ConfMatterE7NE8}Depiction of the toric rays that resolve $\mathcal{T}(\fe_8,\fe_7^1)$ and $\mathcal{T}(\fe_7^M,\fe_8)$ conformal matter
as well as those for the $(M-1)\times$ $\mathcal{T}(\fe_7,\fe_7)$'s between the $M\times $ $\fe_7$ gauge factors.}
\end{table} 

\subsubsection*{T-dual LST}
The T-dual configuration is given as 
\begin{align}
\begin{array}{ccccccc}
&&&&    \overset{\fsp_{2M-2}}{1^*}   &&  \overset{\fsp_{M-3}}{1^*}     \\ 
\lbrack \fso_{32}\rbrack & \overset{\fsp_{9+M}}{1 }       & \overset{\fso_{20+4M}}{4 }    &  \overset{\fsp_{3+3M}}{1 }      &\overset{\fso_{8+8M}}{4^* }       & \overset{\fsp_{3M-1}}{1 }      &\overset{\fso_{4+4M}}{4^* }    \, ,
\end{array}
\end{align}
where we note that the whole flavor algebra is concentrated on the left node. The T-dual fiber type is again $F_{13}$ as before with an $\mathbb{Z}_2$ finite MW group.  
For $M\leq 2$ the $1$ curves are absent and the touching curve admits simply a self-intersection of  $3$. In particular, when $M=2$, the $\fso_{12}$ admits a spinor representation with localized $SO(1)$ flavor symmetry required by the anomaly. For $M=1$, the respective algebra is further reduced to $\fso_7$. The tensor branch dimension of the above LST is given as
\begin{align}
T=5+1^*+1^* \, .
\end{align}
The shape of the quiver coincides with affine $\fe_7$ and also the LS charges are given by its Kac labels, modulo division by two, when an $\fso$ gauge algebra is present 
 \begin{align}
\begin{array}{ccccccc}
&&&2^*& & 1^*&  \\ 
\vec{l}_{LS}=(1,& 1,& 3,& 2 , &3,  &1   ) \, .\end{array}
\end{align}   
The T-dual systems described in the above content, as well as those in the subsequent subsections, are verified through the consistent matching of the 2-group structure constants and the dimension of Coulomb branch, as summarized in \ref{table:T-duals}.  

\subsection{ The $[\fe_8]-\fe_6^M-[\fe_8]$ LST } 
Next we gauge the compact curves with $\fe_6$ while keeping the flavor group. Explicitly for  $M=1$ the full resolved chain is then given as   
\begin{align}
[\fe_8] \, \,  1 \, \, 2\, \,  \overset{\fsp_{1}}{2 } \, \,    \overset{\fg_{2}}{3 }       \, \,  1 \, \,   \overset{\ff_{4}}{5 }     \, \, 1\, \,  \overset{\fsu_{2}}{3 }    \, \,  1 \, \,  \overset{\fe_{6}}{6 }    \, \, 1 \, \, \overset{\fsu_{2}}{3 }  \, \,  1\, \,   \overset{\ff_{4}}{5 }   \, \, 1 \, \, \overset{\fg_{2}}{3 }   \, \, \overset{\fsp_{1}}{2 }\, \,  2\, \,  1\, \,    [\fe_8] \, .
\end{align}
From the above basic data we can compute the LS charges as the following
\begin{align} 
	\vec{l}_{LS}=(1, 1, 1, 1, 2, 1, 3, 2, 3, 1, 3, 2, 3, 1, 2, 1, 1, 1, 1) \, .
\end{align}  
This can be generalized by decorating with $M\times \fe_6$ gauge algebra factors to obtain
\begin{align}
[\fe_8] \, \,  \underbrace{ \overset{\fe_{6}}{1 }  \, \, \overset{\fe_{6}}{2 } \ldots \overset{\fe_{6}}{2 } \ldots \overset{\fe_{6}}{2 }\, \, \overset{\fe_{6}}{1 }  }_{\times M} \, \, [\fe_8]\, .
\end{align}
We then include the $\mathcal{T}(\fe_6,\fe_6)$ superconformal matter factors discussed in \eqref{eq:E6E6SCM} with the general LS charges 
 \begin{align}
N=(1, 1, 1, 1, 2, 1, 3, 2, 3,  \underbrace{1,   3,2,3,   1 \ldots 1    \ldots , 1, 3,2,3,1}_{\times M} 3, 2, 3, 1, 2, 1, 1, 1, 1,) \, .
\end{align} 
The resolution of the above model can again be composed out of the toric rays that resolve the $\fe_6^M$ gauge algebra and $\mathcal{T}(\fe_6^m,\fe_6^{m+1})$ superconformal matter factors depicted in Table~\ref{tab:e6e6ConfMatter}. The $\mathcal{T}(\fe_8, \fe_6)$ superconformal matter is resolved via the rays in Table~\ref{tab:e8e6e7ConfMatter}. 
  
From the above data we can readily deduce the general number of tensor multiplets and the dimension of the 5D Coulomb branch. We use the following ingredients for the superconformal matter theories
\begin{align}
\{\text{T},\text{Dim(CB)}\} (\mathcal{T}(\fe_8, \fe_6))=\{ 9,18\} \, , \qquad  \{ \text{T},\text{Dim(CB)}\} (\mathcal{T}(\fe_6, \fe_6))=\{ 3,5\} \, ,  
\end{align}
and then deduce
\begin{align}
\{\text{T} ,   \text{Dim(CB)} \}( M\cdot \fe_6 + 2 \cdot\mathcal{T}(\fe_8, \fe_6) + (M-1)\cdot\mathcal{T}(\fe_6, \fe_6) -1 )=\{14+4M,30+12M\}\, . 
\end{align}  
\subsubsection*{The T-dual LST}
The T-dual configuration is given as 
\begin{align}
[\fso_{32}]\,\,   \overset{\fsp_{9+M}}{1 }        \,\,   \overset{\fso_{20+4M}}{4 }     \,\,   \overset{\fsp_{3+3M}}{1 }                \,\,     \overset{\fsu_{4+4M}}{2 }          \,\,      \overset{\fsu_{2+2M}}{2 }    \, . 
\end{align}
We find the expected $F_{13}$ fibre type and the whole flavor group is concentrated on the left, inherrited from the T-dual of the $\fe_8$ flavor group. The $\mathbb{Z}_2$ MW group again highlights the presence of an diagonal gauging across the center of flavor and gauge group factors. The LS charge is again simply that of the universal $\fe_6^M$ gauge group dual, i.e.
\begin{align} 
\vec{l}_{LS}=(1, 1, 3, 2, 1) \, ,
\end{align} 
which is consistent with the reduced Kac labels of affine $\ff_4$. From this we can compute and match the 2-group structure constants, as summarized in \ref{table:T-duals}.  

\subsection{The $[\fe_7]-\fe_8^M-[\fe_7] $ LST }
We construct the $[\fe_7]-\fe_8^M-[\fe_7] $ LST which first leads to the chain 
\begin{align} 
\begin{array}{clccclc}
 & 1 & &&&1 & \\
 \lbrack \fe_7\rbrack    &       \overset{\fe_{8}}{2 }        &     \overset{\fe_{8}}{2 }      &\ldots &    \overset{\fe_{8}}{2 }     &   \overset{\fe_{8}}{2 }    & \lbrack \fe_7\rbrack  \, .
\end{array}
\end{align}
Note the necessity to include two $1$ curves for the first and last $\fe_8$ gauge factor. This is also clearly required when including the superconformal matter factors to obtain an anomaly free $\fe_8$. For $M=1$ this chain looks like
\begin{align}
\begin{array}{cccccccccclcccccccccc}
&&&&&&&&&&1&&&&&&&&& \\ 
\lbrack \fe_7\rbrack \,\,  1   &   \overset{\fsu_{2}}{2 }        &    \overset{\fg_{2}}{3 }     & 1&  \overset{\ff_{4}}{5} & 1 & \overset{\fg_{2}}{3 } &  \overset{\fsp_{1}}{2 }  &  2 & 1&   \overset{\fe_{8}}{12 }  & 1& 2& \overset{\fsp_{1}}{2 } &   \overset{\fg_{2}}{3 }     & 1& \overset{\ff_{4}}{5}& 1 &   \overset{\fg_{2}}{3 }     &  \overset{\fsu_{2}}{2 }    & 1 \,\, \lbrack \fe_7\rbrack   \\
&&&&&&&&&&1&&&&&&&&& \\ 
\end{array} \, .
\end{align}
Note that above, we have two $1$ curves touching the middle $\fe_8$ which we will need to resolve torically by an non-flat fibers just as in the $\fe_8$ case discussed in Section~\ref{sec:non-flat}.  The LS charges for those are given as
\begin{align} 
	\vec{l}_{LS}=(1, 1, 1, 2, 1, 3, 2, 3, 4, 5, \begin{array}{c}1\, \\  1, \\ 1\, \end{array} 5, 4, 3, 2, 3, 1, 2, 1, 1, 1, 1, 1) \, .
\end{align}  
When $M>1$ we simply include more $\fe_8^M$ factors and its superconformal matter in between. This results in all $\fe_8$ curves to be $12$ curves with the first and the last having the above mentioned $1$ curve, analogous to the $\fe_8$ flavor case. The LS charges are then given as
\begin{align}
	\vec{l}_{LS} = ( 1, 1, 1, 2, 1, 3, 2, 3, 4, 5, &\begin{array}{c}1 \\ 1, \\ \\ \end{array} 
6, 5, 4, 3, 5, 2, 5, 3, 4, 5, 6, 1 \ldots \times M  \nonumber \\ &\ldots
\begin{array}{c}1 \\ 1 \\ \\ \end{array} , 5, 4, 3, 2, 3, 1, 2, 1, 1, 1) \,.
\end{align} 
The resolution of the elliptic threefolds is analogous to the cases before: Table~\ref{tab:E7E7E7fibre}  includes the $\fe_7$ rays, \eqref{eq:tab:e8i} those of the $\fe_8$'s and Table~\ref{tab:ConfMatter} its superconformal matter. New is only the $\mathcal{T}(\fe_7,\fe_8)$ superconformal matter which requires the removal of one $1$ curve only. The number of tensors and CB parameters is that of the  $[\fe_8]-\fe_8^M-[\fe_8]$ theories minus two.

\subsubsection*{The T-dual LST }
Projecting onto the second sub-polytope in the ray configuration of the resolved model above, leads to an $F_{13}$ type of fibration with the chain for an $M$.
\begin{align}
\begin{array}{cccccccccc}
	&& &&&&\overset{\fsp_{3M-5}}{1^*}&& 
	\\ 
	\lbrack \fso_{24}\rbrack & \overset{\fsp_{7+M}}{1 }    &       
	\overset{\fso_{4M+20}}{\underset{ \lbrack  \fsu_2^2 \rbrack }{4}}
    &  \overset{\fsp{3M+3}}{1 } & \overset{\fso_{8M+8}}{4 }   & \overset{\fsp_{5M-3}}{1 }  &   \overset{\fso_{12M-4}}{4^*} & \overset{\fsp_{4M-4}}{1 }    & \overset{\fso_{4M+4}}{4  } & 
\end{array} \, .
\end{align}
with the familiar LS charge  
\begin{align}
\begin{array}{cccccccc}
&&& &&3^*& & \\ 
\vec{l}_{LS}=(1,& 1,& 3,& 2 , &5,  &3, &4,&1   )\,.
\end{array}   
\end{align}  
The $F_{13}$ fibre type includes two $\fsu_2$ flavor factors in the generic fiber, just as in the $[\fe_7]-\fe_7^M-[\fe_7]$ theories, which intersect the first $\fso_{4M+20}$ factor and lead to bi-fundamental matter, consistent with the gauge anomalies. The shape and the general structure of $\fe_8^M$ T-dual theories has not changed, i.e. the base is that of an $\fe_8$ affine Dynkin diagram. Indeed we might interpret this phase as the T-dual of the Higgs branch deformation of the $E_8\times E_8$ theory, which comes at the cost of two tensors. On the T-dual side, the $SO(32)$ flavor group breaks to $SO(24) \times SU(2)^2$ at the cost of two ranks in the middle $\fsp_{N+9} \rightarrow \fsp_{N+7}$. Note that this specific Higgs branch deformation in the $SO$ side also preserves the $\mathbb{Z}_2$ center gauging. In Appendix~\ref{app:e8e8e7} we discuss the intermediate model with $\fso_{28} \times \fsu_2$ flavor group in the T-dual model, which admits similar features. 

\subsection{ The $[\fe_8]-\fsu_4^M-[\fe_6]$ LST }
Here we construct the deformed $\fsu_4^M$ theory. This one is easy as it does not require superconformal matter between the $\fsu_4$ factors. The chain is given as \footnote{This quiver coincides with $\mathcal K_N(1+1+1+1;1+3; \mathfrak{su}_4)$ quiver presented in Eqn.(4.49) of  \cite{DelZotto:2022ohj} when setting the notation $M=N+1$.}
\begin{align}
	[\fe_8]\, \,  1 \, \, 2 \, \underset{[N_F=1]}{\overset{\mathfrak{su}_2}{2 }}\, \,  \overset{\mathfrak{su}_3}{2 }\, \,  \underbrace{\overset{\mathfrak{su}_4}{2 } \ldots \overset{\mathfrak{su}_4}{2 }}_{\times M } \, \,  \underset{[N_F=2]}{\overset{\mathfrak{su}_3}{2 }} \,\, 1\, \,  [\fe_6] \, .
\end{align}
Note that also the first and last $\fsu_4$ at the edge of the ramp admits one more additional fundamental each. The first $\fsu_2$ and the last $\fsu_3$ admit extra fundamental(s), which does not come from the neighbouring gauge node.  

\subsubsection*{The T-dual LST}
The resolved geometry can readily be constructed via $\fsu_k$ tops along the lines of the sections before. 
We can readily flip to the T-dual configuration. First we note as before, that we obtain a fibration with $F_{11}$ fibre type with an MW group and $\fsu_2$ factor being effective, which are not compatible with a $\mathbb{Z}_2$ center gauging. Indeed, we find
\begin{align}
[\mathfrak{u}_1 \times \fso_{26}]\,    \overset{\mathfrak{sp}_{M+5}}{1 } \, \, 
\underset{[\fsu_2  ]}{  \overset{\mathfrak{su}_{2M+5}}{2 }}
 \, \, \underset{N_F=1}{ \overset{\mathfrak{sp}_{M-1}}{1 }}\,  \, .
\end{align}  
Note that the base topology is that of affine $\fsp_2$, i.e. the $\mathbb{Z}_2$ folding of $\fsu_4$. As there are no $\fso$ groups, also the LS charges are all $1$. The middle $\fsu_{2M+5}$ factor explicitly breaks the $\mathbb{Z}_2$ center gauging, which is consistent with the $F_{11}$ fibre type.  

\subsubsection*{Comparison to $[\fe_8]-\fsu_4^M-[\fe_8]$ LST}
As discussed in the main text, the $[\fe_8]-\fsu_4^M-[\fe_6]$ LST can be seen as a deformed $[\fe_8]-\fsu_4^M-[\fe_8]$ theory with holonomies at infinity via the following chain
 \begin{align}
	[\fe_8] \, \,  1 \, \, 2 \, \overset{\mathfrak{su}_2}{2 }\, \,  \overset{\mathfrak{su}_3}{2 }\, \,  \underbrace{\overset{\mathfrak{su}_4}{2 } \ldots \overset{\mathfrak{su}_4}{2 }}_{\times M } \, \,  \overset{\mathfrak{su}_3}{2 }  \,\,  \overset{\mathfrak{su}_2}{2 }  \,\, 2 \,\, 1\, \,  [\fe_8] \, .
\end{align}
The T-dual of this theory is well known \cite{Aspinwall:1997ye,DelZotto:2020sop} and is given as
\begin{align}
[\fso_{32}]\,    \overset{\mathfrak{sp}_{M+7}}{1 } \, \, 
  \overset{\mathfrak{su}_{2M+6}}{2 } 
 \, \,  \overset{\mathfrak{sp}_{M-1}}{1 }  \,. 
\end{align}
The above theory admits a $\mathbb{Z}_2$ center gauging, obtained from the order two MW group. Also in this model, the LS charge is simply one for each tensor.  

\subsection{The $[\fe_7]-\fe_6^M-[\fe_7]$ LST }
The starting chain is given as
\begin{align}
	[\fe_7]  \, \,   \underbrace{ \overset{\mathfrak{e}_{6}}{1 }  \, \,  \overset{\mathfrak{e}_{6}}{2}\ldots  \overset{\mathfrak{e}_{6}}{2} \ldots\overset{\mathfrak{e}_{6}}{2}   \, \,  \overset{\mathfrak{e}_{6}}{1 } }_{\times M}    \, \,  [\fe_7 ]\, .
\end{align} 
The superconformal matter insertions have been given in the sections before together with their resolutions. The general Little string charge is then composed as
 \begin{align}
	\vec{l}_{LS}=(1, 1, 1, 2, 3,  \underbrace{1,   3,2,3,   1 \ldots 1   \ldots , 1, 3,2,3,1}_{\times M} 3,2,1,1,1) \, .
\end{align}
\subsubsection*{The T-dual LST}
The T-dual theory has an $F_{13}$ fibre type with a $\mathbb{Z}_2$ MW group as well as two non-toric $\fsu_2$ flavor factors.  The dual quiver is given as  
\begin{align}
\lbrack \fso_{24} \rbrack     \, \,   \overset{\mathfrak{sp}_{5+M}}{1 }    \, \,   \overset{\mathfrak{so}_{12+4M}}{4}    \, \,     \overset{\mathfrak{sp}_{3M-1}}{1 }     \, \,      \overset{\mathfrak{su}_{4M}}{2 }       \, \,    \overset{\mathfrak{su}_{2M+2}}{2 }    \lbrack \fsu_2^2 \rbrack   \, ,
\end{align} 
with bifundamentals under the $\fsu_{2M+2}$ and both $\fsu_2$ flavor factors. The LS charges can readily be computed and are given as
\begin{align}
	\vec{l}_{LS}=(1, 1, 3, 2, 1) \, .
\end{align}

\subsection{The $[\fe_6]-\fe_8^M-[\fe_6]$ LST }
At next we start with an $\fe_6$ flavor group and gauge the internal chain by an $\fe_8$ gauge group leading to the chain
\begin{align} 
\begin{array}{clccclc}
 & 1 & &&&1 & \\
\lbrack \fe_6\rbrack   &       \overset{\fe_{8}}{2 }        &     \overset{\fe_{8}}{2 }      &\ldots &    \overset{\fe_{8}}{2 }     &   \overset{\fe_{8}}{2 }     &  \lbrack \fe_6\rbrack \, .
\end{array}
\end{align} 
We again require a $1$ curve for the first and last $\fe_8$ to ensure a $12$ curve upon inserting all conformal matter, given in Eqn.~\eqref{eq:e8e8SCM}.  Upons inserting those, we obtain a regular model, with the following LS charges 
\begin{align}
	\vec{l}_{LS} = (1, 1, 2, 1, 3, 2, 3, 4, 5, &\begin{array}{c}1 \\ 1 \\ \\ \end{array}, 
6, 5, 4, 3, 5, 2, 5, 3, 4, 5, 6  ,1 \ldots \times M  \nonumber \\ &\ldots
\begin{array}{c}1 \\ 1 \\ \\ \end{array} , 5, 4, 3, 2, 3, 1, 2, 1, 1, 1, 1) \, .
\end{align}
\subsubsection*{The T-dual LST}
Since all $\fe_6$ T-duals the generic fiber is of $F_{9}$ type and admits a rank one MW group, we expect a $\mathfrak{u}_1^2$ flavor group. The base topology is that of an affine $\fe_8$ given as 
\begin{align}
\begin{array}{cccccccccc}
&&&  &&&\overset{\fsp_{3M-5}}{1^*}&& 
\\ 
\lbrack \mathfrak{u}_1^2 \times \fso_{20}\rbrack & \overset{\fsp_{M+5}}{1 }    &        \overset{\fso_{ 4M+16 }}{4 }&  \overset{\fsp_{3M+3 }}{ \underset{ \lbrack  N_F=2 \rbrack }{1}  }         & \overset{\fso_{8M+8}}{4 }   & \overset{\fsp_{5M-3}}{1 }       &        \overset{\fso_{12M-4}}{4^*} & \overset{\fsp_{4M-4}}{1 }    & \overset{\fso_{4M+4}}{4  } \, .&  
\end{array} 
\end{align}
There is no $\mathbb{Z}_2$ center gauging, consistent with the absence of a $\mathbb{Z}_2$ MW group. For $M=1$ the $1^*$ curve is absent and the $4^*$ is just a $3$ curve and the respective gauge group is just $\fso_7$. The little string charges are the universal ones for $E_8$ gaugings, given as  
\begin{align}
\begin{array}{cccccccc}
&&& &&3^*& & \\ 
\vec{l}_{LS}=(1,& 1,& 3,& 2 , &5,  &3, &4,&1   )
\end{array}   \, .
\end{align}

\subsection{The $[\fe_6]-\fe_7^M-[\fe_6]$ LST }
We again start by gauging by an $\fe_7$ group, which leads to 
\begin{align}
[\fe_6] \,\,   \underbrace{  \overset{\fe_{7}}{1 }    \,\, \overset{\fe_{7}}{2 }\ldots \overset{\fe_{7}}{2 } \ldots \overset{\fe_{7}}{2 } \, \, \overset{\fe_{7}}{1 } }_{\times M} \,\,  [\fe_6] \, ,
\end{align}
upon inserting the superconformal matter. For $M=1$ this then looks like
\begin{align}   
	[\fe_6]\, \, 1    \, \, \overset{\fsu_{2}}{2 }     \, \,   \overset{\fso_{7}}{3 } \, \,  \overset{\fsu_{2}}{2 } \, \, 1 \, \,    \overset{\fe_{7}}{6 } \, \,        1  \, \,  \overset{\fsu_{2}}{2 } \, \,  \overset{\fso_{7}}{3 } \, \,  \overset{\fsu_{2}}{2 } \, \,  1  \, \,[\fe_6]\, .
\end{align} 
This case has two $\frac12 \mathbf{56}$-plets over the 6 curve. For higher $M$ the first and last $\fe_7$ sit over a $7$ curve with an $\frac12 \mathbf{56}$ plet over each of them.
The toric insertion of $\mathcal{T}(\fe_7,\fe_7)$ conformal matter are given in Section~\ref{ssec:e7e7Me7}. The LS charges are then given as
\begin{align}
	\vec{l}_{LS}=(1, 1, 1, 2, 3, \underbrace{1, 4, 3, 2, 3, 4,   1 \ldots 1  \ldots , 1,4, 3, 2, 3, 4,1}_{\times M} 3, 2, 1, 1, 1) \, .
\end{align}

\subsubsection*{T-dual}
The T-dual admits a generic $F_9$ fiber type with a rank two MW group, with the base again being an affine $\fe_7$ shape and is given as   
\begin{align}
\begin{array}{ccccccc}
&&&&    \overset{\fsp_{2M-4}}{1^*}   &&  \overset{\fsp_{M-3}}{1^*}     \\ 
\lbrack \mathfrak{u}_1^2 \times  \fso_{20}\rbrack  & \overset{\fsp_{M+5 }}{\underset{ \lbrack  N_F=2 \rbrack }{1}}   & \overset{\fso_{4M+12}}{4 }    &  \overset{\fsp_{3M-1}}{1 }      &\overset{\fso_{8M}}{4^* }       & \overset{\fsp_{3M-3}}{1 }    & \overset{\fso_{4M+4 }}{\underset{ \lbrack  N_F=1 \rbrack }{4^*}} 
\end{array}
\end{align}
where the first $1^*$ curve is only present for $M>1$ and the second one for $M>2$. Their respective touching curves are then only $3$ curves. Note that there are two fundamentals for the first $\fsp$ group and another vector for the last $\fso$ group require by anomalies. 
The LS charge is than of a universal $\fe_7$ T-dual, that is
 \begin{align}
\begin{array}{ccccccc}
&&&2^*&&1^*&  \\ 
\vec{l}_{LS}=(1,& 1,& 3,& 2 , &3,  &1   )\,. \end{array}
\end{align}   

\subsection{The $[\fe_7]-\fe_6^M-[\fe_6]$ LST }
We start by an $\fe_7$ and $\fe_6$ flavor group and gauge the $N$ compact curves by $\fe_6$, which yields the following chain
\begin{align}
	[\fe_7]   \, \,   \underbrace{ \overset{\mathfrak{e}_{6}}{1 }  \, \,  \overset{\mathfrak{e}_{6}}{2}\ldots  \overset{\mathfrak{e}_{6}}{2} \ldots\overset{\mathfrak{e}_{6}}{2}   \, \,  \overset{\mathfrak{e}_{6}}{1 } }_{\times M}    \, \,[\fe_6] \, .
\end{align} 
Including the $\mathcal{T}(\fe_7,\fe_6)$ conformal matter increases the negative self-intersection on the $\fe_6$ by three, while the $\mathcal{T}(\fe_6,\fe_6)$ increases it by two. Hence the rightmost $\fe_6$ gauge factor sits on a $5$ curve and hence has one $\mathbf{27}$-plet while the other ones live on $6$ curves. 
The LS charge is given by
 \begin{align}
	\vec{l}_{LS}=(1, 1, 1, 2, 3,  \underbrace{1,   3,2,3,   1 \ldots 1   \ldots , 1, 3,2,3,1}_{\times M}  2, 1,1) \, .
\end{align}   
\subsubsection*{The T-dual LST}
The dual elliptic fibration is of $F_{11}$ fiber type with an $\fsu_2$ and a $\mathfrak{u}_1$ generator. This model
admits no explicit  $\mathbb{Z}_2$ MW generator and therefore no global quotient group. This is consistent with the following T-dual chain that admits two $\fsu$ gauge factors with odd centre symmetries, given as
\begin{align}
\begin{array}{c }
[\fso_{22}] \, {\overset{\mathfrak{\fsp}_{M+4}}{1}} \,  {\overset{\fso_{4M+10}}{4}} \, {\overset{\mathfrak{\fsp}_{3M-2}}{1}} \,  \underset{[N_F=1]}  {\overset{\mathfrak{\fsu}_{4M-1}}{2}} \,  \underset{[N_F=1]}  {\overset{\mathfrak{\fsu}_{2M+1}}{2}} \, [\mathfrak{u}_{2}]
\end{array}
\end{align}  
We have the $\fsu_2$ flavor brane to only intersect the last $\fsu_{2M+1}$ factor contributing a bifundamental which we have combined with the $\mathfrak{u}_1$ MW factor. The LS charges are
\begin{align}
	\vec{l}_{LS}=(1, 1, 3, 2, 1) \, .
\end{align} 

\subsection{The $[\fe_7]-\fe_7^M-[\fe_6]$ LST}
We use the same flavor group as before but gauge by an $\fe_7$ group, leading to the chain
\begin{align}
	[\fe_7]  \, \,   \underbrace{ \overset{\mathfrak{e}_{7}}{1 }  \, \,  \overset{\mathfrak{e}_{7}}{2}\ldots  \overset{\mathfrak{e}_{7}}{2} \ldots\overset{\mathfrak{e}_{7}}{2}   \, \,  \overset{\mathfrak{e}_{7}}{1 } }_{\times M}    \, \,  [\fe_6] \, .
\end{align} 
The inclusion of each conformal matter factor increases the self-intersection by three. Hence the first and last $\fe_7$ gauge factor admit one $\frac12 \mathbf{56}$-plet over a $7$ curve, while the other ones do not. In the $M=1$ case the $\fe_7$ lives over a $6$ curve and admits two $\frac12 \mathbf{56}$-plets. From the configuration, we can then compute the LS charges given by
 \begin{align}
	\vec{l}_{LS}=(1, 1, 1, 2, 3,  \underbrace{1,   4,3,2,3,4,   1   \ldots 1 ,   \ldots , 4,3,2,3,4, 1}_{\times M} 3, 2,1, 1,1) \, .
\end{align}  
\subsubsection*{The T-dual LST}
The dual model is of $F_{11}$ fiber type where an $\fsu_2$ and a $\mathfrak{u}_1$ generator. Similar to the model before, this one admits no $\mathbb{Z}_2$ MW torsion and therefore we do not expect a center gauging of gauge group and flavor factors.  The dual chain is given as 
\begin{align}\begin{array}{c }
\qquad \qquad \qquad  \qquad \overset{\mathfrak{\fsp}_{2M-4}}{1^*}  \qquad \,\, \, \, \overset{\mathfrak{\fsp}_{M-3}}{1^*} \\ 
\lbrack \fso_{22}\rbrack  \, \underset{[N_F=1]} {\overset{\mathfrak{\fsp}_{M+5}}{1}} \,  {\overset{\fso_{4M+12}}{4}} \, {\overset{\mathfrak{\fsp}_{3M-1}}{1}} \,  {\overset{\mathfrak{\fso}_{8M}}{4^*}} \, \, {\overset{\mathfrak{\fsp}_{3M-3}}{1}} \,  {\overset{\mathfrak{\fso}_{4M+4}}{4^*}} [\mathfrak{u}_{2}] \, .\end{array} 
\end{align} 
The $\mathfrak{u}_1$ intersections above explicitly along codimension two component in the discriminant of the singular model. Similarly we find the intersection with the $\fsu_2$ flavor brane intersects at the end of the chain explicitly. The LS charges are given as  
 \begin{align}
\begin{array}{ccccccc}
&&&2^* &&1^*&  \\ 
\vec{l}_{LS}=(1,& 1,& 3,& 2 , &3,  &1   ) \,.\end{array}
\end{align}   

\subsection{The $[\fe_7]-\fe_8^M-[\fe_6]$  LST}
We gauge the tensors along the $E_7\times E_6$ flavor factors with an $\fe_8$. These
require additional $1$ curves in order to be consistent with anomalies such that the chain is given as
\begin{align} 
\begin{array}{clccclc}
 & 1 & &&&1 & \\
 \lbrack \fe_7\rbrack   &       \overset{\fe_{8}}{2 }        &     \overset{\fe_{8}}{2 }      &\ldots &    \overset{\fe_{8}}{2 }     &   \overset{\fe_{8}}{2 }     & \lbrack \fe_6\rbrack  \, .
\end{array}
\end{align}
This is due to the fact, that the superconformal matter insertions, increase the $\fe_8$ curve self-intersection by 5. 
The little string charge is then given as  

\begin{align}
	\vec{l}_{LS} = ( 1, 1, 1, 2, 1, 3, 2, 3, 4, 5, &\begin{array}{c}1 \\ 1 \\ \\ \end{array}, 
6, 5, 4, 3, 5, 2, 5, 3, 4, 5, 6  ,1 \ldots \times M  \nonumber \\ &\ldots
\begin{array}{c}1 \\ 1, \\ \\ \end{array} 5, 4, 3, 2, 3, 1, 2, 1, 1 ) \, .
\end{align}  
\subsubsection*{The T-dual LST}
The fiber ambient for the T-dual is $F_{11}$ with an $\fsu_2$ and $\mathfrak{u}_1$ enhanced flavor symmetry.  The dual base is that of $\fe_{8}$ affine base given as 
\begin{align}
\begin{array}{c} \, \qquad \qquad \qquad \qquad \qquad  \overset{\mathfrak{\fsp}_{3M-5}}{1^*} \\
  \lbrack \fso_{22}\rbrack  \, {\overset{\mathfrak{\fsp}_{M+6}}{1}} \,  \underset{[\mathfrak{u}_2]} {\overset{\fso_{4M+18}}{4}} \,\underset{[N_F=1]}  {\overset{\mathfrak{\fsp}_{3M+3}}{1}} \,  {\overset{\mathfrak{\fso}_{8M+8}}{4}} \, {\overset{\mathfrak{\fsp}_{5M-3}}{1}} \,  {\overset{\mathfrak{\fso}_{12M-4}}{4^*}} \,   {\overset{\mathfrak{\fsp}_{4M-4}}{4}} \,  {\overset{\mathfrak{\fso}_{4M+4}}{4}} \, ,
\end{array}
\end{align}
where the $1^*$ curve exists only for $M>1$, when for $M=1$, the $4^*$ is just a $3$ curve. The little string charges are 
\begin{align}
\begin{array}{cccccccc}
&&& &&3^*& & \\ 
\vec{l}_{LS}=(1,& 1,& 3,& 2 , &5,  &3, &4,&1   )\, .
\end{array}  
\end{align}   
Again we expect no $\mathbb{Z}_2$ gauging of the centre factors since the fundamental of $\fsp_{3N+3}$ with the $\mathfrak{u}_1$ flavor factor breaks it.  

\subsection{The $[\fe_8]-\fe_6^M-[\fe_6]$ LST}
Gauging the $E_8 \times E_6$ theory with $\fe_6$ over the additional tensors, we obtain the chain
\begin{align}
[\fe_8]   \, \,   \underbrace{ \overset{\mathfrak{e}_{6}}{1 }  \, \,  \overset{\mathfrak{e}_{6}}{2}\ldots  \overset{\mathfrak{e}_{6}}{2} \ldots\overset{\mathfrak{e}_{6}}{2}   \, \,  \overset{\mathfrak{e}_{6}}{1 } }_{\times M}    \, \,  [\fe_6] \, .
\end{align} 
The inclusion of the conformal matter factors enhances the self-intersection number of the left-most $\fe_6$ gauge factor by 5 and for all other ones by 4. Hence the rightmost $\fe_6$ sits on a 5 curve and hosts a single $\mathbf{27}$-plet. We then obtain the LS charges as
  \begin{align}
	\vec{l}_{LS}=(1, 1, 1, 1, 2, 1, 3, 2, 3,    \underbrace{1,   3,2,3,   1 \ldots 1   \ldots , 1, 3,2,3,1}_{\times M}  2, 1,1) \, .
\end{align}   
 
\subsubsection*{The T-dual LST}
The T-dual fibration is of $F_{11}$ fiber type  but only the rank one MW group generator is effective. Hence the fibre model is known as $\mathfrak{u}_1$ restricted Tate model, where the $a_6$ Tate-coefficient is absent globally. The dual LST chain is given as 
\begin{align}
\begin{array}{c }
[\mathfrak{u}_1 \times \fso_{26}] \, \,  {\overset{\mathfrak{\fsp}_{M+6}}{1}} \, \,  {\overset{\fso_{4M+14}}{4}} \, \,  {\overset{\mathfrak{\fsp}_{3M }}{1}} \, \,   
\underset{[N_F=1]} {\overset{\fsu_{4M+1 }}{2}}  \, \,  \underset{[N_F=1]} {\overset{\fsu_{2M+1 }}{2}}  \, , 
\end{array} 
\end{align} 
where the $\mathfrak{u}_1$ flavor factor introduces an additional fundamental in the last two gauge group factors. These two factors also highlight the absence of an $\mathbb{Z}_2$ center gauging in this model. The LS charges are
\begin{align}
	\vec{l}_{LS}=(1, 1, 3, 2, 1) \, .
\end{align}

\subsection{The $[\fe_8]-\fe_7^M-[\fe_6]$ LST}
Gauging the $E_8 \times E_6$ flavor factors with $M\times$ $\fe_7$'s, we obtain 
\begin{align}
	[\fe_8]   \, \,   \underbrace{ \overset{\mathfrak{e}_{7}}{1 }  \, \,  \overset{\mathfrak{e}_{7}}{2}\ldots  \overset{\mathfrak{e}_{7}}{2} \ldots\overset{\mathfrak{e}_{7}}{2}   \, \,  \overset{\mathfrak{e}_{7}}{1 } }_{\times M}    \, \,  [\fe_6]\, .
\end{align} 
Introducing the conformal matter factors, enhances the self-intersection of the leftmost $\fe_7$ gauge factor by seven and all the other ones by six. Hence only the rightmost $\fe_7$ admits a matter in the $\frac12 \mathbf{56}$.   The LS charges are then given as
 \begin{align}
	\vec{l}_{LS}=(1, 1, 1, 1, 2, 1, 3, 2, 3,4, \underbrace{1,   4,3,2,3,4,   1   \ldots 1 ,   \ldots , 4,3,2,3,4, 1,}_{\times M} 3, 2,1, 1,1) \, .
\end{align}  

\subsubsection*{The T-dual LST}
The T-dual fibration has $F_{11}$ fibre type and supports a rank one MW group that corresponds to a $\mathfrak{u}_1$ flavor group and no finite MW torsion factor. The dual chain is 
\begin{align}\begin{array}{c }
\qquad \qquad \qquad \qquad \qquad \qquad \,\,\, \, \,\, \overset{\mathfrak{\fsp}_{2M-3}}{1^*} \quad  \qquad \, \, \overset{\mathfrak{\fsp}_{M-3}}{1^*} \\ 
\lbrack \mathfrak{u}_1 \times \fso_{26} \rbrack \, \, \underset{\lbrack N_F =1 \rbrack } {\overset{\mathfrak{\fsp}_{M+7}}{1}} \, \, {\overset{\fso_{4M+16}}{4}} \,\, {\overset{\mathfrak{\fsp}_{3M+1}}{1}} \, \, {\overset{\mathfrak{\fso}_{8M+4}}{4^*}} \,\, {\overset{\mathfrak{\fsp}_{3M-2}}{1}} \, \,\underset{\lbrack N_F=1 \rbrack}  {\overset{\mathfrak{\fso}_{4M+4}}{4^*}}\, \end{array} \, .
\end{align}
We have added an additional $\mathfrak{u}_1$ flavor factor that appears due to the MW group generator in the quiver above. There are also additional fundamentals for the $\fsp$ and last $\fso$ gauge factor.   
The LS  charge is given as  
 \begin{align}
\begin{array}{ccccccc}
&&&2^*&&1^*&  \\ 
\vec{l}_{LS}=(1,& 1,& 3,& 2 , &3,  &1   ) \, .\end{array}
\end{align}     
   
\subsection{The $[\fe_8]-\fe_8^M-[\fe_6]$ LST}
Gauging the flavor factors $E_8\times E_6$ by the $\fe_8^M$ factors yields the following chain
\begin{align} 
\begin{array}{clccclc}
 & 1 & &&&1 & \\
|\fe_8|  &       \overset{\fe_{8}}{2 }        &     \overset{\fe_{8}}{2 }      &\ldots &    \overset{\fe_{8}}{2 }     &   \overset{\fe_{8}}{2 }     &  |\fe_6| \, .
\end{array}
\end{align}
The insertion of the superconformal matter increases the self-intersection by $10$ for all curves. The explicit form and resolutions have been given before. The LS charge of this type of model are then given as 
\begin{align}
	\vec{l}_{LS} = (1,  1, 1, 1, 2, 1, 3, 2, 3, 4, 5, &\begin{array}{c}1 \\ 1 \\ \\ \end{array}, 
6, 5, 4, 3, 5, 2, 5, 3, 4, 5, 6  ,1 \ldots \times M  \nonumber \\ &\ldots
\begin{array}{c}1 \\ 1, \\ \\ \end{array}  5, 4, 3, 2, 3, 1, 2, 1, 1 ) \, .
\end{align}  
 
\subsubsection*{The T-dual LST}
The T-dual fibration is of $F_{11}$ fibre type and supports a rank one MW group that corresponds to a $\mathfrak{u}_1$ flavor group and no finite MW torsion factor. The dual chain is given by  
\begin{align}
\begin{array}{c} \, \qquad \qquad \qquad \qquad \qquad \qquad\,\,\,\,\, \overset{\mathfrak{\fsp}_{3M-5}}{1^*} \\
\lbrack \mathfrak{u}_1 \times \fso_{26}\rbrack \, \,{\overset{\mathfrak{\fsp}_{M+7}}{1}} \, \, {\overset{\fso_{4M+18}}{4}} \, \, \underset{\lbrack N_F=1 \rbrack }  {\overset{\mathfrak{\fsp}_{3M+3}}{1}} \, \, {\overset{\mathfrak{\fso}_{8M+8}}{4}} \, \,{\overset{\mathfrak{\fsp}_{5M-3}}{1}}  \,\, {\overset{\mathfrak{\fso}_{12M-4}}{4^*}} \, \,{\overset{\mathfrak{\fsp}_{4M-4}}{1}}  \,\, {\overset{\mathfrak{\fso}_{4M+4}}{4}} 
\end{array}
\end{align}
Where we find an additional fundamental for $\fsp_{3M+3}$ as required by anomalies. The LS charge is again the same as for all $\fe_8$ gauged T-duals with 
\begin{align}
	\vec{l}_{LS}=(1, \, \,1, \, \,3,\, \, 2,\, \, 5, \begin{array}{c}3^* \\ 3, \\ \\ \end{array} 4,\, \, 1) \, .
\end{align}

\subsection{The $[\fe_8]-\fe_7^M-[\fe_7]$ LST} 
Gauging the $E_8\times E_7$ flavor factors with $M\times$ $\fe_7$'s, we obtain the chain 
\begin{align}
	[\fe_8]   \, \,   \underbrace{ \overset{\mathfrak{e}_{7}}{1 }  \, \,  \overset{\mathfrak{e}_{7}}{2}\ldots  \overset{\mathfrak{e}_{7}}{2} \ldots\overset{\mathfrak{e}_{7}}{2}   \, \,  \overset{\mathfrak{e}_{7}}{1 } }_{\times M}    \, \, [\fe_7] \, .
\end{align} 
The superconformal matter insertions increases the self-intersections for all curves by six but for the rightmost $\fe_7$ curve where it is increased by 7. Thus only the rightmost $\fe_7$ admits on  $\frac12 \mathbf{56}$-plet while the other ones have none.
We have the following LS charges as 
 \begin{align}
	\vec{l}_{LS}=(1, 1, 1, 1, 2, 1, 3, 2, 3, 4, \underbrace{1,   4,3,2,3,4,   1,   \ldots 1 ,   \ldots , 4,3,2,3,4, 1}_{\times M} 3, 2,1, 1,1) \, .
\end{align}   
\subsubsection*{The T-dual LST}
Here the generic ambient space is of $F_{14}$ type and hence admits an $\mathbb{Z}_2$ MW group with a single  $\fsu_2$ flavor group generator. The T-dual chain is given as  
\begin{align}\begin{array}{c }
\qquad \qquad \qquad \qquad  \overset{\mathfrak{\fsp}_{2M-3}}{1^*}\qquad  \quad \overset{\mathfrak{\fsp}_{M-3}}{1^*} \\ 
\lbrack \fso_{28}\rbrack \,  \,{\overset{\mathfrak{\fsp}_{M+7}}{1}} \,\, {\overset{\fso_{4M+16}}{4}} \,  \, {\overset{\mathfrak{\fsp}_{3M+1}}{1}} \, \, {\overset{\mathfrak{\fso}_{8M+4}}{4^*}} \,\, {\overset{\mathfrak{\fsp}_{3M-2}}{1}}  \,\, {\overset{\mathfrak{\fso}_{4M+4}}{4^*}} \,\, [\fsu_2]\end{array} \, .
\end{align}  

According to the anomaly, we find a half-hypermultiplet in the bifundamental representation of $\fso_{4M+4} \times \fsu_2$ on the right. The whole structure is also consistent with the diagonal $\mathbb{Z}_2$ gauings in flavor and gauge group factors, implemented by the $\mathbb{Z}_2$ MW group. The LS charges are the universal ones for a $\fe_7$ gauging and we repeat them as
\begin{align}
	\vec{l}_{LS}=(1, \, \,1, \, \,3, \begin{array}{c} 2^* \\ 2^*, \, \, \\ \\  \end{array} 3, \begin{array}{l}  1^* \\  1 \\ \\ \end{array} ) \, .
\end{align}

\subsection{The $[\fe_8]-\fe_8^M-[\fe_7]$ LST}\label{app:e8e8e7}
Starting with $E_8\times E_7$ flavor symmetry and gauging by $M\times$ $\fe_8$'s leads to the chain
\begin{align} 
\begin{array}{clccclc}
 & 1 & &&&1 & \\
\lbrack \fe_8\rbrack   &       \overset{\fe_{8}}{2 }        &     \overset{\fe_{8}}{2 }      &\ldots &    \overset{\fe_{8}}{2 }     &   \overset{\fe_{8}}{2 }     &  \lbrack \fe_7\rbrack  \, ,
\end{array}
\end{align}
After the insertion of the superconformal matter, we compute the LS charges, which are given by the repeating pattern
\begin{align}
	\vec{l}_{LS} = (1,  1, 1, 1, 2, 1, 3, 2, 3, 4, 5, &\begin{array}{c}1 \\ 1 \\ \\ \end{array}, 
6, 5, 4, 3, 5, 2, 5, 3, 4, 5, 6  ,1 \ldots \times M  \nonumber \\ &\ldots
\begin{array}{c}1 \\ 1 \\ \\ \end{array} , 5, 4, 3, 2, 3, 1, 2, 1, 1 ,1) \, .
\end{align}   
\subsubsection*{The T-dual LST}
The T-dual LST admits a fibration of $F_{13}$ fiber type, which admits an $\fsu_2$ flavor group and a $\mathbb{Z}_2$ MW group. The dual LST chain looks like  
\begin{align}
\begin{array}{c} \, \qquad \qquad \qquad \qquad \qquad  \overset{\mathfrak{\fsp}_{3M-5}}{1^*} \\
\lbrack \fso_{28}\rbrack  \,  \,{\overset{\mathfrak{\fsp}_{M+8}}{1}} \,\,   \underset{\lbrack \fsu_2\rbrack }   {\overset{\fso_{4M+20}}{4}} \, \,   {\overset{\mathfrak{\fsp}_{3M+3}}{1}} \, \,  {\overset{\mathfrak{\fso}_{8M+8}}{4}} \, {\overset{\mathfrak{\fsp}_{5M-3}}{1}}  \,\,  {\overset{\mathfrak{\fso}_{12M-4}}{4^*}} \, {\overset{\mathfrak{\fsp}_{4M-4}}{1}}\,   \, {\overset{\mathfrak{\fso}_{4M+4}}{4}}
\end{array} 
\end{align}
The intersection of the $\fsu_2$ flavor divisor can explicitly be computed and the bifundamentals are consistent with anomaly cancellation as well as the global $\mathbb{Z}_2$ modding in flavor and gauge group factors. The LS charge is then explicitly given as  
\begin{align}
	\vec{l}_{LS}=(1,\, \, 1,\, \, 3,\, \, 2,\, \, 5, \begin{array}{c}3 \\ 3, \\ \\ \end{array} 4, \, \,1) \, .
\end{align}  

\subsection{Second $[\fe_6]-\fe_6^M-[\fe_6]$ LST}
In this section we engineer the $\fe_6$ flavor deformations, as given in \cite{DelZotto:2022ohj} of an $\fe_6$ theory. The minimal tensor branch of this theory is given by the quiver
\begin{align}
    [\fe_6] \, \, 1 \, \,  \overset{\fsu_3}{3} \,\, 1  \, \, \overset{\ff_4}{5}\, \, 1\, \, \overset{\fsu_3}{3}\, \, 1\, \, \overset{\fe_6}{6} \, \,1\, \, \overset{\fsu_3}{3}\, \, 1\, \, [\fe_6] \, .
\end{align}
We therefore consider the class of models as given as 
\begin{align}
    [\fe_6]\, \, \overset{\ff_4}{1} \, \,\underbrace{ \overset{\fe_6}{2}  \ldots \overset{\fe_6}{2} \ldots \overset{\fe_6}{2}}_{\times M} \, \,     \overset{\ff_4}{1}\, \, [\fe_6]
\end{align}
and fill in with the conformal matter, i.e. chains of $1\, \, \overset{\fsu_3}{3}\, \, 1$\, .
The LS charges of such a chain, upon inserting the conformal matter is then given as
\begin{align}
	\vec{l}_{LS} = ( 1,1,2, 1 , 3, 2, 3, \underbrace{1 , 3,2,3, \ldots ,1, \ldots 3,2,3 ,1}_{\times M},3,2,3,1,2,1,1,1  ) 
\end{align}
The resolution follows simply from the the $\fe_6$ rays given in the sections before. 
\subsubsection*{The T-dual LST}
The dual LST admits again a $\mathfrak{u}_1^2$ MW group group, given by the $F_9$ ambient space. The T-dual model can then be obtained from
\begin{align}
    \lbrack \fso_{20}\rbrack \overset{\fsp_{5+M}}{1}\, \, \overset{\fso_{16+4M}}{4}\, \, \underset{[N_F=2]}{\overset{\fsp_{3+3M}}{1}}\, \, \overset{\fsu_{4+4M}}{2}\, \, \overset{\fsu_{2+2M}}{2}
\end{align}
Note that from \cite{DelZotto:2022ohj} we expect the two additional fundamentals to arise from an $\mathfrak{u}_1^2 \rightarrow SO(4)$ enhancement. This model coincides with the dual in Table~5 in \cite{DelZotto:2022ohj} upon switching $M=N+1$. 

\subsection{Second $[\fe_7]-\ff_4-\fe_7^M-\ff_4-[\fe_7]$ LST}
Similarly, there is another deformation of the orbi-instanton theory of an $\fe_7$ singularity to have an $\fe_7$ flavor group, given as
\begin{align}
     [\fe_7] \, \, 1 \, \, \overset{\fsu_2}{2} \, \,  \overset{\fg_2}{3} \,\, 1  \, \,\overset{\ff_4}{5}\, \, 1 \, \,    \overset{\fg_2}{3}\, \, \overset{\fsu_2}{2}\, \, 1 \, \,  \overset{\fe_7}{7} \, \, 1\, \,\overset{\fsu_2}{2} \, \,  \overset{\fso_7}{3}\, \, \overset{\fsu_2}{2}\, \,  1\, \,   \, [\fe_7] \, .
\end{align}
We glue the above pieces together and obtain a chain, modulo conformal matter as
\begin{align}
    [\fe_7] \, \, \textcolor{red}{\overset{\ff_4}{1}} \, \,   \underbrace{\overset{\fe_7}{2}\ldots \overset{\fe_7}{2} \ldots \overset{\fe_7}{2}}_{M\times}\, \, \textcolor{red}{\overset{\ff_4}{1}}  \, \,  [\fe_7]
\end{align}
The LS charges are again easily been read of as
\begin{align}
	\vec{l}_{LS} = (1,1,1,2, \textcolor{red}{1},  3,2,3,4, \underbrace{1,4,3,2,3,4 , \ldots     1, \ldots,4,3,2,3,4 ,1}_{M\times},4,3,2,3,\textcolor{red}{1},2,1,1,1  )
\end{align}
\subsubsection*{The T-dual LST}
The T-dual LST can again be obtained from the usual methods. In the enumeration we have, two more interesting cases appear, i.e. those for $M=1$ and $M=2$ before going to the general case. These cases yield incomplete $\fe_7^{(1)}$ base topologies. In the first case it is
\begin{align}
    \begin{array}{cl}& \\
    \lbrack \fso_{24}\rbrack \overset{\fsp_{8}}{1}\, \, \underset{[\fso_4]}{\overset{\fso_{24}}{4}}  \, \,  \overset{\fsp_{6}}{1} \, \, &\overset{\fso_{16}}{4}\, \, \overset{\fsp_{2 }}{1}  \, \, \overset{\fso_{7}}{3}
    \end{array}\qquad 
    \begin{array}{cl}& \overset{\fsp_2}{1}\\
    \lbrack \fso_{24}\rbrack \overset{\fsp_{9}}{1}\, \, \underset{[\fso_4]}{\overset{\fso_{28}}{4}}  \, \,  \overset{\fsp_{9}}{1} \, \, &\overset{\fso_{24}}{4}\, \, \overset{\fsp_{5 }}{1}  \, \, \underset{[N_S=\frac12]}{\overset{\fso_{12}}{3}}
    \end{array}
\end{align}
and for general $M>2$ 
\begin{align} 
    \begin{array}{cl}& \overset{\fsp_{2M-2}}{1^*}\\
    \lbrack \fso_{24}\rbrack \overset{\fsp_{7+M}}{1}\, \, \underset{[\fso_4]}{\overset{\fso_{20+4M}}{4}}  \, \,  \overset{\fsp_{3+3M}}{1} \, \, &\overset{\fso_{8+8M}}{4}\, \, \overset{\fsp_{3M-1 }}{1}  \, \, \overset{\fso_{4+4M}}{4} \, \, \overset{\fsp_{M-3 }}{1^*} 
    \end{array}
\end{align}
with the LS charges
\begin{align}
\begin{array}{cl}
& 2^*\\ 
\vec{l}_{LS}= (1,\, \,1\, \,3,&2,\, \,3,\, \,1,\, \,1^*) \, ,
    \end{array}
\end{align}
where the stared curves are only present for $M>1$ and $M>2$ respectively. This quiver coincides with that of \cite{DelZotto:2022ohj} when setting $M=N+1$. 

\subsection{The $[\fe_7] - \ff_4 -\fe_6^M - \ff_4-[\fe_7]$ LST}
We turn to the second type of deformations possible for $\fe_6$ singularities that deform the $[\fe_8]$ flavor groups to $[\fe_7]$ , given by 
\begin{align}
\label{eq:e7breakign}
     [\fe_7] \, \, 1 \, \, \overset{\fsu_2}{2} \, \,  \overset{\fg_2}{3} \,\, 1  \, \,\overset{\ff_4}{5}\, \, 1 \, \,    \overset{\fsu_3}{3}\, \,1 \, \,   \overset{\fe_6}{6}\, \, 1 \, \,    \overset{\fsu_3}{3}\, \,1 \, \,   [\fe_6] \, .
\end{align}
which we can fuse along the $\fe_6$. In short, this yields quivers of the type 
\begin{align}
    [\fe_7]\, \, \textcolor{red}{\overset{\ff_4}{1}} \, \,\underbrace{ \overset{\fe_6}{2}  \ldots \overset{\fe_6}{2} \ldots \overset{\fe_6}{2}}_{\times M} \, \,     \textcolor{red}{\overset{\ff_4}{1}}\, \, [\fe_7] \, ,
\end{align}
modulo insertion of superconformal matter which can be directly read off from \eqref{eq:e7breakign}. The LS charges is then given as
\begin{align}
	\vec{l}_{LS} = (1,1,1,2,\textcolor{red}{1}, 3, 2 , 3,\underbrace{1, \ldots ,1, \ldots ,1,}_{M\times} , 3, 2, 3,
    \textcolor{red}{1}, 2,1,1,1  )
\end{align}
The resolution of the above quiver again simply follows from those of $\ff_4,\fe_6$ and $\ff_7$ fibral singularities and the additions of their conformal matter. 

\subsubsection*{The T-dual LST}
The T-dual LST, admits again a $\mathbb{Z}_2$ finite MW group as well as an $\fsu_2^2 \sim \fso_4$ flavor rays in the generic fibre.
\begin{align}
    [\fso_{24}]\, \,  \overset{\fsp_{7+M}}{1} \, \,  \underset{[\fso_4]}{\overset{\fso_{20+4M}}{4}} \, \,  \overset{\fsp_{3+3M}}{1}  \overset{\fsu_{4+4M}}{2}  \overset{\fsu_{2+2M}}{1}\, . 
\end{align}
Note that the above configuration is also consistent for $M=0$. Upon sending $M=N+1$ this configuration coincides with \cite{DelZotto:2022ohj}. The LS charges are simply the universal ones for an $\fe_6$ gauging i.e.
\begin{align}
	\vec{l}_{LS} = (1,1,3,2,1) \, ,
\end{align}

\subsection{The $[\fe_6]-\ff_4 - \fe_7^M - \ff_4 - [\fe_6] $ LST}
This configuration again yields the $\fe_6$ deformation of $\fe_8$ due to an $\fe_7$ singularity, which yields the quiver
\begin{align}
        [\fe_6] \, \, 1 \, \,\overset{\fsu_3}{3} \, \,1\, \, \overset{\ff_4}{5}\, \, 1 \, \,
       \overset{\fg_2}{3} \, \,  \overset{\fsu_2}{2}\, \, 1\, \, \overset{\fe_7}{7} \, \, 1 \, \, \overset{\fsu_2}{2}\, \, \overset{\fso_7}{3} \, \, \overset{\fsu_2}{2} \, \, 1   \, \, [\fe_7] \, .
\end{align}
Thus generalizing, fusing by the $\fe_7$ factors leads to a generalized quiver of LSTs which looks like
\begin{align}
    [\fe_6]\, \, \textcolor{red}{\overset{\ff_4}{1}} \, \,\underbrace{ \overset{\fe_7}{2}  \ldots \overset{\fe_7}{2} \ldots \overset{\fe_7}{2}}_{\times M} \, \,     \textcolor{red}{\overset{\ff_4}{1}}\, \, [\fe_6] \, ,
\end{align}
modulo the conformal matter. The LS charges above are given by
\begin{align}
    \vec{l}_{LS} = (1, 1,2, \textcolor{red}{1} , 3, 2, 3,4, \underbrace{1 , 4,3,2,3,4,  \ldots ,1, \ldots ,4,3,2,3,4  ,1,}_{\times M}4,3,2,3,\textcolor{red}{1},2,1,1  ) 
\end{align}

\subsubsection{The T-dual LST}
The T-dual LST admits again an $\mathfrak{u}_1^2$ generic flavor group and $\fso_{20}$ flavor rays. There are again the two special cases $M=1$ and $M=2$ that are an incomplete fusion.  They are given as
\begin{align}
    \begin{array}{cl}& \\
    \lbrack \fso_{20}\rbrack \overset{\fsp_{6}}{1}\, \,  \overset{\fso_{20}}{4}   \, \,  \underset{[N_F=2]}{\overset{\fsp_{6}}{1}} \, \, &\overset{\fso_{16}}{4}\, \, \overset{\fsp_{2 }}{1}  \, \, \overset{\fso_{7}}{3}
    \end{array}\qquad 
    \begin{array}{cl}& \overset{\fsp_2}{1}\\
    \lbrack \fso_{20}\rbrack \overset{\fsp_{7}}{1}\, \,  \overset{\fso_{24}}{4}   \, \,  \underset{[N_F=2]}{\overset{\fsp_{9}}{1}} \, \, &\overset{\fso_{24}}{4}\, \, \overset{\fsp_{5 }}{1}  \, \, \underset{[N_S=\frac12]}{\overset{\fso_{12}}{3}}
    \end{array}
\end{align}
and for general $M>2$ 
\begin{align} 
    \begin{array}{cl}& \overset{\fsp_{2M-2}}{1^*}\\
    \lbrack \fso_{20}\rbrack \overset{\fsp_{5+M}}{1}\, \, \overset{\fso_{16+4M}}{4}   \, \,  \underset{ [ N_F=2] }{\overset{\fsp_{3+3M}}{1}} \, \, &\overset{\fso_{8+8M}}{4}\, \, \overset{\fsp_{3M-1 }}{1}  \, \, \overset{\fso_{4+4M}}{4^*} \, \, \overset{\fsp_{M-3 }}{1^*} 
    \end{array}
\end{align}
with the LS charges
\begin{align}
\begin{array}{cl}
& 2^*\\ 
\vec{l}_{LS}= (1,1,3,&2,3,1,1^*) \, ,
    \end{array}
\end{align}
where the stared curves are only present for $M>1$ and $M>2$ respectively. This quiver coincides with that of \cite{DelZotto:2022ohj} when setting $M=N+1$. 
 
\subsection{The $[\fe_6] -\fe_6 -\fe_8^M-\fe_6-[\fe_6]$ LST}
At next we discuss  the $E_6'$ deformation of  $\fe_8$ probing an $\fe_8$ singularity given by the quiver
\begin{align}
    [\fe_6]  \, \, 1 \, \,\overset{\fsu_3}{3} \, \,1\, \,  \overset{\fe_6}{6}   \, \, 1 \, \,\overset{\fsu_3}{3} \, \,1\, \,\overset{\ff_4}{5}\, \,   1 \, \,  
    \overset{\fg_2}{3} \, \, \overset{\fsu_2}{2} \, \, 2 \, \, 1\, \, [\fe_8] 
\end{align}
In order to engineer those, we start off with quivers of the following type 
\begin{align}
    [\fe_6]\, \, \textcolor{red}{\overset{\fe_6}{1}} \, \,\underbrace{ \overset{\fe_8}{2}  \ldots \overset{\fe_8}{2} \ldots \overset{\fe_8}{2}}_{\times M} \, \,     \textcolor{red}{\overset{\fe_6}{1}}\, \, [\fe_6] \, ,
\end{align}
modulo the insertion of the respective conformal matter. The LS charges are given as
\begin{align}
\begin{array}{l }
    \vec{l}_{LS} = (1,1,2,\textcolor{red}{1},      
    4,3,5,2,5,3,4,5,6,  
    1, 6,5,4,3,5,2,5,3,4,5,6,1,\ldots 1 \ldots \\ \qquad \qquad  \ldots   1 
   \ldots,
  ,1, 6,5,4,3,5,2,5,3,4,5,6, 1,6,5,4,3,5,2,5,3,4
    ,\textcolor{red}{1}, 2,1,1) \end{array}
\end{align}
The resolution can be done as before, using the toric rays of $\mathcal{T}(\fe_6,\fe_6)$, $\mathcal{T}(\fe_6,\fe_8)$ as well as $\mathcal{T}(\fe_8,\fe_8)$ conformal matter given in the tables above.
\subsubsection*{The T-dual LST}
The T-dual LST is given via the usual flip of the toric rays. The dual fibre is in the $F_9$ toric ambient space which admits an $\mathfrak{u}^2_1$ MW group. The dual quiver is of the usual $\fe_8^{(1)}$ form given as
\begin{align}
\begin{array}{c} \, \, \, \qquad \qquad \qquad \qquad \qquad \quad \overset{[N_F=2]}{\overset{\mathfrak{\fsp}_{3M-1}}{1  }} \\
\lbrack \fso_{20}\rbrack  \,  \,{\overset{\mathfrak{\fsp}_{5+M}}{1}} \,\,    \overset{\fso_{16+4M}}{4}  \, \,    \overset{\mathfrak{\fsp}_{3+3M  }}{1}  \, \,  {\overset{\mathfrak{\fso}_{8M+12}}{4}} \, {\overset{\mathfrak{\fsp}_{5M+1}}{1}}  \,\,  {\overset{\mathfrak{\fso}_{12M+8}}{4 }} \, {\overset{\mathfrak{\fsp}_{4M }}{1}}\,   \,  \overset{\mathfrak{\fso}_{4M+8}}{4} 
\end{array} 
\end{align}
The LS charge is as usual given as
\begin{align}
	\vec{l}_{LS}=(1, 1, 3, 2, 5, \begin{array}{c}3 \\ 3, \\ \\ \end{array} 4, 1) \, .
\end{align} 

\subsection{The $[\fe_7] - \fe_7 -\fe_8^M-\fe_7-[\fe_7]$ LST}
The above configuration corresponds to the fusion of two identical orbi-instanton theories of an $\fe_8$ singularity, with a deformation to $\fe_7$ flavor given by the following quiver
\begin{align}
    \lbrack  \fe_7\rbrack  \, \, 1 \, \overset{\fsu_2}{2} \,   \,\overset{\fso_7}{3} \,\, \overset{\fsu_2}{2} \,   \,1\, \,  \overset{\fe_7}{8}   
    \, \, 1 \, \, \overset{\fsu_2}{2} \, \,  \overset{\fg_2}{3} \,\, 1  \, \,\overset{\ff_4}{5}\, \, 1 \, \,     \overset{\fg_2}{3}\, \,\overset{\fsu_2}{2} \, \,2 \, \, 1\, \,    \lbrack \fe_8\rbrack \, .
\end{align} 
 This configuration corresponds to the $E_7'$ flavor configuration given in \cite{DelZotto:2022ohj}. 
To engineer this, we can equally start with the quiver
\begin{align}
    [\fe_7]\, \, \textcolor{red}{\overset{\fe_7}{1}} \, \,\underbrace{ \overset{\fe_8}{2}  \ldots \overset{\fe_8}{2} \ldots \overset{\fe_8}{2}}_{\times M} \, \,     \textcolor{red}{\overset{\fe_7}{1}}\, \, [\fe_7] \, ,
\end{align}
and insert the conformal matter theories as given above. The LS charges are then given as
\begin{align}
\begin{array}{l}
    \vec{l}_{LS} =(1,1,1,2,3,
    \textcolor{red}{1},
     5, 4, 3, 5, 2, 5, 3, 4, 5, 6,
     1,6,5, 4, 3, 5, 2, 5, 3, 4, 5, 6,1 \ldots  \\
  \qquad  \ldots 1
   \ldots 1,  6,5, 4, 3, 5, 2, 5, 3, 4, 5, 6 , 1,
     6, 5, 4, 3, 5, 2, 5, 3, 4, 5 
    ,\textcolor{red}{1},
    3, 2, 1, 1, 1)\, .
    \end{array}
\end{align}
The resolution can again be done with the usual rays for the conformal matter theories that is given in the bulk of this work.
\subsubsection*{The T-dual LST}
The T-dual LST has as before an $\mathbb{Z}_2$ MW group and two $\fsu_2^2\sim \fso_4$ non-toric rays. The T-dual quiver is given as
\begin{align}
\begin{array}{c}   \quad  \qquad \qquad \qquad \qquad  \overset{\mathfrak{\fsp}_{3M-1}}{1  } \\
\lbrack \fso_{24}\rbrack  \,  \,{\overset{\mathfrak{\fsp}_{7+M}}{1}} \,\,    \overset{\fso_{20+4M}}{4}  \, \,    \overset{\mathfrak{\fsp}_{5+3M  }}{1}  \, \,  {\overset{\mathfrak{\fso}_{8M+16}}{4}} \, {\overset{\mathfrak{\fsp}_{5M+3}}{1}}  \,\,  {\overset{\mathfrak{\fso}_{12M+12}}{4 }} \, {\overset{\mathfrak{\fsp}_{4M+2}}{1}}\,   \, {\overset{\mathfrak{\fso}_{4M+12}}{4}}[\fso_4]
\end{array} 
\end{align}
The LS charges are given as
\begin{align}
	\vec{l}_{LS}=(1, 1, 3, 2, 5, \begin{array}{c}3, \\ 3, \\ \\ \end{array} 4, 1) \, .
\end{align} 
The above theory coincides with the $\widehat{\mathcal{T}}_N (E_7', E_7', \mathfrak{g}=\fe_8)$ T-dual theory upon setting $M=N-1$\, . 
Note that in geometry we find an $\fsu_2^2= \fso_4$ flavor group on the last note, instead of the $\fsp_2$ that one might have expected field theoretically.

\bibliographystyle{ytphys}
\bibliography{refs.bib}

\end{document}